\documentclass[a4paper,12pt]{article}
\usepackage[margin=1in, paperwidth=8.5in, paperheight=11in]{geometry}
\usepackage{amsmath}
\usepackage{graphicx}%
\usepackage{amsfonts}%
\usepackage{amssymb}
\usepackage{color}
\usepackage{bm}
\usepackage{float}
\usepackage{caption}
\usepackage{subcaption}
\usepackage{epstopdf}
\usepackage{setspace}
\usepackage{cite}
\usepackage{amsthm}
\usepackage{comment}
\usepackage{undertilde}

\makeatletter
\renewcommand*\env@matrix[1][*\c@MaxMatrixCols c]{%
  \hskip -\arraycolsep
  \let\@ifnextchar\new@ifnextchar
  \array{#1}}
\makeatother

\renewcommand{\bibitem}{\item[]}
\newenvironment{biblist}
   {\begin{list}{}
    {\setlength{\leftmargin}{4ex}   
     \setlength{\itemindent}{-4ex}  
     \setlength{\itemsep}{0.6ex}      
     \setlength{\parsep}{0ex}       
   }}{\end{list}}

\begin{document}

\thispagestyle{empty}
\setcounter{page}{0}

\begin{center}
{\Large \textbf{Toward a diagnostic toolkit for linear models with Gaussian-process distributed random effects}}

\bigskip

\textbf{Maitreyee Bose$^{1*}$, James S. Hodges$^{2}$, and Sudipto Banerjee$^{3}$}\\
$^{1}$ Department of Biostatistics, University of Washington, Seattle, Washington 98195. \\
$^{2}$ Division of Biostatistics, University of Minnesota, Minneapolis, Minnesota 55455. \\
$^{3}$ Department of Biostatistics, University of California, Los Angeles, California 90095. \\
*email: bosem2@uw.edu \\

\end{center}

\textsc{Summary.}
Gaussian processes (GPs) are widely used as distributions of random effects in linear mixed models, which are fit using the restricted likelihood or the closely-related Bayesian analysis.  This article addresses two problems.  First, we propose tools for understanding how data determine estimates in these models, using a spectral basis approximation to the GP under which the restricted likelihood is formally identical to the likelihood for a gamma-errors GLM with identity link. Second, to examine the data's support for a covariate and to understand how adding that covariate moves variation in the outcome $y$ out of the GP and error parts of the fit, we apply a linear-model diagnostic, the added variable plot (AVP), both to the original observations and to projections of the data onto the spectral basis functions.  The spectral- and observation-domain AVPs estimate the same coefficient for a covariate but emphasize low- and high-frequency data features respectively and thus highlight the covariate's effect on the GP and error parts of the fit respectively.  The spectral approximation applies to data observed on a regular grid;  for data observed at irregular locations, we propose smoothing the data to a grid before applying our methods.  The methods are illustrated using the forest-biomass data of Finley et al.~(2008).

\textsc{Key words:} Gaussian process, linear mixed model, spectral approximation, lack of fit, missing predictor, added variable plot

\newpage

\section{Introduction}\vspace*{-0.2cm}\label{intro}
Gaussian processes (GPs) are widely used in longitudinal, functional, and spatial data analysis because the properties they inherit from the Normal distribution make them easy to work with. Fitting a GP to data involves estimating the process parameters, most commonly the process variance and range, along with an error variance. One way to do this is by writing the GP as a component of a linear mixed model and maximizing the restricted likelihood, which is identical to the marginal posterior from a Bayesian analysis with particular priors. It is, however, unclear how the resulting parameter estimates are influenced by features in the data like outliers or non-stationarities in the mean or covariance function. Fuglstad et al. (2014) argue that even if non-stationarity is present, it is difficult to model properly and fitting a stationary model usually gives satisfactory predictions. Also, GPs are now easily accessible to non-specialists (e.g., in SAS Inc.'s JMP package), so it is useful to know how a given form of non-stationarity affects the fit of a stationary isotropic GP.

To elucidate further, consider an example from Finley et al.~(2008). Data on forest biomass and some covariates were available over a specific region;  prediction of forest biomass at unmeasured locations was of interest.  For over 30 years, tools have been available that completely characterize independent-errors linear models fit to such datasets, but analogous tools do not exist for models with spatially correlated random effects, which are commonly modeled using GPs.  To better understand fits of the latter models, we need a simple, interpretable form of the restricted likelihood.  This paper proposes such a form, which leads to tools for examining fits of linear mixed models with GP-distributed random effects and for choosing covariates.  Section~\ref{sec:real data} demonstrates the tools using this example.

This article addresses two challenging and hitherto untackled problems. First, using the spectral approximation to a GP, we propose tools to help understand exactly \emph{how} the data determine estimates of variance-structure parameters in a mixed linear model with a GP-distributed random effect (MLM/GP).  Phenomena like spatial confounding (e.g., Paciorek, 2010; Hodges \& Reich, 2011) make it clear that we cannot simply assume that a model, in its role as a likelihood, behaves according to its face-value interpretation as a probability model;  rather, our tools must directly display the influence of data on estimates, as do tools for linear models.  Some methods exist for MLM/GPs but they are weak.  One approach, popular in geostatistics, is informally examining residuals using exploratory tools such as variograms (see, e.g., Cressie, 2015; Chiles and Delfiner, 2009; Banerjee et al., 2014), which describe the degree of dependence (spatial range) and extent of variability (sill and nugget) in the data.  Variograms are useful but do not provide specific information about how functions of the data determine estimates. Also, variograms are most useful for stationary, even isotropic, processes and will not help much in ascertaining the effects of nonstationarity on stationary isotropic GP fits.  Exploratory analysis of residuals themselves generally does not provide specific information about how functions of the data determine estimates, and residuals in MLM fits are biased (e.g., Hodges 2014, Chapter~8), with the largest bias in parts of the fit most affected by shrinkage/smoothing.

Our second objective is to help analysts understand how adding a fixed effect to an MLM/GP moves variation in the outcome $y$ \emph{into} the fixed-effect part of the fit and \emph{out of} the GP and error parts of the fit. As we will see, the GP and error variance fits are determined mostly by, respectively, low- and high-frequency data features not captured in fixed effects.  Adding a fixed effect to a MLM/GP can ``take" variation mostly from the GP part of the fit, mostly from the error part of the fit, or substantially from both. The methods developed to understand an MLM/GP fit suggest using a diagnostic tool from linear models, added variable plot (AVPs), which show the data's information, observation-by-observation, about the coefficient of a fixed effect, enabling a modeler to understand whether the information about that fixed effect's coefficient is broadly distributed through the data or arises from a few functions of the data.  The spectral basis we use permits an AVP on the spectral scale;  one can also make an AVP using the original observations.  The two AVPs estimate the same coefficient for the added variable (modulo the spectral approximation) but the spectral-domain AVP emphasizes the contribution of low-frequency data features and thus highlights the effect a candidate predictor will have on the GP part of the fit, while the observation-domain AVP gives more emphasisis to the contribution of high-frequency data features (while avoiding the spectral approximation).  These two AVPs thus adapt a linear-model diagnostic to MLM/GPs in a way that provides information about why the GP and error parts of the fit change the way they do when a fixed effect is added.

In pursuing these goals, we hew to Weisberg's (1983) principles for diagnostics, in particular looking at the data as directly as possible and providing a plot to go with each diagnostic so the effect of individual observations can be assessed.  To meet our objectives, we need a tractable form of the restricted likelihood, which we obtain using a basis approximation to the GP-distributed random effect, the spectral approximation (Wikle, 2002;  Paciorek, 2007). With this choice, the restricted likelihood for the MLM/GP's variance-structure unknowns becomes formally identical to the likelihood arising from a gamma-errors GLM with identity link, so familiar data-analytic intuition and tools can be brought to bear.  The spectral approximation applies to data observed on a regular grid; for data observed at irregular locations, we will assume the data have been mapped to a regular grid (Paciorek, 2007; Reich et al.~2011). We do not see this as a major drawback because our focus is understanding GP fits rather than enhancing the model's richness and flexibility.

We emphasize that the choice of the spectral approximation is predicated on properties of its basis functions that serve our purposes.  This article is \emph{not} about the spectral approximation \emph{per se} or its properties \emph{as a model}.  Others have proposed the spectral approximation to speed computation (e.g., Fuentes 2006, Paciorek 2007), but that is also not our purpose.  We know of no other approximation
(e.g., Karhunen-Lo\`{e}ve expansions, wavelet basis, kernel convolutions, or predictive processes) that would work in the subsequent development.

The rest of this section describes an approach to fitting linear mixed models when the random effect is a one-dimensional GP.  Section~\ref{section2} then details the spectral approximation for intercept-only GPs in one dimension observed at equally spaced locations, and derives the simple restricted likelihood, which Section~\ref{section3} then uses to make conjectures about how data features affect parameter estimates, which simulation experiments support.  Section \ref{sec:cov_apprx1d} extends the approach to models with covariates.  Section~\ref{sec:sim_data} then proposes tools for model building based on the foregoing.  Section~\ref{sec:2dmodel} extends the tools to data observed on two-dimensional regular grids, and Section~\ref{sec:real data} applies them to the forest-biomass data.  The GP has been well investigated as a probability model and as an interpolator given parameter values; we focus on the GP as part of a likelihood used to estimate parameters.  Finally, we discuss only finite sample inferences.

\vspace*{-0.4cm}
\subsection{One dimensional Gaussian process fitting}\label{sec:modeldef1d}\vspace*{-0.2cm}
Given data $y(s)$ at location $s$ for $s \in \{s_1,s_2,...,s_M\}$, we want to fit the model\vspace*{-0.75cm}
\begin{equation} \label{eq:model1d} y(s)=x(s)\beta+w(s)+\epsilon(s)\vspace*{-0.75cm}\end{equation} where $w(s)$ is a stationary GP with mean $0$ and isotropic covariance function $\sigma_\mathfrak{s}^2 K(d; \rho)$, and $\epsilon(s)$ is Normal with mean $0$ and variance
$\sigma_e^2$, independent between locations $s$ and independent of $w(s)$.  $K(d; \rho)$ is a correlation function;  $d$ is the distance between two locations $s$; $\rho$ is an unknown range parameter;  and $d$, $\rho$, and $s$ have the same units (distance).  The row $p$-vector $x(s)$ contains covariates including the intercept and the column $p$-vector $\beta$ contains fixed effects.

Parameters that need to be estimated are $\beta$, $\sigma_\mathfrak{s}^2$ (process variance), $\rho$ (range),
and $\sigma_e^2$ (error variance). One way to fit this model is to write it as a linear mixed model \vspace*{-0.6cm}
\begin{equation}\label{eq:initial model1d}
y=X\beta+I_M \gamma +\epsilon, \vspace*{-0.6cm}
\end{equation}
where $y=(y(s_1), y(s_2),...,y(s_M))'$, $X$'s rows are the $x(s)$, $\gamma \sim N(0,\Sigma)$ with $\Sigma$ = $\sigma_\mathfrak{s}^2 K(d; \rho)$, and $\epsilon \sim N(0,R)$ with $R=\sigma_e^2\hspace*{1mm}I_M$.  Defining $V=\Sigma+R$, the unknowns in $\Sigma$ and $R$ are commonly  estimated by maximizing the log restricted likelihood
\vspace*{-0.6cm}
\begin{equation}\label{eq:lik1d}
\mbox{const} -0.5 (\log|V|+\log|X'V^{-1}X|+y'[V^{-1}-V^{-1}X(X'V^{-1}X)^{-1}X'V^{-1}]y).\vspace*{-0.6cm}
\end{equation}

The restricted likelihood (\ref{eq:lik1d}) has non-closed form terms involving the GP covariance matrix $\Sigma$, so it is a black box. The key to the desired simple form of (\ref{eq:lik1d}) is to diagonalize $V$, leading to a simple matrix-free form that can be used to develop intuition about how the GP model is fit to data.  (Closed form expressions for $V^{-1}$ exist for the Ornstein-Uhlenbeck process (Finley et al.~2009, sec.~2.1.1) but they are not diagonal.) To this end, we approximate the GP using orthogonal basis functions.  The result {\it is} an approximation but it can be used to conjecture about how the exact GP behaves, and the conjectures can be tested in simulations and used as a basis for diagnostic tools.  The approximation's accuracy is not of {\it inherent} interest but rather only to the extent that less accuracy means poorer understanding of the exact GP and less useful tools.  In this regard we note that all diagnostics for non-normal generalized linear models and for Cox regression are based on approximations.

\vspace*{-0.7cm}
\section{Approximating the Gaussian process}\label{section2} \vspace*{-0.2cm}
This section develops a simple approximate form of the log restricted likelihood (\ref{eq:lik1d}) using spectral basis functions. Section~\ref{section3} then interprets that approximate restricted likelihood as the likelihood for a particular generalized linear model and uses it to make conjectures about how features in the data, like outliers or mean-shifts, affect GP fits.

\vspace*{-0.4cm}
\subsection{Linear mixed model representation}\label{subsec2.1}\vspace*{-0.1cm}
An intercept-only GP model, with $\beta$ the intercept, can be approximated as \vspace*{-0.8cm}
\begin{equation}\label{eq:intercept only}
y\approx1_M\beta+Zu+\epsilon \vspace*{-0.75cm}
\end{equation}
where two key conditions hold:  $Z$ is an $M \times (M-1)$ matrix of basis functions that is not a function of any unknown parameters, and $u$ is a zero-mean Normal random vector with a diagonal covariance matrix, $G=\sigma_\mathfrak{s}^2 \hspace*{0.5mm} \mbox{Diag}(b_j(\rho))$, where the $b_j(\rho)$'s are known functions of $\rho$.  $Z$ and $b_j(\rho)$ are chosen so $\mbox{Cov} ( Zu + \epsilon)$ $\approx$ $\mbox{Cov}(y)$.

To yield the desired simplified restricted likelihood, $Z$ must have these properties:  $Z'1_M$ = \textbf{0} and $Z'Z$ is diagonal with diagonal entries $c_1$, $c_2$, ..., $c_{M-1}$.
For such a $Z$, premultiplying (\ref{eq:intercept only}) by $({Z'Z)}^{-1/2}Z'$ gives ${v}=\delta+\varepsilon,$ where
\\ \hspace*{0.9cm}
$\delta$ = $ ({Z'Z)}^{1/2}u\sim N(0, \sigma_\mathfrak{s}^2\hspace*{1mm} \mbox{Diag}\{b_j(\rho)\hspace*{1mm}c_j\}) \quad \mbox{for } j=1,2,..,M-1, \mbox{ and} $ \\
\hspace*{1cm}$\varepsilon$ = $({Z'Z)}^{-1/2}Z'\epsilon \sim N(0,\sigma_e^2\mbox{I}_{M-1}).$

\noindent Then ${v}={(Z'Z)}^{-1/2}Z' y$ is Normal with $\mbox{E}({v})=0$ and diagonal covariance $\sigma_\mathfrak{s}^2 \mbox{Diag}\{b_j(\rho)\hspace*{1mm}c_j\} +\sigma_e^2 I_{M-1}$.
The distribution of the $v_j$'s gives the log restricted likelihood for $(\sigma_\mathfrak{s}^2,\sigma_e^2,\rho)$:\vspace*{-0.5cm}
\begin{equation}\label{eq:lik vj1d}
\mbox{const}-\frac{1}{2}\sum\limits_{j=1}^{M-1}\left(\log(\sigma_\mathfrak{s}^2 a_j(\rho)+\sigma_e^2)+{v}_j^2{(\sigma_\mathfrak{s}^2
a_j(\rho)+\sigma_e^2)}^{-1}\right), \vspace*{-0.5cm}
\end{equation}
where $a_j(\rho)=b_j(\rho)\hspace*{1mm}c_j.$ The columns of $Z{(Z'Z)}^{-0.5}$ are orthonormal and the ${v}_j^2$'s are the squared lengths of projections of $y$ onto these columns. Thus the $ v_j$ decompose the data into components corresponding to these orthonormal predictors; in the spectral approximation described in Section \ref{specapprox1d}, these components correspond to frequencies.

\vspace*{-0.75cm}
\subsection{The generalized linear model form}\label{section2.2}\vspace*{-0.1cm}
Given $\rho$, the approximate restricted likelihood (\ref{eq:lik vj1d}) is identical to the likelihood arising from a gamma-errors generalized linear model with identity link, as in Hodges (2014, Ch.~15) and Henn \& Hodges (2014). As such, the ${v}_j^2$ are the data, the gamma shape parameter is $1/2$, $E({v}_j^2)=\sigma_\mathfrak{s}^2 a_j(\rho) + \sigma_e^2$, and
 $\mbox{Var}({v}_j^2)=2\hspace*{1mm} {(\sigma_\mathfrak{s}^2 a_j(\rho) + \sigma_e^2)}^2$.
Thus the ${v}_j^2$'s and $a_j$'s in the approximate restricted likelihood are the keys to
understanding how the GP's parameters are fit to data, giving a way to examine the model's fit that is immune to the fact that a GP can fit any $y$ perfectly. The ${v}_j^2$'s and $a_j$'s are, in effect, the data and predictors in a regression model that provides the
information about the unknowns $\sigma_\mathfrak{s}^2$, $\sigma_e^2$, and $\rho$.

\vspace*{-0.6cm}
\subsection{The spectral approximation}\label{specapprox1d}\vspace*{-0.2cm}The spectral basis
is a powerful tool, widely used for correlated processes. We develop the spectral approximation of a GP following Royle \& Wikle (2005) and Paciorek (2007).

Assume observations have been made at locations $s_j\in S=\{1,2,...,M\}, j=1,2,...,M$, where $M$ is a multiple of $2$. Define \vspace{-0.5cm}
\begin{equation}
g(s_j)=\sum\limits_{m=0}^{M-1} {\varphi_m(s_j)u_m}\vspace{-0.5cm}
\end{equation}
where $u_m=a_m+i$$b_m,$ $m=0,1,...,M-1$, are the $M$ spectral coefficients. The $\varphi_m(s_j)=\exp(i2\pi\omega_ms_j)$ are basis functions having frequency $\omega_m\in\{0,\frac{1}{M},...,\frac{1}{2},-\frac{1}{2}+\frac{1}{M},...,-\frac{1}{M}\},$ $m=0,1,...,M-1.$ To apply this approximation to real-valued Gaussian processes, assume $u_0,u_1,...,u_{\frac{M}{2}}$ are jointly independent; $u_0$ and $u_{\frac{M}{2}}$ are real valued ($b_0=b_{\frac{M}{2}}=0$); and $u_{\frac{M}{2}+1}=\bar{u}_{\frac{M}{2}-1},...,u_{M-1}=\bar{u}_1.$  This makes $g(s_j)$ real valued: \vspace*{-0.5cm}
\begin{equation}\label{eq:spectral}
g(s_j)=a_0+2\sum\limits_{m=1}^{\frac{M}{2}-1}{(a_m \cos(2\pi \omega_ms_j)-b_m \sin(2 \pi \omega_ms_j))}+a_{\frac{M}{2}} \cos(2 \pi \omega_{\frac{M}{2}}s_j),\vspace*{-0.5cm}
\end{equation}
where the $a_m$'s and $b_m$'s have independent mean zero Gaussian distributions with variances
V($a_0$) = $\frac{1}{M}\sigma_\mathfrak{s}^2\phi(\omega_0;\rho)$; V($a_{\frac{M}{2}}$) = $\frac{1}{M}\sigma_\mathfrak{s}^2\phi(\omega_{\frac{M}{2}};\rho)$; and V($a_m$) = V($b_m$) = $\frac{1}{2M}\sigma_\mathfrak{s}^2\phi(\omega_m;\rho)$ for  $m\neq0$ or $M/2$, where $\sigma_\mathfrak{s}^2 \phi(.;\rho)$ is the spectral density of the covariance function $\sigma_\mathfrak{s}^2 K(d; \rho)$. For large $M$, this approximate process is a Gaussian process with mean zero and covariance function close to that of the GP it approximates (Web Supplement Appendix A).

For data observed at locations $s_1=1,s_2=2,...,s_M=M$, then, model (\ref{eq:initial model1d}) becomes
\vspace*{-0.75cm}\begin{equation}\label{eq:lmmspec}
y= 1_M\beta+Zu+\epsilon  \vspace*{-0.75cm}
\end{equation}
where $\epsilon\sim N(0,R), R=\sigma_e^2 I_M$, \vspace*{-3mm}
\begin{itemize}
\item $\beta$ is the coefficient for the intercept, the only fixed effect, \vspace*{-3mm}
\item $u=(a_1,b_1,a_2,b_2,...,a_{\frac{M}{2}-1},b_{\frac{M}{2}-1},a_{\frac{M}{2}})'$ is the vector of random effects, and\vspace*{-3mm}
\item $Z$ is an $M\times {(M-1)}$ matrix with $j^{th}$ column $z_j$ given by\vspace*{-1mm}
\begin{itemize}
\item[] $\left(2\cos(\omega_{\frac{j+1}{2}}2\pi),2\cos(\omega_{\frac{j+1}{2}}2\pi2),...,2\cos(\omega_{\frac{j+1}
{2}}2\pi M)\right)' ; \hspace*{2mm} j\in\{1,3,...,M-3\},$
\item[] $\left(-2\sin(\omega_{\frac{j}{2}}2\pi),-2\sin(\omega_{\frac{j}{2}}2\pi2),...,-2\sin(\omega_{\frac{j}
{2}}2\pi M)\right)' ;\hspace*{2mm} j\in\{2,4,...,M-2\},$
\item[] $\left(\cos(\omega_{\frac{j+1}{2}}2\pi),\cos(\omega_{\frac{j+1}{2}}2\pi2),...,\cos(\omega_{\frac{j+1}
{2}}2\pi M)\right)' ; \hspace*{2mm} j=M-1.$
\end{itemize}\vspace*{-3mm}
\end{itemize}
 Finally, $u\sim N(0,G)$ for $G=\sigma_\mathfrak{s}^2\hspace*{1mm}\mbox{Diag}
(\frac{1}{2M}\phi(\omega_{m(1)};\rho),\frac{1}{2M}\phi(\omega_{m(2)};\rho)$,...,
$\frac{1}{2M}\phi(\omega_{m(M-2)};\rho),$\\$\frac{1}{M}\phi(\omega_{m(M-1)};\rho)
)$, with $m(j)=j/2$ for even $j$ and $m(j)=(j+1)/2$ for odd $j$.
The coefficient $a_0$ in (\ref{eq:spectral}) is not identified if an intercept is included in the model, so it has been omitted.

$Z$ does not depend on unknowns;  $Z'Z=\mbox{Diag}(2M,2M,....,2M,M)$ and $Z'1_M=$\textbf{0}, i.e., $Z$'s columns are orthogonal to each other and to the constant vector (proofs are in the Web Supplement Appendix A).  The successive columns capture trends in the data corresponding to increasing frequencies, with the elements of $u$ being the weights for these trends.

\begin{figure}[H]
\centering\begin{subfigure}[b]{0.3\textwidth}
                 \includegraphics[width=\textwidth]{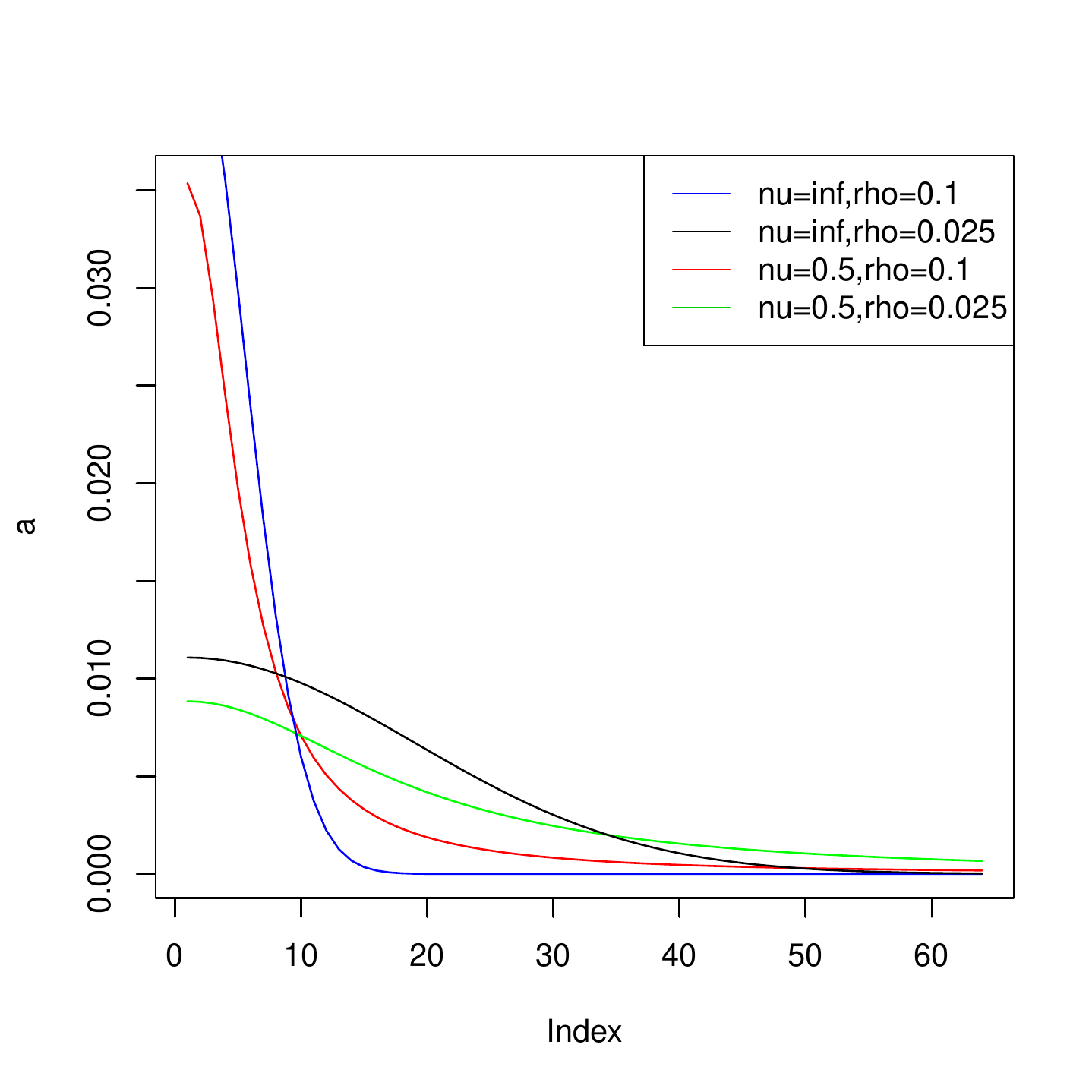}
                 \caption{\\$a_j(\rho)$ for Mat\'{e}rn $\nu$=0.5 and Mat\'{e}rn $\nu$$=$$\infty$, for two values of $\rho$, for $j\in$[1,2,...,63,64].}\label{figure1a}
         \end{subfigure}
         \quad
         \begin{subfigure}[b]{0.3\textwidth}
                          \includegraphics[width=\textwidth]{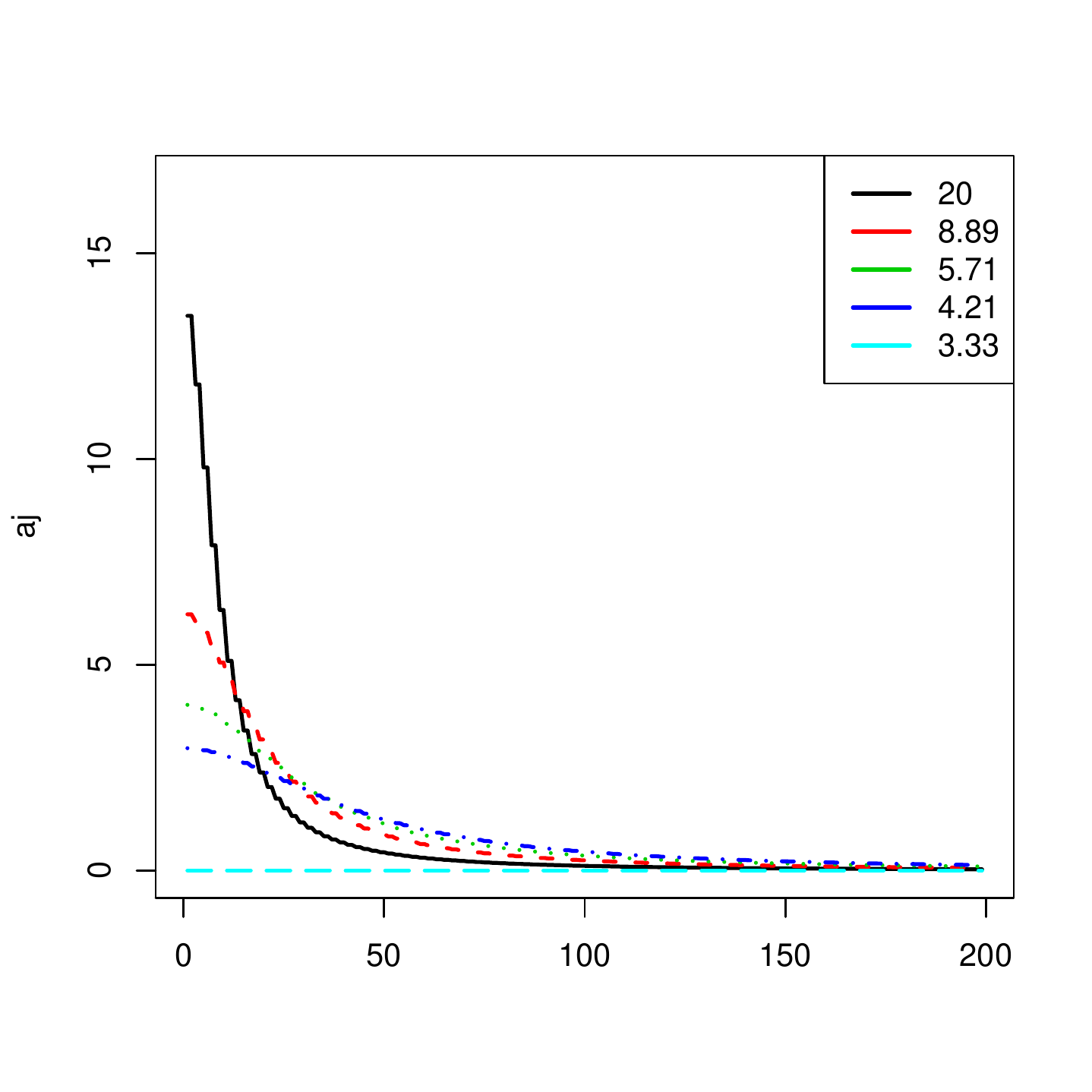}
                 \caption{$a_j(\rho)$ for Mat\'{e}rn $\nu$=0.5 for different values of $\rho$ for $j\in$[1,2,...,199,200].\\}\label{figure1b}
                 \end{subfigure}
                 \quad
                 \begin{subfigure}[b]{0.3\textwidth}
                 \includegraphics[width=\textwidth]{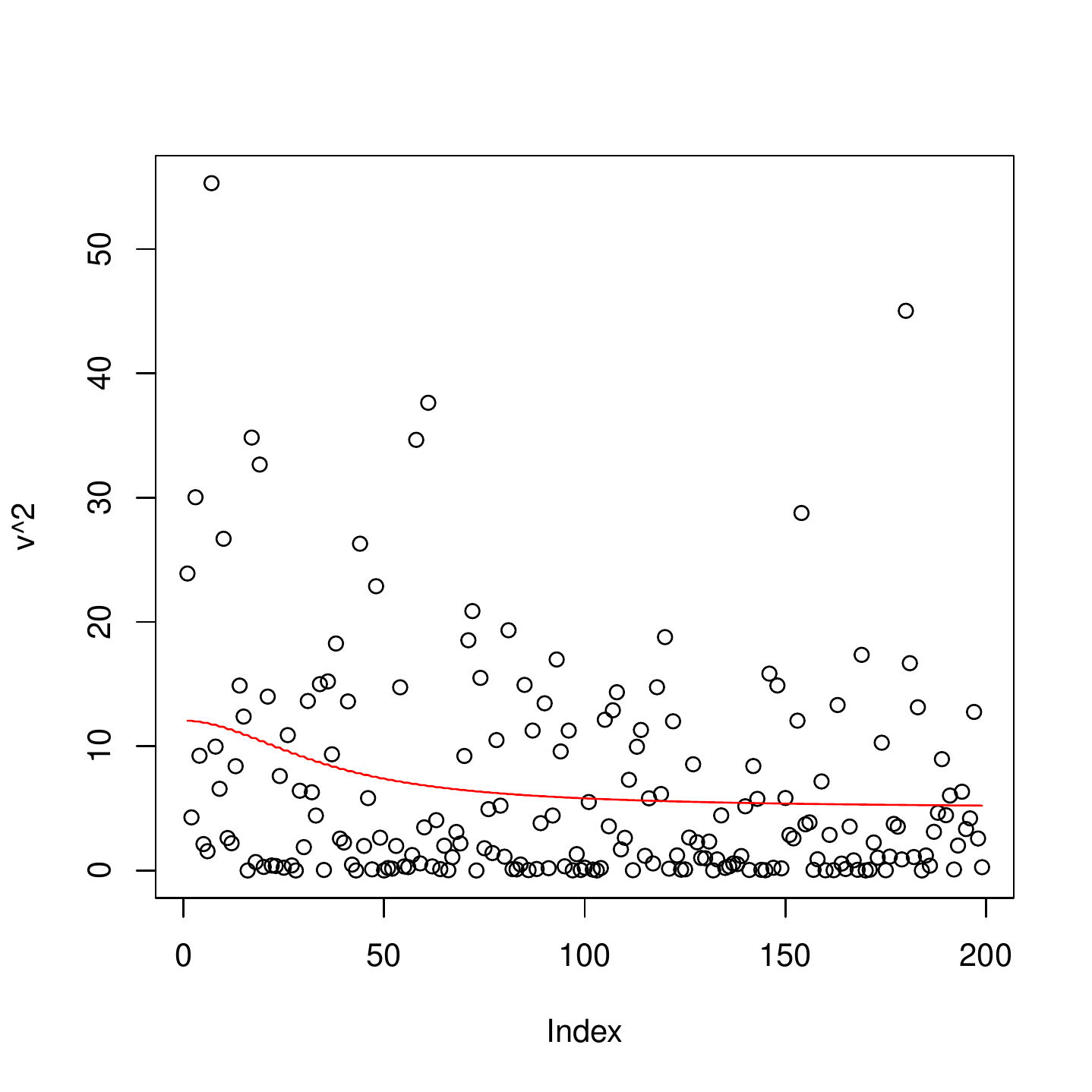}
                 \caption{Black circles: ${v_j}^2$'s for data simulated from GP with $\sigma_\mathfrak{s}^2$=2, $\sigma_e^2$=5 and $\rho$=5. Red line: $\sigma_\mathfrak{s}^2 a_j (\rho) + \sigma_e^2$. }
                 \label{figure4}
                 \end{subfigure}
         \caption{}
         \end{figure}

In the spectral approximation, the $a_j(\rho)$'s defined in Section~\ref{subsec2.1} are given by\ \vspace*{-0.8cm}
$$a_j(\rho)=\phi(\omega_{m(j)};\rho),\vspace*{-0.8cm}$$
with $m(j)=j/2$ for even $j$ and $m(j)=(j+1)/2$ for odd $j$, $j=1,2,..,M-1$. For example, for the exponential correlation function (Mat\'{e}rn with smoothness parameter $\nu=0.5$),\vspace*{-0.7cm}\begin{equation}\label{eq:exp cov}K(s_i,s_j;\rho)=
\exp(-\sqrt{2} |s_i-s_j|/\rho),\vspace*{-0.5cm}\end{equation} and $\phi(\omega;\rho)$ has the form of a
Cauchy density\vspace*{-0.7cm}
\begin{equation}\label{eq:spdensity_exp}
\phi(\omega;\rho)=\frac{1}{\sqrt{2}}\rho {\left(1+\frac{{(\pi\rho)}^2}{2}\omega^2\right)}^{-1}.\vspace*{-0.5cm}
\end{equation}
For any $\nu$, the Mat\'{e}rn($\nu$) correlation function corresponds to a particular function $a_j(\rho)$. Figure~\ref{figure1a} shows $a_j(\rho)$ for the Mat\'{e}rn($\nu$) for $\nu=0.5$ and $\infty$. Given $\rho$, the $a_j(\rho)$ for different $\nu$ hardly differ, indicating how little information about $\nu$ the data can provide. This corroborates the well-known fact that $\nu$ is generally difficult to estimate from data.

A known aspect of the spectral approximation is that $a_j(\rho)$ is non-increasing in $j$ and aproaches zero for large $j$.  The $a_j$'s start higher and decline faster as $\rho$ increases (Figure~\ref{figure1b}).

Empirically, we observe that the spectral approximation causes the correlation, as a function of the distance between observations, to decrease to zero at a faster rate than it should. In one dimension, maximizing the approximate restricted likelihood compensates by making the estimate of $\rho$ (range) larger than the estimate from the exact restricted likelihood. Also, the approximate process is periodic: $g(0)=g(2 \pi)$ (Figures 1 and 2 in Web Supplement Appendix B show examples).  To mitigate this, Royle \& Wikle (2005) use a grid larger than the observation domain, known as padding. Paciorek (2007) pads by mapping the periodic domain $(0,2\pi)$ to $(0,2)$ and then mapping the observation domain onto $(0,1)$. For our purpose, $v_j$ must be a function of the data, so we cannot pad in this way.  This is, admittedly, a weakness of the approximation but does not imply that it is useful only for analyses of periodic functions;  as noted, the approximation's utility arises from its ability to provide insight, which is reasonably unimpaired, as we now argue.

\vspace*{-0.5cm}
\section{Conjectures about parameter estimates}\label{section3}\vspace*{-0.4cm}

Recall from Section~\ref{section2.2} that for fixed $\rho$, the approximate restricted likelihood (\ref{eq:lik vj1d}) is identical to the likelihood from a gamma-errors GLM with identity link. This gives a way to generate conjectures about how parameter estimates are fit to data.

The matrix $Z$ is the same for all GPs on a given location set so given $y$, the $v_j^2$'s are also the same for all GPs;  the only thing distinguishing GP models for $u$ is their $a_j(\rho)$'s. The ${v}_j^2$'s are the ``data" and the parameters are fit for a model with $\mbox{E}({{v}_j^2}|\sigma_\mathfrak{s}^2,\rho,\sigma_e^2)=\sigma_\mathfrak{s}^2a_j(\rho)+\sigma_e^2$ and
$\mbox{Var}({v}_j^2|\sigma_\mathfrak{s}^2,\rho,\sigma_e^2)=2\hspace*{1mm} {(\sigma_\mathfrak{s}^2 a_j(\rho) + \sigma_e^2)}^2$.

Because  $a_j(\rho)$ approaches 0 for large $j$,  $\mbox{E}({v}_j^2|\sigma_\mathfrak{s}^2,\rho,\sigma_e^2) \approx \sigma_e^2$ for large $j$;  heuristically, ${v}_j^2$'s for large $j$ are more informative about ${\sigma}_e^2$ and ${v}_j^2$'s for small $j$ are more informative about ${\sigma}_\mathfrak{s}^2$ and $\rho$. Because $a_j(\rho)$ is non-increasing in $j$, the ``data" ${v}_j^2$ have higher variance for smaller $j$, so the data provide more information about ${\sigma}_e^2$ than about $\sigma_\mathfrak{s}^2$ or $\rho$.  For Mat\'{e}rn correlation functions like (\ref{eq:spdensity_exp}), $\sigma_\mathfrak{s}^2 a_j(\rho)$ has the form $\sigma_\mathfrak{s}^2 \rho f(\rho, \omega)$;  $\sigma_\mathfrak{s}^2$ and $\rho$ are identified in the approximate restricted likelihood (\ref{eq:lik vj1d}) only by $f(\rho, \omega)$, which describes how $a_j(\rho)$ declines with $j$, and $\hat{\rho}$ and $\hat{\sigma}_\mathfrak{s}^2$ are chosen to fit this rate of decline to the $v_j^2$'s.  The noise in $ v_j^2$ is a function of $j$ so $\sigma_\mathfrak{s}^2$ and $\rho$ are not always well identified;  this and lack of consistency in joint estimates of $\sigma_\mathfrak{s}^2$ and $\rho$ on a fixed domain are well-known problems (Ying 1991, Zhang 2004).

Figure~\ref{figure4} shows the ${v}_j^2$'s and $\sigma_\mathfrak{s}^2 a_j(\rho)+\sigma_e^2$ for a dataset simulated with exponential correlation function (\ref{eq:exp cov}) and true $\sigma_\mathfrak{s}^2=2$, $\sigma_e^2=5$ and $\rho=5$.  As described, $\sigma_\mathfrak{s}^2$, $\sigma_e^2$ and $\rho$ are estimated so that the $\hat\sigma_\mathfrak{s}^2 a_j(\hat\rho)+\hat\sigma_e^2$ fit the ${v}_j^2$ as best they can. (The estimates are in Web Supplement Appendix C, Table 1.) Thus, the GLM formulation (\ref{eq:lik vj1d}) allows us to visualize how features in the data produce the parameter estimates. Sections~\ref{outliersec} to~\ref{meanshiftsec} present and test some conjectures about how features of the data affect the parameter estimates.

To do this, data were simulated from a GP with mean 0 and correlation function (\ref{eq:exp cov}) and Normal(0, $\sigma_e^2$) errors using each of Table~\ref{table2}'s eight combinations of true parameter values, with observations at locations \{1,2,...,199,200\}.  100 datasets were simulated for each combination.  We call these simulated datasets uncontaminated data.  Parameter estimates were obtained by maximizing the exact log restricted likelihood; estimates were also obtained by maximizing the approximate log restricted likelihood. Table~\ref{table2} presents averages of the estimates over these 100 datasets with Monte Carlo standard errors.

We then re-fit the GPs to two kinds of contaminated data, contaminating by:\vspace*{-0.3cm}
\begin{itemize}
\item an outlier:  the $100^{th}$ observation was replaced by 18 (Section \ref{outliersec});\vspace{-0.2cm}
\item a mean shift:  5 was added to the last 100 observations (Section \ref{meanshiftsec}).\vspace{-0.2cm}
\end{itemize}

\noindent Web Supplement Appendix C, also shows simulation results for a contamination in which the GP's range parameter $\rho$ was changed halfway through the series.


\subsection{Outlier}\label{outliersec}
How does an ordinary outlier affect the estimates of the GP parameters?

We conjecture that an outlier will inflate the $v_j^2$'s for larger $j$'s, corresponding to high frequencies. Because $\hat\sigma_e^2$ is driven largely by those ${v}^2_j$'s, $\hat\sigma_e^2$ will be inflated. The $v_j^2$'s for smaller $j$'s (low frequencies) will be comparatively unaffected, so the outlier will have little effect on $\hat{\sigma}_\mathfrak{s}^2$ and $\hat{\rho}$. These two effects lead to a smoother fit.

For all eight true parameter combinations, the most striking effect of the outlier contamination is in fact an inflated $\hat{\sigma}_e^2$ (Table~\ref{table2}), as conjectured (Figure 19a in Web Supplement Appendix C shows this for one dataset). Figure 19b in Appendix C shows that the fit is indeed smoother when the outlier is present (Appendix C, Table 1 gives the estimates).  An outlier at the end of the data series has a similar effect (not shown).

\subsection{Mean shift}\label{meanshiftsec}

\begin{figure}[H]
         \centering
         \begin{subfigure}[b]{0.3\textwidth}
                 \includegraphics[width=\textwidth]{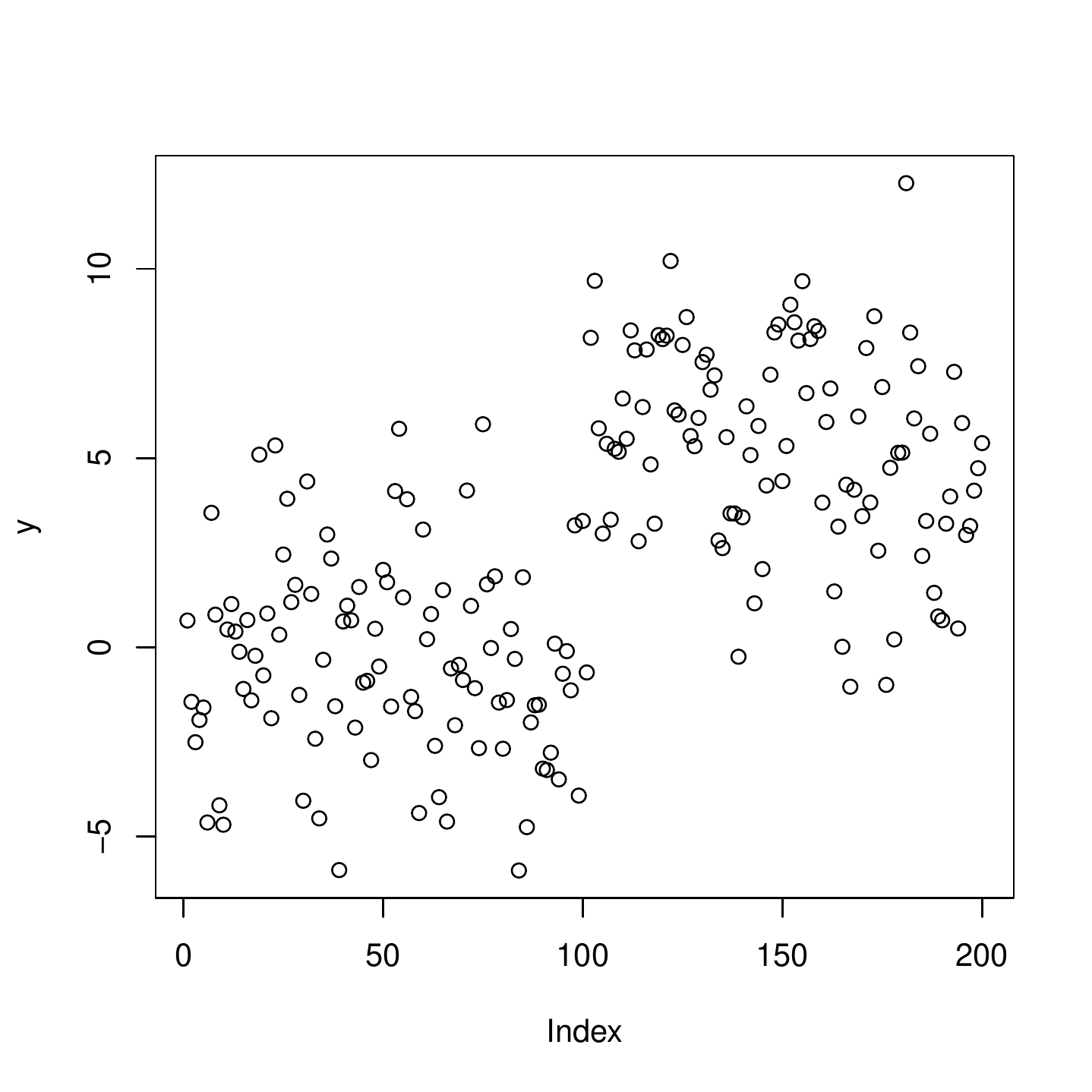}
                 \caption{Data simulated from GP with $\sigma_\mathfrak{s}^2$=2, $\sigma_e^2$=5 and $\rho$=5 with mean shift from 0 to 5 midway.}\label{figure6a}
         \end{subfigure}
         \quad
         \begin{subfigure}[b]{0.3\textwidth}
                 \includegraphics[width=\textwidth]{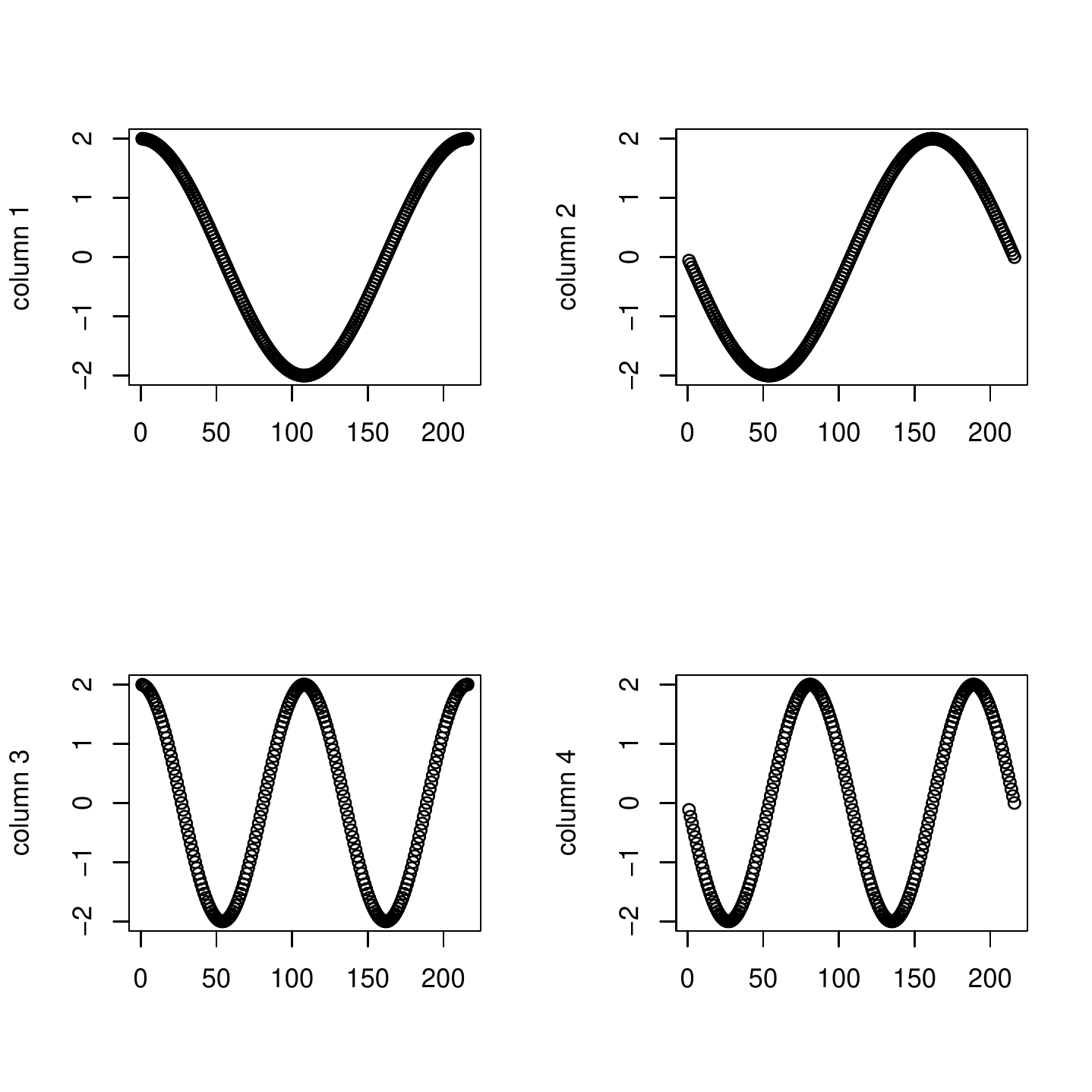}
                 \caption{First four columns of $Z$, the spectral basis matrix, on the domain [1,2,...,199,200].\\}\label{figure6b}
         \end{subfigure}
         \quad
         \begin{subfigure}[b]{0.3\textwidth}
                 \includegraphics[width=\textwidth]{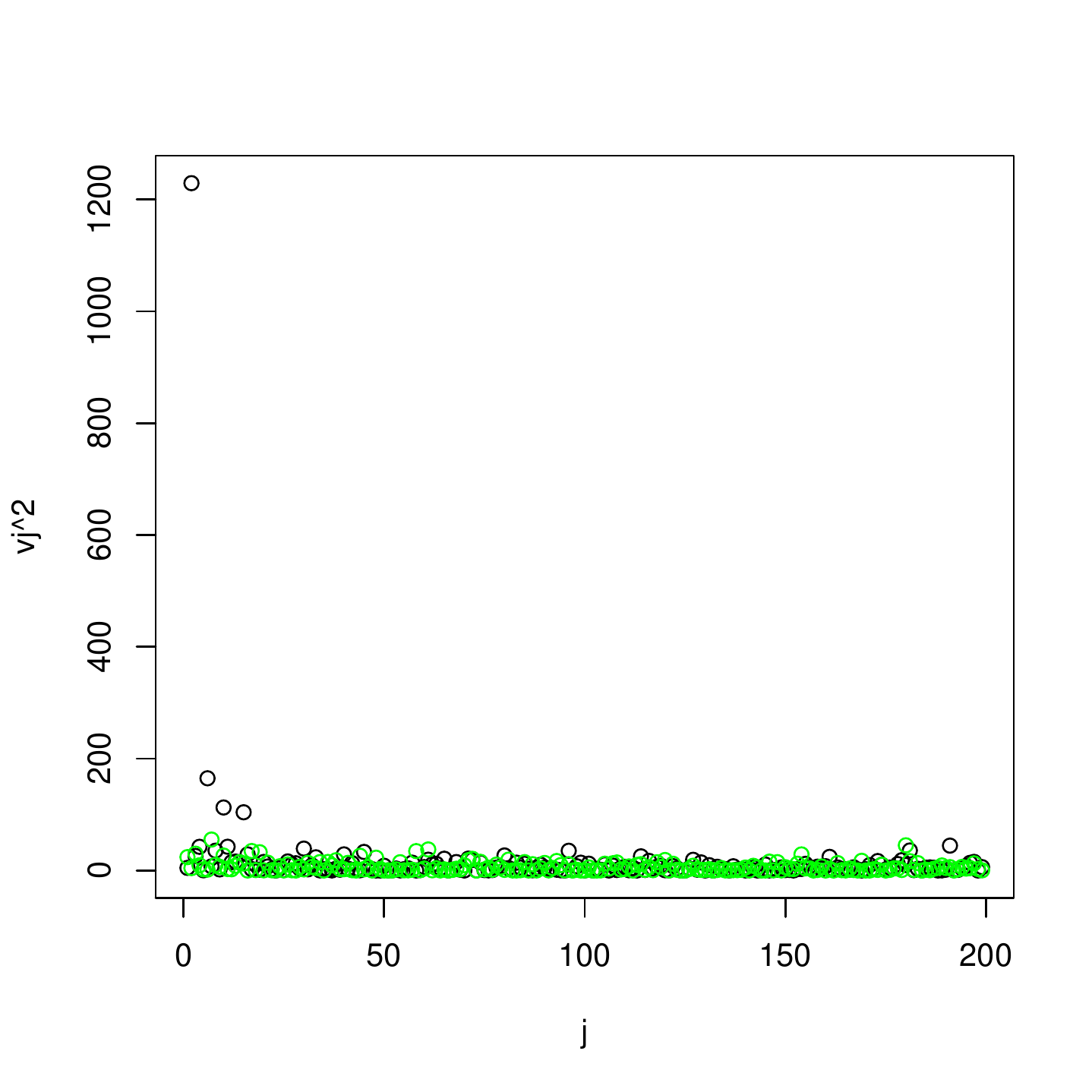}
                 \caption{Green circles: ${v_j}^2$'s from uncontaminated simulated data. Black circles: ${v_j}^2$'s from data with mean shift. }\label{figure7a}
         \end{subfigure}
         \caption{}
         \end{figure}

Consider the data in Figure~\ref{figure6a}, a draw from a GP with a shift in the mean halfway through the series. How does this mean shift affect the stationary GP's parameter estimates?

The contaminated data look most like the $2^{nd}$ column of the $Z$ matrix (Figure~\ref{figure6b}'s upper right), so we conjecture that the ${v}_j^2$ arising from this column, ${v}_2^2$, will be greatly increased, which in turn will inflate $\hat{\sigma}_\mathfrak{s}^2$. The large ${v}_2^2$ will cause the ${v}_j^2$'s to decline more sharply in $j$, so $\hat{\rho}$ will be inflated to capture that decline (recall Figure~\ref{figure1b}).  The mean shift, a low frequency data feature, will not affect the ${v}_j^2$'s for large $j$'s, so $\hat{\sigma}_e^2$ should change little.

The ${v}_j^2$ arising from $Z$'s second column, ${v}_2^2$, is indeed affected most by the mean-shift contamination (Figure~\ref{figure7a} shows this for one simulated dataset) leading, as conjectured, to inflated $\hat{\sigma}_\mathfrak{s}^2$ and $\hat{\rho}$ (Tables 1 and 2 in Web Supplement Appendix C).

\vspace*{-0.4cm}
\section{Regressing out covariates}\label{sec:cov_apprx1d}
The spectral representation above is for an intercept-only GP model. If the fixed-effect design matrix $X$ is not just a vector of ones, we propose first regressing it out as follows, and then applying Section \ref{section2}'s spectral representation. Let $P_X= X{(X'X)}^{-1}X'$ be the orthogonal projector onto $X$'s column space. Premultiply both sides of (\ref{eq:initial model1d}) by ($I_{M}-P_X$) to give \vspace*{-0.75cm}
\begin{equation}\label{eq:lmm cov1d}
y^*= \gamma^* +  \epsilon^*,\vspace*{-0.75cm}
\end{equation}
where $y^*=(I_{M}-P_X)\hspace*{1mm}y$ is the residual from a regression on $X$, $\gamma^*=(I_{M}-P_X)\hspace*{1mm}\gamma$, and $\epsilon^*= (I_{M}-P_X)\hspace*{1mm}\epsilon$, with $\mbox{Cov}(\gamma^*)=(I_{M}-P_X)\hspace*{1mm}\Sigma\hspace*{1mm} (I_{M}-P_X)$ and $\mbox{Cov}( \epsilon^*)=\sigma_e^2 (I_{M}-P_X)$. The likelihood arising from (\ref{eq:lmm cov1d}) is the restricted likelihood of the original model (\ref{eq:initial model1d}).

If rank(${X}$) is small compared to $M$, these approximations are reasonable:  Cov$(\gamma^*)\approx \Sigma$, and Cov($\epsilon^*)\approx  \sigma_e^2 I_{M} = R $, i.e., ignore changes in Cov$(y^*)$ induced by the residual projection, as standard linear-model diagnostics do.  Priestley (1981, Ch.~7) discusses fitting stationary processes to residuals from least-squares fits. If the residuals arise from a polynomial fit, then the spectral densities estimated from $y^*$ and $y$ have the same asymptotic properties.

Thus we assume the residuals $\gamma^*$ after regressing on the covariates can be approximately modeled by a GP having the same covariance form as $\gamma$ and the errors $\epsilon^*$ approximately modeled by the same form as $\epsilon$, \textit{with possibly different parameter values}. With the approximation Cov($y^*)\approx \Sigma+R$, the model becomes
\vspace*{-0.75cm}\begin{equation}\label{eq:resmodel}y^*= I_M \gamma^{*}+\epsilon^*,\vspace*{-0.2cm}\end{equation}
 where $y^*={(I_M-P_X)}\hspace*{1mm}y$, $\gamma^{*}={(I_M-P_X)}\hspace*{1mm}\gamma\overset{approx}{\sim} \mbox{GP}(0,\Sigma)$, $\epsilon^*=(I_M-P_X) \hspace*{1mm}\epsilon\overset{approx}{\sim} N(0,R)$.

It is helpful to see what this approximation does in practice. When $X$ is a column of $1$'s and $Z$ is  Section~\ref{specapprox1d}'s spectral basis matrix, $v_j$ is the unshrunk projection of $y$ onto the $j^{th}$ column of $W = Z (Z'Z)^{1/2}$. If we add fixed effects to $X$ and proceed as proposed, replacing $y$ with $y^*$ = $(I_M - P_X) y$ but keeping the same $Z$, then the unshrunk projections of $y^*$ onto the columns of $W$, i.e., the $W'y^*$, are
\vspace*{-0.75cm}
\begin{equation}
W'(I_M - P_X) y =
(Z'Z)^{1/2}Z'y - (Z'Z)^{1/2}Z'P_X y  =  v - (Z'Z)^{1/2}Z'P_X y \label{eq:just 1}\vspace*{-1cm}\end{equation}
 \vspace*{-0.75cm}\begin{equation} \hspace*{9.8cm} =  v - [P_X Z (Z'Z)^{1/2}]' y. \label{eq:just_2}\vspace*{-0.5cm}\end{equation}
Each $v_j$ is reduced by an amount depending on how much of it ``goes away" when $y$ is projected onto the orthogonal complement of $X$'s column space, as in (\ref{eq:just 1}), or by an amount determined by the projection of $Z$'s $j^{th}$ column onto $X$, as in (\ref{eq:just_2}). The approximation Cov($y^*$) $\approx$ $\Sigma+ R$ reduces $y$'s projection onto $Z$'s $j^{th}$ column to an extent depending on how $X$ (collectively) is correlated with $Z$'s $j^{th}$ column. Note that in computing the exact restricted likelihood, all of $y$'s variation in $X$'s column space is attributed to $X$;  the approximation retains this key feature.

For another view, consider the Kullback-Leibler distance between the densities $N(0, \Sigma+R)$ and $N(0, P\Sigma P + R)$, \vspace*{-0.6cm}$$0.5 \log[\mbox{det}(P\Sigma P + R)/\mbox{det}(\Sigma + R)] - M + \mbox{trace}[(P\Sigma P +R)^{-1}(\Sigma +R)],\vspace*{-0.6cm}$$ where $P$ is an $M\times M$ projector of rank $M-p$.  If $(P\Sigma P + R)^{-1}(\Sigma +R)$ is approximately the identity, this distance is small, as is the case here.

Figures 3 to 18 in Web Supplement Appendix B show Cov($y^*$) and Cov($y$) for various $M$, $\sigma_\mathfrak{s}^2/\sigma_e^2$, $\rho$, correlation functions, and $X$.  These figures show that the structure of Cov($y^*$) is (for our purposes) satisfactorily approximated by the functional form of Cov($y$).

\section{A small toolkit for assessing goodness of fit and considering covariates}\label{sec:sim_data}
Conventional residuals can highlight a few extreme outliers in data space but cannot identify lack of fit, especially in the GP part of the model. Indeed, any model of this type can be made to fit any $y$ arbitrarily well by setting $\rho$ small and $\sigma_\mathfrak{s}^2$ large. This section shows how to use the tools from earlier sections to avoid this problem, in particular highlighting lack of fit in the GP part of the model. When covariates are available, we present tools for considering which covariates to add. If no covariates are available, the tools identify properties of potential covariates, to aid in seeking them.

When covariates (potential fixed effects) are available, a modeler must make a choice:  either let the fitting machinery interpret strong low-frequency data features as evidence of stationary GP errors with large $\sigma_\mathfrak{s}^2$ and $\rho$, or attribute those features to covariates to the extent possible.  Whatever your view on this matter, it is essential to know whether such features are present so a well-informed choice can be made.  A plot with $v_j^2$ on the vertical axis and $j$ on the horizontal axis (henceforth ``the $v_j^2$ plot") shows such prominent low-frequency data features as large $v_j^2$ for small $j$.  If, in the same plot, some $v_j^2$ are large for large $j$, that is evidence of outliers (high frequency trends).  Generally, a large $v_j^2$ suggests a missing covariate with high power at the frequency corresponding to $j$.  A polynomial or sinusoidal curve of that frequency could be added and this may be defensible in some cases, e.g., a linear term parallel to a coordinate axis or an annual cycle. However, adding substantively meaningful covariates is generally more satisfactory.

The $v_j^2$ plot is visually dominated by low frequency $j$;  outlying $y_i$ or high frequency trends in $y$ will be spread out among high-frequency $v_j^2$ and may not be visible in the $v_j^2$ plot.  If potential covariates are available, however, a modest adaptation of the familiar added variable plot can be used to examine covariates irrespective of their prominent frequencies.  We now describe added variable plots in the observation and spectral (frequency) domains.
\vspace*{-0.7cm}
\subsection{Added variable plots}\label{sec:add_var}
 In an ordinary linear model, the added variable plot for a candidate predictor $C$ is drawn as follows (Cook \& Weisberg 1982, p.~44; Atkinson 1985, Section 5.2):
 \vspace*{-0.3cm}
\begin{enumerate}
\item Compute residuals from regressing the outcome $y$ on all predictors except $C$.\vspace*{-0.4cm}
\item Compute residuals from regressing $C$ on all predictors other than $C$.\vspace*{-0.4cm}
\item Plot the residuals from steps 1 and 2 on the vertical and horizontal axes respectively. \vspace*{-0.4cm}
\item Fit a regression through the origin to the plotted data;  this estimates the coefficient of $C$ if it were included in the model.
    \vspace*{-0.3cm}
\end{enumerate}
This usual added variable plot assumes errors are independent with constant variance. A linear mixed model with a GP random effect has neither property;  for this model, we describe how to adapt added variable plots in both the observation and spectral domains.  Both plots estimate the same slope for the covariate $C$.
\\ \textbf{Observation domain:}
Consider adding $C$ to give the model
$y=X\beta+C\alpha+\gamma+\epsilon,$ where $X$ contains predictors already in the model including the intercept, $\gamma$ is a GP-distributed random effect, and $\epsilon$ is iid Normal errors. Pre-multiply both sides of the model equation by $\hat V^{-0.5}$, where $\hat{.}$ denotes estimates from fitting the model without $C \alpha$. Then pre-multiply both sides of the model equation by $\hat P=I - \hat V^{-0.5} X {(X'\hat V^{-1}X)}^{-1}X'\hat V^{-0.5}$ to give
$\hat P\hat V^{-0.5}y=\hat P \hat V^{-0.5} C\alpha + \hat P \hat V^{-0.5} (u + \epsilon)$.
The added variable plot shows $\hat P \hat V^{-0.5} y $ vs $\hat P \hat V^{-0.5}C$.
\\ \textbf{Spectral domain:}
For the same model equation, pre-multiply both sides by $(I-P_X)$, then pre-multiply by ${(Z'Z)}^{-0.5}Z'$, then pre-multiply by $\hat{D}=\mbox{Diag}(1/\sqrt{\hat\sigma_\mathfrak{s}^2 a(\hat \rho) + \hat \sigma_e^2})$, to give
$\hat{D}v^* = \hat{D}v_C^* \alpha +\hat{D}{(Z'Z)}^{-0.5}Z' (\gamma+  \epsilon) ,$ where $v^*={(Z'Z)}^{-0.5}Z'(I-P_X)y$ are the $v_j$ from the residuals $y^*$ and $v_C^*={(Z'Z)}^{-0.5}Z'(I-P_X)C$ are $v_j$ from the residuals for $C$.  This added variable plot shows $\hat{D}v^*$ vs $\hat{D}v^*_C$.

A common method for assessing the effect of a predictor is to fit the model with and without the predictors and compare the fits; the model is fit twice, which may be inefficient depending on the cost of a model fit. In contrast, added variable plots do not require re-fitting the model for each predictor and also show the effect of individual observations.

In the example below and the Web Supplement, we demonstrate the model building tools described here, i.e., the $v_j^2$ plot and the added variable plots. (Web Supplement Appendix G uses a simulated example;  Appendices H, I, and J use Finley et al's (2008) forest biomass data.) These examples show how a low-frequency covariate can have a weak signal in the observation-domain added variable plot but a strong signal in the spectral-domain added variable plot; similarly, the two added variable plots may have signals of differing strength for a covariate with power mainly in high frequencies.

\section{Gaussian process on two dimensions}\label{sec:2dmodel}
Section \ref{section2} described the spectral basis and GLM interpretation of the approximate restricted likelihood for observations at locations in one dimension. Spatial data are commonly observed at two-dimensional locations;  this section outlines derivation of the 2-dimensional (2-D) approximate restricted likelihood;  Web Supplement Appendix D gives details.

For observations on an equally-spaced $M_1$ $\times$ $M_2$ grid, the restricted likelihood has the same form as (\ref{eq:lik1d}) and as in the 1-D case, we approximate the intercept-only GP using spectral basis functions.  If the model includes fixed effects $X$, we proceed as in Section \ref{sec:cov_apprx1d}'s 1-D case, i.e., assume (approximately) that the residuals follow a linear mixed model with a GP-distributed random effect, then approximate this GP using the spectral basis. For the 2-D model, the approximate log restricted likelihood has the matrix-free form \vspace*{-0.5cm}
\begin{equation}\label{eq:lik_vj}
\mbox{ALR}(\sigma_\mathfrak{s}^2,\sigma_e^2, \rho)\mbox{ = const}-\frac{1}{2}\sum\limits_{j=1}^{M_1M_2-1}\left(\log(\sigma_\mathfrak{s}^2 a_j(\rho)+\sigma_e^2)+{v}_j^2{(\sigma_\mathfrak{s}^2 a_j(\rho)+\sigma_e^2)}^{-1}\right); \vspace*{-0.3cm}\end{equation} the $a_j(\rho)$ are sorted to be non-increasing in $j$;  it is easy to prove this order is invariant to $\rho$.

We can now ask for the 2-D case how GP fits respond to features in the data. As in the 1-D case, by construction sin/cos column pairs of the spectral basis matrix $Z$ decompose the data into frequency components. Thus a high frequency feature, e.g., an outlier, falls in the space spanned by columns of $Z$ with large $j$ and inflates $v_j^2$ for larger $j$, which in turn inflates $\hat{\sigma}_e^2$.  A low frequency feature in $y$, e.g., a linear trend, inflates $v_j^2$ for smaller $j$, which in turn inflates $\hat{\rho}$ and $\hat{\sigma}_\mathfrak{s}^2$.

For data observed on a regular grid, we have made two approximations: approximating $(I_{M_1 M_2}-P_X)\hspace*{1mm}\Sigma\hspace*{1mm} (I_{M_1 M_2}-P_X)$ + $\sigma_e^2 (I_{M_1M_2} - P_X)$ by $\Sigma$ + $R$, and the spectral approximation. If the observation locations are not on a regular grid, we suggest first smoothing the data onto such a grid, as follows.

\subsection{Smoothing observed data onto a regular grid}\label{sec:irregular grid}
Construct a rectangular uniformly spaced grid on the observation domain; the grid size must be a multiple of 2 in each dimension.  Label the grid locations $\{1,2,...,M_1\}$ $\times$ $\{1,2,..,M_2\}$ and re-scale the actual observation domain to $[1,M_1]$ $\times$ $[1,M_2]$. To minimize space in the grid with no observations, the map of observation locations may need to be rotated to make it more nearly rectangular, and $M_1/M_2$ should be close to the aspect ratio of the observation locations.
Then for each location on the grid, we suggest constructing an artificial datum using inverse distance weighting (IDW) (Shepard 1968, Zimmerman 1999):  if the data are $(y_1, y_2, .., y_{n})'$ at 2-D locations $(s_1,s_2,..s_{n})'$, the value at grid location $\varrho$ is
\vspace*{-0.5cm}
\begin{equation}\label{ref:idw}
\frac{\sum_{k=1}^{n}t_k\hspace*{1mm} y_k}{\sum_{k=1}^{n}t_k}, \quad \quad \mbox{where} \quad \quad t_k=\frac{1}{d(\varrho,s_k)^\lambda},
\vspace*{-0.5cm}
\end{equation} where $d$ is Euclidean distance and $\lambda$ is a tuning constant.

In Section \ref{sec:real data}'s example, we chose $M_1$, $M_2$, and $\lambda$ so $\sigma_\mathfrak{s}^2$, $\sigma_e^2$, and $\rho$ would have estimates as similar as possible from\ the artificial and actual data. Other choices of $M_1$, $M_2$, and $\lambda$ would give results qualititatively similar to those we present.  Web Supplement Appendix K describes a simulation experiment showing the consequences of different $M_1$, $M_2$, and $\lambda$. Broadly, IDW largely preserves $y$'s low-frequency trends while sacrificing power at high frequencies, less so for larger $\lambda$. Thus $\hat{\sigma}_e^2$ is smaller for the artificial data than for the actual data, $\hat{\rho}$ is inflated slightly, and $\hat{\sigma}_\mathfrak{s}^2$ is affected but not in a systematic way. Added variable plots in the observation domain preserve $y$'s high-frequency information;  spectral-domain plots lose some of it.  IDW's main virtue for our purpose is computing speed;  an ideal method, if one exists, would minimally affect power at all frequencies.

\section{Application to data}\label{sec:real data}
We illustrate use of the tools with the 2002 forest inventory data analyzed by Finley et al.~(2008) and included with the R package spBayes (BEF.dat;  Finley et al.~2007). The outcome $y$ is red maple total basal area (RM\_02BAREA $\times$ BAREA02\_TOT) and potential predictors are ELEV, SLOPE, SPR\_02\_TC2, SPR\_02\_TC3, SUM\_02\_TC1, SUM\_02\_TC3, and FALL\_02\_TC2, all measured at 437 locations. The data were smoothed to a grid using IDW with grid size 28 $\times$ 20, with $\lambda$ = 7 for $y$ and $\lambda$ = 9 for the predictors. We fit an intercept-only model in which the GP had the exponential covariance function, i.e., Mat\'{e}rn with $\nu = 0.5$, then used the tools to examine the fit and consider adding predictors. Section \ref{sec:real_data_part1} considers the spectral-transformed data and spectral-domain added variable plots for selecting predictors, Section \ref{sec:real_data_yaddvar} considers observation-domain added variable plots and Section \ref{sec:real_data_exact} shows fits and tests for added variables using the exact restricted likelihood and the original data.  Here we show just the first of a sequence of model-building steps;  Web Supplement Appendices H, I, and J give all of the steps.  We present this only to illustrate how the tools could be used;  we make no claim that these steps are optimal.

\subsection{Model-building in the spectral domain}\label{sec:real_data_part1}
The questions are:  does the intercept-only model suffice to explain variation in $y$ or should covariates be added and if so, which ones? A stationary GP model is flexible enough that the fit to $y$ is never bad but we presume that apparent deviations from stationarity are better modeled using covariates than with the GP.

Consider Figure \ref{fig:step1_vjsq}, the $v_j^2$ plot from the intercept-only fit; the plotting symbol is $j$. The point $j = 1$ corresponds to a cubic or linear north-south trend (the spectral basis has no linear-like component), so the huge $v_1^2$ indicates a north-south trend.   Adding a covariate could remove this trend.  Figure \ref{fig:step1_addvar2} shows spectral-domain added variable plots for the candidate covariates, each of which (after pre-smoothing to the grid) has been standardized by subtracting its average and dividing by its standard deviation, so the covariates' slopes are comparable.  Table \ref{table:table_step1} suggests that adding covariates will improve the fit.  Note that for Elevation and Slope, Figures~\ref{fig:step1_addvar2}b and c, the signal for adding the covariate is concentrated in few transformed observations with extreme values on the horizontal axis, including the lowest frequencies $j$ = 1 and 2, while the signal for the other covariates is more diffuse in $j$.

The natural impulse is to add the covariate with the largest slope, which is Elevation. However, because $v_1^2$, a north-south trend, is the most prominent $v_j^2$, we also want the added covariate to explain some of this trend, the more the better. To this end we calculated Cook's distance, describing the influence of each point on the regression slope; we want $j = 1$ to have a large Cook's distance. Among the candidate covariates, Elevation has the smallest p-value for its added variable plot's slope and $j = 1$ has the largest Cook's distance. Therefore we add Elevation to the model.  (Slope has results similar to Elevation so we could have added Slope instead. Had we done so, then applying the same considerations in the next step would lead to adding Elevation next.)

Web Supplement Appendices H, I, and J show further model-building steps.  Again, we make no claim this is an optimal sequence of steps but merely intend to illustrate how the tools can be used.

\subsection{Added variable plots in the observation domain}\label{sec:real_data_yaddvar}
Added variable plots in both domains estimate the same slope (apart from effects of the approximations used for the spectral-domain plot) but they emphasize different aspects of the outcome $y$ and the candidate predictors.  Table~\ref{table:table_step2_obs} shows slopes and p-values for observation-domain added variable plots, comparable to Table~\ref{table:table_step1}.  The most striking differences are for Elevation and Slope:  their strong low-frequency components are emphasized by the spectral-domain added variable plot but de-emphasized by the observation-domain plot.  Thus, although in the observation-domain plots these two predictors have the largest slopes in absolute value (Table~\ref{table:table_step2_obs}), they have much larger p-values than in the spectral-domain plots, reflecting lower power in the observation domain.  Among the other predictors, Spring TC3, Summer TC1, and Summer TC3 have somewhat larger slopes and somewhat smaller p-values in the observation domain, reflecting their relative strength in high frequencies.

\subsection{Fits using the raw data and exact restricted likelihood}\label{sec:real_data_exact}
Table~\ref{table:dumbfit_first} shows estimates of coefficients of the candidate covariates using the original data $y$ and maximizing the exact restricted likelihood to estimate $\sigma_\mathfrak{s}^2$, $\rho$, and $\sigma_e^2$.  These estimates are almost exactly equal to the slopes of the observation-domain added variable plots (Section~\ref{sec:real_data_yaddvar}), as might be expected given that the latter plot involves less approximation.  The p-values in Table~\ref{table:dumbfit_first} are from Wald tests that treat estimates of $\sigma_\mathfrak{s}^2$, $\rho$, and $\sigma_e^2$ as if they are known to be true, as is typical in non-Bayesian analyses, which ignores variability accounted for in the p-values for both added variable plots.

Table~\ref{table:dumbfit_first} also shows $\hat{\sigma}_\mathfrak{s}^2$, $\hat{\sigma}_e^2$, and $\hat{\rho}$ for the intercept-only model and for models adding each candidate covariate.  Elevation and Slope, which have strong low-frequency components and low p-values in the spectral-domain added variable plots, have the biggest impact on $\hat{\sigma}_\mathfrak{s}^2$ and $\hat{\rho}$.  Entering either moves substantial variation out of the GP part of the fit and into the fixed effects, leaving the GP part of the fit with less variation ($\sigma_\mathfrak{s}^2$) and spatial correlation range ($\rho$).  The other candidate covariates have much smaller effects on estimates of the GP parameters.  Elevation is the only candidate covariate with a noteworthy impact on $\hat{\sigma}_e^2$.  Web Supplement Appendices H, I, and J show analogous estimates for later steps in the model-building process.

\section{Discussion}\vspace*{-0.2cm}
This paper is a step toward tools for doing data-analysis with linear mixed models with a random effect distributed as a stationary Gaussian process, specifically tools for model-building and understanding model fits.  We used the spectral approximation for stationary isotropic GPs on regularly-spaced grids to give a linear mixed model in which the random effect has a design matrix with orthogonal columns not depending on any unknowns and a diagonal covariance matrix.  The resulting approximate restricted likelihood is formally identical to the likelihood from a GLM with gamma errors and identity link.  The transformed observations $v^2_j$ and the functions $a_j(\rho)$ --- the spectral density of the GP's covariance --- are the keys to understanding the restricted likelihood as a function of the unknowns.  The spectral approximation could also contribute to understanding spatial confounding;  this is secondary to the present purpose but is discussed in Web Supplement Appendix L.

The approximate restricted likelihood fits the $v_j^2$'s to $\sigma_\mathfrak{s}^2 a_j(\rho) + \sigma_e^2$, so a prominent $v_j^2$, especially for small $j$, suggests a missing covariate with substantial power at the corresponding frequency. If covariates are available, added variable plots in the observation and spectral domains can be used to examine the support in the data for adding each potential covariate.

The spectral representation requires data observed on an regular grid;  we suggested a way to apply these methods to data observed at irregularly spaced locations but other approaches are possible.  We view the present work as a beginning, not a definitive approach. Finally, we used a dataset to sketch how to use the model building tools developed here. A future publication will include a more fully-worked example.

\vspace*{-0.6cm}
\section{References}\vspace*{-0.2cm}


\section{Figures and Tables}

\vspace*{-0.5cm}

\begin{table}[H]
\small
\caption{Average over 100 simulated datasets of estimates maximizing the exact and approximate restricted likelihoods; contamination by an outlier. Standard errors are in parentheses.} \label{table2}
\centering
\begin{tabular}{c c}

\begin{tabular}{ c c c c } \hline \hline
 & \multicolumn{3}{c}{exact RL} \\
            \cline{2-4}
&$\sigma_\mathfrak{s}^2$ &  $\sigma_e^2$& $\rho$ \\ \hline\hline
actual values &2 &5&5\\
uncontaminated & 2.29 (0.11) &4.75 (0.11)&6.89 (0.82)\\
contaminated & 2.44 (0.10)& 5.99 (0.11)&7.25 (0.60)\\
\hline
actual values & 2&5 &16.67\\
uncontaminated  & 2.16 (0.09) & 4.89 (0.07)& 26.12 (4.32)\\
 contaminated & 2.23 (0.11)& 6.14 (0.09)&20.11 (1.84)\\
\hline
actual values &2 & 0.1&5\\
uncontaminated & 1.94 (0.03) &0.099 (0.01)&4.97 (0.13) \\
contaminated  &1.98 (0.05) &1.36 (0.03) & 5.46 (0.24)\\
\hline
actual values& 2& 0.1&16.67\\
uncontaminated & 2.09 (0.07) &0.093 (0.00)& 18.07 (0.84)\\
contaminated & 2.08 (0.06)&1.33 (0.02) & 17.69 (1.04)\\
\hline
actual values& 10& 5&5\\
uncontaminated & 10.17 (0.20)& 4.99 (0.13)& 5.58 (0.18)\\
contaminated & 10.02 (0.22)& 6.01 (0.15) & 5.59 (0.24)\\
\hline
actual values& 10& 5&16.67\\
uncontaminated &  10.44 (0.37)& 4.80 (0.06)& 17.57 (0.83)\\
contaminated &9.65 (0.29) &  6.24 (0.09)&17.99 (1.01) \\
\hline
actual values& 10& 0.1&5\\
uncontaminated & 9.70 (0.17) &0.21 (0.03) & 5.26 (0.13)\\
contaminated &10.17 (0.19) & 1.36 (0.08) &5.19 (0.13) \\
\hline
actual values& 10& 0.1&16.67\\
uncontaminated &  9.99 (0.34)& 0.11 (0.01)&17.29 (0.67) \\
contaminated & 10.52 (0.31)&1.39 (0.05)  & 18.14 (0.76)\\
\hline

\end{tabular}
&

\begin{tabular}{ c c c} \hline \hline
  \multicolumn{3}{c}{approximate RL} \\
            \cline{1-3}
$\sigma_\mathfrak{s}^2$ &  $\sigma_e^2$& $\rho$ \\ \hline\hline
2 &5&5\\
 2.39 (0.16)&4.90 (0.09)&13.51 (1.74)\\
 2.48 (0.18)  &6.04 (0.11) &12.76 (1.21)\\
\hline
2&5 &16.67\\
 2.17 (0.09)&4.91 (0.06)& 38.11 (2.53)\\
  2.39 (0.11) & 6.07 (0.08)&33.99 (2.74)\\
\hline
2 & 0.1&5\\
 1.97 (0.04)&0.232 (0.01)& 10.24 (0.25)\\
2.03 (0.05)& 1.45 (0.03)& 10.94 (0.94)\\
\hline
 2& 0.1&16.67\\
  2.01 (0.08)&0.144 (0.00)&34.84 (1.67) \\
2.03 (0.08)& 1.39 (0.02)& 34.57 (1.76)\\
\hline
10& 5&5\\
  9.98 (0.21)&5.63 (0.10)& 10.87 (0.41)\\
 10.12 (0.22) &6.70 (0.13) &10.24 (0.40) \\
\hline
 10& 5&16.67\\
 10.54 (0.43) &5.08 (0.06) &35.60 (2.05) \\
10.27 (0.39)&6.59 (0.08) & 37.32 (2.64)\\
\hline
 10& 0.1&5\\
 9.65 (0.17)& 0.76 (0.03)& 10.34 (0.28)\\
 9.55 (0.19)& 2.09 (0.07)&10.12 (0.32) \\
\hline
 10& 0.1&16.67\\
 10.02 (0.37) & 0.32 (0.01)&32.70 (1.37) \\
 10.30 (0.35)&1.59 (0.05) & 34.29 (1.29)\\
\hline
\end{tabular}
\end{tabular}
\end{table}

\vspace*{-0.2cm}
\begin{figure}[H]\label{fig:cov}
         \centering\vspace*{-1.5cm}
                          \includegraphics[width=0.5\textwidth]{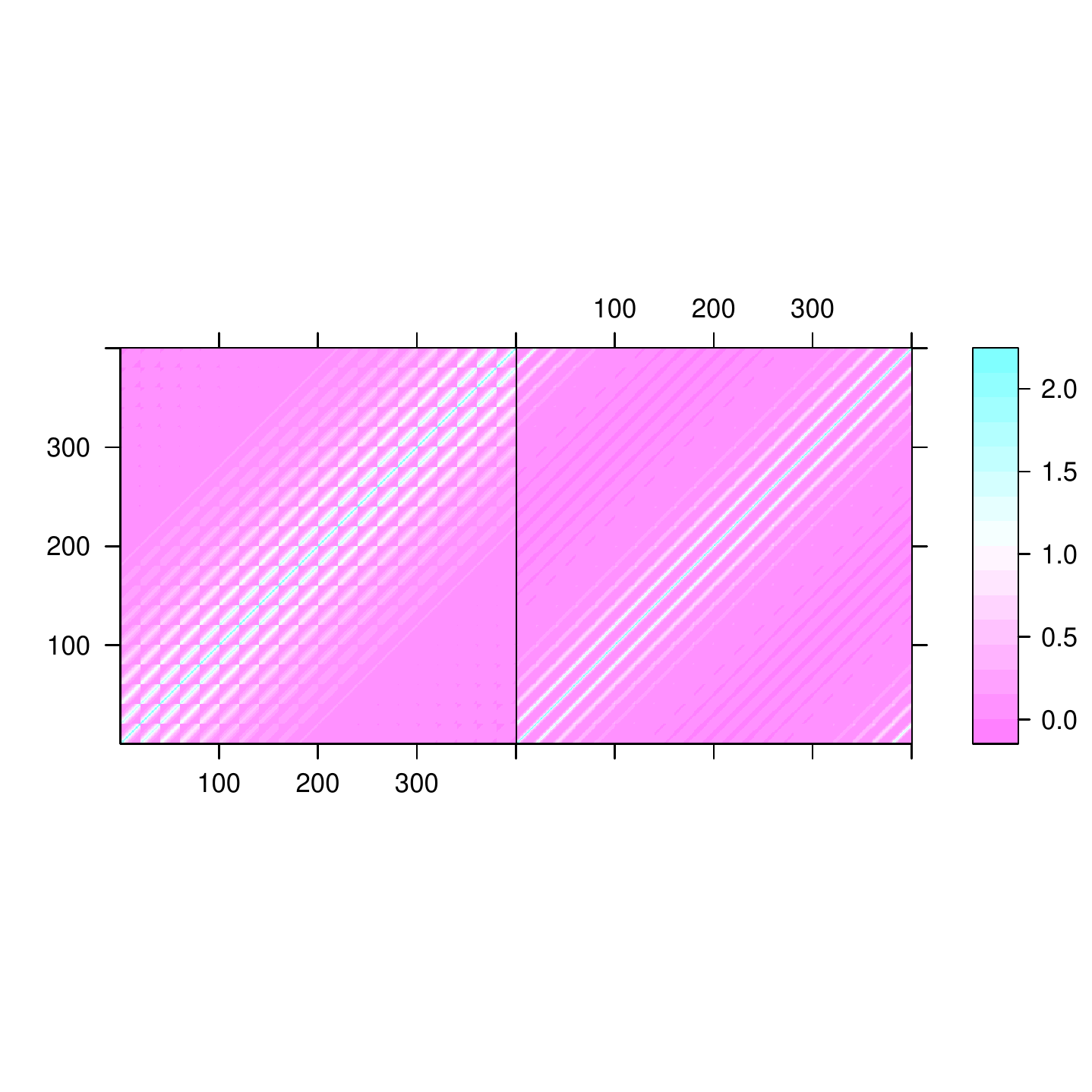}\vspace*{-1.5cm} \vspace*{-0.5cm}
                 \caption{Left: exact 2-D exponential covariance (including the iid errors). Right: approximate covariance. $\sigma_\mathfrak{s}^2=2, \sigma_e^2=0.1, \rho=5$.  Observation domain [1,2,..,20]$\times$[1,2,..,20].}
                 \label{fig:cov}
         \end{figure}

\begin{figure}[H]
         \centering \begin{subfigure}[b]{0.22\textwidth}
 \includegraphics[width=\textwidth]{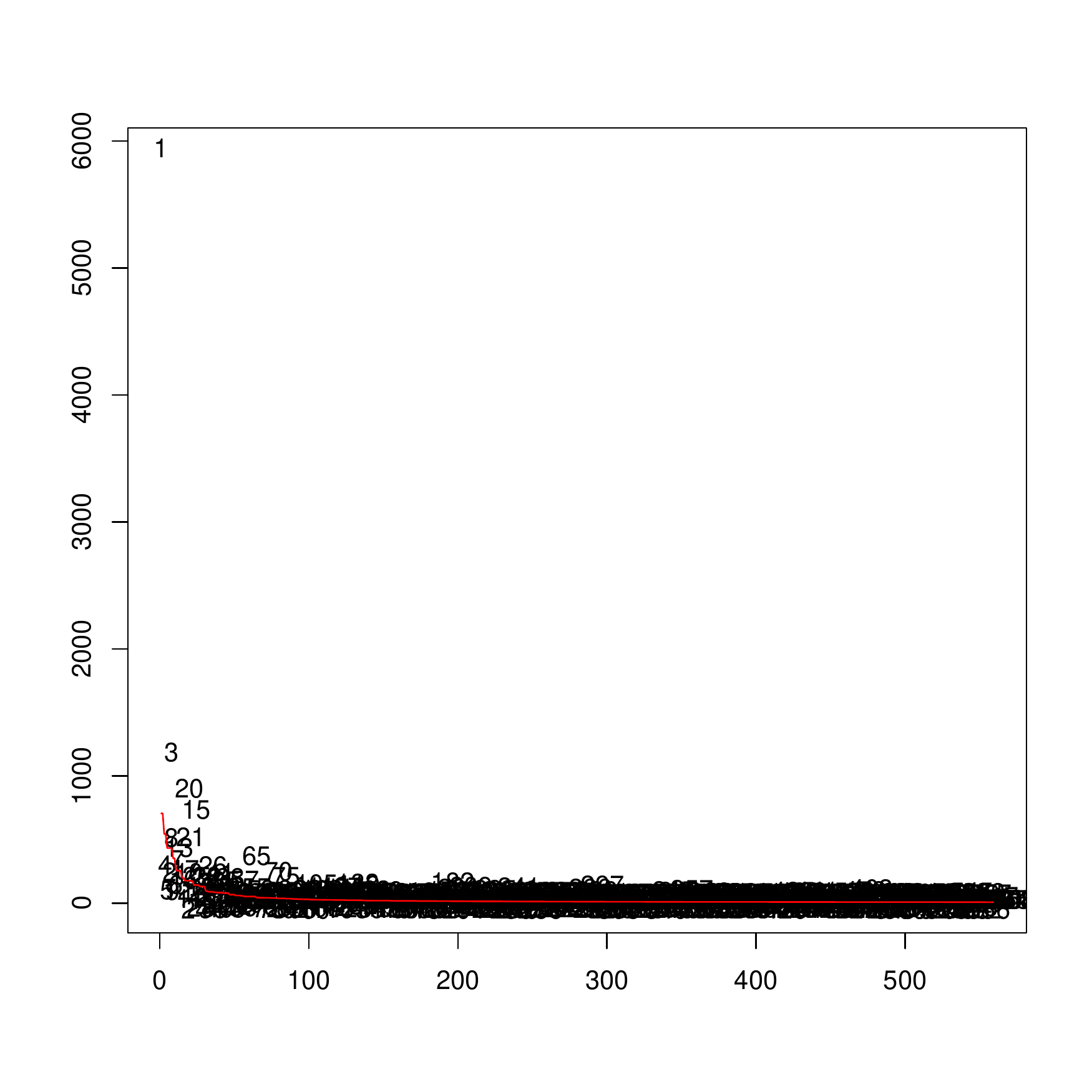}
                 \caption{$v_j^2$ vs $j$.}
                                  \label{fig:step1_vjsq}
\end{subfigure}
\quad
 \begin{subfigure}[b]{0.22\textwidth}
 \includegraphics[width=\textwidth]{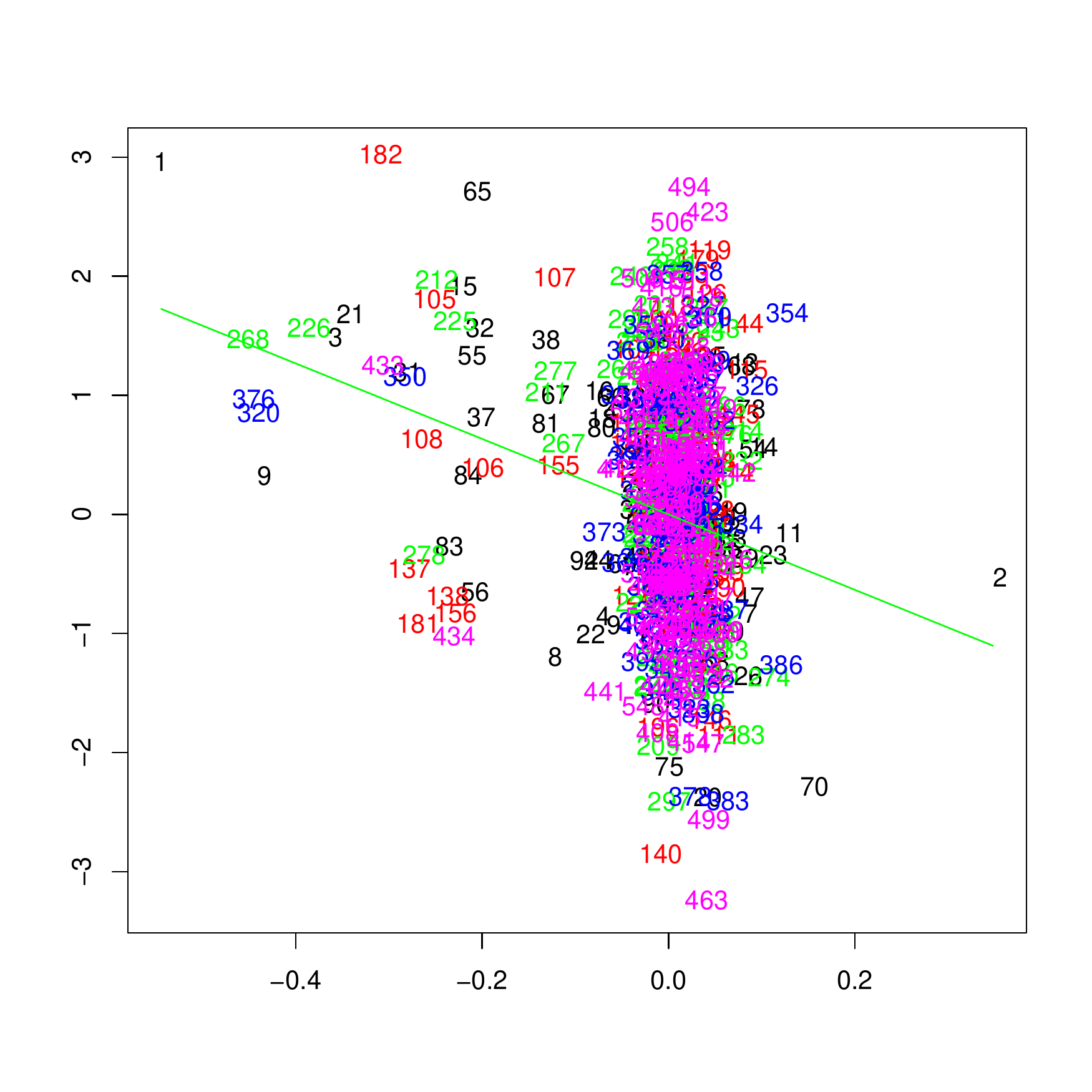}
                 \caption{Elevation.}
                                  \label{fig:RM_add_var1}
\end{subfigure}
         \quad
\begin{subfigure}[b]{0.22\textwidth}
                                                  \includegraphics[width=\textwidth]{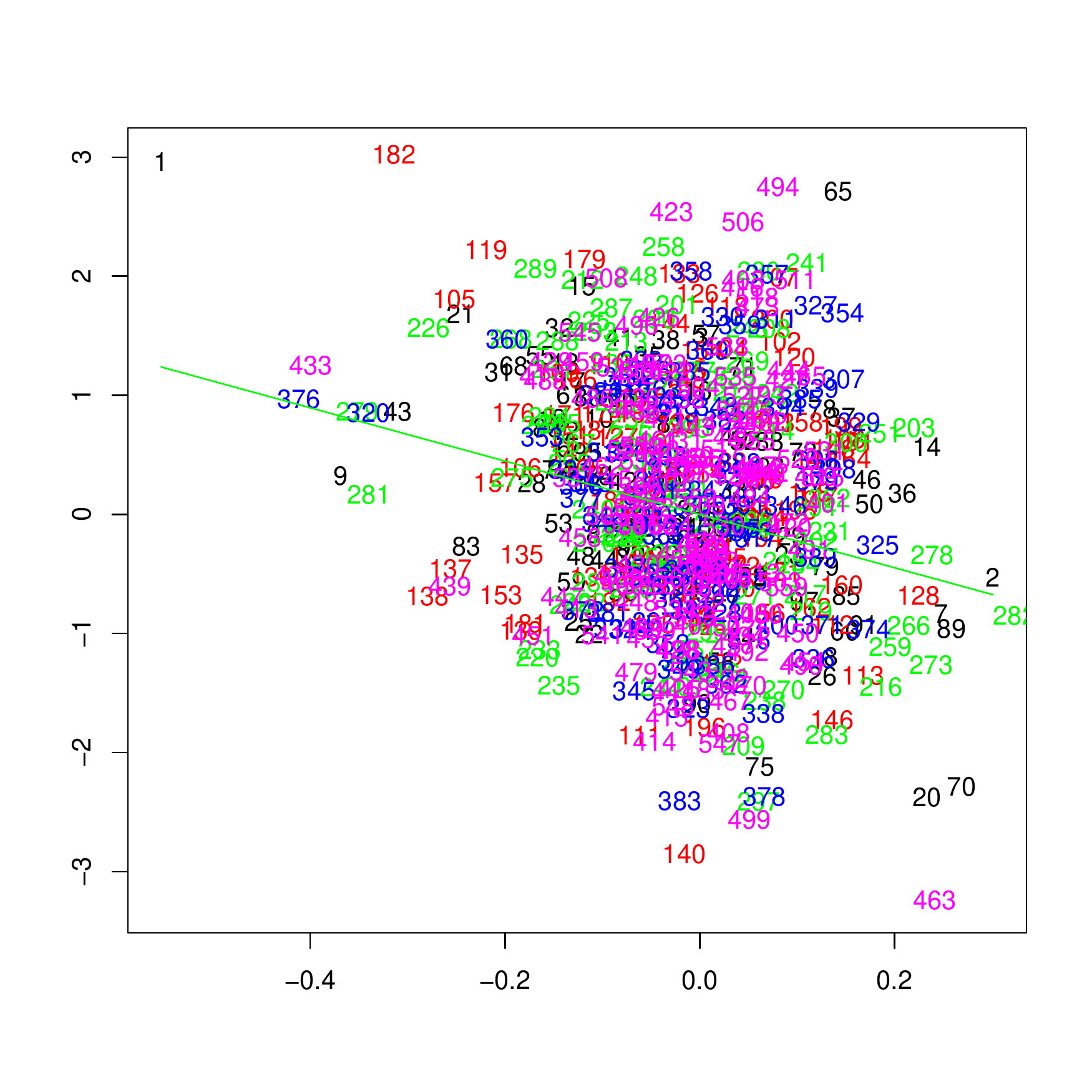}
                 \caption{Slope. }
                 \label{fig:RM_add_var2}
\end{subfigure}
         \quad
\begin{subfigure}[b]{0.22\textwidth}
                                                  \includegraphics[width=\textwidth]{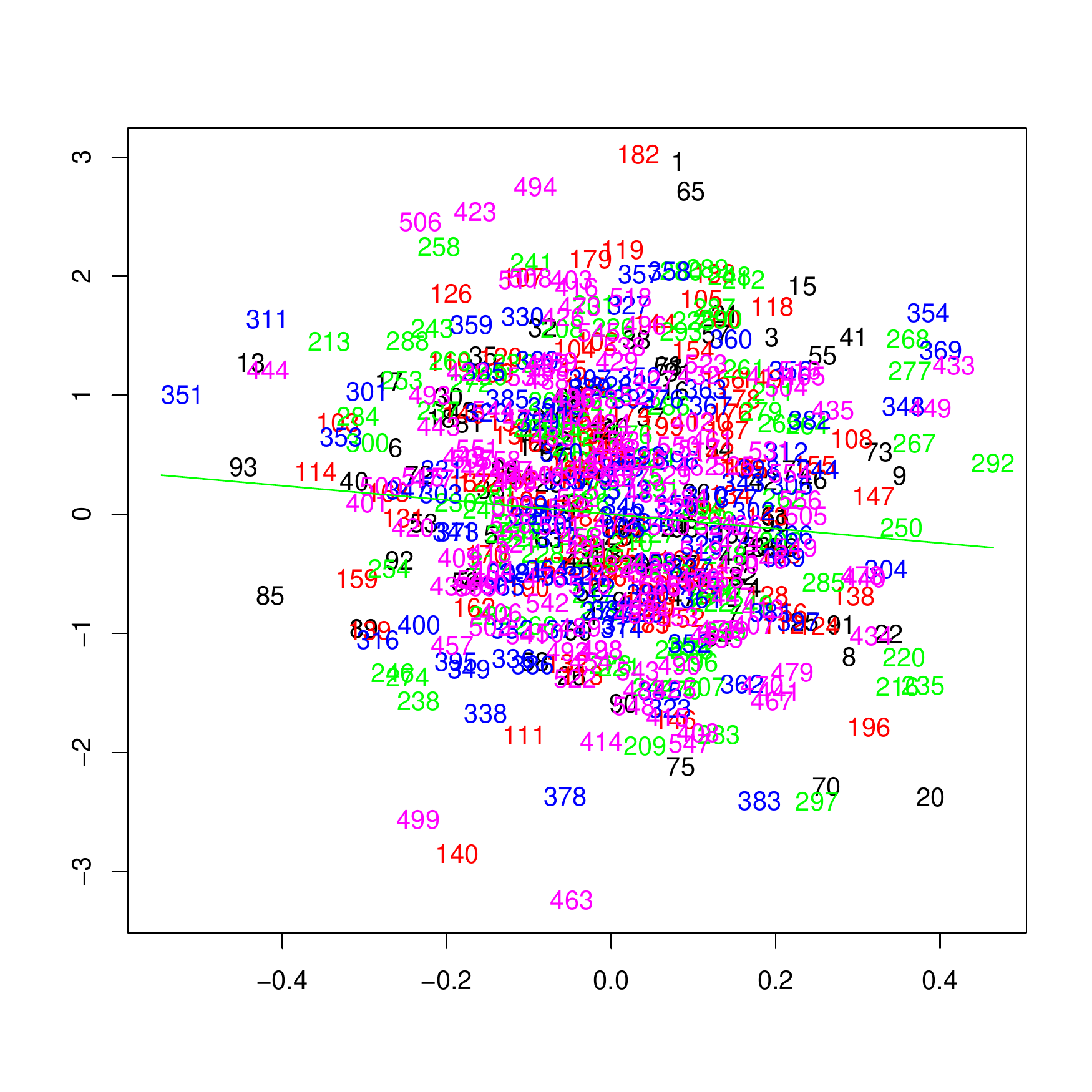}
                 \caption{SpringTC2.}
                 \label{fig:RM_add_var3}
\end{subfigure}
\quad

\begin{subfigure}[b]{0.22\textwidth}
                                                  \includegraphics[width=\textwidth]{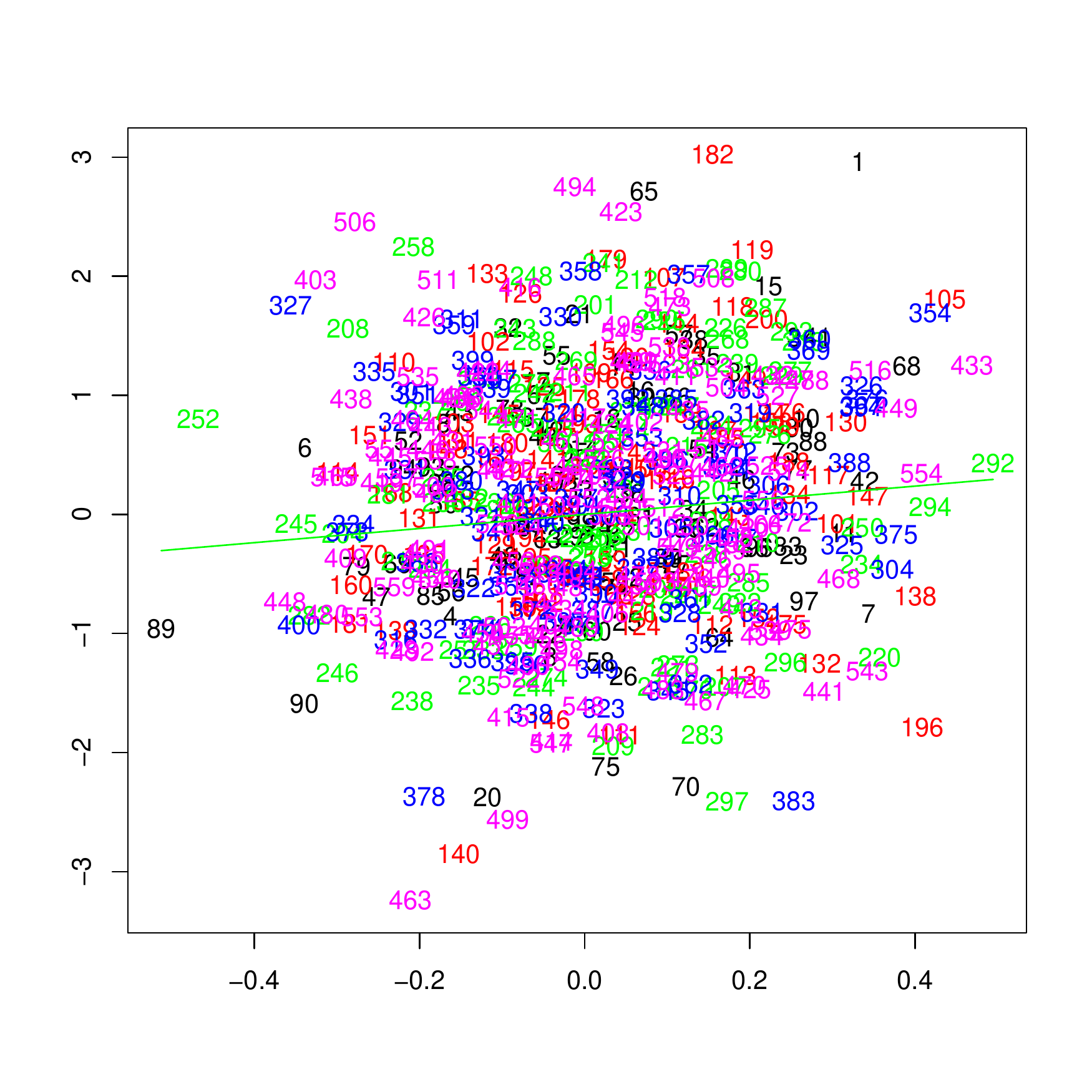}
                 \caption{SpringTC3.}
                 \label{fig:RM_add_var4}
\end{subfigure}
\quad
         \begin{subfigure}[b]{0.22\textwidth}
 \includegraphics[width=\textwidth]{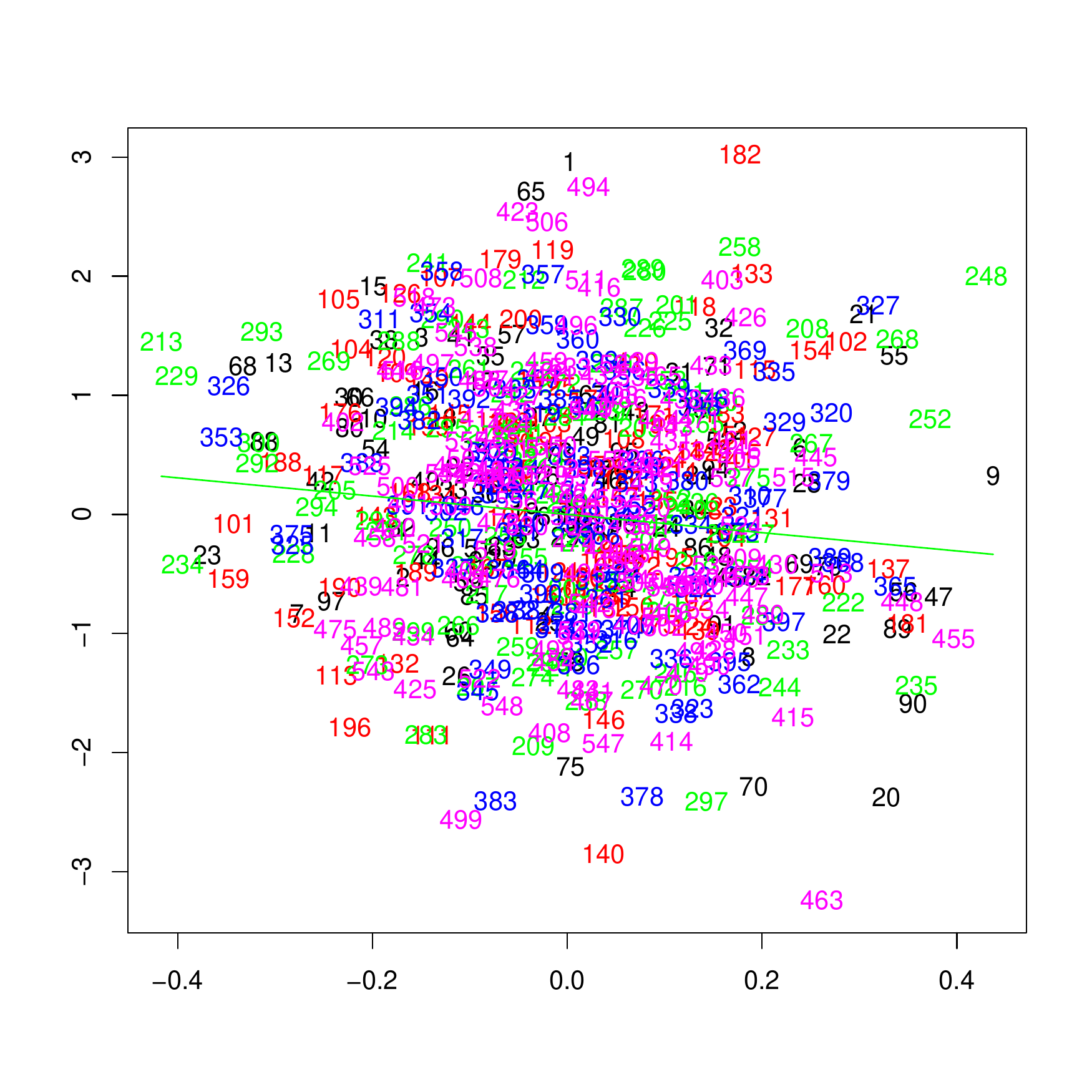}
                 \caption{SummerTC1.}
                                  \label{fig:RM_add_var1}
\end{subfigure}
         \quad
\begin{subfigure}[b]{0.22\textwidth}
                                                  \includegraphics[width=\textwidth]{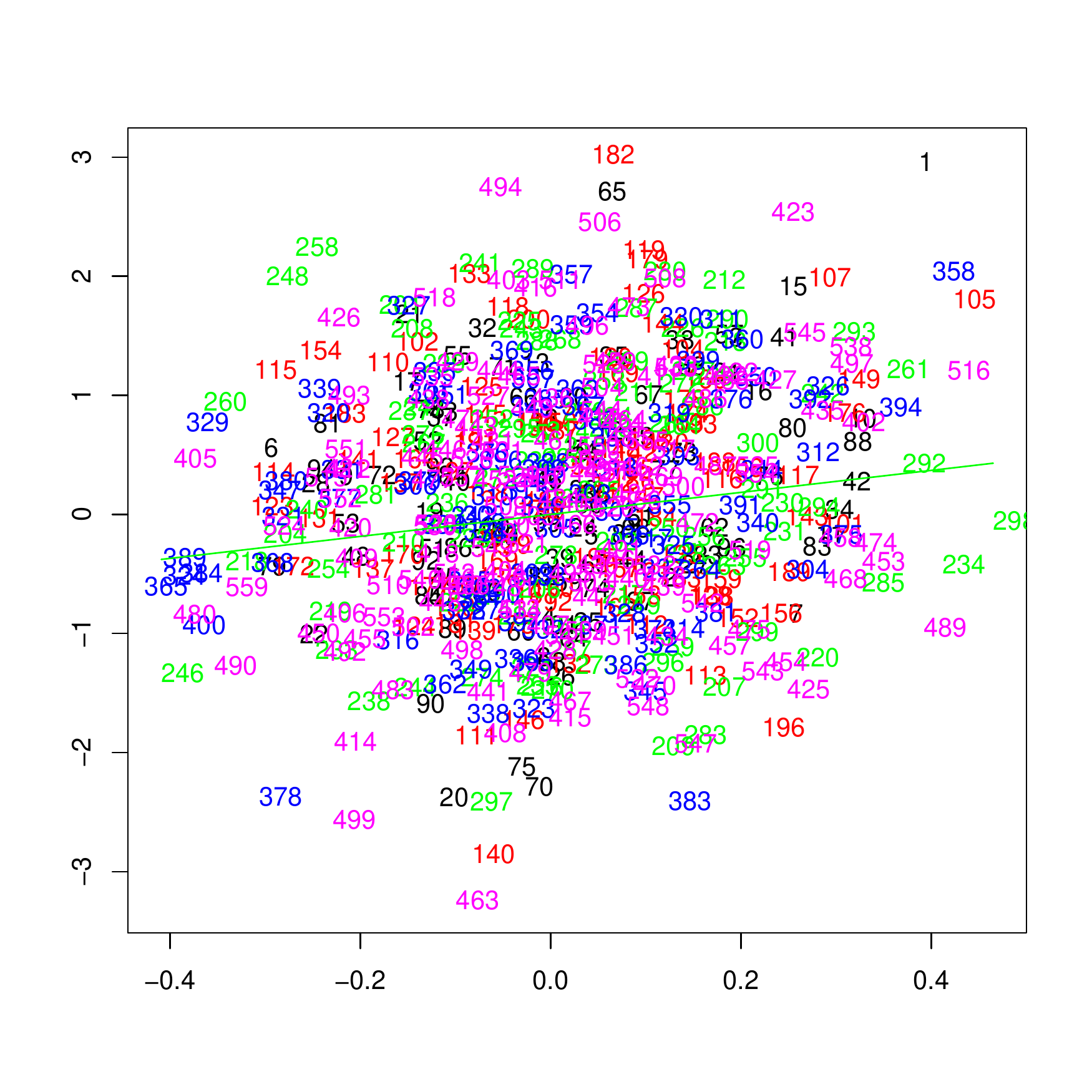}
                 \caption{SummerTC3. }
                 \label{fig:RM_add_var2}
\end{subfigure}
         \quad
\begin{subfigure}[b]{0.22\textwidth}
                                                  \includegraphics[width=\textwidth]{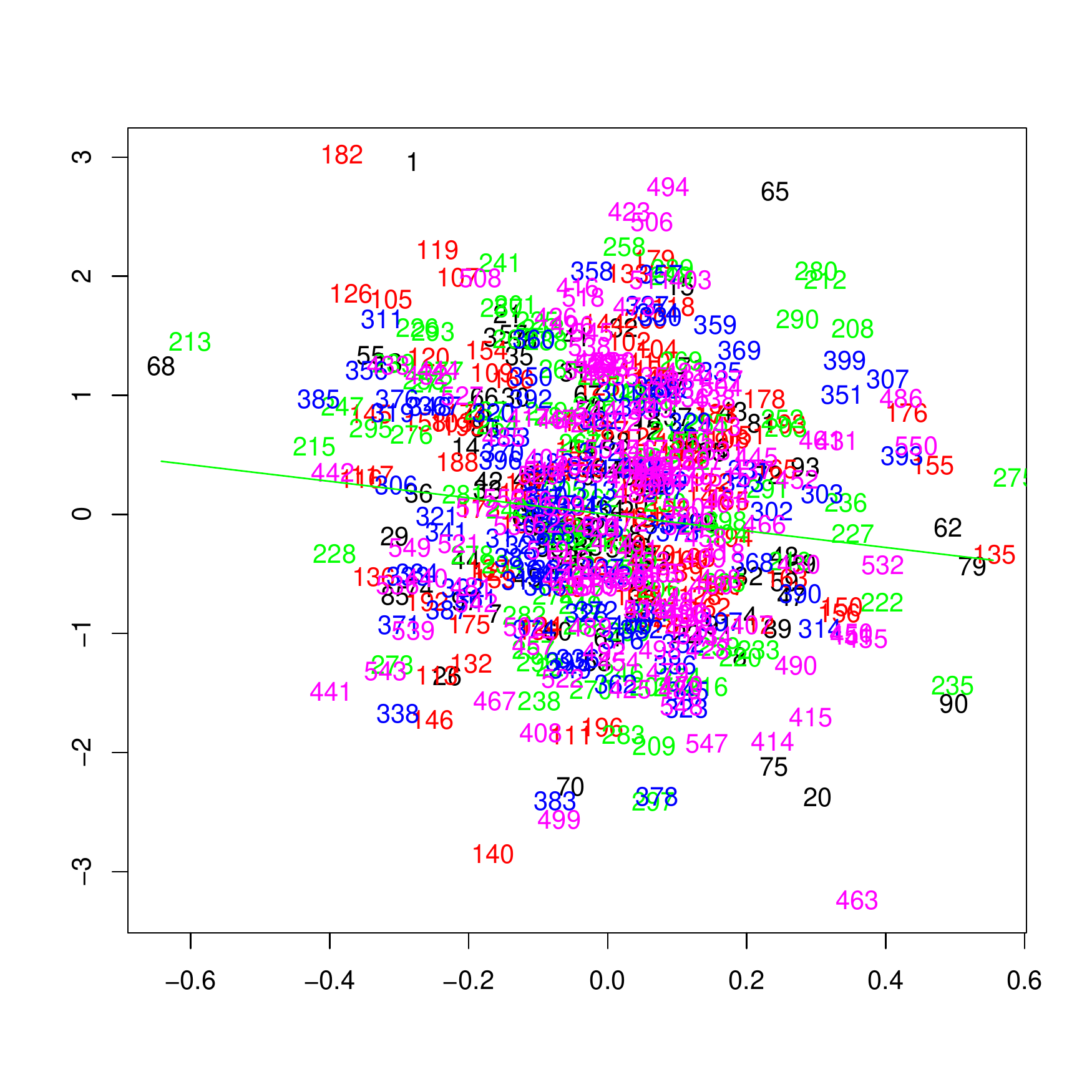}
                 \caption{FallTC2.}
                 \label{fig:RM_add_var3}
\end{subfigure}
 \caption{After intercept-only fit; (b) to (h) are spectral-domain added variable plots.}\label{fig:step1_addvar2}
         \end{figure}

              \begin{table}[H]
\centering
  \begin{tabular}{ |c || c |c|c|}
    \hline
    Candidate covariate &  Slope & p-value&$j$ with top 5 Cook's dist\\ \hline
    Elevation& -3.17  &$10^{-10}$&1,182,9,181,434\\ \hline
    Slope & -2.24 &$10^{-9}$&1,182,463,70,65\\ \hline
    SpringTC2&-0.60 &0.03&20,499,354,268,369\\ \hline
    SpringTC3 & 0.59&0.02&1,506,196,403,463\\ \hline
    SummerTC1 &-0.77&0.007&248,463,20,268,327\\ \hline
    SummerTC3 & 0.92&0.0004&1,258,378,358,248\\ \hline
    FallTC2 &-0.69&0.004&182,463,1,212,280\\ \hline  \hline
    \end{tabular}
  \caption{Slopes of spectral-domain added variable plots, after the intercept-only fit. } \label{table:table_step1}
\end{table}

\begin{table}[H]
 \small
\centering
  \begin{tabular}{ |c || c |c||}
    \hline
    Candidate covariate &Slope&p-value \\ \hline
    Elevation&-2.02 &0.07\\ \hline
    Slope  &-1.45 & 0.004 \\ \hline
    SpringTC2 &-0.28 &0.42\\ \hline
    SpringTC3 &0.96 &0.002\\ \hline
    SummerTC1 &-0.98&0.002\\ \hline
    SummerTC3  & 1.25&$10^{-5}$\\ \hline
    FallTC2 &-0.83&0.004\\ \hline  \hline
    \end{tabular}
  \caption{Slopes of observation-domain added variable plots, after the intercept-only fit.}\label{table:table_step2_obs}
\end{table}

\begin{table}[H]
\small
\centering
  \begin{tabular}{ |c ||c|c||c|c|c|}
    \hline
    & \multicolumn{2}{c||}{Coefficient}& \multicolumn{3}{c|}{Variance parameters}\\ \hline
    Candidate covariate &  slope &p-value&  \textbf{$\hat{\sigma}_\mathfrak{s}^2$}& \textbf{$\hat{\sigma}_e^2$}& \textbf{$\hat{\rho}$} \\ \hline
    Intercept-only & --   & --         &29.62 &16.20  &5.96  \\ \hline
    Elevation      &-2.52 &$<10^{-12}$ &21.96 &13.82  &2.85  \\ \hline
    Slope          &-1.63 &$10^{-6}$   &20.31 &16.11  &3.97  \\ \hline
    SpringTC2      &-0.28 &0.16        &29.69 &16.35  &6.13  \\ \hline
    SpringTC3      &0.99  &$10^{-5}$   &26.80 &17.15  &6.93  \\ \hline
    SummerTC1      &-1.03 &$10^{-5}$   &30.98 &17.54  &8.88  \\ \hline
    SummerTC3      &1.25  &$10^{-7}$   &26.91 &16.19  &6.14  \\ \hline
    FallTC2        &-0.87 &0.0001      &26.50 &17.33  &6.98  \\ \hline  \hline
    \end{tabular}
  \caption{Exact restricted likelihood maximized for the raw data:  Estimated coefficient and Wald-test p-value for each candidate covariate, added individually (one at a time) to the model, and resulting estimates of $\sigma_\mathfrak{s}^2$, $\sigma_e^2$, and $\rho$. }\label{table:dumbfit_first}

\end{table}

\newpage
\begin{center}
{\Large \textbf{Web-based Supplementary Materials for \\Understanding Gaussian Process Fits Using an Approximate Form of the Restricted Likelihood}}

\bigskip

\textbf{Maitreyee Bose, James S. Hodges, and Sudipto Banerjee}

\end{center}

\section{{Web Appendix A. Spectral approximation in one dimension: technical details}}
\subsection{Distribution of $u_m$ from $a_m$ and $b_m$}
Given the assumed normal prior distributions for the ${a_m}$'s and ${b_m}$'s, it can be shown that the $u_m$'s have complex normal distributions as follows.

Assume\vspace*{-1.5cm}
$$a_m\sim N(0,\frac{1}{2M}\sigma_\mathfrak{s}^2\phi(\omega_m;\rho)),\mbox{ }b_m\sim N(0,\frac{1}{2M}\sigma_\mathfrak{s}^2\phi(\omega_m;\rho))\hspace*{2mm} \forall m\neq0 \hspace*{1mm}\mbox{or} \hspace*{1mm}\frac{M}{2}$$
\vspace*{-1.2cm}$$a_0\sim N(0,\frac{1}{M}\sigma_\mathfrak{s}^2\phi(\omega_0;\rho)),\mbox{ } a_{\frac{M}{2}}\sim N(0,\frac{1}{M}\sigma_\mathfrak{s}^2\phi(\omega_{\frac{M}{2}};\rho)), \mbox{ }b_0=b_{\frac{M}{2}}=0.\vspace*{-0.2cm}$$

Therefore, $u_m=a_m+ib_m$ has a complex normal distribution with\vspace*{-0.2cm}
\begin{enumerate}
\item $E(u_m)=E(a_m)+iE(b_m)=0;$\vspace*{-0.3cm}
\item covariance matrix: $V(u_m)=V_{aa}+V_{bb}+i(V_{ab}-V_{ba})=\frac{1}{M}\sigma_\mathfrak{s}^2\phi(\omega_m;\rho) \hspace*{2mm}\forall m;$\vspace*{-0.3cm}
\item relation matrix: $C(u_m)=V_{aa}-V_{bb}+i(V_{ba}+V_{ab})=0.$\vspace*{-0.3cm}
\end{enumerate}

So, $E(u_m)=0$, $E(u_mu_n)=0$ for $m\neq n$ and $E{\mid u_m \mid}^2=\frac{1}{M}\sigma_\mathfrak{s}^2\phi(\omega_m;\rho).$
\\
\\
Note that in all the following proofs, the specific choices of the values of the $s_j$'s and the $\omega_m$'s have a crucial role.

\subsection{Proof that the spectral approximation is valid }
We prove the approximate equality $\mbox{Cov}(g(s_j),g(s_j+d))\simeq
\sigma_\mathfrak{s}^2 K(d;\rho)$.

\textbf{Definition of spectral density }(Priestley 1981, p.~199, 211). Let $R(d)$ be the covariance function of a continuous parameter stationary process.
If $R(d)$ is absolutely integrable and isotropic, then its spectral density is defined as\vspace*{-0.5cm}
$$h(\omega)=\int_{-\infty}^\infty \exp(-2\pi i\omega d)R(d)\mbox{ d}d.\vspace*{-0.5cm}$$
When the process is observed only at a discrete set of integer locations, then $R(d)$ is defined only for integer values
of $d$ and the above integral has to be replaced by a discrete sum (Priestley 1981, p.~222):\vspace*{-0.5cm}
$$\hspace*{0.4cm} h(\omega)=\sum\limits_{d=-\infty}^ \infty \exp(-2\pi i\omega d)R(d),\hspace*{0.8cm} -\frac{1}{2}\leq\omega\leq\frac{1}{2}.\vspace*{-0.2cm}$$

Now, $\exp(-2\pi i\omega d)$ is a periodic function of $\omega$ with period $1$, so the components in the observed discrete parameter process with
frequencies $\omega-1$, $\omega+1$, $\omega-2$, $\omega+2$,.... will all appear to have frequency $\omega$. The frequency $\omega$ is then said to be the
alias of the frequencies $\omega\pm1$, $\omega\pm2$,..... Since every frequency outside $[-\frac{1}{2}, \frac{1}{2}]$ has an alias inside this range,
the spectral density is defined only
for $\omega$ in the range $[-\frac{1}{2}, \frac{1}{2}]$.

Then \vspace*{-0.2cm}$$\int_{-\frac{1}{2}}^\frac{1}{2} \exp(2\pi i\omega d)h(\omega)\mbox{ d}\omega=R(d).\vspace*{-0.2cm}$$
Proof:\vspace*{-0.4cm}$$\int_{-\frac{1}{2}}^\frac{1}{2} \exp(2\pi i\omega d)h(\omega)\mbox{ d}\omega
=\int_{-\frac{1}{2}}^\frac{1}{2} \sum\limits_{\varsigma=-\infty}^ \infty \exp(-2\pi i\omega (\varsigma-d))R(\varsigma)\mbox{ d}\omega\vspace*{-0.2cm}$$
using the definition of spectral density
$$=\sum\limits_{\varsigma=-\infty}^ \infty R(\varsigma)\int_{-\frac{1}{2}}^\frac{1}{2} \exp(-2\pi i\omega (\varsigma-d))\mbox{ d}\omega$$
$$=R(d)+\sum\limits_{\varsigma\neq d}^{} \frac{R(\varsigma)}{2\pi i(\varsigma-d)}
(\exp(\pi i(\varsigma-d))-\exp(-\pi i(\varsigma-d)))$$
$$=R(d)+\sum\limits_{\varsigma\neq d}^{} \frac{R(\varsigma)\sin(\pi (\varsigma-d))}{\pi(\varsigma-d)}=R(d)$$
because $\sin$ of any integer multiple of $\pi$ is 0.

Thus we have \vspace*{-0.5cm}$$\sigma_\mathfrak{s}^2 K(d;\rho)=\int_{-\frac{1}{2}}^\frac{1}{2} \exp(2\pi i\omega d)\sigma_\mathfrak{s}^2 \phi(\omega;\rho)\mbox{ d}\omega,
$$ where $d$ is the distance between two locations belonging to $S$ and our proof of $\mbox{Cov}(g(s_j),g(s_j+d))\simeq \sigma_\mathfrak{s}^2
K(d;\rho)$  is complete if we can show \vspace*{-0.5cm}
$$\mbox{Cov}(g(s_j),g(s_j+d))\simeq \int_{-\frac{1}{2}}^\frac{1}{2} \exp(2\pi i\omega d)\sigma_\mathfrak{s}^2 \phi(\omega;\rho)\mbox{ d}\omega.\vspace*{-0.2cm}$$

\textbf{Spectral representation theorem }(Gelfand et al.~2010, p.~60). Suppose $Y_0,Y_1,...,Y_{M-1}$ are mean zero complex random variables with
$E(Y_lY_m)=0$ for $l\neq m=0,1,...,M$ and $E{\mid Y_m\mid}^2=f_m$ for each $m$, and suppose
$\omega_0,\omega_1,...,\omega_{M-1}\in \Re, -\frac{1}{2}\leq\omega_m\leq\frac{1}{2}\hspace*{1mm}  \forall\hspace*{1mm}  m.$
\\Consider \vspace*{-0.4cm}$$Z(s)=\sum\limits_{m=0}^{M-1}{\exp(i 2 \pi \omega_ms)Y_m} \hspace*{1cm}s\in S \in \aleph.$$
Then $Z(s)$ are realizations of a weakly stationary process in $\Re$ with covariance function \vspace*{-0.4cm}$$C(d)=\sum\limits_{m=0}^{M-1}{\exp(i 2 \pi
\omega_m d)f_m} \vspace*{-0.2cm}$$ where $d$ is the distance between two locations belonging to the set of locations
$S$.

Thus, for our approximation, by the spectral representation theorem, $\{g(s_j);s_j\in S\}$ is a weakly stationary process in $\Re$ with
covariance function $\mbox{Cov}(g(s_j),g(s_j+d))$ given by\vspace*{-0.5cm}
          $$C(d)=\sum\limits_{m=0}^{M-1}{\exp( i 2 \pi \omega_m d)E{\mid u_m \mid}^2}\vspace*{-0.2cm}$$                                                                                                $$=\frac{1}{M}\sum\limits_{m=0}^{M-1}{\exp(i2 \pi\omega_m d)\sigma_\mathfrak{s}^2 \phi(\omega_m;\rho)}$$
$$\hspace*{1.4cm}\simeq \int_{-\frac{1}{2}}^\frac{1}{2} \exp(i2\pi\omega d)\sigma_\mathfrak{s}^2 \phi(\omega;\rho)\mbox{ d}\omega\hspace*{2mm}
\mbox{as}\hspace*{2mm} M\rightarrow\infty,$$ where the last step follows from the Riemann sum formula for the definite integral.

\subsection{Orthogonality of columns of Z}For $z_j$ the $j^{th}$ column of $Z$, we need three results:
\vspace*{-0.2cm}
\begin{enumerate}
\item $z_j'z_j=2M,\hspace*{2mm}\mbox{for}\hspace*{2mm} j\in \{1,2,...,M-2\},$  \hspace*{5mm}2. $z_j'z_j=M,\hspace*{2mm}\mbox{for}\hspace*{2mm} j={M-1}$ \vspace*{-0.2cm}
\item $z_j'z_l=0,\hspace*{2mm}\mbox{for}\hspace*{2mm} l\neq j\in\{1,2,...,M-1\}.$\vspace*{-0.2cm}
\end{enumerate}
We prove 1 below; 2 and 3 can be proved similarly. \\
Proof:
$\mbox{For}\hspace*{2mm} j\in\{1,3,...,M-3\}$\vspace*{-0.5cm}
\begin{equation}\label{eqn1}
z_j'z_j
=2\sum\limits_{k=1}^{M}(\cos 0+\cos(\omega_\frac{j+1}{2}\frac{2\pi}{M}2k))\vspace*{-0.5cm}
\end{equation}
because $\cos(mx)\cos(nx)=\frac{1}{2}(\cos(m-n)x+\cos(m+n)x).$ Then equation (\ref{eqn1})\vspace*{-0.5cm}
\begin{eqnarray}&&=2M+2\sum\limits_{k=1}^{M}\cos(\omega_\frac{j+1}{2}\frac{4\pi k}{M})
=2M+\sum\limits_{k=1}^{M}(e^{ck}+e^{-ck}),\hspace*{1mm}\mbox{for}\hspace*{1mm}c=i\hspace{1mm}\omega_{\frac{j+1}{2}}\frac{4\pi}{M},
\nonumber\\ &&=2M+\frac{e^{c}(e^{Mc}-1)}{e^{c}-1}+\frac{e^{-Mc}(e^{Mc}-1)}{e^{c}-1}
=2M+\frac{(e^{Mc}-1)(e^{c}+e^{-Mc})}{e^{c}-1}
=2M \nonumber\end{eqnarray} because $e^{Mc}=e^{i\omega_{\frac{j+1}{2}}4\pi}=\cos(\omega_\frac{j+1}{2}4\pi)+i\hspace{0.5mm}
\sin(\omega_\frac{j+1}{2}4\pi)=1+0=1.$

$\mbox{For}\hspace*{2mm} j\in\{2,4,...,M-2\}$\vspace*{-1cm}
\begin{equation}\label{eqn2}z_j'z_j
=2\sum\limits_{k=1}^{M}(\cos 0-\cos(\omega_\frac{j}{2}\frac{2\pi}{M}2k))\vspace*{-1cm}
\end{equation}
\\ because $\sin(mx)\sin(nx)=\frac{1}{2}(\cos(m-n)x-\cos(m+n)x)$. Then equation (\ref{eqn2})
\begin{eqnarray}
&&
=2M-2\sum\limits_{k=1}^{M}\cos(\omega_\frac{j}{2}\frac{4\pi k}{M})
=2M-\frac{(e^{Mc}-1)(e^{c}+e^{-Mc})}{e^{c}-1}, \hspace*{1mm}c=i\hspace{1mm}\omega_{\frac{j}{2}}\frac{4\pi}{M},
=2M\nonumber
\end{eqnarray}because $e^{Mc}=e^{i\omega_{\frac{j}{2}}4\pi}=\cos(\omega_\frac{j}{2}4\pi)+i\hspace{0.5mm} \sin(\omega_\frac{j}{2}4\pi)=1+0=1.$

\subsection*{Orthogonality of X=1 and Z}
We show below that the odd-numbered columns of $Z$ sum to $0$. It can be shown similarly that the even-numbered columns also sum to 0.
\\Proof: 
$\mbox{For}\hspace*{2mm} j\in\{1,3,...,M-3\},$ the
sum of $j^{th}$ column is
\begin{eqnarray}
&&2\sum\limits_{k=1}^{M}\cos(\omega_\frac{j+1}{2}\frac{2\pi}{M}k)
=\sum\limits_{k=1}^{M}(e^{ck}+e^{-ck}),\hspace*{1mm}\mbox{for}\hspace*{1mm}c=i\hspace{1mm}\omega_{\frac{j+1}{2}}\frac{2\pi}{M},
\nonumber\\ &&=\frac{e^{c}(e^{Mc}-1)}{e^{c}-1}+\frac{e^{-Mc}(e^{Mc}-1)}{e^{c}-1}
=\frac{(e^{Mc}-1)(e^{c}+e^{-Mc})}{e^{c}-1}
=0 \nonumber\end{eqnarray} because $e^{Mc}=e^{i\omega_{\frac{j+1}{2}}2\pi}=\cos(\omega_\frac{j+1}{2}2\pi)+i\hspace{0.5mm} \sin(\omega_\frac{j+1}{2}2\pi)=1+0=1.$

$\mbox{For}\hspace*{2mm} j=M-1,$ the
sum of $j^{th}$ column is \vspace*{-0.3cm}
\begin{eqnarray}
&&=2\sum\limits_{k=1}^{M}\cos(\omega_\frac{M}{2}\frac{2\pi}{M}k)
=2\sum\limits_{k=1}^{M}\cos(\pi k), \hspace*{1mm}\mbox{since}\hspace*{1mm}\omega_\frac{M}{2}=\frac{M}{2}\nonumber\\&&
=\sum\limits_{k=1}^{M}(e^{\pi k}+e^{-\pi k}), =\frac{(e^{Mc}-1)(e^{c}+e^{-M})}{e^{c}-1},\hspace*{1mm}\mbox{for}\hspace*{1mm}c=i\hspace{1mm}\pi,
=0 \nonumber\vspace*{-0.5cm}\end{eqnarray}
because $e^{Mc}=e^{i M \pi}=\cos(M \pi)+i\hspace{0.5mm} \sin(M \pi)=1+0=1.$

\section{Web Appendix B. Covariance matrices}
\begin{figure}[H]
 \centering
\begin{subfigure}[b]{0.44\textwidth}
                 \includegraphics[width=\textwidth]{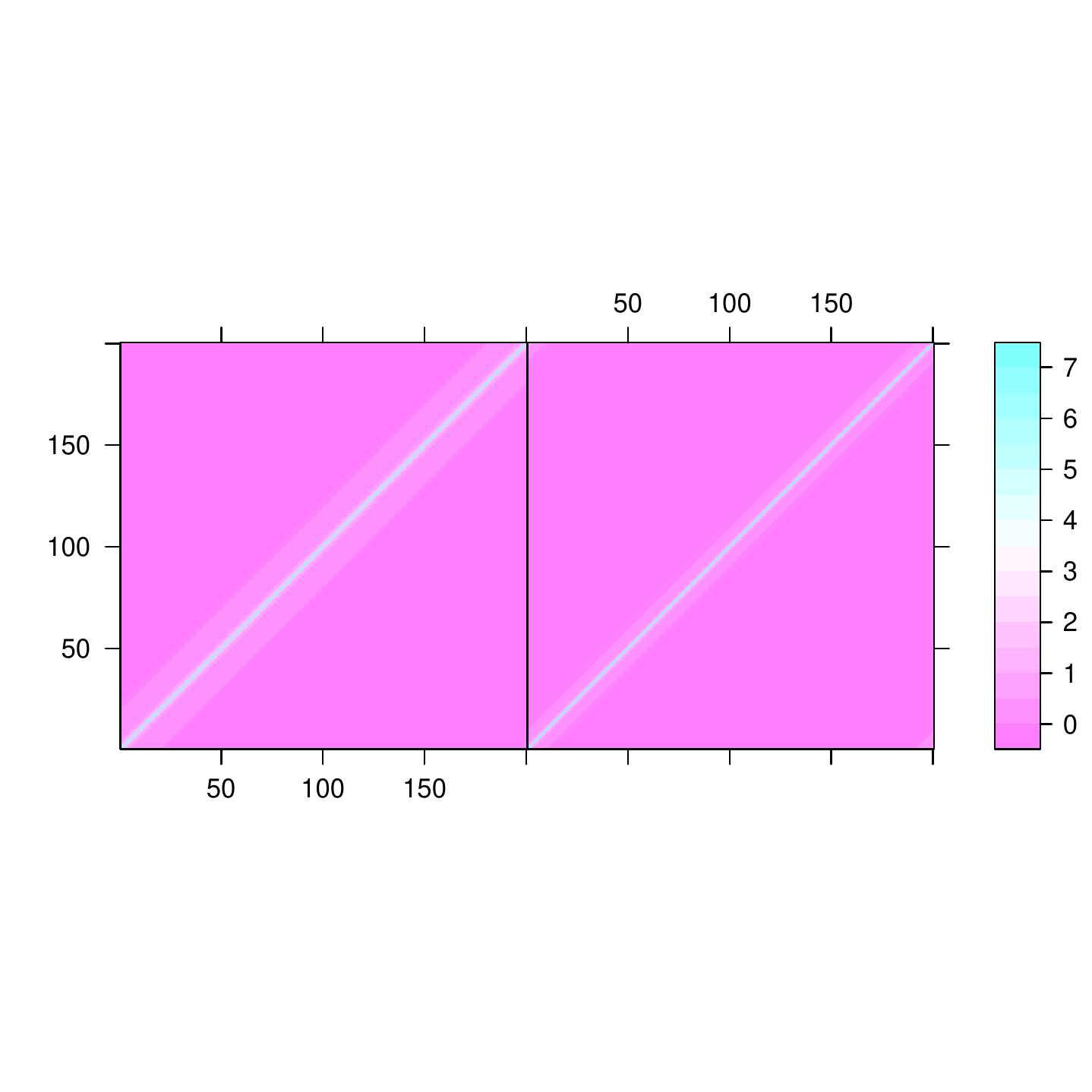}\vspace*{-1.8cm}
                 \caption{Left: Exact covariance matrix of [GP + N error] with $\sigma_\mathfrak{s}^2$=2, $\sigma_e^2$=5 and $\rho=5$ on the domain [1,2,...,199,200]. Right:
                 Approximate covariance matrix of the same model.}
                 \label{figure2a}
\end{subfigure}
         \quad
 \begin{subfigure}[b]{0.44\textwidth}
                 \includegraphics[width=\textwidth]{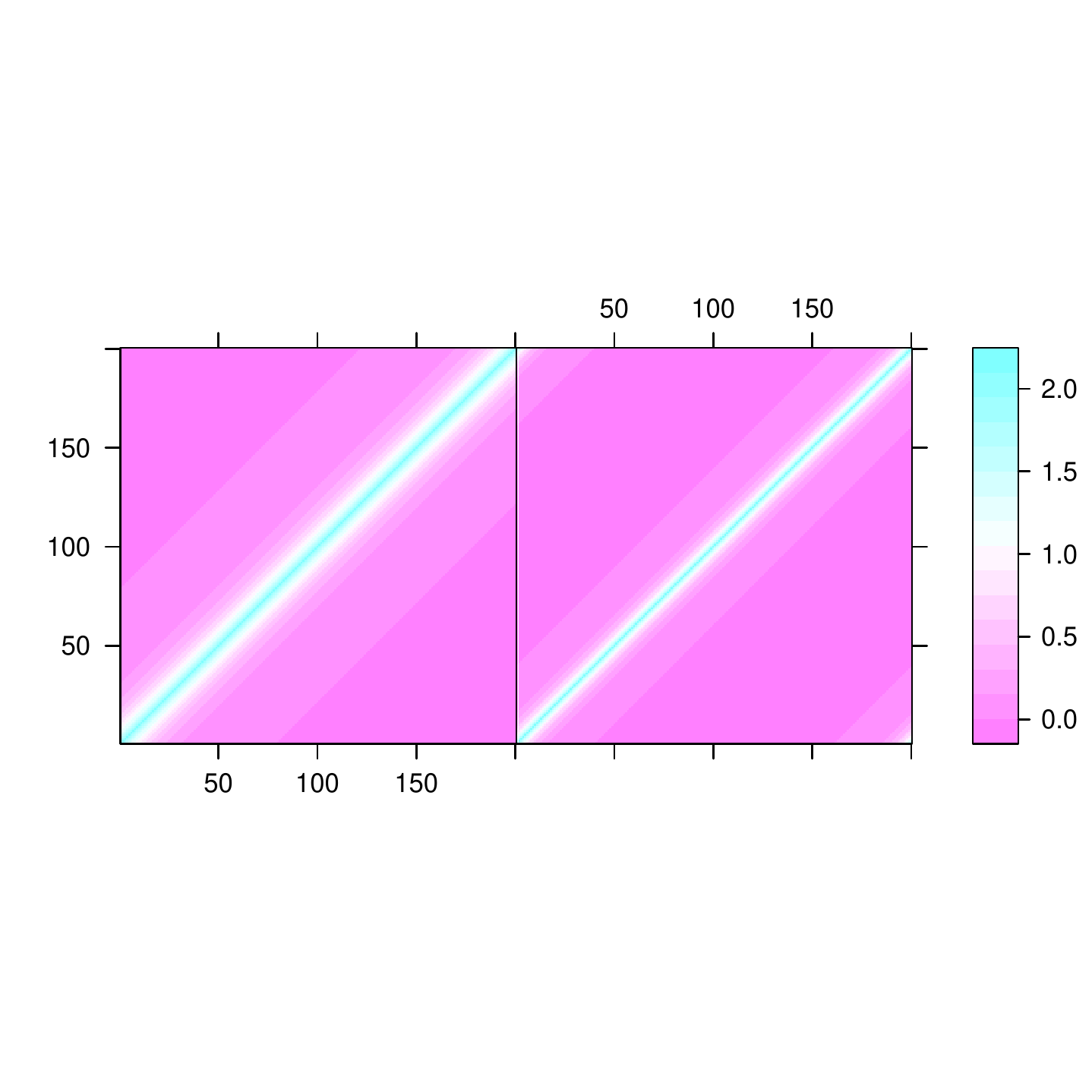}\vspace*{-1.8cm}
                 \caption{Left: Exact covariance matrix of [GP + N error] with $\sigma_\mathfrak{s}^2$=2, $\sigma_e^2$=0.1 and $\rho=16.67$ on the domain [1,2,...,199,200]. Right:
                 Approximate covariance matrix of the same model.}
                 \label{figure2b}
\end{subfigure}
                \caption{}\label{figure2}
         \end{figure}

\begin{figure}[H]
 \centering
\begin{subfigure}[b]{0.44\textwidth}
                 \includegraphics[width=\textwidth]{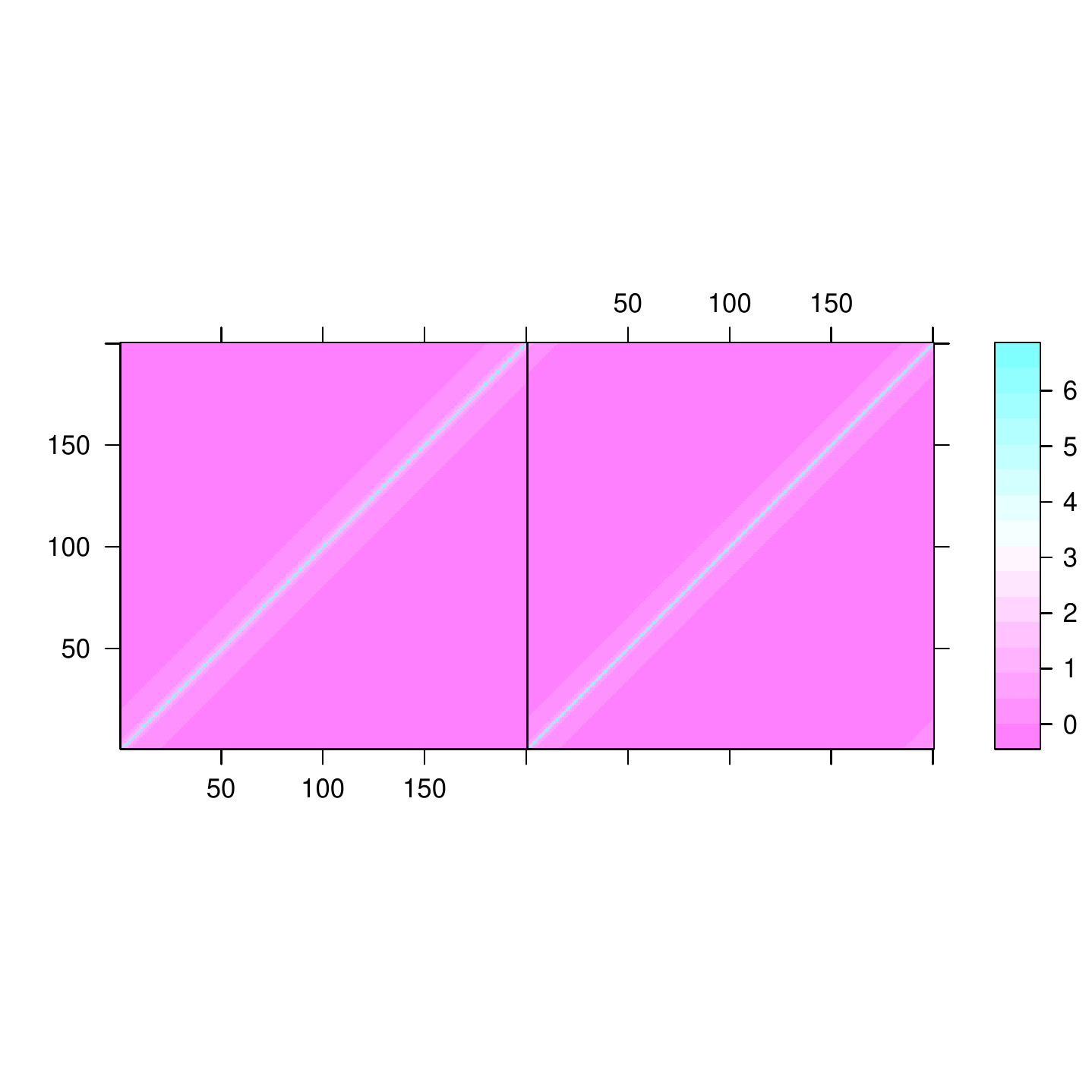}\vspace*{-1.8cm}
                 \caption{Left: Fitted exact covariance matrix for data with $\sigma_\mathfrak{s}^2$=2, $\sigma_e^2$=5 and $\rho$=5 using estimates from the exact restricted likelihood (RL). Right: Fitted approximate covariance using estimates from the approximate RL.}
                 \label{figure3a}
\end{subfigure}
         \quad
 \begin{subfigure}[b]{0.44\textwidth}
                 \includegraphics[width=\textwidth]{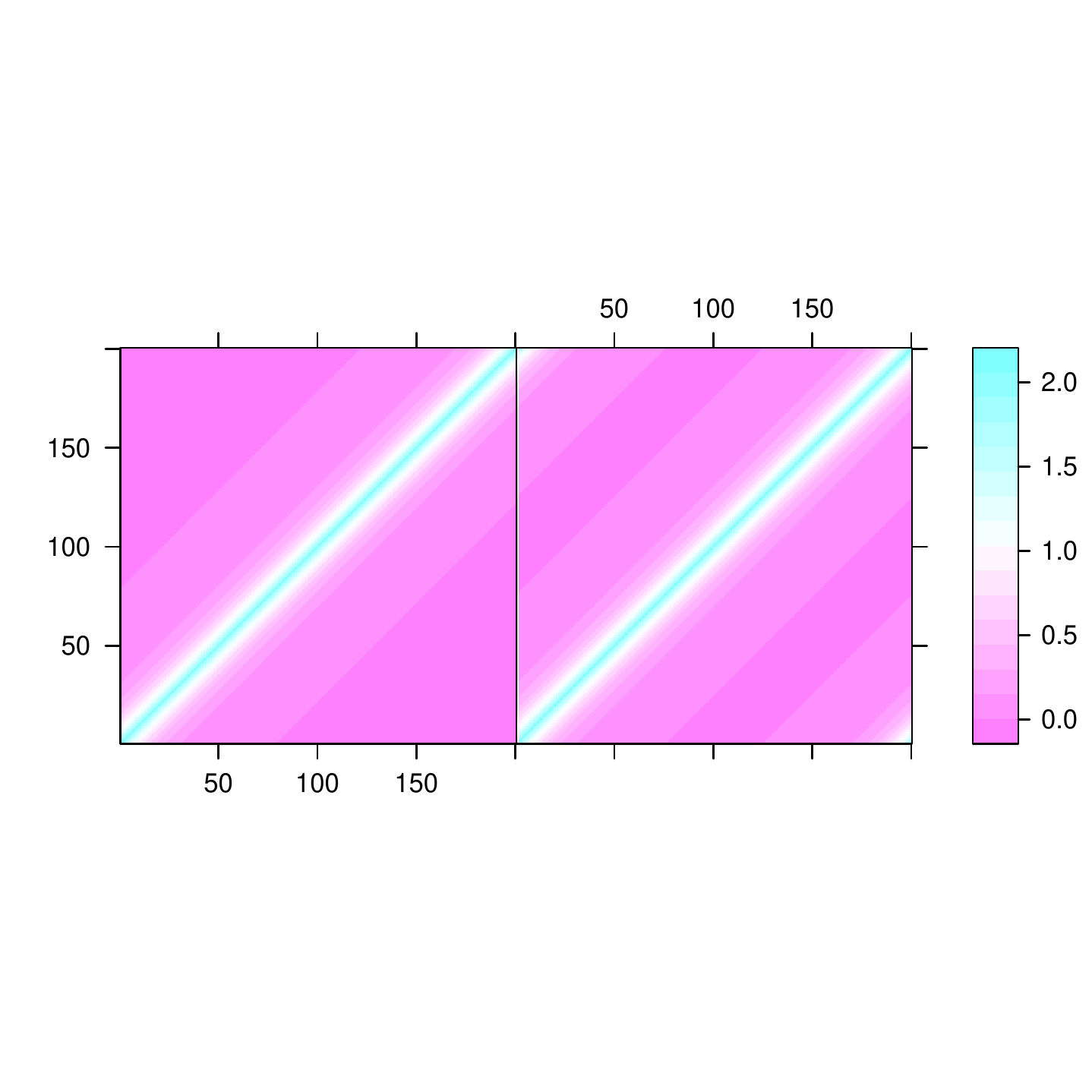}\vspace*{-1.8cm}
                 \caption{Left: Fitted exact covariance matrix for data with $\sigma_\mathfrak{s}^2$=2, $\sigma_e^2$=0.1 and $\rho$=16.67 using estimates from the exact restricted likelihood (RL). Right: Fitted approximate covariance using estimates from the approximate RL.}
                 \label{figure3b}
\end{subfigure}
\caption{}\label{figure3}
         \end{figure}

Figures \ref{figure2} and \ref{figure3} show some exact, approximate, and estimated covariance matrices for our model [GP + iid Normal error] with the spectral approximation, as discussed in the last paragraph of Section 2.3 in the main paper.
\vspace*{-1.5cm}

\vspace*{-2cm}

\subsection{Cov($y$) and Cov($y^*$) for $M$=100, and $M$=200}
In Section 4 of the main paper, we claim that the covariance of the residuals from a regression of the data on the fixed effects in the model is reasonably well approximated by the assumed covariance structure of the data $y$, and implicitly argue that this approximation should be better, for a given rank of fixed-effect design matrix $X$, if the number of observations $M$ is larger. Denote the residuals by $y^*$. Note that it is enough that Cov($y^*$) resembles the covariance structure of $y$;  it need not be the same in value as Cov($y$), because the values of the parameters $\sigma_\mathfrak{s}^2$, $\sigma_e^2$ and $\rho$ will be different for $y$ and for $y^*$. This appendix shows covariance matrices for $y$ and for $y^*$ (plotted using the same color scale), for the scenarios described below. Figures 3 to 6 use the exponential covariance (Mat\'{e}rn with $\nu$=0.5), and $X$ has two columns, namely a column of ones and a column of random draws from $N(0,1)$. Figures 7 to 10 also use the exponential covariance, but the second column of $X$ is now the first column of the spectral basis matrix $Z$ which is an extreme case because the effect of the residual transform would be concentrated on the column of $Z$ with the largest eigenvalue. Figures 11 to 14 use the Mat\'{e}rn with $\nu$=3.5 covariance, and $X$ has two columns, namely a column of ones and a column of random draws from $N(0,1)$. Figures 15 to 18 also use the Mat\'{e}rn with $\nu$=3.5 covariance, but now the second column of $X$ is the first column of the spectral basis matrix. We consider two sets of parameter values for the two variance parameters: $\sigma_\mathfrak{s}^2=2$, $\sigma_e^2=5$, and
$\sigma_\mathfrak{s}^2=2$, $\sigma_e^2=0.1$, the object being to have two different ratios $\sigma_\mathfrak{s}^2/\sigma_e^2$. The two parameter values considered for the range parameter are
 $\rho$=5, and $\rho$=16.67.

\graphicspath{{exp_N01/}}
\begin{figure}[H]
         \centering
\begin{subfigure}[b]{0.22\textwidth}
                 \includegraphics[width=\textwidth]{11}
                 \caption{Cov($y$). $\sigma_\mathfrak{s}^2=2$, $\sigma_e^2$=5, $\rho$=5, $M$=100.}
                 \label{fig:gull}
\end{subfigure}
         \quad
\begin{subfigure}[b]{0.22\textwidth}

                 \includegraphics[width=\textwidth]{12}
                 \caption{Cov($y^*$). $\sigma_\mathfrak{s}^2=2$, $\sigma_e^2$=5, $\rho$=5, $M$=100.}
                 \label{fig:gull}
\end{subfigure}
 \quad
\begin{subfigure}[b]{0.22\textwidth}
                 \includegraphics[width=\textwidth]{21}
                 \caption{Cov($y$). $\sigma_\mathfrak{s}^2=2$, $\sigma_e^2$=5, $\rho$=5, $M$=200.}
                 \label{fig:gull}
\end{subfigure}
         \quad
\begin{subfigure}[b]{0.22\textwidth}

                 \includegraphics[width=\textwidth]{22}
                 \caption{Cov($y^*$). $\sigma_\mathfrak{s}^2=2$, $\sigma_e^2$=5, $\rho$=5, $M$=200.}
                 \label{fig:gull}
\end{subfigure}
 \caption{}\label{fig:animals}
         \end{figure}

\begin{figure}[H]
         \centering
\begin{subfigure}[b]{0.22\textwidth}
                 \includegraphics[width=\textwidth]{31}
                 \caption{Cov($y$). $\sigma_\mathfrak{s}^2=2$, $\sigma_e^2$=5, $\rho$=16.67, $M$=100.}
                 \label{fig:gull}
\end{subfigure}
         \quad
\begin{subfigure}[b]{0.22\textwidth}

                 \includegraphics[width=\textwidth]{32}
                 \caption{Cov($y^*$). $\sigma_\mathfrak{s}^2=2$, $\sigma_e^2$=5, $\rho$=16.67, $M$=100.}
                 \label{fig:gull}
\end{subfigure}
 \quad
\begin{subfigure}[b]{0.22\textwidth}
                 \includegraphics[width=\textwidth]{41}
                 \caption{Cov($y$). $\sigma_\mathfrak{s}^2=2$, $\sigma_e^2$=5, $\rho$=16.67, $M$=200.}
                 \label{fig:gull}
\end{subfigure}
         \quad
\begin{subfigure}[b]{0.22\textwidth}

                 \includegraphics[width=\textwidth]{42}
                 \caption{Cov$(y^*)$. $\sigma_\mathfrak{s}^2=2$, $\sigma_e^2$=5, $\rho$=16.67, $M$=200.}
                 \label{fig:gull}
\end{subfigure}
 \caption{}\label{fig:animals}
         \end{figure}

\begin{figure}[H]
         \centering
\begin{subfigure}[b]{0.22\textwidth}
                 \includegraphics[width=\textwidth]{51}
                 \caption{Cov($y$). $\sigma_\mathfrak{s}^2=2$, $\sigma_e^2$=0.1, $\rho$=5, $M$=100.}
                 \label{fig:gull}
\end{subfigure}
         \quad
\begin{subfigure}[b]{0.22\textwidth}

                 \includegraphics[width=\textwidth]{52}
                 \caption{Cov($y^*)$. $\sigma_\mathfrak{s}^2=2$, $\sigma_e^2$=0.1, $\rho$=5, $M$=100.}
                 \label{fig:gull}
\end{subfigure}
 \quad
\begin{subfigure}[b]{0.22\textwidth}
                 \includegraphics[width=\textwidth]{61}
                 \caption{Cov($y$). $\sigma_\mathfrak{s}^2=2$, $\sigma_e^2$=0.1, $\rho$=5, $M$=200.}
                 \label{fig:gull}
\end{subfigure}
         \quad
\begin{subfigure}[b]{0.22\textwidth}

                 \includegraphics[width=\textwidth]{62}
                 \caption{Cov($y^*$). $\sigma_\mathfrak{s}^2=2$, $\sigma_e^2$=0.1, $\rho$=5, $M$=200.}
                 \label{fig:gull}
\end{subfigure}
 \caption{}\label{fig:animals}
         \end{figure}

         \begin{figure}[H]
         \centering
\begin{subfigure}[b]{0.22\textwidth}
                 \includegraphics[width=\textwidth]{71}
                 \caption{Cov($y$). $\sigma_\mathfrak{s}^2=2$, $\sigma_e^2$=0.1, $\rho$=16.67, $M$=100.}
                 \label{fig:gull}
\end{subfigure}
         \quad
\begin{subfigure}[b]{0.22\textwidth}

                 \includegraphics[width=\textwidth]{72}
                 \caption{Cov($y^*$). $\sigma_\mathfrak{s}^2=2$, $\sigma_e^2$=0.1, $\rho$=16.67, $M$=100.}
                 \label{fig:gull}
\end{subfigure}
 \quad
\begin{subfigure}[b]{0.22\textwidth}
                 \includegraphics[width=\textwidth]{81}
                 \caption{Cov($y$). $\sigma_\mathfrak{s}^2=2$, $\sigma_e^2$=0.1, $\rho$=16.67, $M$=200.}
                 \label{fig:gull}
\end{subfigure}
         \quad
\begin{subfigure}[b]{0.22\textwidth}

                 \includegraphics[width=\textwidth]{82}
                 \caption{Cov($y^*$). $\sigma_\mathfrak{s}^2=2$, $\sigma_e^2$=0.1, $\rho$=16.67, $M$=200.}
                 \label{fig:gull}
\end{subfigure}
 \caption{}\label{fig:animals}
         \end{figure}

\graphicspath{{exp_z1/}}
\begin{figure}[H]
         \centering
\begin{subfigure}[b]{0.22\textwidth}
                 \includegraphics[width=\textwidth]{11}
                 \caption{Cov($y$). $\sigma_\mathfrak{s}^2=2$, $\sigma_e^2$=5, $\rho$=5, $M$=100.}
                 \label{fig:gull}
\end{subfigure}
         \quad
\begin{subfigure}[b]{0.22\textwidth}

                 \includegraphics[width=\textwidth]{12}
                 \caption{Cov($y^*$). $\sigma_\mathfrak{s}^2=2$, $\sigma_e^2$=5, $\rho$=5, $M$=100.}
                 \label{fig:gull}
\end{subfigure}
 \quad
\begin{subfigure}[b]{0.22\textwidth}
                 \includegraphics[width=\textwidth]{21}
                 \caption{Cov($y$). $\sigma_\mathfrak{s}^2=2$, $\sigma_e^2$=5, $\rho$=5, $M$=200.}
                 \label{fig:gull}
\end{subfigure}
         \quad
\begin{subfigure}[b]{0.22\textwidth}

                 \includegraphics[width=\textwidth]{22}
                 \caption{Cov($y^*$). $\sigma_\mathfrak{s}^2=2$, $\sigma_e^2$=5, $\rho$=5, $M$=200.}
                 \label{fig:gull}
\end{subfigure}
 \caption{}\label{fig:animals}
         \end{figure}

\begin{figure}[H]
         \centering
\begin{subfigure}[b]{0.22\textwidth}
                 \includegraphics[width=\textwidth]{31}
                 \caption{Cov($y$). $\sigma_\mathfrak{s}^2=2$, $\sigma_e^2$=5, $\rho$=16.67, $M$=100.}
                 \label{fig:gull}
\end{subfigure}
         \quad
\begin{subfigure}[b]{0.22\textwidth}

                 \includegraphics[width=\textwidth]{32}
                 \caption{Cov($y^*$). $\sigma_\mathfrak{s}^2=2$, $\sigma_e^2$=5, $\rho$=16.67, $M$=100.}
                 \label{fig:gull}
\end{subfigure}
 \quad
\begin{subfigure}[b]{0.22\textwidth}
                 \includegraphics[width=\textwidth]{41}
                 \caption{Cov($y$). $\sigma_\mathfrak{s}^2=2$, $\sigma_e^2$=5, $\rho$=16.67, $M$=200.}
                 \label{fig:gull}
\end{subfigure}
         \quad
\begin{subfigure}[b]{0.22\textwidth}

                 \includegraphics[width=\textwidth]{42}
                 \caption{Cov$(y^*)$. $\sigma_\mathfrak{s}^2=2$, $\sigma_e^2$=5, $\rho$=16.67, $M$=200.}
                 \label{fig:gull}
\end{subfigure}
 \caption{}\label{fig:animals}
         \end{figure}

\begin{figure}[H]
         \centering
\begin{subfigure}[b]{0.22\textwidth}
                 \includegraphics[width=\textwidth]{51}
                 \caption{Cov($y$). $\sigma_\mathfrak{s}^2=2$, $\sigma_e^2$=0.1, $\rho$=5, $M$=100.}
                 \label{fig:gull}
\end{subfigure}
         \quad
\begin{subfigure}[b]{0.22\textwidth}

                 \includegraphics[width=\textwidth]{52}
                 \caption{Cov($y^*)$. $\sigma_\mathfrak{s}^2=2$, $\sigma_e^2$=0.1, $\rho$=5, $M$=100.}
                 \label{fig:gull}
\end{subfigure}
 \quad
\begin{subfigure}[b]{0.22\textwidth}
                 \includegraphics[width=\textwidth]{61}
                 \caption{Cov($y$). $\sigma_\mathfrak{s}^2=2$, $\sigma_e^2$=0.1, $\rho$=5, $M$=200.}
                 \label{fig:gull}
\end{subfigure}
         \quad
\begin{subfigure}[b]{0.22\textwidth}

                 \includegraphics[width=\textwidth]{62}
                 \caption{Cov($y^*$). $\sigma_\mathfrak{s}^2=2$, $\sigma_e^2$=0.1, $\rho$=5, $M$=200.}
                 \label{fig:gull}
\end{subfigure}
 \caption{}\label{fig:animals}
         \end{figure}

         \begin{figure}[H]
         \centering
\begin{subfigure}[b]{0.22\textwidth}
                 \includegraphics[width=\textwidth]{71}
                 \caption{Cov($y$). $\sigma_\mathfrak{s}^2=2$, $\sigma_e^2$=0.1, $\rho$=16.67, $M$=100.}
                 \label{fig:gull}
\end{subfigure}
         \quad
\begin{subfigure}[b]{0.22\textwidth}
                 \includegraphics[width=\textwidth]{72}
                 \caption{Cov($y^*$). $\sigma_\mathfrak{s}^2=2$, $\sigma_e^2$=0.1, $\rho$=16.67, $M$=100.}
                 \label{fig:gull}
\end{subfigure}
 \quad
\begin{subfigure}[b]{0.22\textwidth}
                 \includegraphics[width=\textwidth]{81}
                 \caption{Cov($y$). $\sigma_\mathfrak{s}^2=2$, $\sigma_e^2$=0.1, $\rho$=16.67, $M$=200.}
                 \label{fig:gull}
\end{subfigure}
         \quad
\begin{subfigure}[b]{0.22\textwidth}

                 \includegraphics[width=\textwidth]{82}
                 \caption{Cov($y^*$). $\sigma_\mathfrak{s}^2=2$, $\sigma_e^2$=0.1, $\rho$=16.67, $M$=200.}
                 \label{fig:gull}
\end{subfigure}
 \caption{}\label{fig:animals}
         \end{figure}

\graphicspath{{mat35_N01/}}
\begin{figure}[H]
         \centering
\begin{subfigure}[b]{0.22\textwidth}
                 \includegraphics[width=\textwidth]{11}
                 \caption{Cov($y$). $\sigma_\mathfrak{s}^2=2$, $\sigma_e^2$=5, $\rho$=5, $M$=100.}
                 \label{fig:gull}
\end{subfigure}
         \quad
\begin{subfigure}[b]{0.22\textwidth}

                 \includegraphics[width=\textwidth]{12}
                 \caption{Cov($y^*$). $\sigma_\mathfrak{s}^2=2$, $\sigma_e^2$=5, $\rho$=5, $M$=100.}
                 \label{fig:gull}
\end{subfigure}
 \quad
\begin{subfigure}[b]{0.22\textwidth}
                 \includegraphics[width=\textwidth]{21}
                 \caption{Cov($y$). $\sigma_\mathfrak{s}^2=2$, $\sigma_e^2$=5, $\rho$=5, $M$=200.}
                 \label{fig:gull}
\end{subfigure}
         \quad
\begin{subfigure}[b]{0.22\textwidth}

                 \includegraphics[width=\textwidth]{22}
                 \caption{Cov($y^*$). $\sigma_\mathfrak{s}^2=2$, $\sigma_e^2$=5, $\rho$=5, $M$=200.}
                 \label{fig:gull}
\end{subfigure}
 \caption{}\label{fig:animals}
         \end{figure}

\begin{figure}[H]
         \centering
\begin{subfigure}[b]{0.22\textwidth}
                 \includegraphics[width=\textwidth]{31}
                 \caption{Cov($y$). $\sigma_\mathfrak{s}^2=2$, $\sigma_e^2$=5, $\rho$=16.67, $M$=100.}
                 \label{fig:gull}
\end{subfigure}
         \quad
\begin{subfigure}[b]{0.22\textwidth}

                 \includegraphics[width=\textwidth]{32}
                 \caption{Cov($y^*$). $\sigma_\mathfrak{s}^2=2$, $\sigma_e^2$=5, $\rho$=16.67, $M$=100.}
                 \label{fig:gull}
\end{subfigure}
 \quad
\begin{subfigure}[b]{0.22\textwidth}
                 \includegraphics[width=\textwidth]{41}
                 \caption{Cov($y$). $\sigma_\mathfrak{s}^2=2$, $\sigma_e^2$=5, $\rho$=16.67, $M$=200.}
                 \label{fig:gull}
\end{subfigure}
         \quad
\begin{subfigure}[b]{0.22\textwidth}
                 \includegraphics[width=\textwidth]{42}
                 \caption{Cov$(y^*)$. $\sigma_\mathfrak{s}^2=2$, $\sigma_e^2$=5, $\rho$=16.67, $M$=200.}
                 \label{fig:gull}
\end{subfigure}
 \caption{}\label{fig:animals}
         \end{figure}

\begin{figure}[H]
         \centering
\begin{subfigure}[b]{0.22\textwidth}
                 \includegraphics[width=\textwidth]{51}
                 \caption{Cov($y$). $\sigma_\mathfrak{s}^2=2$, $\sigma_e^2$=0.1, $\rho$=5, $M$=100.}
                 \label{fig:gull}
\end{subfigure}
         \quad
\begin{subfigure}[b]{0.22\textwidth}

                 \includegraphics[width=\textwidth]{52}
                 \caption{Cov($y^*)$. $\sigma_\mathfrak{s}^2=2$, $\sigma_e^2$=0.1, $\rho$=5, $M$=100.}
                 \label{fig:gull}
\end{subfigure}
 \quad
\begin{subfigure}[b]{0.22\textwidth}
                 \includegraphics[width=\textwidth]{61}
                 \caption{Cov($y$). $\sigma_\mathfrak{s}^2=2$, $\sigma_e^2$=0.1, $\rho$=5, $M$=200.}
                 \label{fig:gull}
\end{subfigure}
         \quad
\begin{subfigure}[b]{0.22\textwidth}

                 \includegraphics[width=\textwidth]{62}
                 \caption{Cov($y^*$). $\sigma_\mathfrak{s}^2=2$, $\sigma_e^2$=0.1, $\rho$=5, $M$=200.}
                 \label{fig:gull}
\end{subfigure}
 \caption{}\label{fig:animals}
         \end{figure}

         \begin{figure}[H]
         \centering
\begin{subfigure}[b]{0.22\textwidth}
                 \includegraphics[width=\textwidth]{71}
                 \caption{Cov($y$). $\sigma_\mathfrak{s}^2=2$, $\sigma_e^2$=0.1, $\rho$=16.67, $M$=100.}
                 \label{fig:gull}
\end{subfigure}
         \quad
\begin{subfigure}[b]{0.22\textwidth}

                 \includegraphics[width=\textwidth]{72}
                 \caption{Cov($y^*$). $\sigma_\mathfrak{s}^2=2$, $\sigma_e^2$=0.1, $\rho$=16.67, $M$=100.}
                 \label{fig:gull}
\end{subfigure}
 \quad
\begin{subfigure}[b]{0.22\textwidth}
                 \includegraphics[width=\textwidth]{81}
                 \caption{Cov($y$). $\sigma_\mathfrak{s}^2=2$, $\sigma_e^2$=0.1, $\rho$=16.67, $M$=200.}
                 \label{fig:gull}
\end{subfigure}
         \quad
\begin{subfigure}[b]{0.22\textwidth}

                 \includegraphics[width=\textwidth]{82}
                 \caption{Cov($y^*$). $\sigma_\mathfrak{s}^2=2$, $\sigma_e^2$=0.1, $\rho$=16.67, $M$=200.}
                 \label{fig:gull}
\end{subfigure}
 \caption{}\label{fig:animals}
         \end{figure}

\graphicspath{{mat35_z1/}}
\begin{figure}[H]
         \centering
\begin{subfigure}[b]{0.22\textwidth}
                 \includegraphics[width=\textwidth]{11}
                 \caption{Cov($y$). $\sigma_\mathfrak{s}^2=2$, $\sigma_e^2$=5, $\rho$=5, $M$=100.}
                 \label{fig:gull}
\end{subfigure}
         \quad
\begin{subfigure}[b]{0.22\textwidth}

                 \includegraphics[width=\textwidth]{12}
                 \caption{Cov($y^*$). $\sigma_\mathfrak{s}^2=2$, $\sigma_e^2$=5, $\rho$=5, $M$=100.}
                 \label{fig:gull}
\end{subfigure}
 \quad
\begin{subfigure}[b]{0.22\textwidth}
                 \includegraphics[width=\textwidth]{21}
                 \caption{Cov($y$). $\sigma_\mathfrak{s}^2=2$, $\sigma_e^2$=5, $\rho$=5, $M$=200.}
                 \label{fig:gull}
\end{subfigure}
         \quad
\begin{subfigure}[b]{0.22\textwidth}

                 \includegraphics[width=\textwidth]{22}
                 \caption{Cov($y^*$). $\sigma_\mathfrak{s}^2=2$, $\sigma_e^2$=5, $\rho$=5, $M$=200.}
                 \label{fig:gull}
\end{subfigure}
 \caption{}\label{fig:animals}
         \end{figure}

\begin{figure}[H]
         \centering
\begin{subfigure}[b]{0.22\textwidth}
                 \includegraphics[width=\textwidth]{31}
                 \caption{Cov($y$). $\sigma_\mathfrak{s}^2=2$, $\sigma_e^2$=5, $\rho$=16.67, $M$=100.}
                 \label{fig:gull}
\end{subfigure}
         \quad
\begin{subfigure}[b]{0.22\textwidth}

                 \includegraphics[width=\textwidth]{32}
                 \caption{Cov($y^*$). $\sigma_\mathfrak{s}^2=2$, $\sigma_e^2$=5, $\rho$=16.67, $M$=100.}
                 \label{fig:gull}
\end{subfigure}
 \quad
\begin{subfigure}[b]{0.22\textwidth}
                 \includegraphics[width=\textwidth]{41}
                 \caption{Cov($y$). $\sigma_\mathfrak{s}^2=2$, $\sigma_e^2$=5, $\rho$=16.67, $M$=200.}
                 \label{fig:gull}
\end{subfigure}
         \quad
\begin{subfigure}[b]{0.22\textwidth}

                 \includegraphics[width=\textwidth]{42}
                 \caption{Cov$(y^*)$. $\sigma_\mathfrak{s}^2=2$, $\sigma_e^2$=5, $\rho$=16.67, $M$=200.}
                 \label{fig:gull}
\end{subfigure}
 \caption{}\label{fig:animals}
         \end{figure}

\begin{figure}[H]
         \centering
\begin{subfigure}[b]{0.22\textwidth}
                 \includegraphics[width=\textwidth]{51}
                 \caption{Cov($y$). $\sigma_\mathfrak{s}^2=2$, $\sigma_e^2$=0.1, $\rho$=5, $M$=100.}
                 \label{fig:gull}
\end{subfigure}
         \quad
\begin{subfigure}[b]{0.22\textwidth}

                 \includegraphics[width=\textwidth]{52}
                 \caption{Cov($y^*)$. $\sigma_\mathfrak{s}^2=2$, $\sigma_e^2$=0.1, $\rho$=5, $M$=100.}
                 \label{fig:gull}
\end{subfigure}
 \quad
\begin{subfigure}[b]{0.22\textwidth}
                 \includegraphics[width=\textwidth]{61}
                 \caption{Cov($y$). $\sigma_\mathfrak{s}^2=2$, $\sigma_e^2$=0.1, $\rho$=5, $M$=200.}
                 \label{fig:gull}
\end{subfigure}
         \quad
\begin{subfigure}[b]{0.22\textwidth}

                 \includegraphics[width=\textwidth]{62}
                 \caption{Cov($y^*$). $\sigma_\mathfrak{s}^2=2$, $\sigma_e^2$=0.1, $\rho$=5, $M$=200.}
                 \label{fig:gull}
\end{subfigure}
 \caption{}\label{fig:animals}
         \end{figure}

        \begin{figure}[H]
         \centering
\begin{subfigure}[b]{0.22\textwidth}
                 \includegraphics[width=\textwidth]{71}
                 \caption{Cov($y$). $\sigma_\mathfrak{s}^2=2$, $\sigma_e^2$=0.1, $\rho$=16.67, $M$=100.}
                 \label{figure8a}
\end{subfigure}
         \quad
\begin{subfigure}[b]{0.22\textwidth}
                 \includegraphics[width=\textwidth]{72}
                 \caption{Cov($y^*$). $\sigma_\mathfrak{s}^2=2$, $\sigma_e^2$=0.1, $\rho$=16.67, $M$=100.}
                 \label{figure8b}
\end{subfigure}
 \quad
\begin{subfigure}[b]{0.22\textwidth}
                 \includegraphics[width=\textwidth]{81}
                 \caption{Cov($y$). $\sigma_\mathfrak{s}^2=2$, $\sigma_e^2$=0.1, $\rho$=16.67, $M$=200.}
                 \label{figure8c}
\end{subfigure}
         \quad
\begin{subfigure}[b]{0.22\textwidth}

                 \includegraphics[width=\textwidth]{82}
                 \caption{Cov($y^*$). $\sigma_\mathfrak{s}^2=2$, $\sigma_e^2$=0.1, $\rho$=16.67, $M$=200.}
                 \label{figure8d}
\end{subfigure}
 \caption{}\label{figure8}
  \end{figure}

\section{Web Appendix C. Simulation results}
This appendix supplements the simulation results reported in the main paper regarding the effects of contaminating data simulated from an intercept-only linear mixed model with GP-distributed random effect and iid Normal errors, for two contaminations:  adding an outlier (Section 3.1 in the main paper;  Table 1 and Figure 19a,b below) and adding a mean shift in the data series (Section 3.2 of the main paper;  Tables 1 and 2 below).  This appendix also presents conjectures and simulation results regarding contamination by replacing a part of the data series by data simulated from a GP with a different range parameter ($\rho)$, which is not discussed in the main paper.  Section 3.0.1 below gives these conjectures and discusses simulation results presented in Figure 19c and Tables 1 and 3 below.

         \begin{figure}[H]
         \centering
\begin{subfigure}[b]{0.3\textwidth}
                 \includegraphics[width=\textwidth]{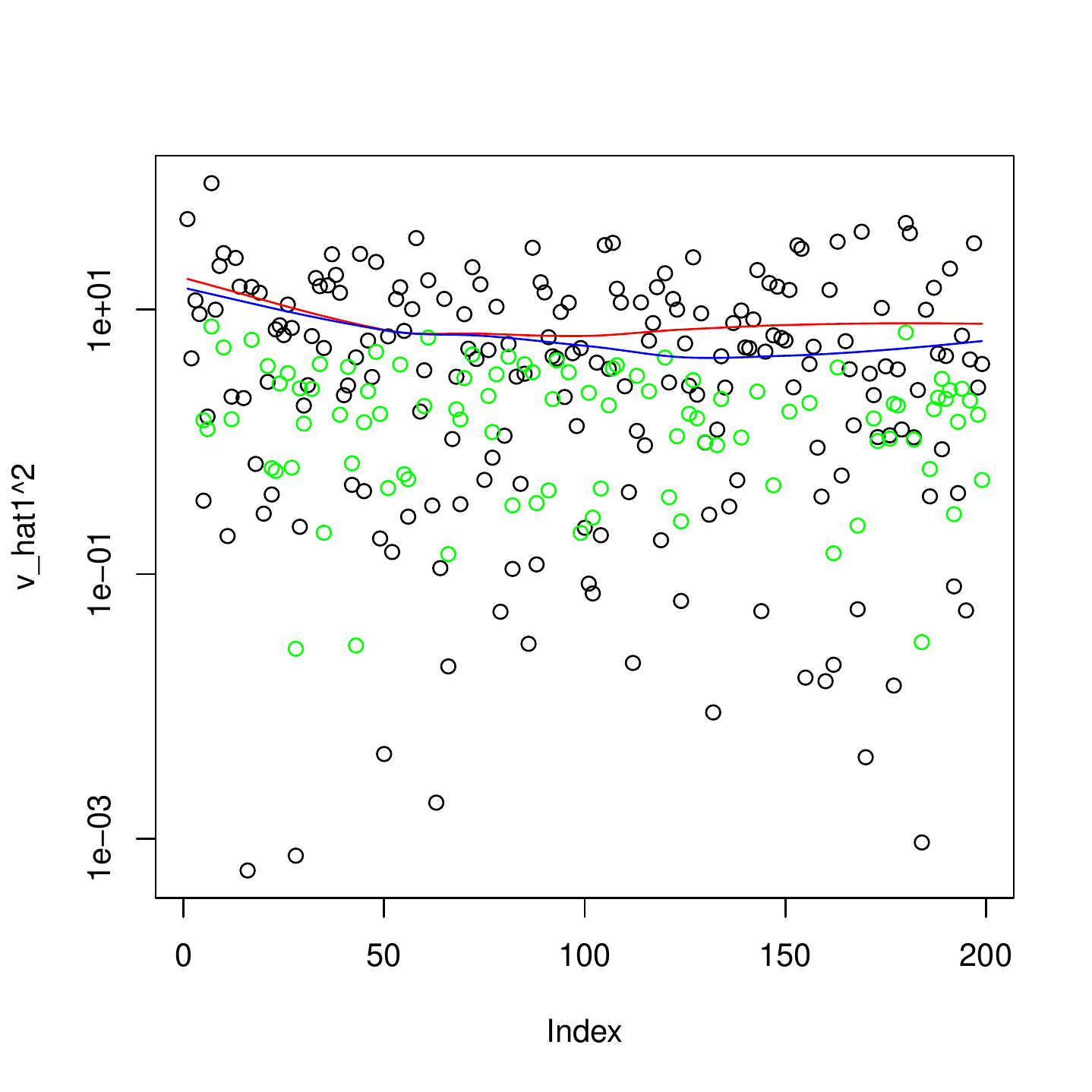}
                 \caption{Green circles:
                 log ${v_j}^2$ from simulated data; blue line: loess smooth of log ${v_j}^2$;  black circles, red line: from data contaminated by outlier.} \label{figure5a}
\end{subfigure}
         \quad
\begin{subfigure}[b]{0.3\textwidth}

                 \includegraphics[width=\textwidth]{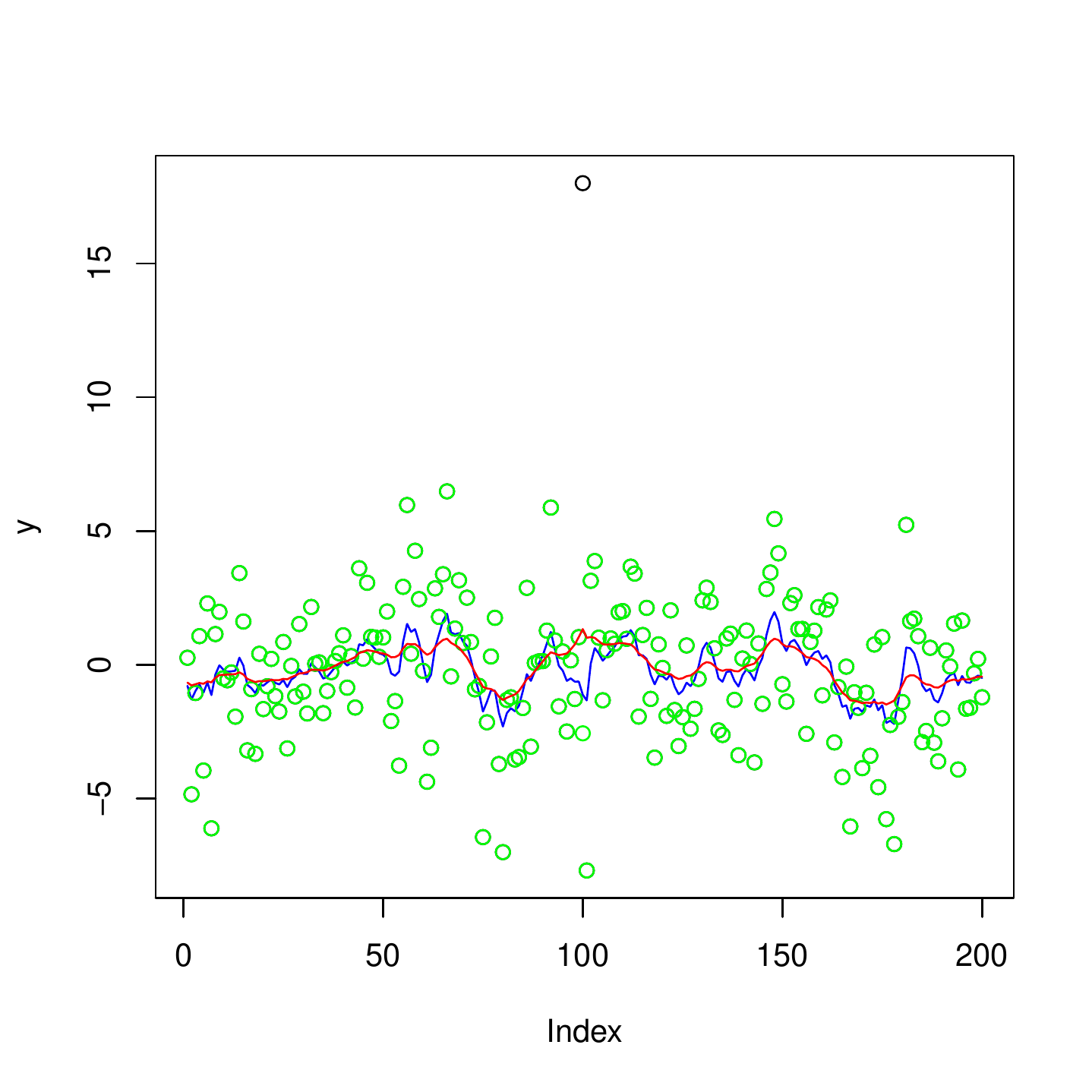}
                 \caption{Green circles: simulated data; blue line: fit;  black circle:
                  outlier added; red line: fit of data contaminated by outlier.\\} \label{figure5b}
\end{subfigure}
 \quad
         \begin{subfigure}[b]{0.3\textwidth}
                 \includegraphics[width=\textwidth]{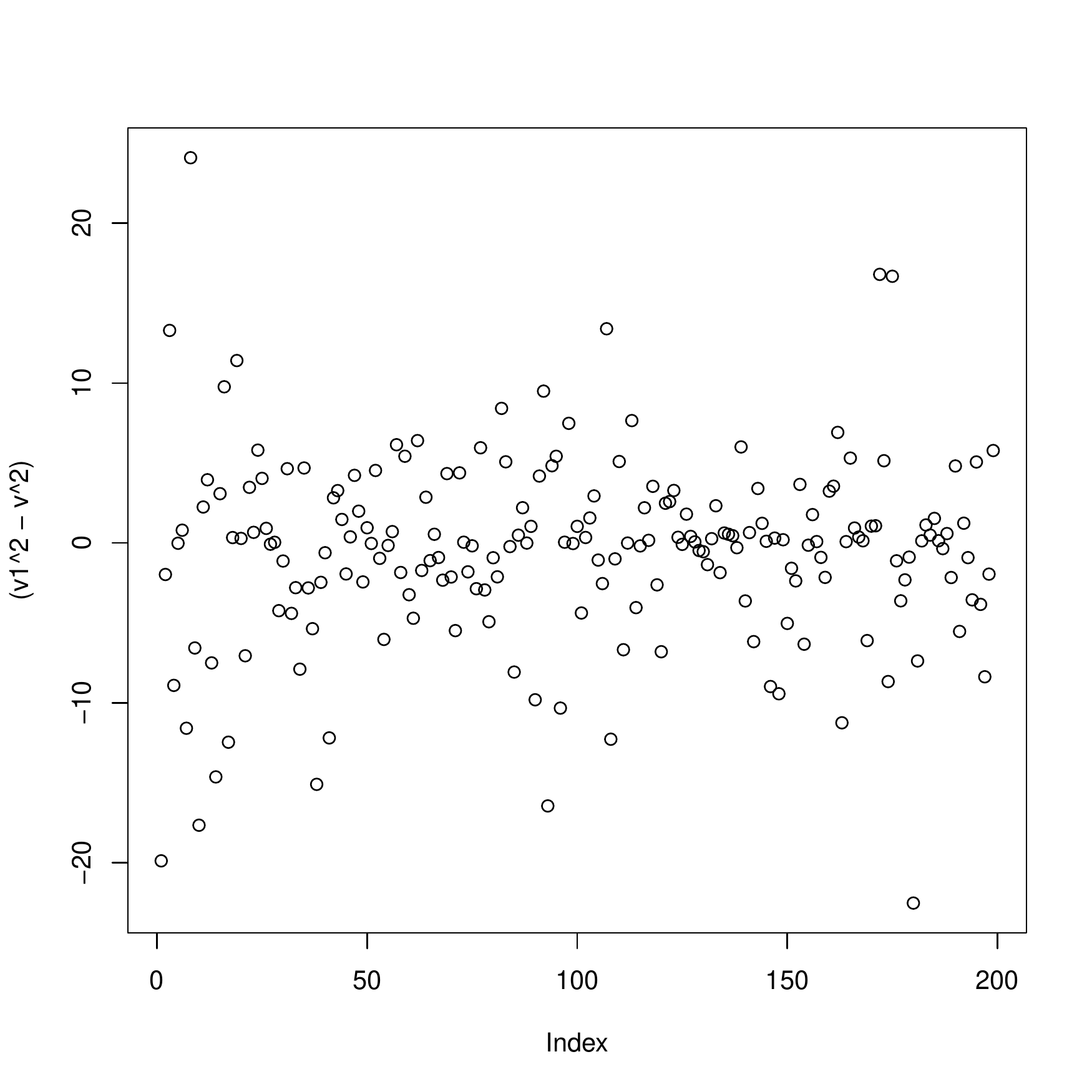}
                 \caption{Difference of ${v_j}^2$'s (contaminated - uncontaminated) from simulated data, contaminated by GP with $\rho$=16.67.\\}\label{figure7b}
        \end{subfigure}
         \caption{(a) and (b) show effects of contamination by an outlier; (c) shows effects of contamination by GP with a different range. Parameter values: $\sigma_\mathfrak{s}^2$=2, $\sigma_e^2$=5 and $\rho$=5. Estimates from exact restricted likelihood maximization. }
         \end{figure}

\begin{table}[H]
\small
\caption{Estimates of GP parameters maximizing the exact and approximate restricted log likelihoods (``exact RL'' and ``apprx RL'' respectively) for two simulated datasets. Also presented are estimates after the data is contaminated by an outlier in the middle, by a mean-shift midway, or by replacing a subseries of observations by observations from a GP with a different $\rho$.}\label{table1}
\begin{tabular}{c c}

\begin{tabular}{ c c c c } \hline \hline

             & \multicolumn{3}{c}{exact RL} \\
            \cline{2-4}
&$\sigma_\mathfrak{s}^2$ &  $\sigma_e^2$& $\rho$ \\ \hline\hline
actual values used in simulation &2 &5&5\\
estimates from simulated data & 1.76 &4.66&4.00\\
\hline
from data contaminated by outlier& 1.07&7.02 &11.11\\
from data contaminated by shift  &13.05& 5.63&58.82 \\
from data contaminated by another GP &1.69& 4.24 & 3.04  \\
\hline
\hline
actual values used in simulation &2 &0.1&16.7\\
estimates from simulated data & 1.95 &0.11&16.7\\
\hline
from data contaminated by outlier& 1.59&1.93 &12.50\\
from data contaminated by shift  &11.51& 0.03&55.56 \\
from data contaminated by another GP &1.60& 0.09 & 8.55  \\
\hline
\end{tabular}
&

\begin{tabular}{ c c c} \hline \hline
  \multicolumn{3}{c}{apprx RL} \\
            \cline{1-3}
$\sigma_\mathfrak{s}^2$ &  $\sigma_e^2$& $\rho$ \\ \hline\hline
2 &5&5\\
1.71 &4.84&8.33\\
 \hline
  1.07&7.06&25.00  \\
 14.51& 5.71 & 106.38  \\
 1.56& 4.51 &  7.09 \\
\hline
\hline
2 &0.1&16.7\\
1.94 &0.15&33.33\\
 \hline
  1.62&1.96&25.00  \\
 10.99& 0.12 & 93.46  \\
 1.56& 0.18 &  17.86 \\
\hline
\end{tabular}
\end{tabular}
\end{table}

\begin{table}[H]
\small
\caption{Average over 100 simulated datasets of estimates maximizing the exact and approximate restricted likelihoods: contamination by mean shift. Monte Carlo standard errors are in parentheses.}\label{table3}
\begin{tabular}{c c}

\begin{tabular}{ c c c c } \hline \hline
& \multicolumn{3}{c}{exact RL} \\
            \cline{2-4}
&$\sigma_\mathfrak{s}^2$ &  $\sigma_e^2$& $\rho$ \\ \hline\hline
actual values &2 &5&5\\
uncontaminated & 2.29 (0.11) &4.75 (0.11)&6.89 (0.82)\\
contaminated & 11.50 (0.36)& 5.61 (0.06)&106.48 (6.06)\\
\hline
actual values & 2&5 &16.67\\
uncontaminated  & 2.16 (0.09) & 4.89 (0.07)& 26.12 (4.32)\\
 contaminated & 12.56 (0.59)&5.22 (0.07) &140.13 (8.31)\\
\hline
actual values &2 & 0.1&5\\
uncontaminated & 1.94 (0.03) &0.099 (0.01)&4.97 (0.13) \\
contaminated  & 9.82 (0.17)& 0.260 (0.01)& 34.89 (1.29)\\
\hline
actual values& 2& 0.1&16.67\\
uncontaminated & 2.09 (0.07) &0.093 (0.00)& 18.07 (0.84)\\
contaminated &10.04 (0.29) &0.117 (0.00)&72.84 (2.96)\\
\hline

actual values& 10& 5&5\\
uncontaminated & 10.17 (0.20)& 4.99 (0.13)& 5.58 (0.18)\\
contaminated &16.26 (0.49) & 5.84 (0.13) & 15.60 (1.32)\\
\hline
actual values& 10& 5&16.67\\
uncontaminated &  10.44 (0.37)& 4.80 (0.06)& 17.57 (0.83)\\
contaminated &18.89 (0.84) &  5.09 (0.07)& 39.18 (3.05)\\
\hline
actual values& 10& 0.1&5\\
uncontaminated & 9.70 (0.17) &0.21 (0.03) & 5.26 (0.13)\\
contaminated &16.09 (0.35) &0.52 (0.04)  & 10.30 (0.34)\\
\hline
actual values& 10& 0.1&16.67\\
uncontaminated &  9.99 (0.34)& 0.11 (0.01)&17.29 (0.67) \\
contaminated & 17.96 (0.77)&0.16 (0.01)  & 29.99 (1.32)\\
\hline
\end{tabular}
&

\begin{tabular}{ c c c} \hline \hline
   \multicolumn{3}{c}{approximate RL} \\
            \cline{1-3}
$\sigma_\mathfrak{s}^2$ &  $\sigma_e^2$& $\rho$ \\ \hline\hline
2 &5&5\\
 2.39 (0.15)&4.90 (0.09)&13.51 (1.74)\\
 11.60 (0.41) &5.33 (0.07)& 122.0 (6.49)\\
\hline
2&5 &16.67\\
 2.17 (0.09)&4.91 (0.06)& 38.11 (2.53)\\
 15.14 (0.83)&5.01 (0.05) &169.7 (11.2)\\
\hline
2 & 0.1&5\\
 1.97 (0.03)&0.232 (0.01)& 10.24 (0.25)\\
10.05 (0.20) & 0.346  (0.01) & 63.89 (2.21)\\
\hline
 2& 0.1&16.67\\
  2.01 (0.08)&0.144 (0.00)&34.84 (1.67) \\
14.16 (0.79)&0.176 (0.00) &153.7 (9.00) \\
\hline
 10& 5&5\\
 9.98 (0.21)&5.63 (0.10)& 10.87 (0.41)\\
17.39 (0.39) & 6.18 (0.11)& 26.22 (1.27)\\
\hline
 10& 5&16.67\\
 10.54 (0.43) &5.08 (0.06) &35.60 (2.05) \\
19.29 (0.87)&5.28 (0.06) &66.12 (4.24) \\
\hline
10& 0.1&5\\
 9.65 (0.17)& 0.76 (0.03)& 10.34 (0.28)\\
 16.66 (0.37) & 1.05 (0.04)&21.14 (0.64) \\
\hline
10& 0.1&16.67\\
10.02 (0.37) & 0.32 (0.01)&32.70 (1.37) \\
19.70 (1.13) &0.37 (0.01)&61.49 (5.08) \\
\hline
\end{tabular}
\end{tabular}
\end{table}

\subsubsection{Contamination by changing $\rho$ partway through the series}\label{changerhosec}

Suppose a draw from an intercept-only mixed linear model with GP-distributed random effect is contaminated by replacing a substantial subseries with a draw from another GP with the same $\sigma_{\mathfrak{s}}^2$ and $\sigma_e^2$ but with a different range $\rho$. How does this contamination affect the estimates of the GP parameters?  Recalling how $a_j(\rho)$ depends on $\rho$, we conjecture that:\\
1. When the contamination is by a GP with higher $\rho$, the ${v}_j^2$'s for small $j$'s (low frequency columns of $Z$) will be inflated, which will inflate $\hat{\sigma}_\mathfrak{s}^2$. The ${v}_j^2$'s for larger $j$'s will be comparatively unaffected, so $\hat{\sigma}_e^2$ will be comparatively unaffected.  Also, the ${v}_j^2$'s will decline more sharply with $j$, inflating $\hat{\rho}$.
\\2.  When the contamination is by a GP with smaller $\rho$, the ${v}_j^2$'s for large $j$'s (high frequencies) will be inflated, leading to an inflated $\hat{\sigma}_e^2$. The ${v}_j^2$'s for smaller $j$ will be comparatively unaffected, so $\hat{\sigma}_\mathfrak{s}^2$ will be unaffected too. $\hat{\rho}$ will be diminished to capture the more gradual decline of the ${v}_j^2$'s with $j$.

In the simulated data, on average contamination by a GP with higher $\rho$ inflates $\hat{\rho}$,  contamination by a GP with a smaller $\rho$ diminishes $\hat{\rho}$, and $\hat{\sigma}_\mathfrak{s}^2$ and $\hat{\sigma}_e^2$ are largely unaffected, all as expected (Table 3 below).  Figure 19c above shows the change in the ${v}_j^2$'s for one simulated dataset, for which it is not clear what the net effect on the estimates will be (Table 1 above).

\begin{table}[H]
\small
\caption{Average over 100 simulated datasets of estimates maximizing the exact and approximate restricted likelihoods: contaminated by GP with $\rho$=5 and $\rho$=16.67 for uncontaminated GP with $\rho$=16.67 and $\rho$=5 respectively. Monte Carlo standard errors are given in parentheses.}\label{table4}
\begin{tabular}{c c}

\begin{tabular}{ c c c c } \hline \hline
& \multicolumn{3}{c}{exact RL} \\
            \cline{2-4}
&$\sigma_\mathfrak{s}^2$ &  $\sigma_e^2$& $\rho$ \\ \hline\hline
actual values &2 &5&5\\
uncontaminated & 2.29 (0.11) &4.75 (0.11)&6.89 (0.82)\\
contaminated & 2.22 (0.11) &4.79 (0.09) & 14.62 (3.45)\\
\hline
actual values& 2&5 &16.67\\
uncontaminated   & 2.16 (0.09) & 4.89 (0.07)& 26.12 (4.32)\\
contaminated  & 2.19 (0.08) &4.90 (0.06)&18.17 (2.42) \\
\hline
actual values&2 & 0.1&5\\
uncontaminated  & 1.94 (0.03) &0.099 (0.01)&4.97 (0.13) \\
contaminated & 1.93 (0.04) &0.116 (0.01)&6.15 (0.18) \\
\hline
actual values & 2& 0.1&16.67\\
 uncontaminated  & 2.09 (0.07) &0.093 (0.00)& 18.07 (0.84)\\
contaminated  & 2.00 (0.06) &0.110 (0.01)& 11.62(0.44)\\
\hline

actual values& 10& 5&5\\
uncontaminated & 10.17 (0.20)& 4.99 (0.13)& 5.58 (0.18)\\
contaminated & 10.11 (0.24)& 4.89 (0.12) &6.74 (0.31) \\
\hline
actual values& 10& 5&16.67\\
uncontaminated &  10.44 (0.37)& 4.80 (0.06)& 17.57 (0.83)\\
contaminated &10.19 (0.31) & 5.09 (0.09) & 12.49 (0.66)\\
\hline
actual values& 10& 0.1&5\\
uncontaminated & 9.70 (0.17) &0.21 (0.03) & 5.26 (0.13)\\
contaminated &9.70 (0.19) & 0.27 (0.03) & 6.66 (0.18)\\
\hline
actual values& 10& 0.1&16.67\\
uncontaminated &  9.99 (0.34)& 0.11 (0.01)&17.29 (0.67) \\
contaminated &9.67 (0.29) & 0.19 (0.02) & 11.29 (0.38)\\
\hline
\end{tabular}
&

\begin{tabular}{ c c c} \hline \hline
  \multicolumn{3}{c}{approximate RL} \\
            \cline{1-3}
$\sigma_\mathfrak{s}^2$ &  $\sigma_e^2$& $\rho$ \\ \hline\hline
2 &5&5\\
 2.39 (0.15)&4.90 (0.09)&13.51 (1.74)\\
 2.28 (0.09)&4.94 (0.08)&13.42 (0.66)\\
\hline
2&5 &16.67\\
2.17 (0.09)&4.91 (0.06)& 38.11 (2.53)\\
 2.28 (0.09) &4.69 (0.07)&18.65 (1.39) \\
\hline
2 & 0.1&5\\
 1.97 (0.03)&0.232 (0.01)& 10.24 (0.25)\\
2.03 (0.04)&0.223 (0.01)&12.99 (0.36) \\
\hline
 2& 0.1&16.67\\
  2.01 (0.08)&0.144 (0.00)&34.84 (1.67) \\
1.95 (0.06)&0.183 (0.00)&22.55 (0.92) \\
\hline
 10& 5&5\\
 9.98 (0.21)&5.63 (0.10)& 10.87 (0.41)\\
10.16 (0.23)&5.48 (0.10) &13.55 (0.49) \\
\hline
10& 5&16.67\\
10.54 (0.43) &5.08 (0.06) &35.60 (2.05) \\
 9.92 (0.32)&5.43 (0.08) &24.39 (1.03) \\
\hline
 10& 0.1&5\\
  9.65 (0.17)& 0.76 (0.03)& 10.34 (0.28)\\
 9.50 (0.20)& 0.73 (0.03)&12.67 (0.38) \\
\hline
 10& 0.1&16.67\\
10.02 (0.37) & 0.32 (0.01)&32.70 (1.37)\\
10.22 (0.29) &0.49 (0.02) & 23.69 (1.01)\\
\hline
\end{tabular}
\end{tabular}
\end{table}

\section{Web Appendix D. Two dimensional Gaussian process model}\label{sec:2dmodel}
This section describes the spectral representation of GPs in two dimensions and develops the 2-dimensional (2-D) approximate restricted likelihood.  Web Appendix E gives proofs that the approximation is valid and that the columns of $Z$ are orthogonal to each other and to a vector of 1s (the fixed-effect design column for the intercept).

Suppose observations are made on a 2-D grid with grid points at $[\{s_{1,1}, s_{1,2},...,s_{1,M_1}\}\times$ \newline $\{s_{2,1},s_{2,2},...,s_{2,M_2}\}]$ and
suppose the vector of observations is given by \vspace*{-0.5cm}\begin{equation}\label{eq:locations}y=(y( s_{1,1},s_{2,1}), y( s_{1,1},  s_{2,2}), ..., y( s_{1,1}, s_{2,M_2}),......., y(s_{1,M_1},  s_{2,M_2}))'.\vspace*{-0.5cm}\end{equation}
Given data $y(s)$ at location $s \in \{( s_{1,1},s_{2,1}), ( s_{1,1},  s_{2,2}), ..., ( s_{1,1}, s_{2,M_2}),......., (s_{1,M_1},  s_{2,M_2})\}$, we want to fit the model\vspace*{-0.5cm}
$$y(s)=x(s)\beta+w(s)+\epsilon(s)\vspace*{-0.5cm}$$  where all variables are defined analogously to the 1-dimensional (1-D) model.  Similarly, the linear mixed model to be fit to $y$ is\vspace*{-0.5cm}
\begin{equation}\label{eq:initial model}
y=X\beta+I_{M_1M_2}\gamma+\epsilon.\vspace*{-0.5cm}
\end{equation}
The restricted likelihood has the same form as in the 1-D case, and we obtain a matrix-free approximation to it by approximating the intercept-only GP using spectral basis functions.  If the model includes fixed effects $X$, we regress them out as in the 1-D case, assume (approximately) that the residuals follow a linear mixed model with a random effect having a GP covariance, and then approximate the GP in this model using spectral basis functions. Section \ref{sec:spec} below describes the spectral approximation following Wikle (2002) and Paciorek (2007);  Section \ref{sec:2dlikelihood} below derives the simple approximate restricted likelihood arising from applying the spectral approximation to the intercept-only GP.

\subsection{The spectral approximation in 2-D}\label{sec:spec}

Define $(\overset{1}{\omega}_{m_1}, \overset{2}{\omega}_{m_2}) \in\{0,\frac{1}{M_1},\frac{2}{M_1},...,\frac{1}{2},$ $-\frac{1}{2}+\frac{1}{M_1},-\frac{1}{2}+\frac{2}{M_1},...,-\frac{1}{M_1}\} \times \{0,\frac{1}{M_2},\frac{2}{M_2},...,\frac{1}{2},-\frac{1}{2}+\frac{1}{M_2},-\frac{1}{2}+\frac{2}{M_2},...,-\frac{1}{M_2}\}$, $m_1=0,1,...,M_1-1;$ $m_2=0,1,...,M_2-1.$ At observation location $(s_{1,\imath}, s_{2,j})$, $s_{1,\imath} \in \{1,2,..,M_1\}$, $s_{2,j} \in \{1,2,..,M_2\}$, $\imath=1,...,M_1,$ $j=1,...,M_2,$ the mean zero stationary isotropic GP can be approximated, using spectral basis functions, by \vspace*{-0.4cm}
$$g(s_{1,\imath}, s_{2,j})=\sum\limits_{m_1=0}^{M_1-1} \sum\limits_{m_2=0}^{M_2-1}{
\varphi_{m_1, m_2}(2\pi s_{1,\imath},2\pi s_{2,j}) u_{m_1,m_2}}\vspace*{-0.4cm}$$ \vspace*{-0.4cm}
$$\overset{\mbox{def}}{=}\sum\limits_{m_1=0}^{M_1-1} \sum\limits_{m_2=0}^{M_2-1}\exp (i 2 \pi (\overset{1}{\omega}_{m_1}s_{1,\imath}+\overset{2}{\omega}_{m_2}s_{2,j}))
 (a_{m_1,m_2}+i b_{m_1,m_2})$$ where the $a_m$'s and $b_m$'s have independent mean zero Gaussian prior distributions with
 V($a_{m_1,m_2}$) = V($b_{m_1,m_2}$) = $\frac{1}{2 M_1 M_2}\sigma_{\mathfrak s}^2\phi(\overset{1}{\omega}_{m_1}, \overset{2}{\omega}_{m_2};\rho)$, where $\phi(\overset{1}{\omega}_{m_1}, \overset{2}{\omega}_{m_2};\rho)$ is the spectral density function of $K(d;\rho)$, described further below. Note that the observation locations $s_{1,\imath} \in \{1,2,..,M_1\}$, $s_{2,j} \in \{1,2,..,M_2\}$ lie on a uniformly spaced rectangular 2-D grid and the grid sizes in the two dimensions, $M_1$ and $M_2$, must be even integers.

 To approximate real valued processes (Wikle 2002, Paciorek 2007), assume \vspace*{-0.7cm}$$u_{m_1,m_2}=\overline{u}_{M_1-m_1,M_2-m_2}\vspace*{-0.8cm}$$ for\vspace*{-0.7cm}
 \begin{itemize}
 \item $m_1=1,2,..,M_1/2-1$, $m_2=1,2,..,M_2/2$\vspace*{-0.5cm}
 \item $m_1=1,2,..,M_1/2$, $m_2=M_2/2+1,M_2/2+2,..,M_2-1$.\vspace*{-0.5cm}
  \end{itemize}Finally, assume \vspace*{-0.8cm}$$u_{m_1,m_2}=\overline{u}_{m_1,M_2-m_2},\mbox{ for }m_1=0,\mbox{ }m_2=1,2,..,M_2/2-1\vspace*{-0.7cm}$$ and \vspace*{-0.8cm}$$u_{m_1,m_2}=\overline{u}_{M_1-m_1,m_2},\mbox{ for }m_1=1,2,..,M_1/2-1,\mbox{ }m_2=0.\vspace*{-0.5cm}$$ Also assume \vspace*{-0.5cm} $$b_{00}=b_{0,M_2/2}=b_{M_1/2,0}=b_{M_1/2,M_2/2}=0.\vspace*{-0.5cm}$$
 Then defining $\bm\omega=(\overset{1}{\omega}_{m_1},\overset{2}{\omega}_{m_2})'$ and $\bm s =(s_{1,\imath}, s_{2,j})'$, the approximation $g(s_{1,\imath},s_{2,j})$ becomes\vspace*{-0.5cm}
 $$g(s_{1,\imath},s_{2,j}) = 2  \sum\limits_{(m_1,m_2) \in Q}^{} \left(
 a_{m_1,m_2} \cos( 2 \pi \bm{\omega}'\bm{s})-b_{m_1,m_2}\sin( 2 \pi \bm{\omega}'\bm{s})\right)\vspace*{-0.5cm}$$\vspace*{-0.9cm}
$$+ \mbox{ } a_{00}\vspace*{-0.7cm}$$ \vspace*{-0.5cm}$$+ \mbox{ } a_{0,M_2/2}\cos(2 \pi (
\overset{1}{\omega}_{0}s_{1,\imath} +\overset{2}{\omega}_{M_2/2}s_{2,j}))+a_{M_1/2,0}\cos(2\pi( \overset{1}{\omega}_{M_1/2} s_{1,\imath}+ \overset{2}{\omega}_{0}s_{2,j}))\vspace*{-0.5cm}$$\vspace*{-0.5cm}\begin{equation}\label{eq:spec}
+ \mbox{ } a_{M_1/2,M_2/2}\cos(2\pi( \overset{1}{\omega}_{M_1/2} s_{1,\imath}+ \overset{2}{\omega}_{M_2/2}s_{2,j})),\vspace*{-0.5cm}\end{equation}
where $Q= Q_1 \cup Q_2 \cup Q_3 \cup Q_4$,  with \\$Q_1=\{(m_1,m_2): m_1=1,2,...,M_1/2-1; m_2=1,2,...,M_2/2\}$, \\$Q_2=\{(m_1,m_2): m_1=1,2,...,M_1/2; m_2=M_2/2+1,M_2/2+2,...,M_2-1\}$, \\$Q_3=\{(m_1,m_2): m_1=0; m_2=1,2,...,M_2/2-1\}$, \\$Q_4=\{(m_1,m_2): m_1=1,2,...,M_1/2-1; m_2=0\}.$

\noindent Then \vspace*{-1cm}$$\left(g(s_{1,1},s_{2,1}), g(s_{1,1},s_{2,2}), ..., g(s_{1,1},s_{2,M_2}),..., g( s_{1,M_1}, s_{2,M_2})\right)'=1_{M_1M_2}\beta^*+Zu,\vspace*{-0.5cm}$$
 where $\beta^*$ is the intercept, $Z$ is an $M_1 M_2 \times (M_1M_2-1)$ matrix described below, $u=(a_{11}, b_{11}, a_{12}, b_{12},$ ..,$a_{1,\frac{M_2}{2}}, b_{1,\frac{M_2}{2}}, $
 $.......,a_{\frac{M_1}{2}-1,1}, b_{\frac{M_1}{2}-1,1},.., a_{\frac{M_1}{2}-1,\frac{M_2}{2}},b_{\frac{M_1}{2}-1,\frac{M_2}{2}},$
 \\ \hspace*{0.9cm}$a_{1,\frac{M_2}{2}+1}, b_{1,\frac{M_2}{2}+1},a_{1,\frac{M_2}{2}+2},b_{1,\frac{M_2}{2}+2},..,  a_{1,M_2-1},b_{1,M_2-1}, $
  $.......,$
\\ \hspace*{1.2cm}$a_{\frac{M_1}{2},\frac{M_2}{2}+1},b_{\frac{M_1}{2},\frac{M_2}{2}+1},.., a_{\frac{M_1}{2},M_2-1},b_{\frac{M_1}{2},M_2-1}$,
  \\ \hspace*{0.9cm}$a_{01},b_{01},a_{02},b_{02},...,a_{0,M_2/2-1}, b_{0,M_2/2-1}$,
  \\ \hspace*{0.9cm}$a_{10},b_{10},a_{20},b_{20},...,a_{M_1/2-1,0}, b_{M_1/2-1,0}$,
  \\ \hspace*{0.9cm}$a_{0,M_2/2}, a_{M_1/2,0}, a_{M_1/2,M_2/2})'$
 , and \begin{enumerate}
 \item V($a_{m_1,m_2}$) = V($b_{m_1,m_2}$) = $\frac{1}{2M_1M_2}\sigma_{\mathfrak s}^2\phi(\overset{1}{\omega}_{m_1}, \overset{2}{\omega}_{m_2};\rho)$  for $m_1=0,1,..,M_1-1$, $m_2=0,1,..,M_2-1$, $(m_1,m_2)\notin \{(0,0),(0,M_2/2),(M_1/2,0),(M_1/2,M_2/2)\}.$
    \item V($a_{0,M_2/2}$) = $\frac{1}{M_1M_2}\sigma_{\mathfrak s}^2\phi(\overset{1}{\omega}_{0}, \overset{2}{\omega}_{M_2/2};\rho)$, V($a_{M_1/2,0}$) = $\frac{1}{M_1M_2}\sigma_{\mathfrak s}^2\phi(\overset{1}{\omega}_{M_1/2}, \overset{2}{\omega}_{0};\rho).$
     \item V($a_{M_1/2,M_2/2}$) = $\frac{1}{M_1M_2}\sigma_{\mathfrak s}^2\phi(\overset{1}{\omega}_{M_1/2}, \overset{2}{\omega}_{M_2/2};\rho).$

\end{enumerate}
\textbf{Constructing $G$}
\\In this construction, the covariance of the random effect vector $u$ is a matrix $G$ with all off-diagonal elements zero and diagonal elements given by the above variances.

The function $\phi(\bm\omega; \rho)$ is the spectral density, in two dimensions, of the correlation function of the GP being approximated.  For example, the Mat\'{e}rn correlation function with smoothness $\nu$, for Euclidean distances $d$, is\vspace*{-0.5cm} \begin{equation}\label{eq:exp_cov}K(d;\rho,\nu)=\frac{1}{\Gamma(\nu) 2^{\nu-1}} {\left(\frac{\sqrt{2\nu} d}{\rho}\right)}^\nu K_\nu \left(\frac{\sqrt{2\nu} d}{\rho}\right),\vspace*{-0.5cm}\end{equation}
and it has spectral density\vspace*{-0.5cm}
$$\phi(\bm\omega;\rho,\nu)=\frac{\Gamma(\nu+D/2){(4 \nu)}^\nu}{\pi^{D/2}\Gamma(\nu){(\pi \rho)}^{2 \nu}} {\left( \frac{4\nu}{(\pi \rho)^2}+\bm\omega'\bm\omega \right)}^{-(\nu+D/2)},\vspace*{-0.5cm}$$where $D$ is the dimension of the process (Paciorek 2007). For the exponential correlation function (Mat\'{e}rn with $\nu=0.5$) in two dimensions, the spectral density is a Cauchy density \vspace*{-0.5cm}
 \begin{equation}\label{eq:exp_cov} \phi(\bm\omega; \rho) = \frac{\pi \rho ^{2}}{4} {\left(1+\frac{(\pi \rho)^2}{2} \bm\omega'\bm\omega\right)}^{-3/2}. \vspace*{-0.5cm}\end{equation}The elements of $u$ are ordered so that $\phi(\bm{\omega_j}; \rho)$ is a non-increasing function of $j$ (this ordering will be the same irrespective of the value of $\rho$).
\\ \textbf{Constructing $Z$}
\\After constructing $u$, the matrix $Z$ is constructed conformably from (\ref{eq:spec}) as follows.

Consider the vector of length $(M_1/2-1)M_2/2$ \vspace*{-0.6cm}$$[2\cos(\overset{1}{\omega}_{m_1}2\pi s_{1,\imath}+\overset{2}{\omega}_{m_2} 2\pi s_{2,j})]_{m_1=1,2,..,M_1/2-1; m_2=1,2,..,M_2/2}, \mbox{ where }m_2\mbox{ varies fastest}.\vspace*{-0.6cm}$$ Construct one such vector for each $(s_{1,\imath}, s_{2,j})$ combination; this gives us $M_1M_2$ such vectors. Let these vectors be $A_1, A_2,...,A_{M_1M_2}$ respectively. Next, similarly to the above, consider the vector of length $(M_1/2-1)M_2/2$ \vspace*{-0.2cm}$$[-2\sin(\overset{1}{\omega}_{m_1}2\pi s_{1,\imath}+\overset{2}{\omega}_{m_2} 2\pi s_{2,j})]_{m_1=1,2,..,M_1/2-1; m_2=1,2,..,M_2/2}, \small{\mbox{ where }m_2\mbox{ varies fastest.}}\vspace*{-0.2cm}$$ Construct one such vector for each $(s_{1,\imath}, s_{2,j})$ combination; this gives us $M_1M_2$ such vectors. Let these vectors be $B_1, B_2,...,B_{M_1M_2}$ respectively. Now construct a matrix $P_1$ with the $r^{th}$ row of $P_1$ given by $(A_{r1}, B_{r1},A_{r2}, B_{r2}, ..., A_{r,(M_1/2-1)M_2/2}, B_{r,(M_1/2-1)M_2/2})$, where $A_{r t}$ denotes the $t^{th}$ element of the vector $A_r$, $r=1,2,...,M_1M_2$, $t=1,2,...,{(M_1/2-1)}M_2/2$. The matrix $P_1$ constitutes the first ${(M_1/2-1)}{M_2}$ columns of $Z$.

Next, consider the vector of length $M_1/2(M_2/2-1)$ \vspace*{-0.6cm}$$[2\cos(\overset{1}{\omega}_{m_1}2\pi s_{1,\imath}+\overset{2}{\omega}_{m_2} 2\pi s_{2,j})]_{m_1=1,2,..,M_1/2; m_2=M_2/2+1,M_2/2+2,..,M_2-1},\vspace*{-0.6cm}$$ where  $m_2$ varies fastest. Construct one such vector for each $(s_{1,\imath}, s_{2,j})$ combination. Let these vectors be $C_1, C_2,...,C_{M_1M_2}$ respectively. Next, similarly to the above, consider the vector of length $M_1/2(M_2/2-1)$ \vspace*{-0.6cm}$$[-2\sin(\overset{1}{\omega}_{m_1}2\pi s_{1,\imath}+\overset{2}{\omega}_{m_2} 2\pi s_{2,j})]_{m_1=1,2,..,M_1/2; m_2=M_2/2+1,M_2/2+2,..,M_2-1},\vspace*{-0.6cm}$$ where $m_2$ varies fastest. Construct one such vector for each $(s_{1,\imath}, s_{2,j})$ combination; this gives us $M_1M_2$ such vectors. Let these vectors be $D_1, D_2,...,D_{M_1M_2}$ respectively. Now construct a matrix $P_2$ with the $r^{th}$ row of $P_2$ given by $(C_{r1}, D_{r1},C_{r2}, D_{r2}, ..., C_{r,M_1/2(M_2/2-1)}, D_{r,M_1/2(M_2/2-1)})$, where $C_{r t}$ denotes the $t^{th}$ element of the vector $C_r$, $r=1,2,...,M_1M_2$, $t=1,2,...,{M_1/2}(M_2/2-1)$. The matrix $P_2$ constitutes the next ${M_1}{(M_2/2-1)}$ columns of $Z$.

Next, consider the vector \vspace*{-0.7cm}$$[2\cos(2\pi(\overset{1}{\omega}_{0}  s_{1,\imath}+\overset{2}{\omega}_{m_2} s_{2,j}))]_{m_2=1,2,..,M_2/2-1}.\vspace*{-0.7cm}$$ Construct one such vector for every $(s_{1,\imath},s_{2,j})$. Call them $F_1, F_2, ..., F_{M_1M_2}$ respectively.
Consider the vector \vspace*{-0.5cm}$$[-2\sin(2\pi(\overset{1}{\omega}_0  s_{1,\imath}+\overset{2}{\omega}_{m_2} s_{2,j}))]_{m_2=1,2,..,M_2/2-1}.\vspace*{-0.2cm}$$ Construct one such vector for every $(s_{1,\imath},s_{2,j})$. Call them ${H}_1, {H}_2, ..., {H}_{M_1M_2}$ respectively.
Construct a matrix $P_3$ with $M_2-2$ columns as: if $r$ is odd the $r^{th}$ column of $P_3$ is given by $({\frac{r+1}{2}}^{th}$ element of $F_1$, ${\frac{r+1}{2}}^{th}$ element of $F_2$,..., ${\frac{r+1}{2}}^{th}$ element of $F_{M_1M_2})'$;
if $r$ is even the $r^{th}$ column of $P_3$ is given by (${\frac{r}{2}}^{th}$ element of $H_1$,
${\frac{r}{2}}^{th}$ element of $H_2$,..., ${\frac{r}{2}}^{th}$ element of $H_{M_1M_2})'.$
Then, the next $M_2-2$ columns of $Z$ are $P_3$.

Consider the vector \vspace*{-1cm}$$[2\cos(2\pi(\overset{1}{\omega}_{m_1} s_{1,\imath}+\overset{2}{\omega}_0  s_{2,j}))]_{m_1=1,2,..,M_1/2-1}.\vspace*{-0.8cm}$$  Construct one such vector for every $(s_{1,\imath},s_{2,j})$. Call them $\Lambda_1, \Lambda_2, ..., \Lambda_{M_1M_2}$ respectively.
Consider the vector \vspace*{-0.5cm}$$[-2\sin(2\pi(\overset{1}{\omega}_{m_1} s_{1,\imath}+\overset{2}{\omega}_0  s_{2,j}))] _{m_1=1,2,..,M_1/2-1}.\vspace*{-0.5cm}$$ Construct one such vector for every $(s_{1,\imath},s_{2,j})$. Call them $\Delta_1, \Delta_2, ..., \Delta_{M_1M_2}$ respectively.
Construct a matrix $P_4$ with $M_1-2$ columns as: if $r$ is odd the $r^{th}$ column of $P_4$ is given by $({\frac{r+1}{2}}^{th}$ element of $\Lambda_1$, ${\frac{r+1}{2}}^{th}$ element of $\Lambda_2$,..., ${\frac{r+1}{2}}^{th}$ element of $\Lambda_{M_1M_2})'$;
if $i$ is even the $r^{th}$ column of $P_4$ is given by (${\frac{r}{2}}^{th}$ element of $\Delta_1$,
${\frac{r}{2}}^{th}$ element of $\Delta_2$,..., ${\frac{r}{2}}^{th}$ element of $\Delta_{M_1M_2})'.$
Then, the next $M_1-2$ columns of $Z$ are $P_4$.

Finally, construct three more vectors $P_5$, $P_6$, and $P_7$ as follows.
Construct the vector $P_5$ \vspace*{-0.8cm}$$[\cos(\overset{1}{\omega}_02\pi s_{1,\imath}+\overset{2}{\omega}_{M_2/2} 2\pi s_{2,j})]_{\imath=1,2,..,M_1;j=1,2,..,M_2}, \mbox {where }j\mbox{  varies fastest.}\vspace*{-0.8cm}$$
Construct the vector $P_6$ \vspace*{-0.8cm}$$[\cos(\overset{1}{\omega}_{M_1/2}2\pi s_{1,\imath}+\overset{2}{\omega}_{0}2\pi  s_{2,j})]_{\imath=1,2,..,M_1;j=1,2,..,M_2}, \mbox {where }j\mbox{  varies fastest.}\vspace*{-0.8cm}$$
Construct the vector $P_7$ \vspace*{-0.8cm}$$[\cos(\overset{1}{\omega}_{M_1/2}2\pi s_{1,\imath}+\overset{2}{\omega}_{M_2/2}2\pi  s_{2,j})]_{\imath=1,2,..,M_1;j=1,2,..,M_2}, \mbox {where }j\mbox{  varies fastest.}\vspace*{-0.8cm}$$
The last $3$ columns of $Z$ are $P_5, P_6,$ and $P_7$ respectively.

Now order the columns of $Z$ so that $\phi(\bm{\omega_j}; \rho)$ is a non-increasing function of $j$ (this ordering will be the same irrespective of the value of $\rho$).

\subsection{The simple approximate restricted likelihood}\label{sec:2dlikelihood}

With the definitions given above, the approximate model to be fit to the data $y$, if $X$ in Section \ref{sec:2dmodel} above is just a column of 1's, or to be fit to the residuals $y^*$, if $X$ in Section \ref{sec:2dmodel} also includes observed covariates,  is\vspace*{-0.5cm}
      \begin{equation}\label{eq:app_2d}y \mbox{ or }y^*\approx1_{M_1M_2} \beta^* + Z u +\epsilon^*,\vspace*{-0.3cm}\end{equation}
       with $\epsilon$ an $M_1M_2 \times 1$ vector of iid N(0,$\sigma_e^2)$ errors.  (If the left-hand side of equation (\ref{eq:app_2d}) is $y^*$, $\beta^*$ is necessarily be zero.)  $Z$ has these properties: (i)$Z' 1_{M_1 M_2}$=\textbf{0}, and (ii)$Z'Z=M_1 M_2 \mbox{Diag}(2,2, ..., 2,$$1,1,1)$, i.e., the columns of $Z$ are orthogonal to each other and to the constant vector (proofs are in Web Appendix E).
      Premultiplying (\ref{eq:app_2d}) by ${(Z'Z)}^{-0.5}Z'$, under the assumed model, the transformed data vector ${v}={(Z'Z)}^{-0.5}Z' y$ or ${(Z'Z)}^{-0.5}Z' y^*$ has a Normal distribution with $\mbox{E}({v})=0$ and $\mbox{Cov}({v})=\sigma_\mathfrak{s}^2 \mbox{Diag} (a_j(\rho)) +\sigma_e^2 I_{M_1M_2-1}$, for $a_j(\rho) = \phi(\bm {\omega_j}; \rho)$.
From the distribution of $ v_j$, the approximate log restricted likelihood has the matrix-free form \vspace*{-0.5cm}
\begin{equation}\label{eq:lik_vj}
\mbox{ALR}(\sigma_\mathfrak{s}^2,\sigma_e^2, \rho)\mbox{ = const}-\frac{1}{2}\sum\limits_{j=1}^{M_1M_2-1}\left(\log(\sigma_\mathfrak{s}^2 a_j(\rho)+\sigma_e^2)+{v}_j^2{(\sigma_\mathfrak{s}^2 a_j(\rho)+\sigma_e^2)}^{-1}\right). \vspace*{-0.2cm}\end{equation}

By convention, we sort the $a_j(\rho)$ so they are in non-increasing order as $j$ increases and the columns of $Z$ in the same order.  It is easy to show that this order does not depend on $\rho$.  By construction, the successive column pairs of the spectral basis matrix $Z$ decompose the data into successive frequency components, as in the 1-D case. Thus a high frequency feature, e.g., an outlier, will fall in the space spanned by the columns of $Z$ corresponding to large $j$ and will affect $v_j^2$ for larger $j$, which in turn will affect $\hat{\sigma}_e^2$.  A low frequency feature, for example, a linear or quadratic trend in the data, will fall in the space spanned by the columns of $Z$ corresponding to small $j$ and thus will affect $v_j^2$ for smaller $j$, which in turn will affect $\hat{\sigma}_\mathfrak{s}^2$ and $\hat{\rho}$.  A prominent low-frequency feature will inflate both $\hat{\sigma}_\mathfrak{s}^2$ and $\hat{\rho}$, which have opposite effects on the fit's smoothness, so the net effect of a low frequency feature will depend on the specifics of the data and model.

Note that to get to this point, we have made two approximations: approximating $(I_{M_1 M_2}-P_X)\hspace*{1mm}\Sigma\hspace*{1mm} (I_{M_1 M_2}-P_X)$ + $\sigma_e^2 (I_{M_1M_2} - P_X)$ by $\Sigma$ + $R$, and the spectral approximation.  Web Appendix F shows some examples of these covariance matrices.

\section{Web Appendix E. 2-D spectral approximation: technical details}
\subsection{Proof that it is a valid approximation} The correlation of the approximate process is approximately equal to that of the actual GP for $M_1\rightarrow \infty, M_2\rightarrow \infty$.

Proof: For two dimensional frequencies $\bm\omega$ and two dimensional locations $\bm s$, the spectral density $\phi_2(\bm\omega;\rho)$ of the correlation function $K(\bm s;\rho)$ (correlation of the actual GP) is defined as\vspace*{-0.2cm}
$$\phi_2(\bm\omega;\rho)=\int_{R^2}^{} \exp(-2 \pi i \bm\omega'\bm s) K(\bm s;\rho) \mbox{ d}\bm s.\vspace*{-0.2cm}$$ For a process observed only a discrete uniform integer grid locations, the integral has to be replaced by a sum. Because every frequency outside $[-\frac{1}{2},\frac{1}{2}]$ has an alias in $[-\frac{1}{2},\frac{1}{2}]$, the spectral density is only defined for $\bm\omega$ in $[-\frac{1}{2},\frac{1}{2}]\times [-\frac{1}{2},\frac{1}{2}]$.

Then\vspace*{-0.2cm}
$$\int_{-\frac{1}{2}}^{\frac{1}{2}}\int_{-\frac{1}{2}}^{\frac{1}{2}} \exp(2\pi i \bm\omega'\bm s)\phi_2(\bm\omega;\rho)\mbox{ d}\bm\omega =K(\bm s;\rho).\vspace*{-0.1cm}$$

By the spectral representation theorem,
$g(\bm s)$ follows a Normal distribution with mean 0 and correlation $K(\bm s;\rho)=\frac{1}{M_1M_2}\sum\limits_{m_1=0}^{M_1-1} \sum\limits_{m_2=0}^{M_2-1}\exp(i 2 \pi(\overset{1}{\omega}_{m_1},\overset{2}{\omega}_{m_2})\bm s)\phi_2((\overset{1}{\omega}_{m_1},\overset{2}{\omega}_{m_2});\rho).$
As $M_1\rightarrow \infty, M_2\rightarrow \infty$, this sum is approximately \vspace*{-0.2cm}$$\int_{-\frac{1}{2}}^{\frac{1}{2}}\int_{-\frac{1}{2}}^{\frac{1}{2}} \exp(2\pi i \bm\omega'\bm s)\phi_2(\bm\omega;\rho)\mbox{ d}\bm\omega,\vspace*{-0.2cm}\mbox{ which is equal to $K(\bm s;\rho)$.}$$

\subsection{Orthogonality of columns of $Z$, and of $X=1$ and $Z$} For ${s}_{1,1}=1,{s}_{1,2}=2, ...,{s}_{1,M_1}=M_1$, and $s_{2,1}=1,{s}_{2,2}=2, ...,{s}_{2,M_2}=M_2$, $Z$ satisfies the properties: (i)$Z'\bm{1}=\bm{0}$, (ii)$Z'Z=\mbox{Diag}(2M_1M_2,2M_1M_2, ...,2M_1M_2,M_1M_2,M_1M_2, M_1M_2).$

Proof of (i):
We show below that the first element of $Z'\bm{1}$ is 0; the other elements are also 0 by similar arguments.
\\The first element of $Z'\bm{1}$ is $A_{11}+A_{21}+...+A_{M_1M_2,1}$
\\$=4 \sum\limits_{\imath=1}^{M_1} \sum\limits_{j=1}^{M_2} \cos(\overset{1}{\omega}_1 2 \pi s_{1,\imath}+\overset{2}{\omega}_1 2 \pi s_{2,j})$
\\$=2 \sum\limits_{\imath=1}^{M_1} \sum\limits_{j=1}^{M_2} \left(e^{i(\overset{1}{\omega}_1 2 \pi s_{1,\imath}+\overset{2}{\omega}_1 2 \pi s_{2,j})}+e^{-i(\overset{1}{\omega}_1 2 \pi s_{1,\imath}+\overset{2}{\omega}_1 2 \pi s_{2,j})}\right)$
\\$=2\sum\limits_{\imath=1}^{M_1} \sum\limits_{j=1}^{M_2} e^{i(\overset{1}{\omega}_1 2 \pi s_{1,\imath}+\overset{2}{\omega}_1 2 \pi s_{2,j})} +2\sum\limits_{\imath=1}^{M_1} \sum\limits_{j=1}^{M_2} e^{-i(\overset{1}{\omega}_1 2 \pi s_{1,\imath}+\overset{2}{\omega}_1 2 \pi s_{2,j})}$
\\$=2\sum\limits_{k=1}^{M_1}  e^{c_1k} \sum\limits_{k=1}^{M_2}  e^{c_2k} +2\sum\limits_{k=1}^{M_1 } e^{-c_1k} \sum\limits_{k=1}^{M_2}  e^{-c_2k}$, $c_1=i\overset{1}{\omega}_12\pi$, $c_2=i\overset{2}{\omega}_12\pi$
\\$=2{\left(\frac{e^{c_1}(e^{M_1c_1}-1)}{e^{c_1}-1}\right)}{\left(\frac{e^{c_2}(e^{M_2c_2}-1)}{e^{c_2}-1}\right)} + 2{\left(\frac{e^{-M_1c_1}(e^{M_1c_1}-1)}{e^{c_1}-1}\right)}{\left(\frac{e^{-M_2c_2}(e^{M_2c_2}-1)}{e^{c_2}-1}\right)}$
\\$=0$, since $e^{M_1c_1}=\cos(M_1\overset{1}{\omega}_1 2\pi)+i \sin(M_1\overset{1}{\omega}_1 2\pi)$ $=1+0=1$, and $e^{M_2c_2}=\cos(M_2\overset{2}{\omega}_1 2\pi)+i \sin(M_2\overset{2}{\omega}_1 2\pi)$ $=1+0=1$.

Proof of (ii):
We show below that the ${(1,1)}$ element of $Z'Z$ is $2M_1M_2$, and the ${(1,2)}$ element of $Z'Z$ is $0$; the other elements of $Z'Z$ will be $2M_1M_2$, or $M_1M_2$, or $0$ by similar arguments.
\\The ${(1,1)}$ element of $Z'Z$ is
$A_{11}^2+A_{21}^2+...+A_{M_1M_2,1}^2$
\\$=2 \sum\limits_{\imath=1}^{M_1} \sum\limits_{j=1}^{M_2} \left(\cos (2 (\overset{1}{\omega}_12\pi s_{1,\imath}+\overset{2}{\omega}_12\pi s_{2,j}))+1\right)$
\\$=2M_1M_2+2\sum\limits_{\imath=1}^{M_1} \sum\limits_{j=1}^{M_2} \cos (2 (\overset{1}{\omega}_12\pi s_{1,\imath}+\overset{2}{\omega}_12\pi s_{2,j}))$
\\$= 2M_1M_2+\sum\limits_{\imath=1}^{M_1}\sum\limits_{j=1}^{M_2} e^{i 2 (\overset{1}{\omega}_12\pi s_{1,\imath}+\overset{2}{\omega}_12\pi s_{2,j})}+ \sum\limits_{\imath=1}^{M_1} \sum\limits_{j=1}^{M_2} e^{-i 2 (\overset{1}{\omega}_12\pi s_{1,\imath}+\overset{2}{\omega}_12\pi s_{2,j})}  $
\\$=2M_1M_2+ \sum\limits_{k=1}^{M_1} e^{c_1k} \sum\limits_{k=1}^{M_2} e^{c_2k}+\sum\limits_{k=1}^{M_1} e^{-c_1k} \sum\limits_{k=1}^{M_2} e^{-c_2k}$, $c_1=2i \overset{1}{\omega}_1 2 \pi$, $c_2=2i \overset{2}{\omega}_1 2 \pi$
\\$=2M_1M_2+{\left(\frac{e^{c_1}(e^{M_1c_1-1})}{e^{c_1}-1}\right)}{\left(\frac{e^{c_2}(e^{M_2c_2-1})}{e^{c_2}-1}\right)}    + {\left(\frac{e^{-M_1c_1}(e^{M_1c_1-1})}{e^{-M_1c_1}-1}\right)} {\left(\frac{e^{-M_2c_2}(e^{M_2c_2-1})}{e^{-M_2c_2}-1}\right)}$
\\$=2M_1M_2$, since $e^{M_1c_1}=e^{M_1i\overset{2}{\omega}_1 4\pi}=1$, and $e^{M_2c_2}=e^{M_2i\overset{2}{\omega}_1 4\pi}=1$.

The ${(1,2)}$ element of $Z'Z$ is
$A_{11}B_{11}+A_{21}B_{21}+...+A_{M_1M_2,1}B_{M_1M_2,1}$ \\=$-4 \sum\limits_{\imath=1}^{M_1} \sum\limits_{j=1}^{M_2} \cos (\overset{1}{\omega}_12\pi s_{1,\imath}+\overset{2}{\omega}_12 \pi s_{2,j})\sin (\overset{1}{\omega_1}2\pi s_{1,\imath}+\overset{2}{\omega}_12 \pi s_{2,j})$
\\$=-2 \sum\limits_{\imath=1}^{M_1} \sum\limits_{j=1}^{M_2} \sin (2 (\overset{1}{\omega}_12\pi s_{1,\imath}+\overset{2}{\omega}_12\pi s_{2,j}))$
\\$=-\frac{1}{i}\sum\limits_{\imath=1}^{M_1} \sum\limits_{j=1}^{M_2} e^{i 2 (\overset{1}{\omega}_12\pi s_{1,\imath}+\overset{2}{\omega}_12\pi s_{2,j})}-\frac{1}{i}\sum\limits_{\imath=1}^{M_1} \sum\limits_{j=1}^{M_2} e^{-i 2 (\overset{1}{\omega}_12\pi s_{1,\imath}+\overset{2}{\omega}_12\pi s_{2,j})}$
\\$=-\frac{1}{i}\sum\limits_{k=1}^{M_1} e^{ic_1k}\sum\limits_{k=1}^{M_2} e^{ic_2k} -\frac{1}{i}\sum\limits_{k=1}^{M_1} e^{-ic_1k}\sum\limits_{k=1}^{M_2} e^{-ic_2k}$
\\=$-\frac{1}{i} {\left(\frac{e^{c_1}(e^{M_1c_1-1})}{e^{c_1}-1}\right)}{\left(\frac{e^{c_2}(e^{M_2c_2-1})}{e^{c_2}-1}\right)}-\frac{1}{i} {\left(\frac{e^{-M_1c_1}(e^{M_1c_1-1})}{e^{-M_1c_1}-1}\right)}{\left(\frac{e^{-M_2c_2}(e^{M_2c_2-1})}{e^{-M_2c_2}-1}\right)}$
$=0$.

\section{Web Appendix F. Covariance matrices in 2-D}
The figures in this appendix show some exact, approximate, and estimated covariance matrices for our model in 2-D with the spectral approximation, referred to in the last paragraph of Web Appendix D.
\vspace*{-1.5cm}
\begin{figure}[H]
 \centering
\begin{subfigure}[b]{0.45\textwidth}
                                  \includegraphics[width=\textwidth]{covs_2_0point1_5}\vspace*{-1.8cm}
                 \caption{\\Left: exact covariance matrix (including the iid errors). Right: approximate covariance matrix. $\sigma_\mathfrak{s}^2=2, \sigma_e^2=0.1, \rho=5$.}
                 \label{fig:cov}
         \end{subfigure}
\quad
        \begin{subfigure}[b]{0.45\textwidth}
                 \includegraphics[width=\textwidth]{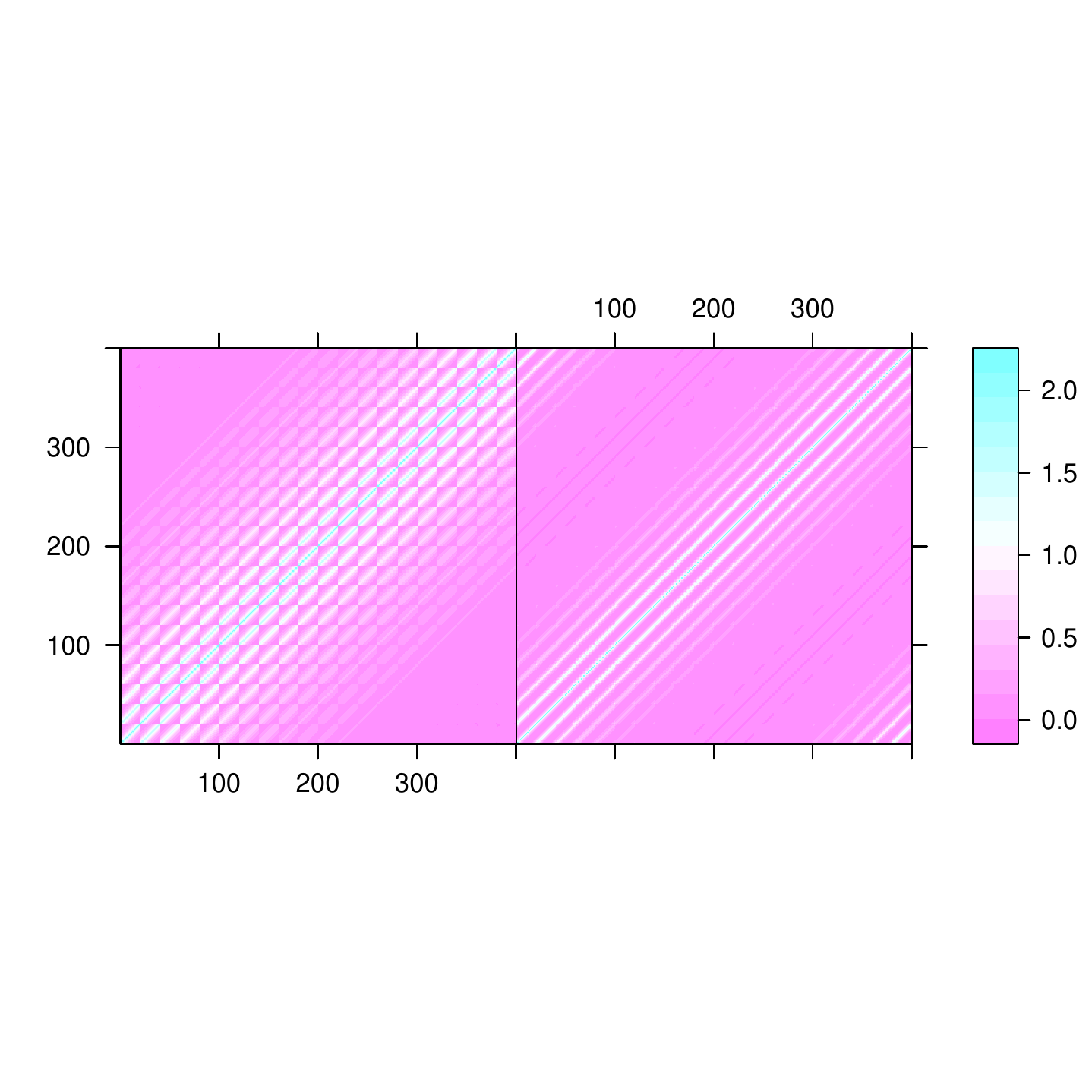}\vspace*{-1.8cm}
                 \caption{Left: Fitted exact covariance matrix using estimates from exact RL. Right: Fitted approximate covariance matrix using estimates from approximate RL.}
               \label{fig:cov_fit}
         \end{subfigure}
         \caption{2-D GP's covariance matrices.}\label{fig:covs_ch2}
\end{figure}

\section{Web Appendix G. Demonstration with simulated data}\label{sec:sim_data_analysis}
This section uses a simulated dataset to show how the plot of $v_j^2$ vs $j$ and added variable plots can be used to examine goodness of fit and potential covariates. We simulate data that depend on covariates, then omit those covariates from the fitted model to see how the missing covariates show themselves in the $v_j^2$ plot and the added variable plots in the observation and spectral domains.

\begin{figure}[H]
         \centering \begin{subfigure}[b]{0.22\textwidth}
 \includegraphics[width=\textwidth]{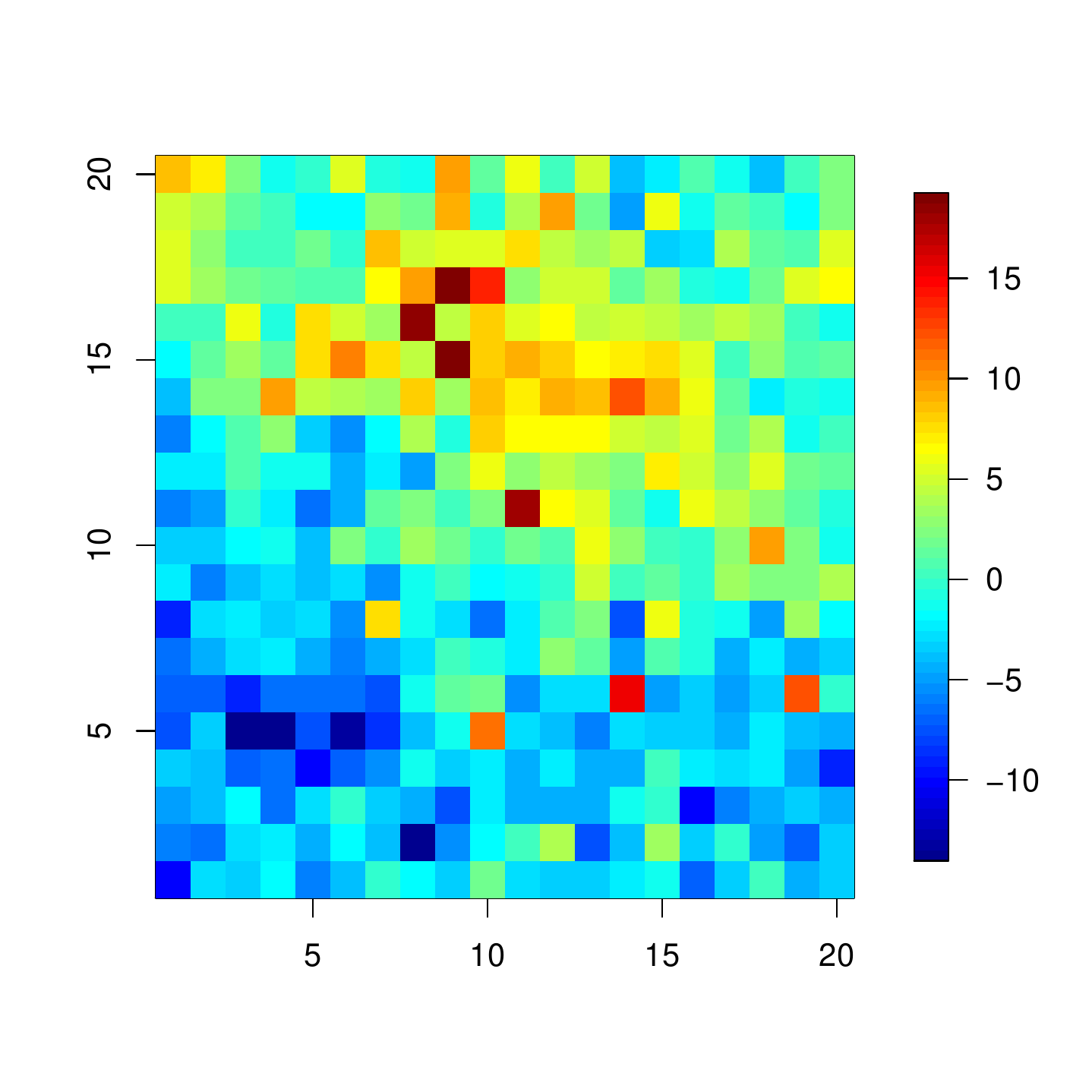}
                 \caption{\\Simulated data with covariates.\\}
                                  \label{fig:sim_fig1}
\end{subfigure}
         \quad
\begin{subfigure}[b]{0.22\textwidth}
                                                  \includegraphics[width=\textwidth]{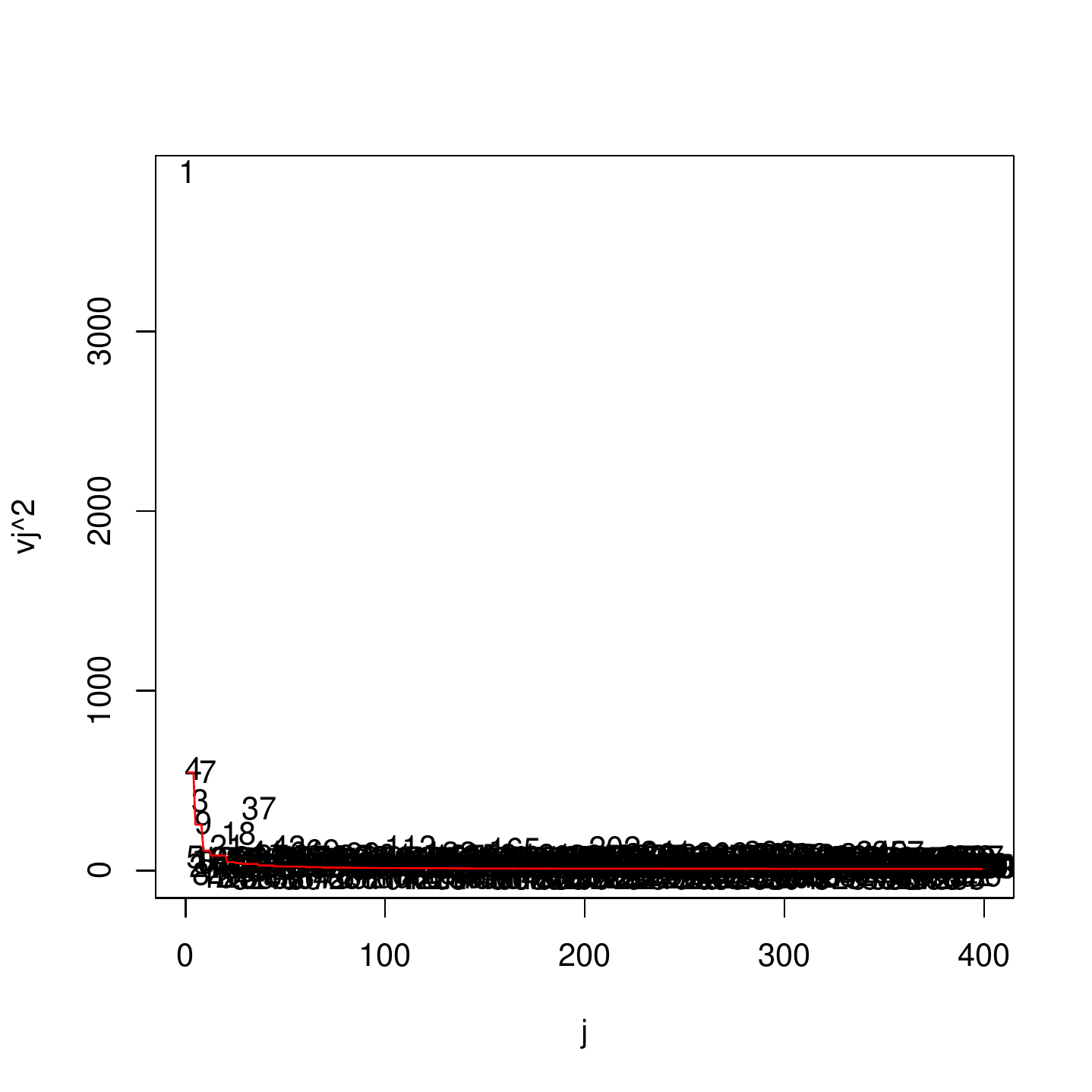}
                 \caption{\\$v_j^2$ vs $j$ for intercept-only fit.\\}
                 \label{fig:sim_fig2}
\end{subfigure}
         \quad
\begin{subfigure}[b]{0.22\textwidth}
                                                  \includegraphics[width=\textwidth]{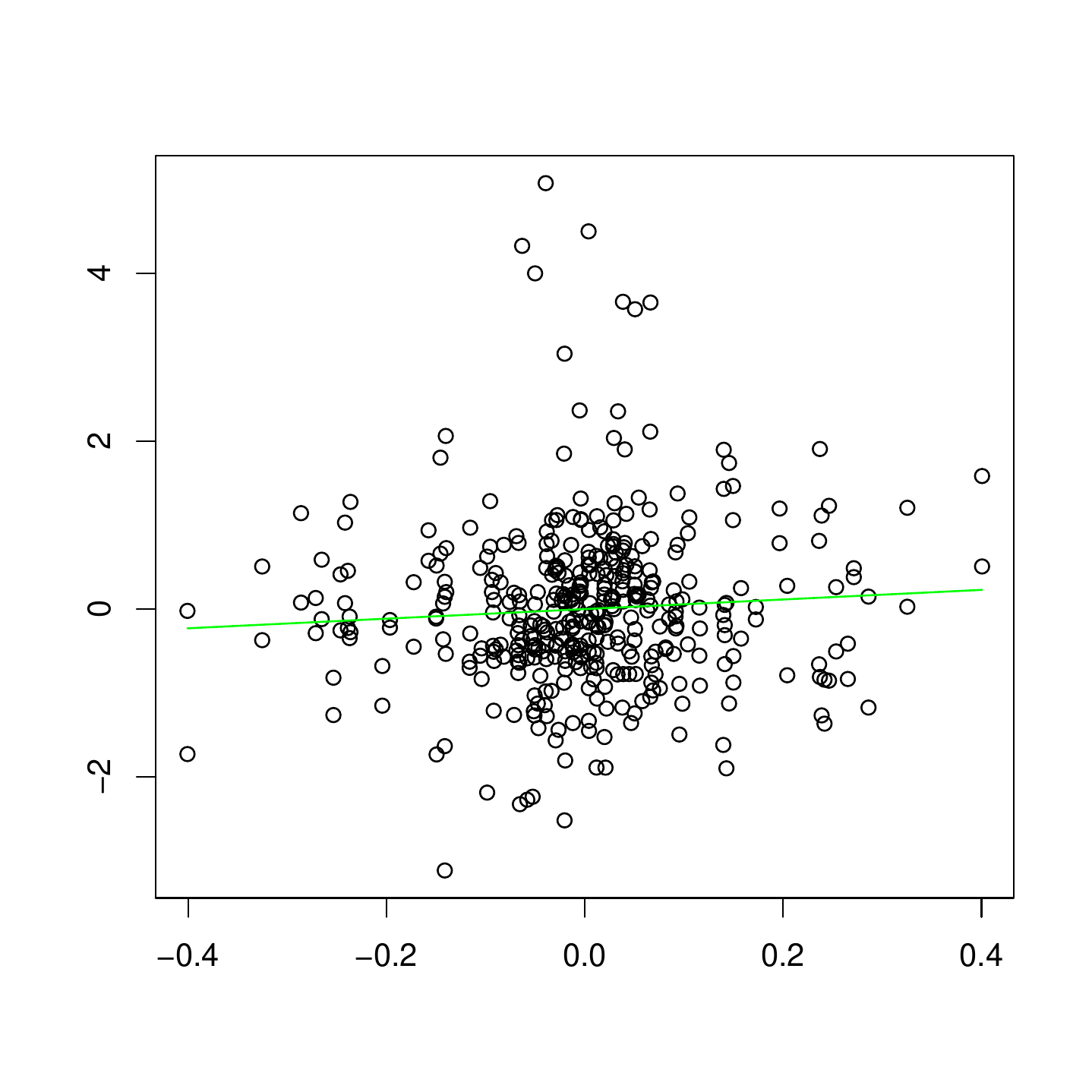}
                 \caption{Added variable plot in observation domain for low frequency covariate.}
                 \label{fig:sim_fig3}
\end{subfigure}
\quad
\begin{subfigure}[b]{0.22\textwidth}
        \includegraphics[width=\textwidth]{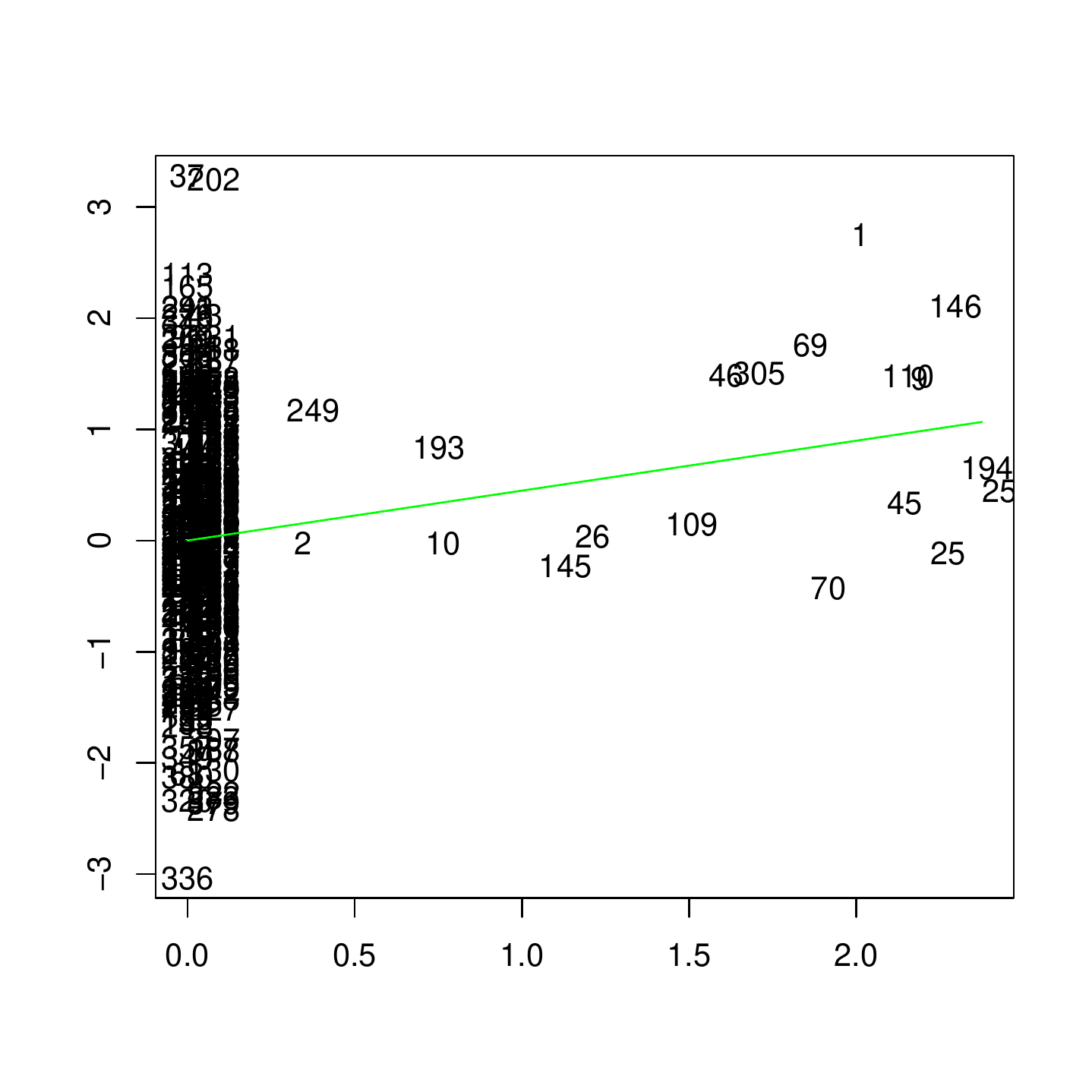}
                 \caption{Added variable plot in spectral domain for low frequency covariate.}
                 \label{fig:sim_fig4}
\end{subfigure}
\quad
\begin{subfigure}[b]{0.22\textwidth}
        \includegraphics[width=\textwidth]{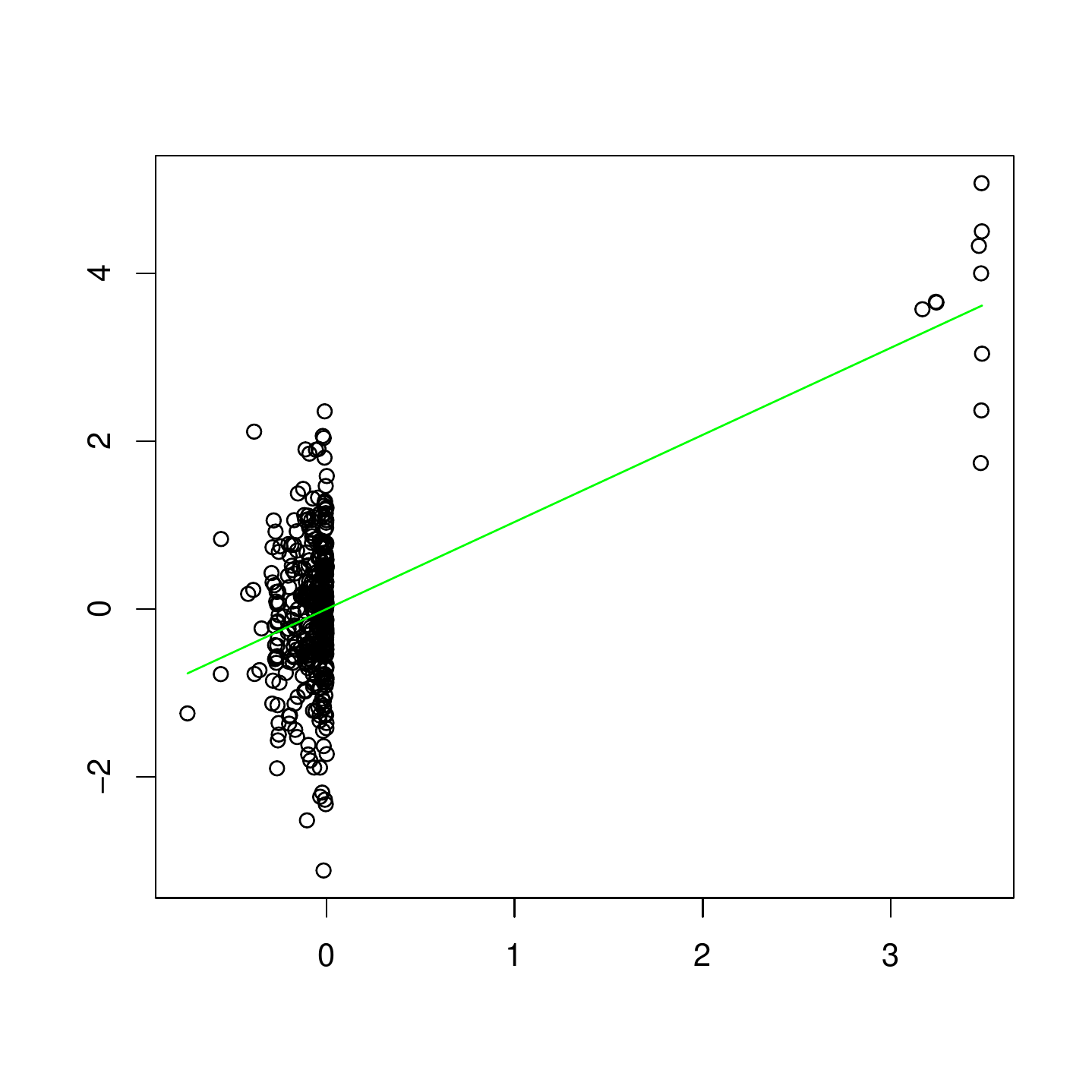}
                 \caption{Added variable plot in observation domain for high frequency covariate.}
                 \label{fig:sim_fig5}
\end{subfigure}
\quad
\begin{subfigure}[b]{0.22\textwidth}
        \includegraphics[width=\textwidth]{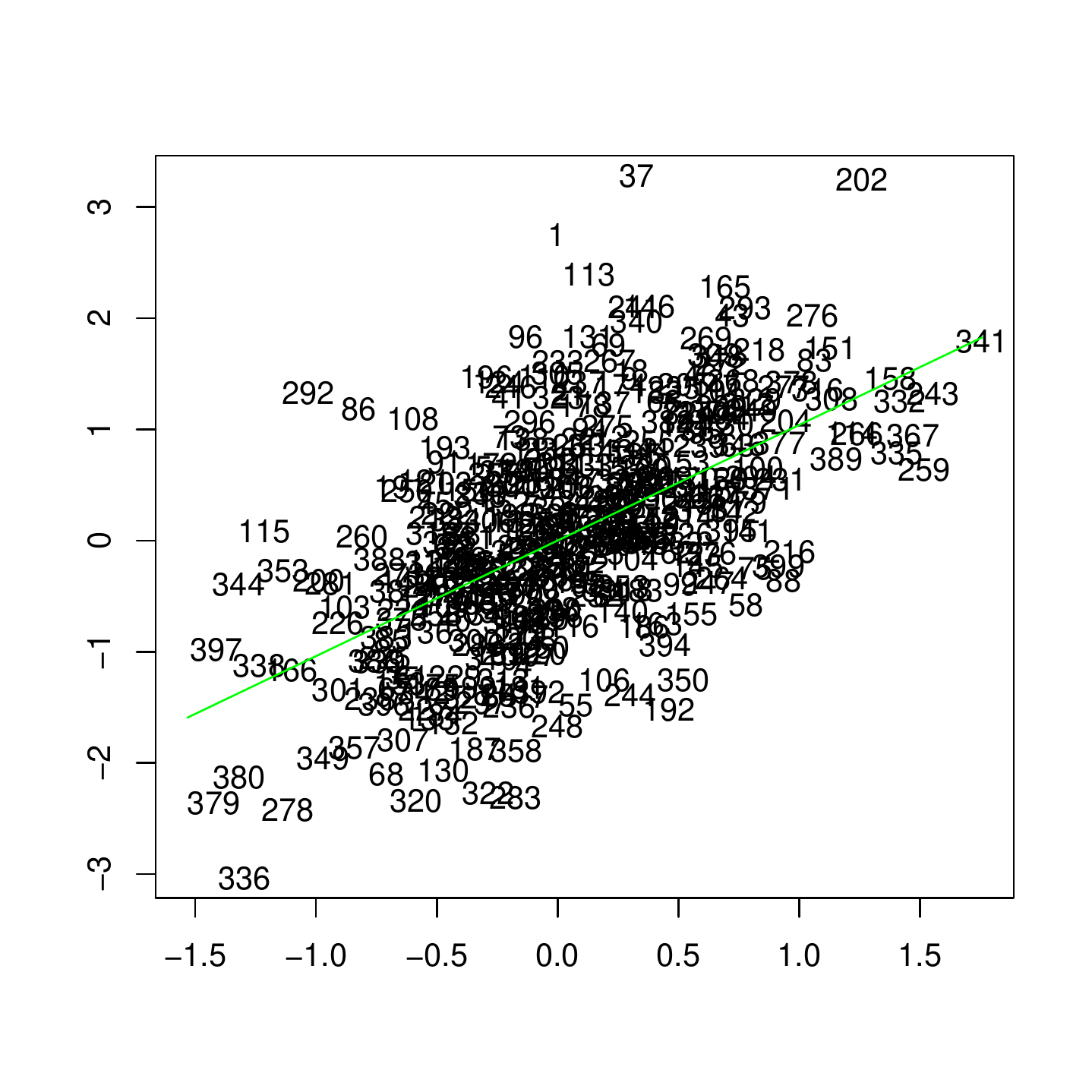}
                 \caption{Added variable plot in spectral domain for high frequency covariate.}
                 \label{fig:sim_fig6}
\end{subfigure}
\quad
\begin{subfigure}[b]{0.22\textwidth}
        \includegraphics[width=\textwidth]{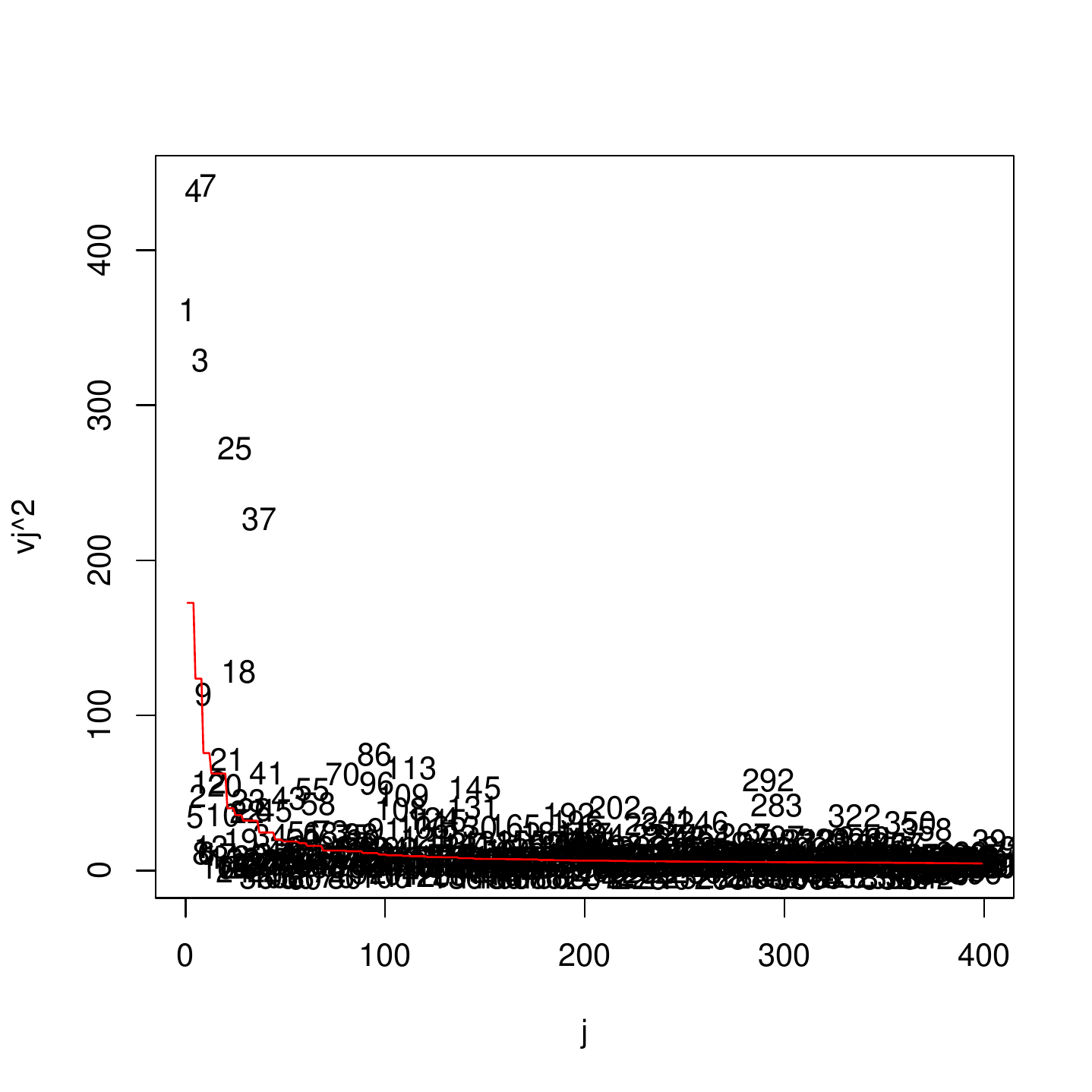}
                 \caption{\\$v_j^2$ vs $j$ for fit with both covariates.\\}
                 \label{fig:sim_fig7}
\end{subfigure}
 \caption{Plots for the analysis of the simulated data.}\label{fig:sim}
         \end{figure}

Consider the data in Figure \ref{fig:sim_fig1}.  Observations were simulated at integer locations on a $20 \times 20$ regularly spaced grid from a GP with the Mat\'{e}rn $\nu=0.5$ correlation function with the GP's variance and range parameters being $\sigma_\mathfrak{s}^2=12$ and $\rho=5$ and with errors distributed as iid Normal(0, $\sigma_e^2=5$). Then two features (covariates) were added as fixed effects to give the simulated data:  The first is a north-south linear trend spanning the entire grid;  the second feature is addition of the value 12 to ten randomly selected locations. Thus the first and second covariates have low and high frequency respectively.

We fit to the data an intercept-only model with the same correlation function used to simulate the data. This fit gave estimates $\hat{\sigma}_\mathfrak{s}^2$, $\hat{\sigma}_e^2$, and $\hat{\rho}$ equal to 21.94, 8.52, and 13.19 respectively. All three estimates are inflated: The omitted low-frequency feature inflates $\hat{\sigma}_\mathfrak{s}^2$ and $\hat{\rho}$ and the omitted high-frequency feature inflates $\hat{\sigma}_e^2$.  Figure \ref{fig:sim_fig2} is the $v_j^2$ plot with $j$ as the plotting character.  In this plot, the smooth line denotes the fit of the $v_j^2$, $\hat{\sigma}_{\mathfrak{s}}^2 a_j(\hat{\rho}) + \hat{\sigma}_e^2$, where the estimates are obtained by maximizing the exact restricted likelihood. In Figure \ref{fig:sim_fig2}, the $v_j^2$ plot has a prominent point for $j=1$, indicating a strong north-south linear trend.  The observation-domain added variable plot for the first covariate (Figure \ref{fig:sim_fig3}), only weakly detects this missing covariate and does not have a significant slope (P = 0.18). In contrast, the spectral-domain added variable plot (Figure \ref{fig:sim_fig4}) gives a strong signal that this covariate belongs in the model (P = 0.0005) and shows the frequencies where the covariate's signal is concentrated.

Now consider the second covariate. This missing high-frequency covariate, with no spatial pattern, is not visible in the $v_j^2$ plot (Figure \ref{fig:sim_fig2}).  However, both added variable plots, in the observation domain (Figure \ref{fig:sim_fig5}) and in the spectral domain (Figure \ref{fig:sim_fig6}), show that the covariate should be included (P $<$ $10^{-15}$ for both), although the former identifies the few observations (outlier locations) that drive the fit while the spectral-domain added variable plot distributes the effect of the outliers diffusely over the $v_j$ corresponding to high frequencies.

Figure \ref{fig:sim_fig7} is the $v_j^2$ plot when both covariates have been included in the model.  This plot shows no sign of lack of fit;  $\hat{\sigma}_\mathfrak{s}^2$, $\hat{\sigma}_e^2$ and $\hat{\rho}$ are now 11.98, 4.03, and 5.28 respectively, close to the values used to simulate the data.

We have shown how a missing high-frequency covariate, i.e., outliers in $y$, can be detected from a observation-domain added variable plot or a spectral-domain added variable plot for that covariate, but is not visible in the $v_j^2$ plot. On the other hand, missing large-scale trends, i.e., non-stationarity in the form of a linear or quadratic trend, can be visible in the $v_j^2$ plot. Once detected, these latter trends may be included as covariates, and if a covariate is available that captures such trends, then an added variable plot for that covariate should have a large slope. The added variable plots in the two domains estimate the same slope for the candidate covariate but have different power for testing the slope. For low-frequency trends, the spectral-domain added variable plot appears to have more power than the observation-domain plot.

For each kind of missing covariate considered above --- which were pure types used for demonstration --- one kind of added variable plot shows how the signal in the data is concentrated in some $y_i$ or $v_j$.  In real datasets, a potential covariate may be a mix of low- and high-frequency features.

\section{Web Appendix H. \hspace*{3mm} Model building in the spectral domain}\label{sec:real_data_part1}

The main paper's Section 7 showed the first of a sequence of model-building steps, illustrating how to use the tools.  This appendix and Appendices I and J show the full sequence of steps considering both the spectral and observation domains.  Again, we make no claim that this is an optimal sequence of steps; we merely intend to illustrate how the tools can be used.

\textbf{Step 1}: First, fit the intercept-only model. Consider the resulting plot of the $v_j^2$ (Figure \ref{fig:step1_vjsq}) in which the plotting symbol is $j$, corresponding to a particular frequency. For example, the point numbered 1 corresponds to a cubic or linear north-south trend in the data (the spectral approximation has no linear-like component), so the prominent $v_1^2$ indicates non-stationarity likely in the form of a north-south trend.   This deviation from stationarity may be addressed by including an appropriate covariate in the model, so we draw spectral-domain added variable plots for all the covariates (Figure \ref{fig:step1_addvar2}).
 In the added variable plots, the colors are: black for $j$ in 1 to 100, \textcolor{red}{red} for $j$ in 101 to 200, \textcolor{green}{green} for $j$ in 201 to 300, \textcolor{blue}{blue} for $j$ in 301 to 400, \textcolor{magenta}{pink} for $j$ in 401 to 559.
 Each covariate (after being pre-smoothed onto the regular grid) has been standardized by subtracting its sample mean and dividing by its sample standard deviation, so the slopes for the covariates are comparable. The p-values for the added variable plot slopes suggest that the fit can be improved by adding covariates (Table \ref{table:table_step1}).
The natural impulse is to include first the covariate having the largest slope in its added variable plot, in this case Elevation. However, since the $v_j^2$ plot for the data has $v_1^2$ as the most prominent point,  corresponding to a north-south trend, we also want the added covariate to explain at least some of this trend, the more the better. For this purpose we calculate Cook's distance for points in the added variable plot, describing their respective influence on the regression slope; we want the point numbered 1 to have one of the largest Cook's distances. Among the potential covariates, Elevation has the smallest p-value for the added variable plot's slope, and point number 1 has the largest Cook's distance for the coefficient of Elevation, i.e., it is most highly influential in determining the slope of Elevation's added variable plot. Therefore we add Elevation as a covariate.

Slope has results similar to Elevation's, so we could have added Slope to the model at this stage. If we had, it turns out that applying the same considerations we would add Elevation in the second step, so it does not matter if we add Elevation or Slope first.

\textbf{Step 2}: Now fit the model with one covariate, Elevation. In Figure \ref{fig:step2_vjsq}'s plot of the $v_j^2$, the $v_j^2$ for low $j$ (low frequences) are not as striking as for the intercept-only model, but some are outstanding, suggesting deviation from stationarity; it seems the model still attributes some low-frequency data components to the GP and perhaps they can be captured using explicit covariates instead. The spectral-domain added variable plots (Figure \ref{fig:step2_addvar2}) have large slopes for several candidate covariates (Table \ref{table:table_step2}).  Because $v_8^2$ is the largest $v_j^2$, we prefer to include a covariate for which the added variable plot's slope is large and which is also influenced substantially by the point $j = 8$, corresponding to a quartic ($4^{th}$ degree polynomial) trend. The covariate SummerTC1 has the smallest p-value for its slope, but the point $j = 8$ does not have a large Cook's distance for SummerTC1. The point $j = 8$ does have a large Cook's distance for the covariate Slope.  Thus, in this step we add both covariates, SummerTC1 and Slope, to the model that includes Elevation.

\textbf{Step 3}: Now fit the model including Elevation, Slope, and SummerTC1.  The $v_j^2$ plot again shows low-frequency trends suggesting deviation from stationarity (Figure \ref{fig:step3_vjsq}); the point $j = 4$, corresponding to a quadratic north-south trend, is most prominent.  Added variable plots for the remaining covariates (Figures \ref{fig:step3_addvar}) show that SpringTC2 has the smallest p-value but $j = 4$ is not particularly influential for its slope (Table \ref{table:table_step3}). The indication of non-stationarity cannot be explained by SpringTC2, so we add a quadratic north-south trend as a covariate along with SpringTC2.  This is tolerable because a quadratic tend has very low frequency; were we collaborating with subject matter experts, such a simple trend might suggest a potential covariate that could be used instead of the non-substantive quadratic.

An alternative may be to add FallTC2 instead of the north-south quadratic trend. If we do that, the point $j = 4$ comes down in the result $v_j^2$ plot but is still prominent (Figure \ref{fig:step3_vjsqfaltc2}), so we add the north-south quadratic trend instead of FallTC2.

\textbf{Step 4}:  Fit the model to the smoothed data including Elevation, Slope, SummerTC1, SpringTC2, and the north-south quadratic trend.  The $v_j^2$ plot now has no prominent points (Figure \ref{fig:step4_vjsq}).  Figure \ref{fig:step4_addvar} shows added variable plots for the remaining covariates;  based on the p-values (Table \ref{table:table_step4}), we add the covariate SummerTC3.

\textbf{Step 5}:  The best-fitting model includes Elevation, Slope, SummerTC1, SpringTC2, a north-south quadratic trend, and SummerTC3.  The $v_j^2$ plot has no prominent points (Figure \ref{fig:step5_vjsq}) and the remaining covariates appear unable to absorb variation currently relegated to the GP or error parts of the model (Table \ref{table:table_step5}, Figure \ref{fig:step5_addvar}).

\newpage

\graphicspath{{gridded/}}
\vspace*{-0.4cm}
\begin{figure}[H]
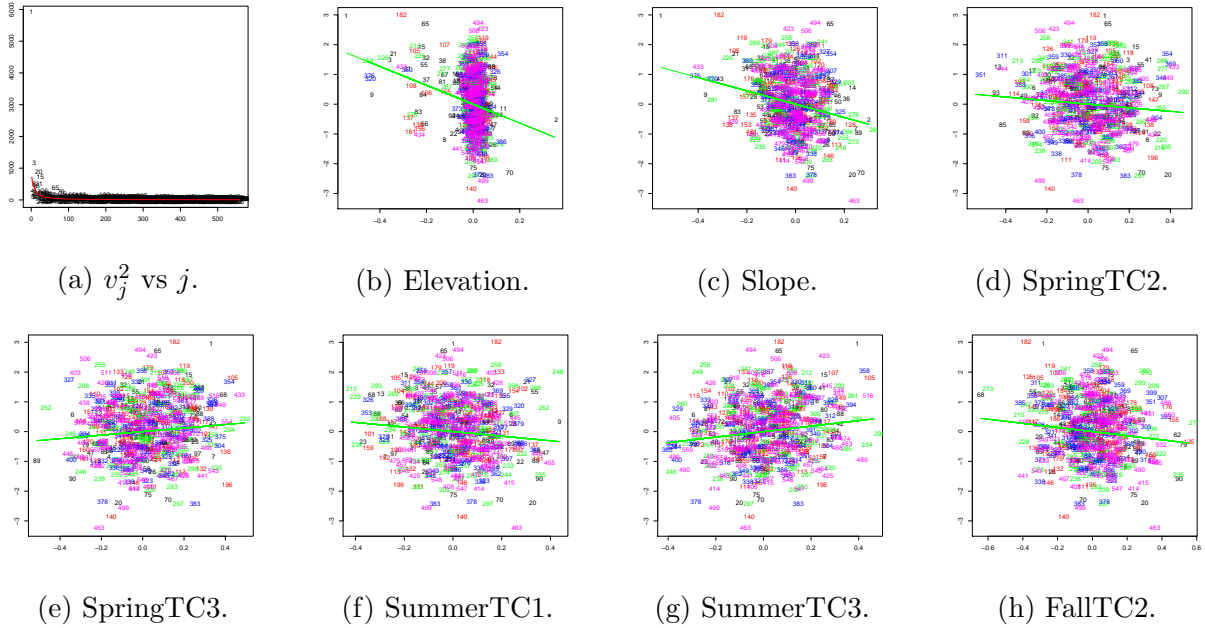

         \centering \begin{subfigure}[b]{0.22\textwidth}
 \includegraphics[width=\textwidth]{RM_new_intonly_vj_gridded}
                 \caption{$v_j^2$ vs $j$.}
                                  \label{fig:step1_vjsq}
\end{subfigure}
\quad
 \begin{subfigure}[b]{0.22\textwidth}
 \includegraphics[width=\textwidth]{RM_new_intonly_vjadvarfor_elevation_gridded}
                 \caption{Elevation.}
                                  \label{fig:RM_add_var1}
\end{subfigure}
         \quad
\begin{subfigure}[b]{0.22\textwidth}
                                                  \includegraphics[width=\textwidth]{RM_new_intonly_vjadvarfor_slope_gridded}
                 \caption{Slope. }
                 \label{fig:RM_add_var2}
\end{subfigure}
         \quad
\begin{subfigure}[b]{0.22\textwidth}
                                                  \includegraphics[width=\textwidth]{RM_new_intonly_vjadvarfor_sprtc2_gridded}
                 \caption{SpringTC2.}
                 \label{fig:RM_add_var3}
\end{subfigure}
\quad

\begin{subfigure}[b]{0.22\textwidth}
                                                  \includegraphics[width=\textwidth]{RM_new_intonly_vjadvarfor_sprtc3_gridded}
                 \caption{SpringTC3.}
                 \label{fig:RM_add_var4}
\end{subfigure}
\quad
         \begin{subfigure}[b]{0.22\textwidth}
 \includegraphics[width=\textwidth]{RM_new_intonly_vjadvarfor_sumtc1_gridded}
                 \caption{SummerTC1.}
                                  \label{fig:RM_add_var1}
\end{subfigure}
         \quad
\begin{subfigure}[b]{0.22\textwidth}
                                                  \includegraphics[width=\textwidth]{RM_new_intonly_vjadvarfor_sumtc3_gridded}
                 \caption{SummerTC3. }
                 \label{fig:RM_add_var2}
\end{subfigure}
         \quad
\begin{subfigure}[b]{0.22\textwidth}
                                                  \includegraphics[width=\textwidth]{RM_new_intonly_vjadvarfor_faltc2_gridded}
                 \caption{FallTC2.}
                 \label{fig:RM_add_var3}
\end{subfigure}
 \caption{\textbf{Step 1} After intercept-only fit; (b) to (h) are spectral-domain added variable plots.}\label{fig:step1_addvar2}
         \end{figure}

              \begin{table}[H]
\centering
  \begin{tabular}{ |c || c |c|c|}
    \hline
    Candidate covariates &  slope & p-value&$j$ with top 5 Cook's dist\\ \hline
    \textbf{Elevation}& -3.17  &$10^{-10}$&1,182,9,181,434\\ \hline
    Slope & -2.24 &$10^{-9}$&1,182,463,70,65\\ \hline
    SpringTC2&-0.60 &0.03&20,499,354,268,369\\ \hline
    SpringTC3 & 0.59&0.02&1,506,196,403,463\\ \hline
    SummerTC1 &-0.77&0.007&248,463,20,268,327\\ \hline
    SummerTC3 & 0.92&0.0004&1,258,378,358,248\\ \hline
    FallTC2 &-0.69&0.004&182,463,1,212,280\\ \hline  \hline
    \end{tabular}
  \caption{ \textbf{Step 1} Slopes of spectral-domain added variable plots iafter the intercept-only fit. The covariate chosen to be added at the next step is in bold.  } \label{table:table_step1}
\end{table}

\begin{figure}[H]
         \centering  \begin{subfigure}[b]{0.22\textwidth}
 \includegraphics[width=\textwidth]{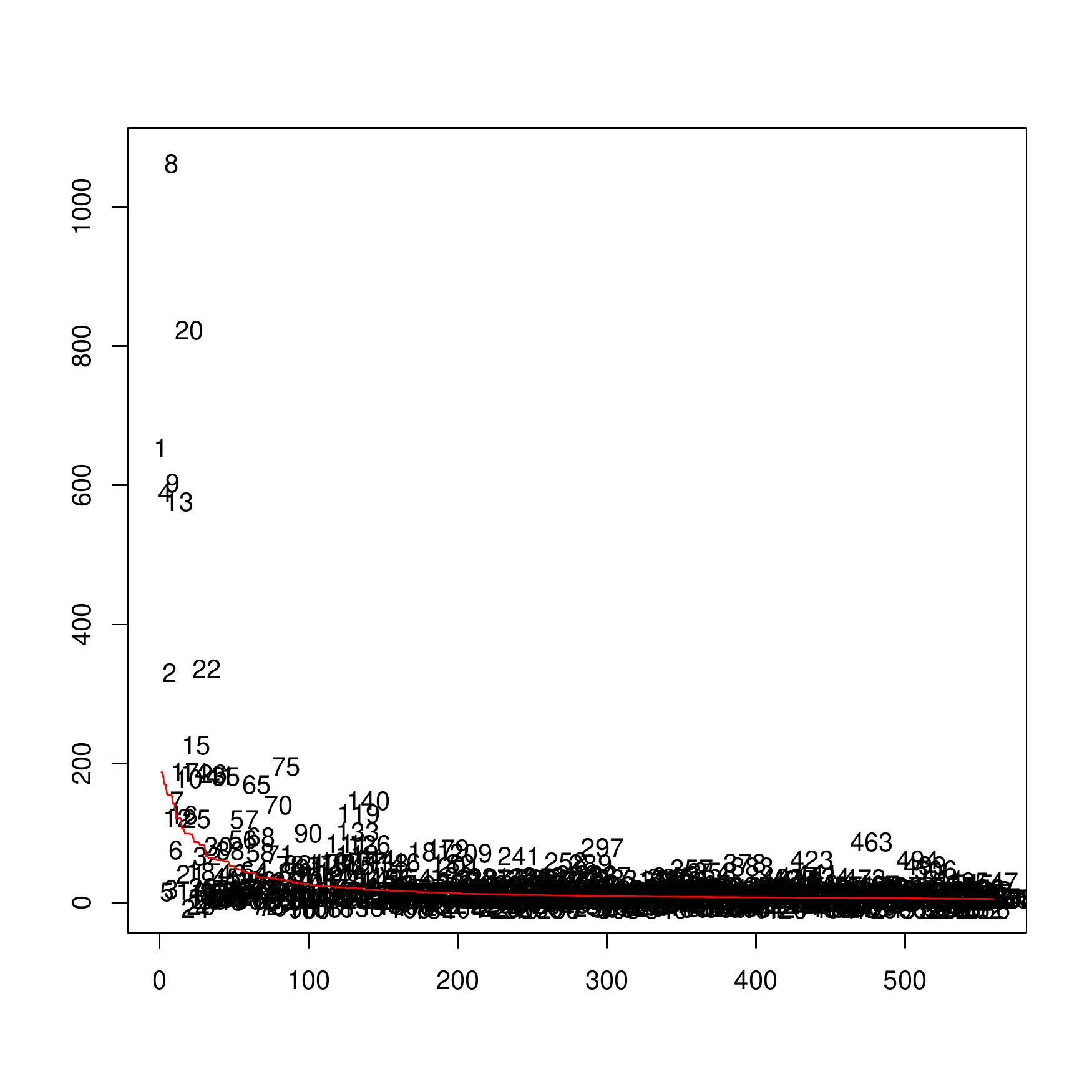}
                 \caption{$v_j^2$ vs $j$.}
                                 \label{fig:step2_vjsq}
\end{subfigure}
 \quad
\begin{subfigure}[b]{0.22\textwidth}
          \includegraphics[width=\textwidth]{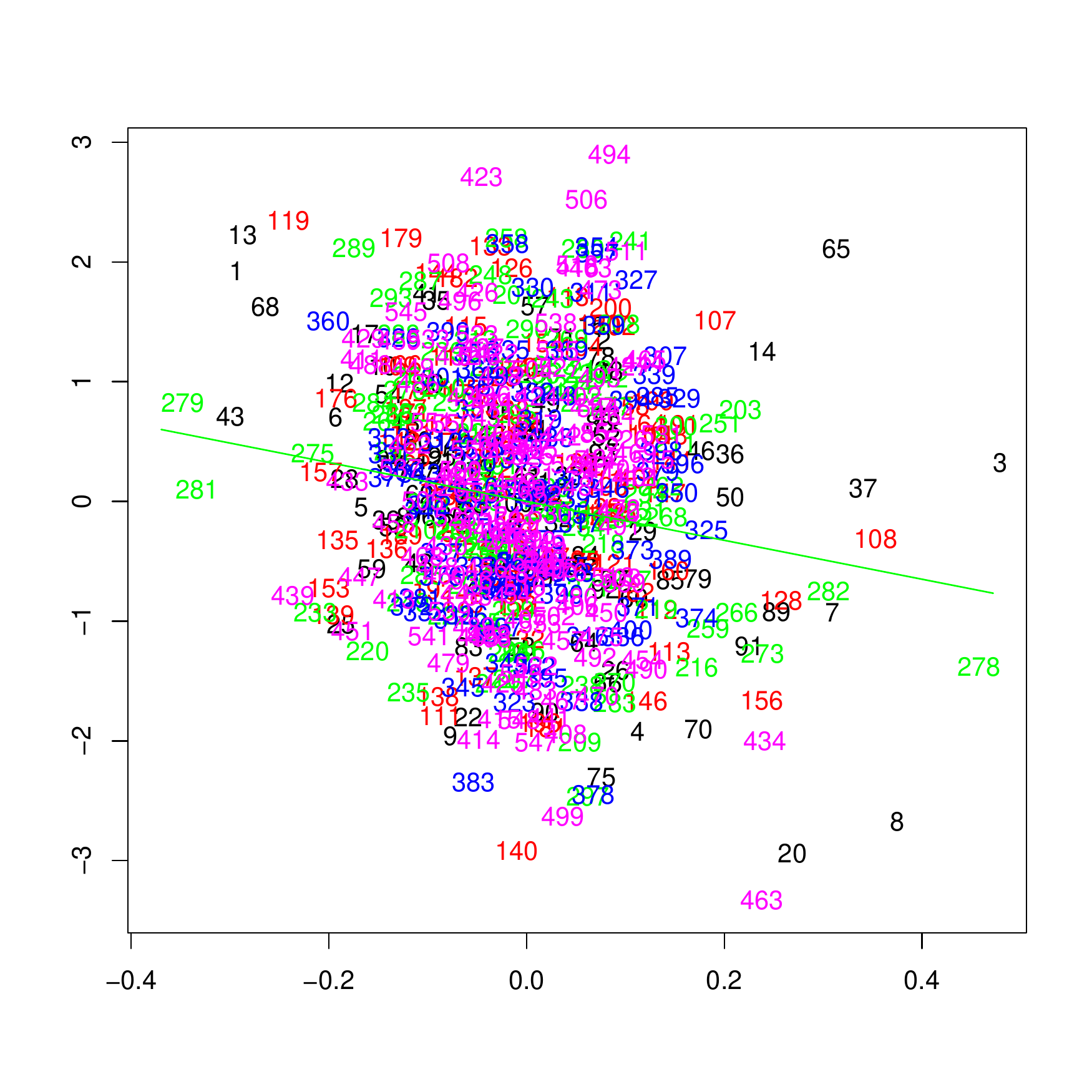}
                 \caption{Slope.}
                                  \label{fig:RM_add_var1}
\end{subfigure}
                  \quad
\begin{subfigure}[b]{0.22\textwidth}
                                                  \includegraphics[width=\textwidth]{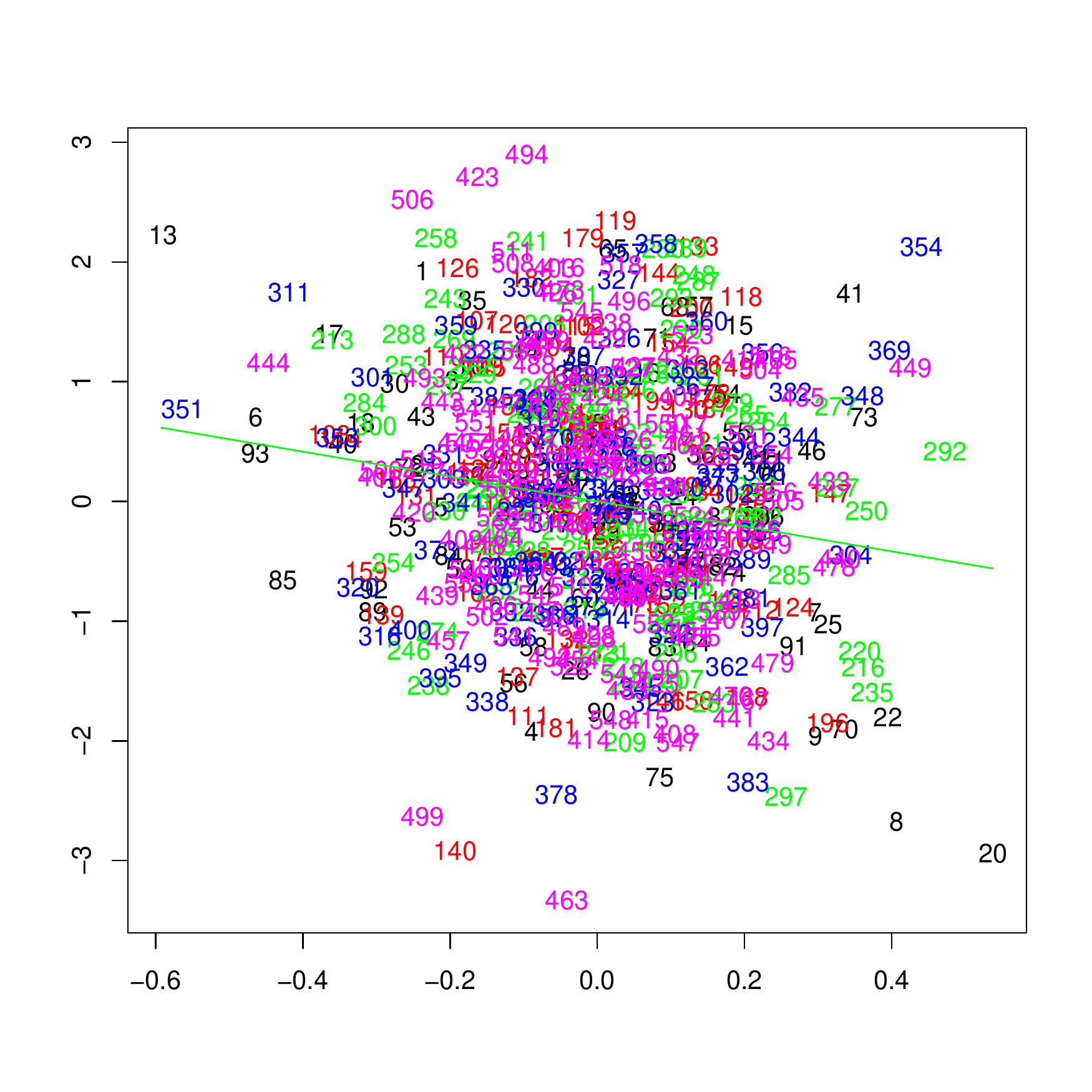}
                 \caption{SpringTC2.}
                 \label{fig:RM_add_var3}
\end{subfigure}
\quad
\begin{subfigure}[b]{0.22\textwidth}
                                                  \includegraphics[width=\textwidth]{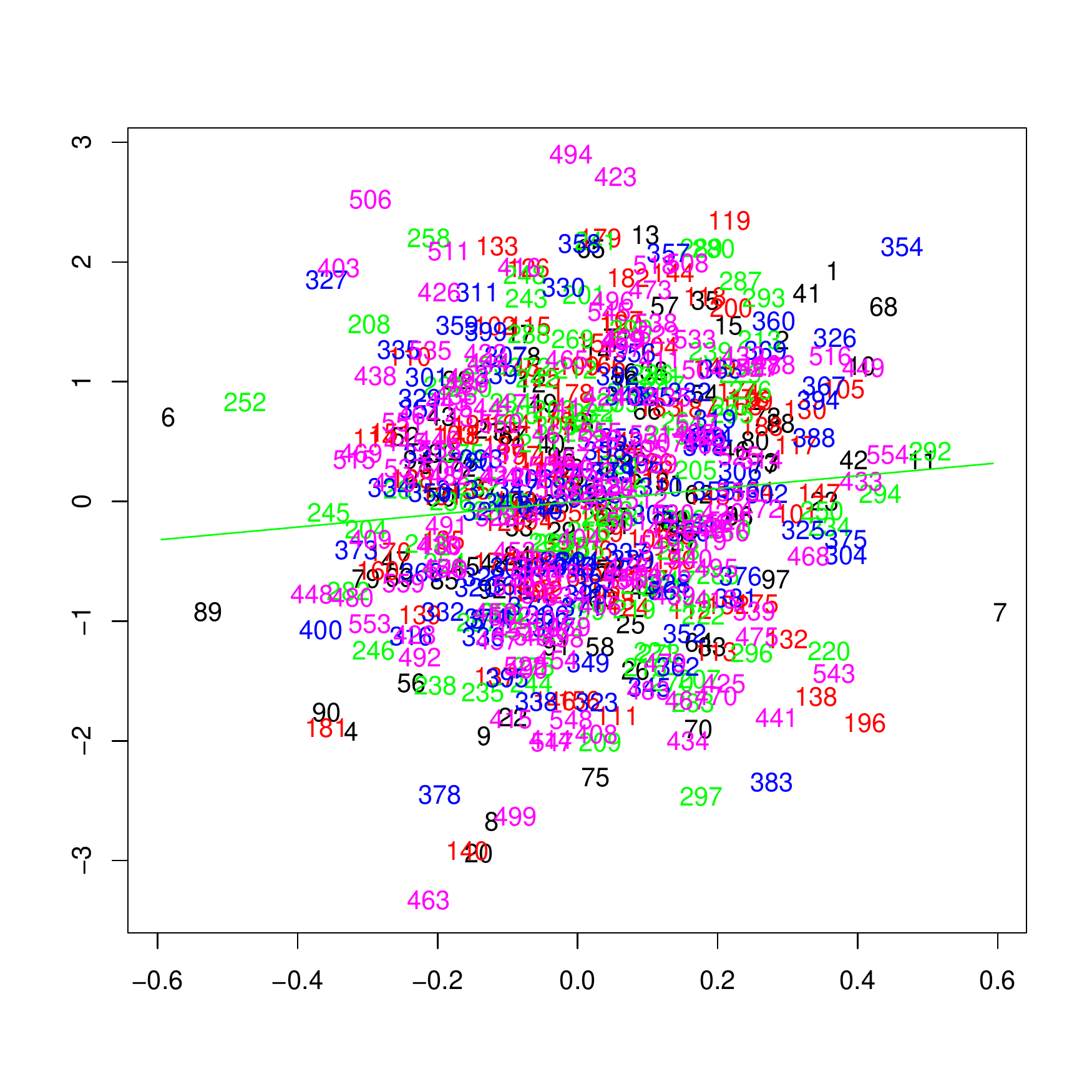}
                 \caption{SpringTC3.}
                 \label{fig:RM_add_var4}
\end{subfigure}
 \quad
  \begin{subfigure}[b]{0.22\textwidth}
 \includegraphics[width=\textwidth]{with_elev_vjadvarfor_sprtc3}
                 \caption{SummerTC1.}
                                  \label{fig:RM_add_var1}
\end{subfigure}
         \quad
\begin{subfigure}[b]{0.22\textwidth}
                                                  \includegraphics[width=\textwidth]{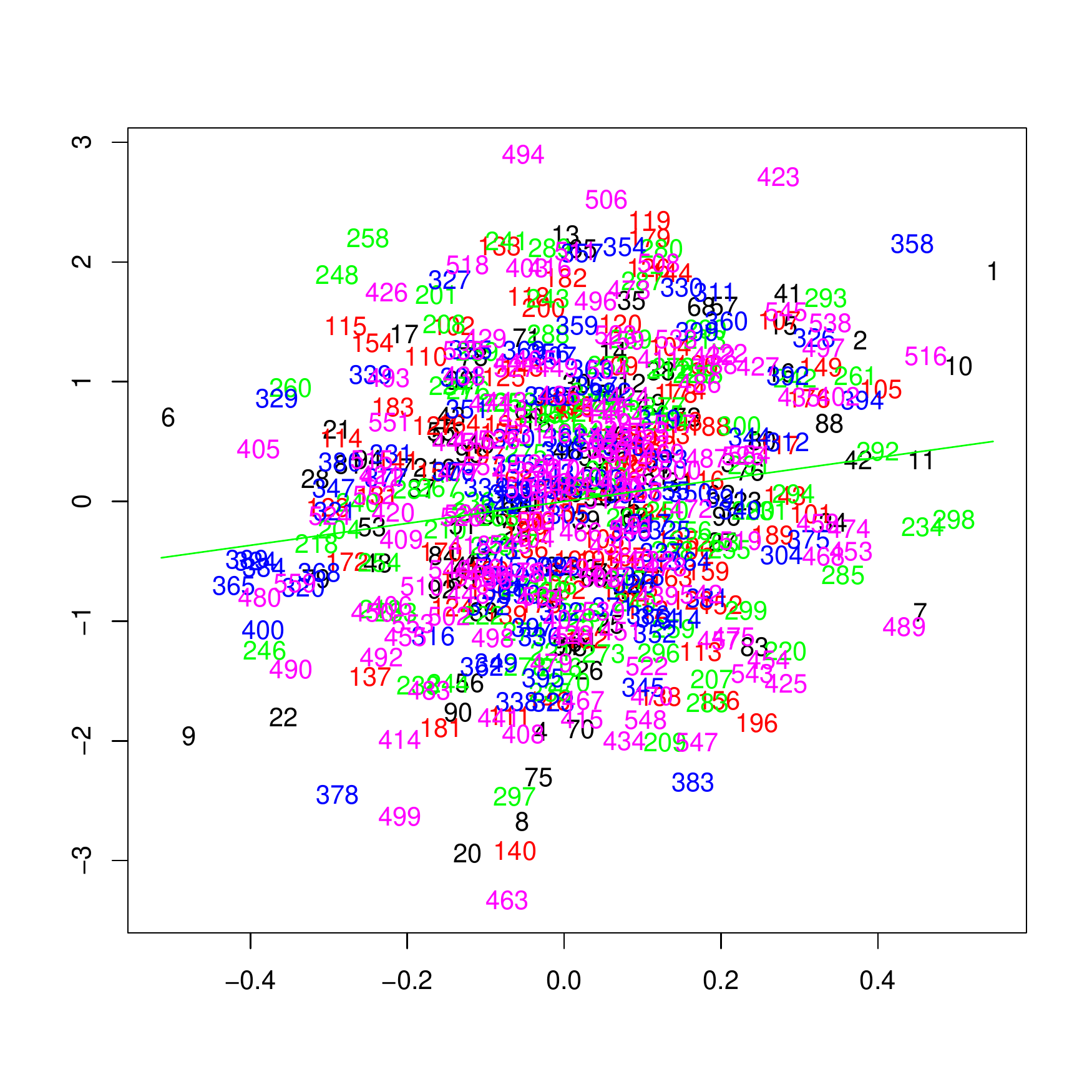}
                 \caption{SummerTC3. }
                 \label{fig:RM_add_var2}
\end{subfigure}
         \quad
\begin{subfigure}[b]{0.22\textwidth}
                                                  \includegraphics[width=\textwidth]{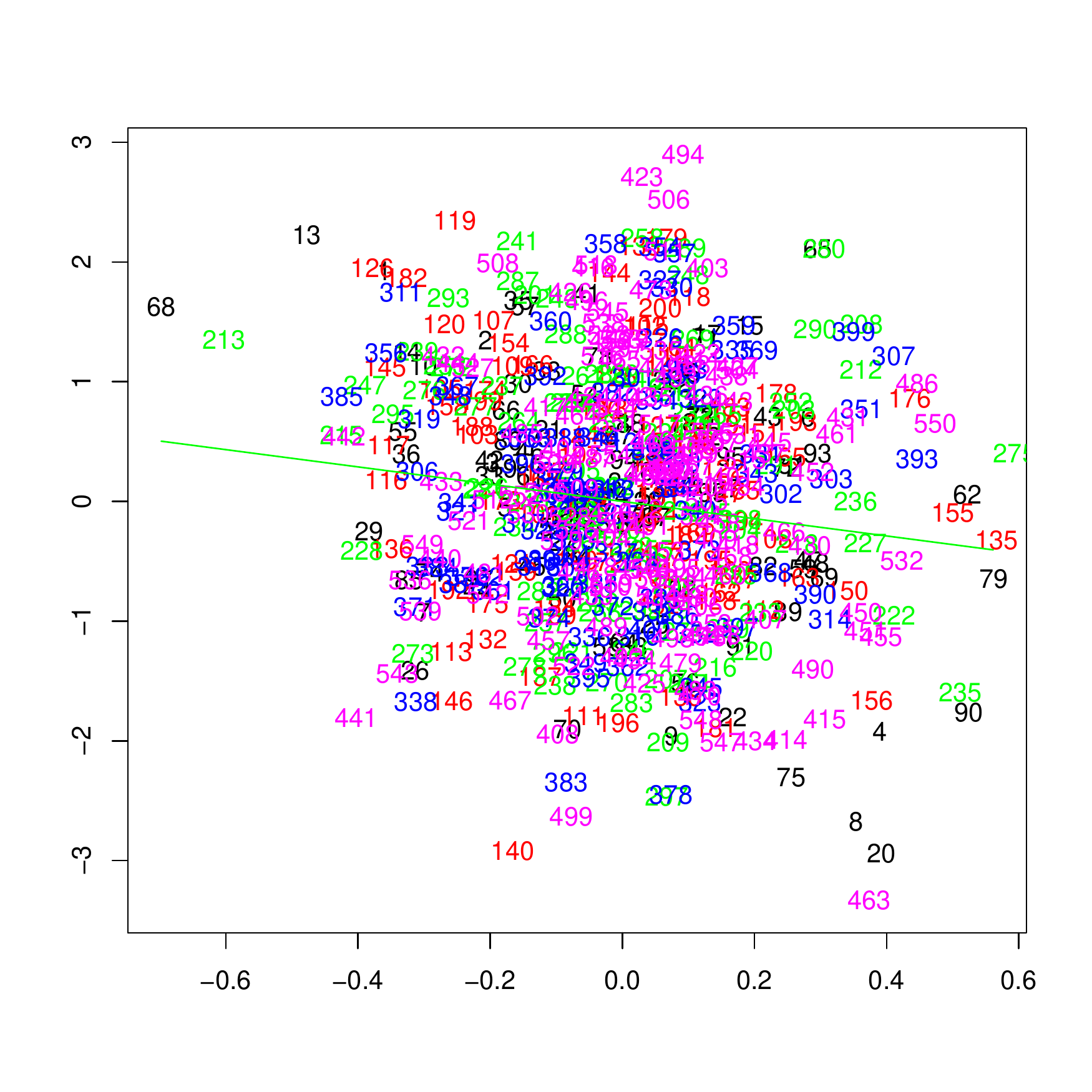}
                 \caption{FallTC2.}
                 \label{fig:RM_add_var3}
\end{subfigure}
 \caption{\textbf{Step 2} After fit with covariate Elevation; (b) to (g) are spectral-domain added variable plots. }\label{fig:step2_addvar2}
         \end{figure}

              \begin{table}[H]\centering
\begin{tabular}{ |c || c |c|c|}
    \hline
    Candidate covariates &  slope & p-value&$j$ with top 5 Cook's dist \\ \hline
    Elevation& --  &--&--\\ \hline
    \textbf{Slope} &-1.63  &$10^{-5}$&65,8,20,463,3\\ \hline
    SpringTC2&-1.04 &$10^{-5}$&20,354,8,13,499\\ \hline
    SpringTC3 & 0.54&0.03&506,196,463,327,403\\ \hline
    \textbf{SummerTC1} &-1.27&$10^{-6}$&20,248,463,9,327\\ \hline
    SummerTC3 & 0.92&0.0003&9,1,378,378,358,489\\ \hline
    FallTC2 &-0.72&0.002&463,20,441,8,13\\ \hline  \hline
    \end{tabular}
  \caption{ \textbf{Step 2} Slopes of spectral-domain added variable plots after the fit with covariate Elevation. The covariates chosen to be added at the next step are in bold. The symbol ``--" denotes covariates already in the model.}\label{table:table_step2}
\end{table}

\begin{figure}[H]\centering \begin{subfigure}[b]{0.17\textwidth}
 \includegraphics[width=\textwidth]{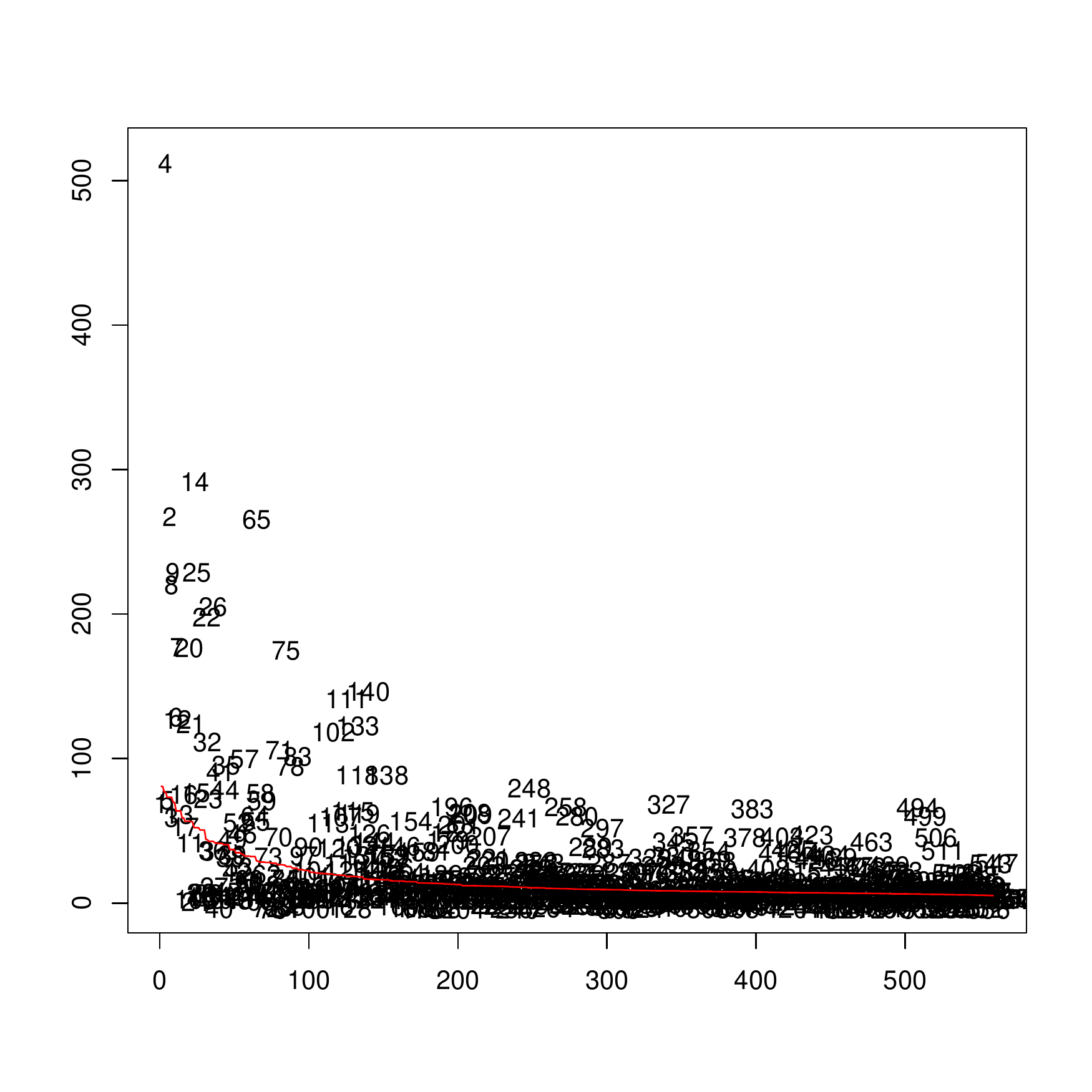}
                 \caption{$v_j^2$ vs $j$.}
                                 \label{fig:step3_vjsq}
\end{subfigure}
\quad
\begin{subfigure}[b]{0.17\textwidth}
                                                  \includegraphics[width=\textwidth]{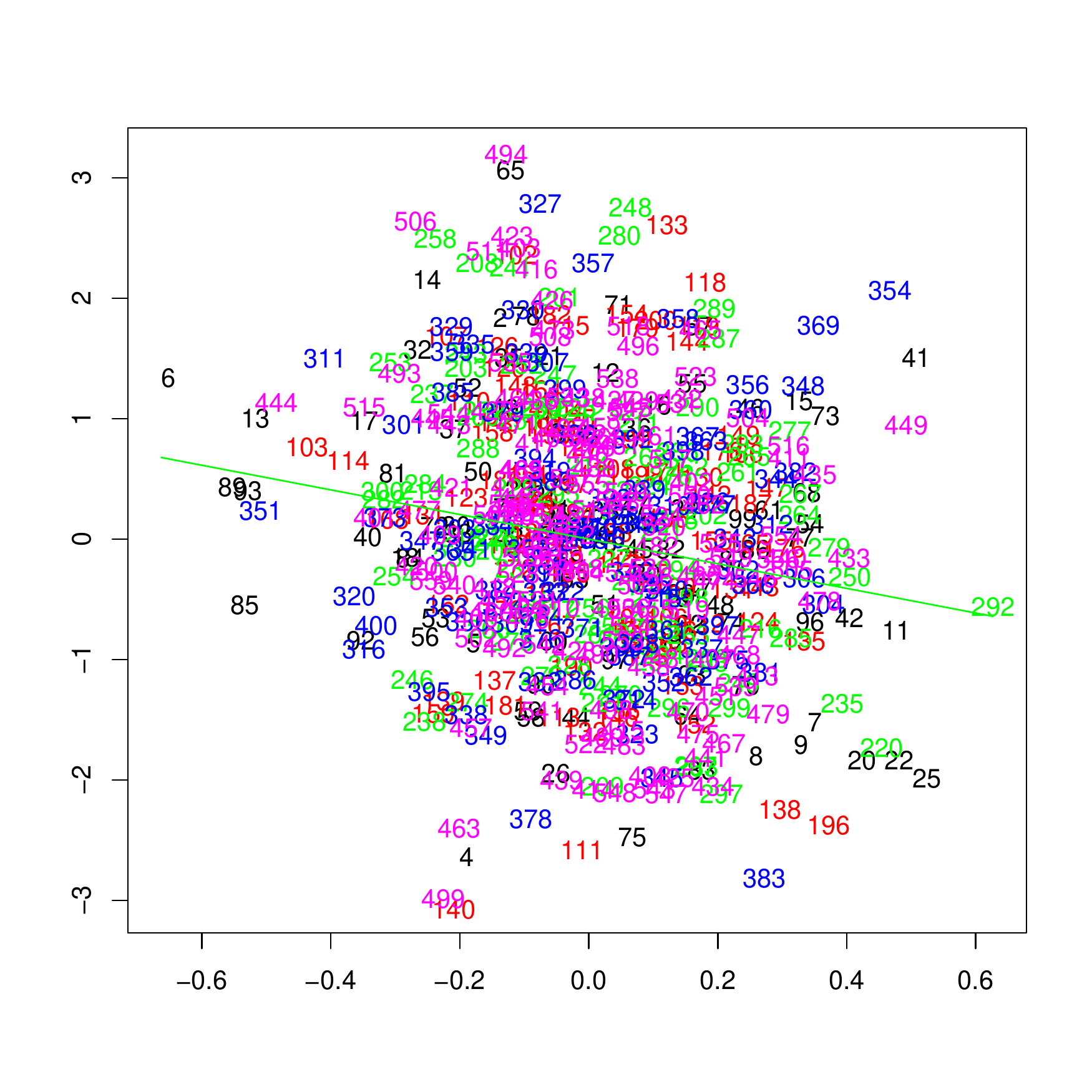}
                 \caption{SpringTC2.}
                 \label{fig:RM_add_var4}
\end{subfigure}\quad
 \begin{subfigure}[b]{0.17\textwidth}
 \includegraphics[width=\textwidth]{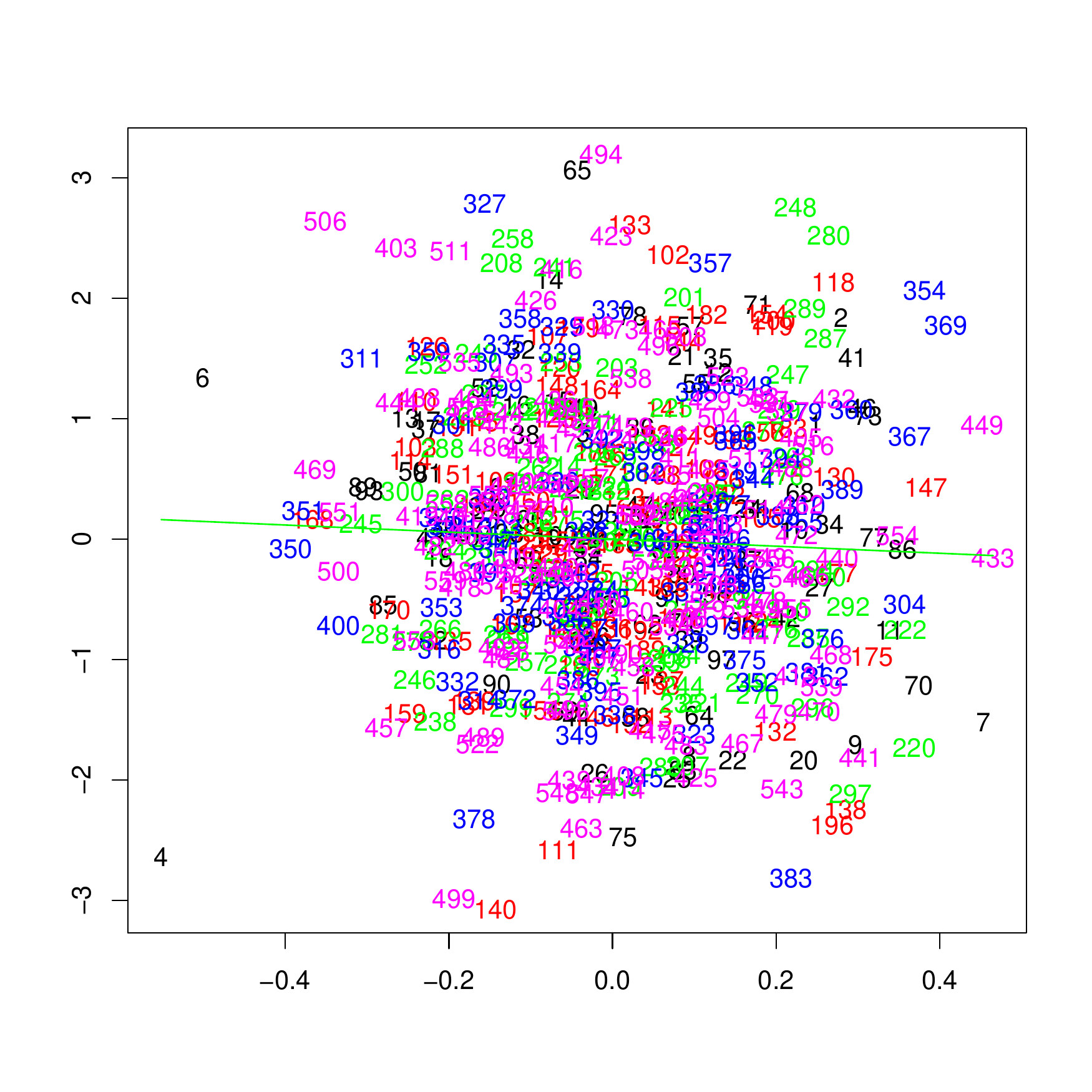}
                 \caption{SpringTC3.}
                                  \label{fig:RM_add_var1}
\end{subfigure}
         \quad
\begin{subfigure}[b]{0.17\textwidth}
                                                  \includegraphics[width=\textwidth]{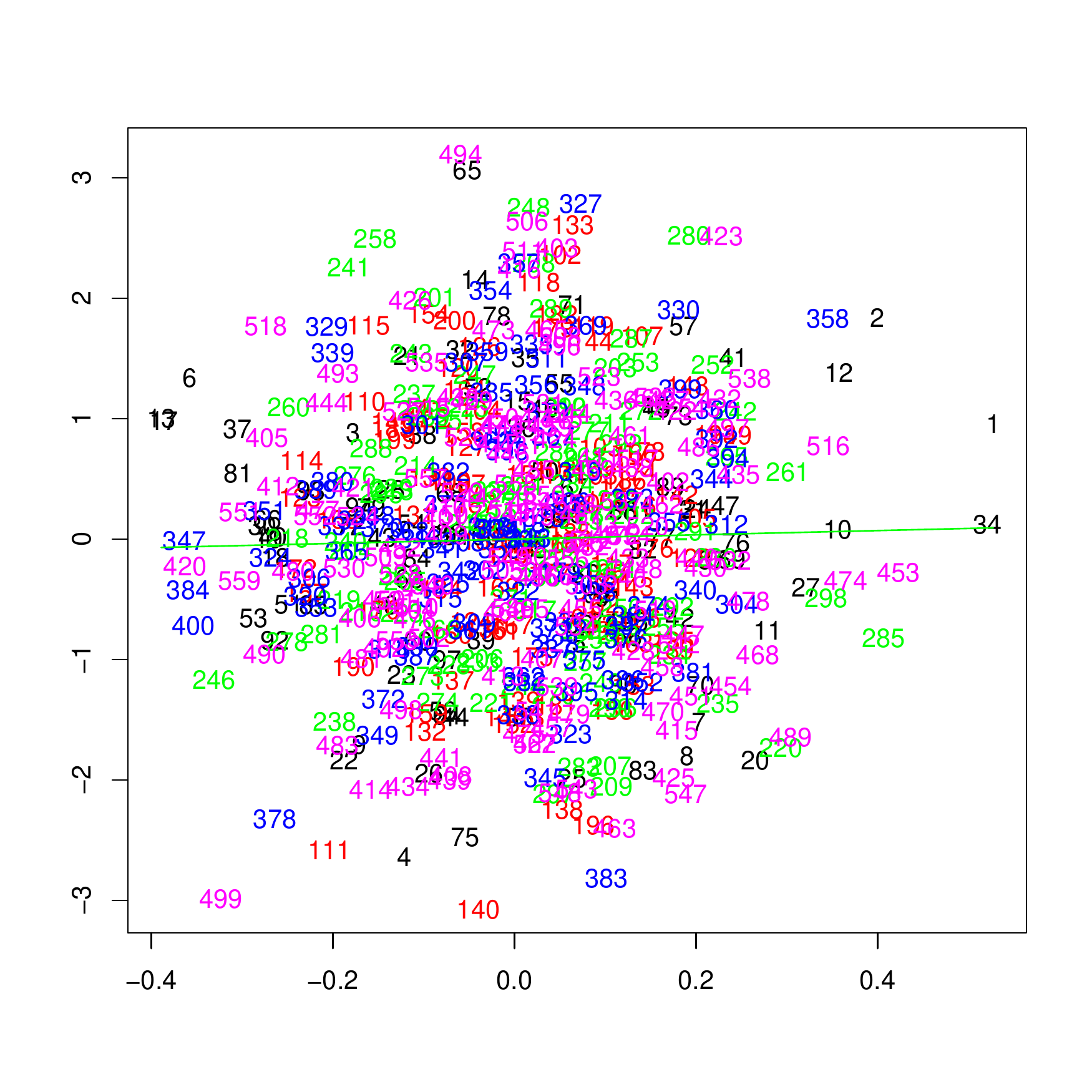}
                 \caption{SummerTC3. }
                 \label{fig:RM_add_var2}
\end{subfigure}
         \quad
\begin{subfigure}[b]{0.17\textwidth}
                                                  \includegraphics[width=\textwidth]{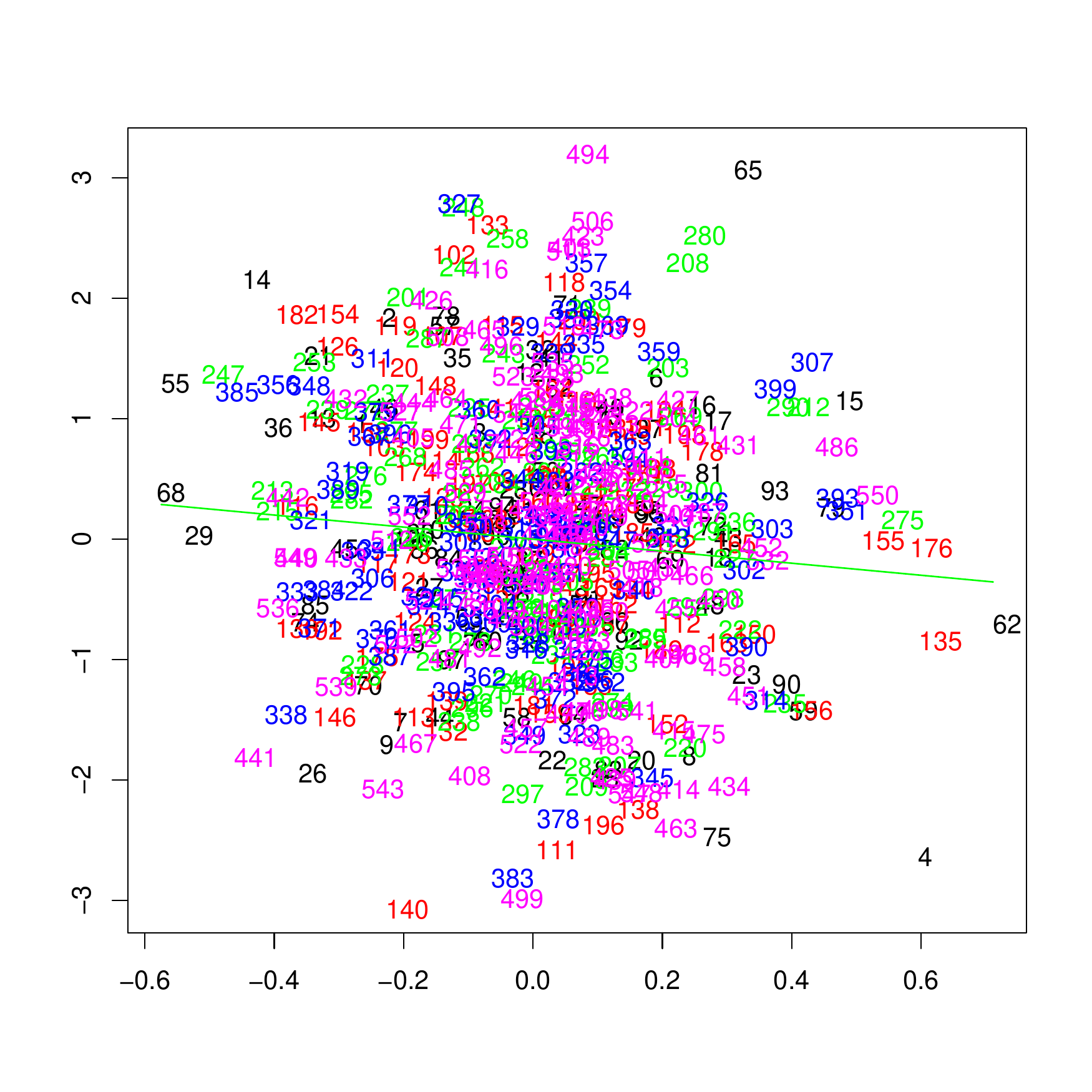}
                 \caption{FallTC2.}
                 \label{fig:RM_add_var3}
\end{subfigure}
 \caption{\textbf{Step 3} After the fit with covariate Elevation, Slope, SummerTC1; (b) to (e) are spectral-domain added variable plots.}\label{fig:step3_addvar}
         \end{figure}

                 \begin{table}[H]
\centering
  \begin{tabular}{ |c || c |c|c|}
    \hline
    Candidate covariates &  slope & p-value&pts with top 5 Cook's dist\\ \hline
    Elevation&  -- &--&--\\ \hline
    Slope & -- &--&--\\ \hline
    \textbf{SpringTC2}&-1.02 &$10^{-5}$&354,41,499,25,369\\ \hline
    SpringTC3 & -0.29&0.29&4,506,354,369,280\\ \hline
    SummerTC1 &--&--&--\\ \hline
   SummerTC3 &0.18&0.57 &499,2,378,111,358\\ \hline
    FallTC2 &-0.49&0.03&4,65,441,14,26\\ \hline  \hline
    \end{tabular}
  \caption{\textbf{Step 3} Slopes of spectral-domain added variable plots after the fit with covariates Elevation, Slope, SummerTC1. The covariate chosen to be added at the next step is in bold. The symbol ``--" denotes covariates already in the model.}\label{table:table_step3}
\end{table}

\begin{figure}[H]
         \centering
 \includegraphics[width=0.3\textwidth]{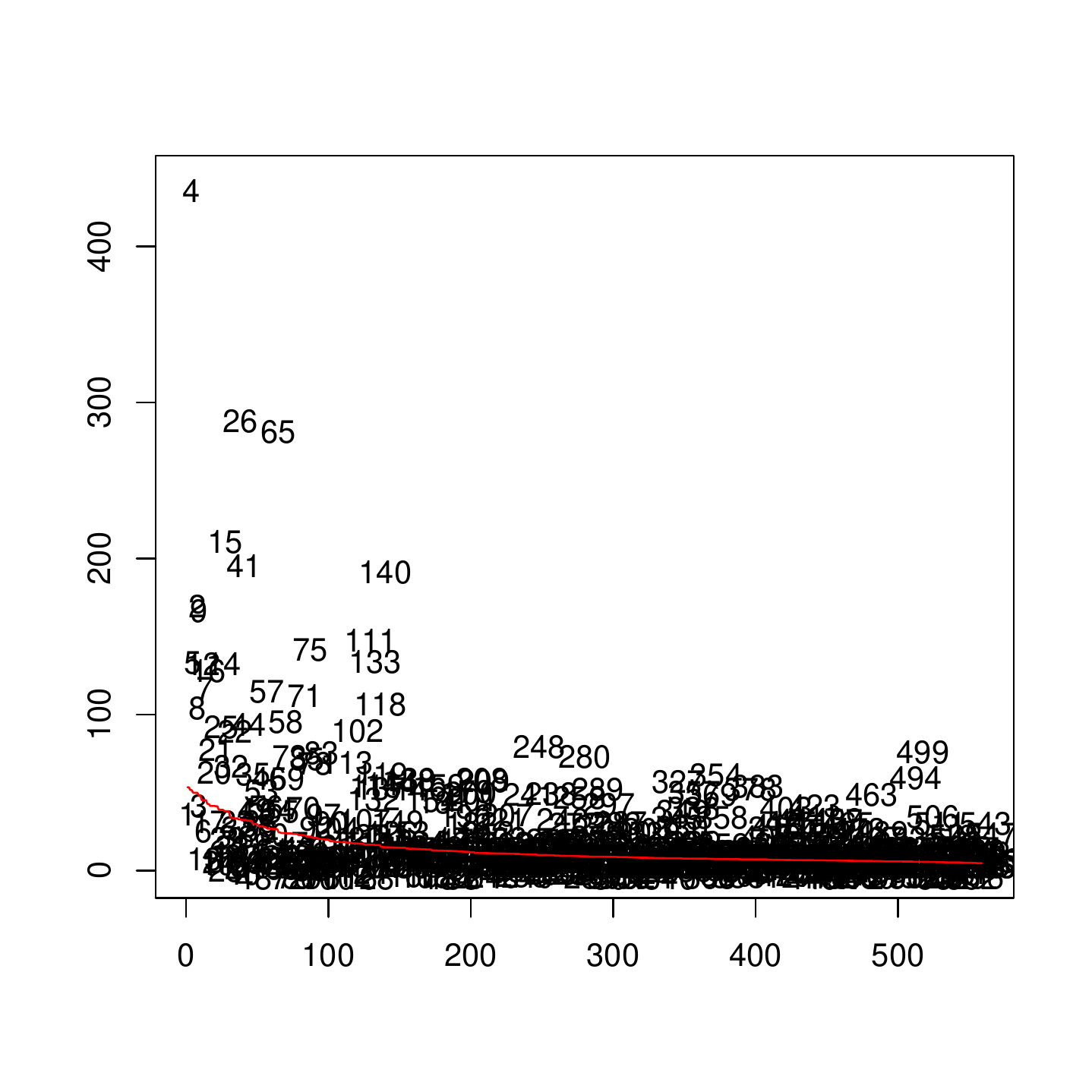}
                 \caption{$v_j^2$ vs $j$ for the fit with covariates Slope, Elevation, SpringTC2, SummerTC1, and FallTC2.}
                                 \label{fig:step3_vjsqfaltc2}
\end{figure}

\begin{figure}[H]
         \centering
\begin{subfigure}[b]{0.22\textwidth}
 \includegraphics[width=\textwidth]{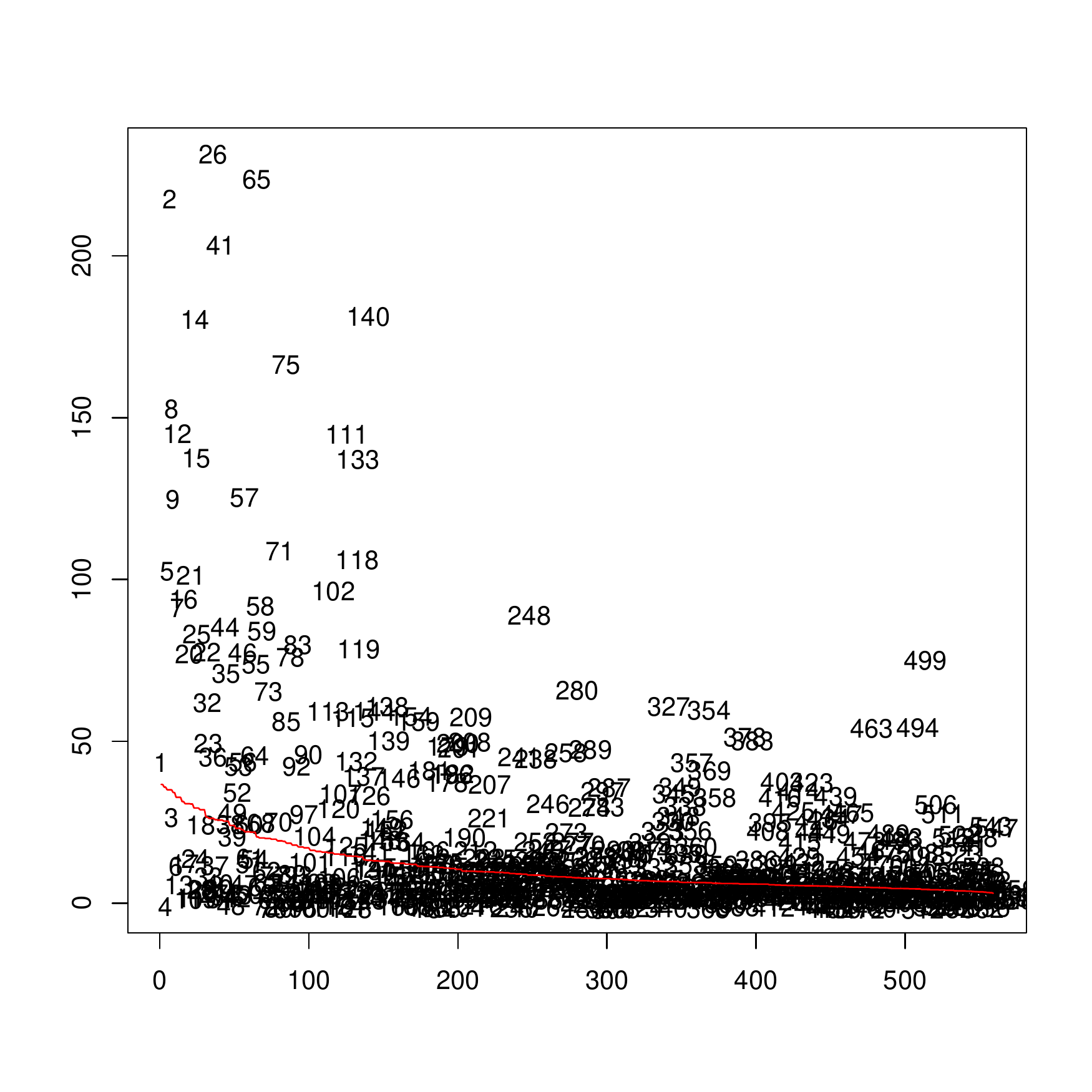}
                 \caption{$v_j^2$ vs $j$.}
                                 \label{fig:step4_vjsq}
\end{subfigure}
 \quad
\begin{subfigure}[b]{0.22\textwidth}
 \includegraphics[width=\textwidth]{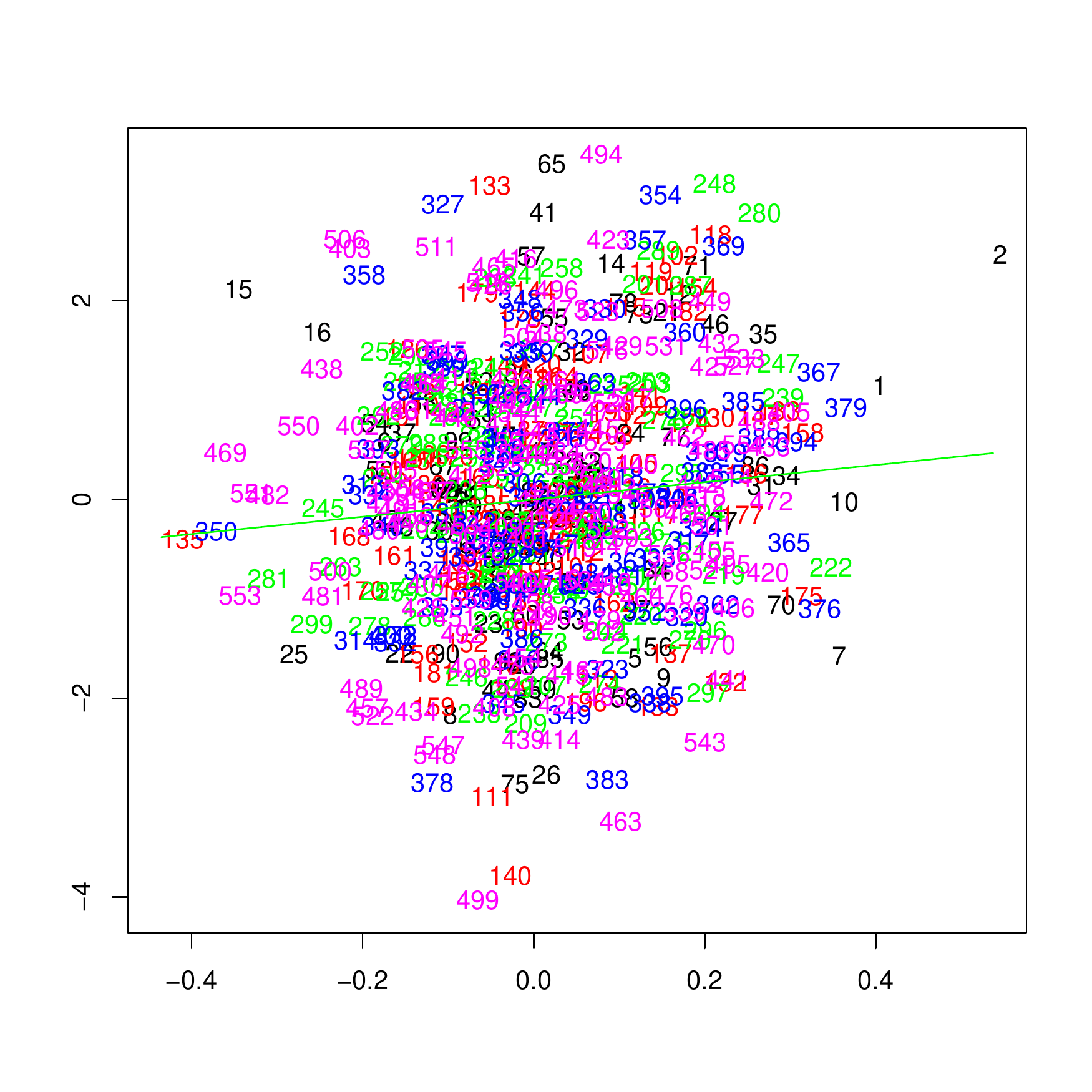}
                 \caption{SpringTC3.}
                 \label{fig:RM_add_var4}
\end{subfigure}
 \quad \begin{subfigure}[b]{0.22\textwidth}
 \includegraphics[width=\textwidth]{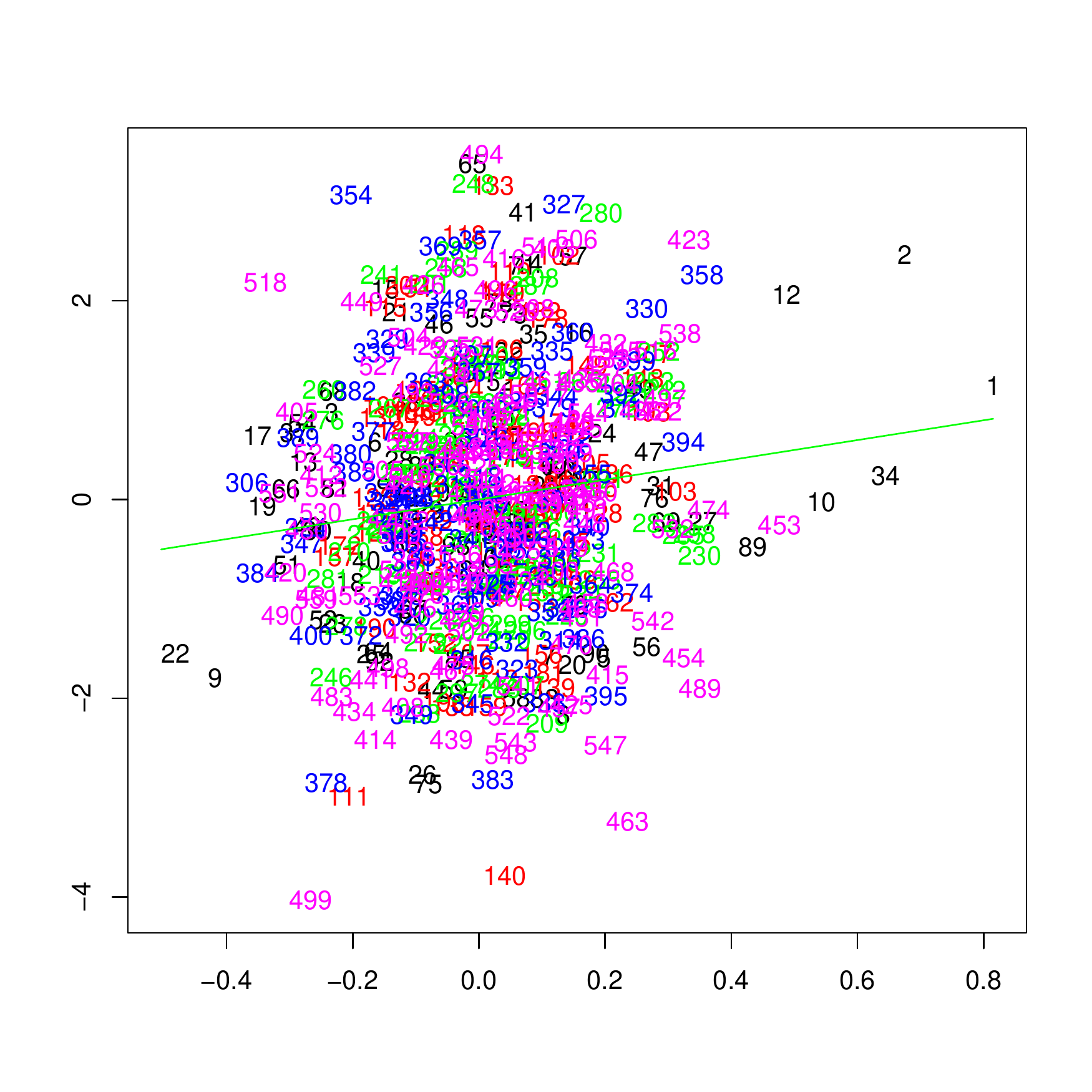}
                 \caption{SummerTC3.}
                                  \label{fig:RM_add_var1}
\end{subfigure}
         \quad
\begin{subfigure}[b]{0.22\textwidth}
                                                  \includegraphics[width=\textwidth]{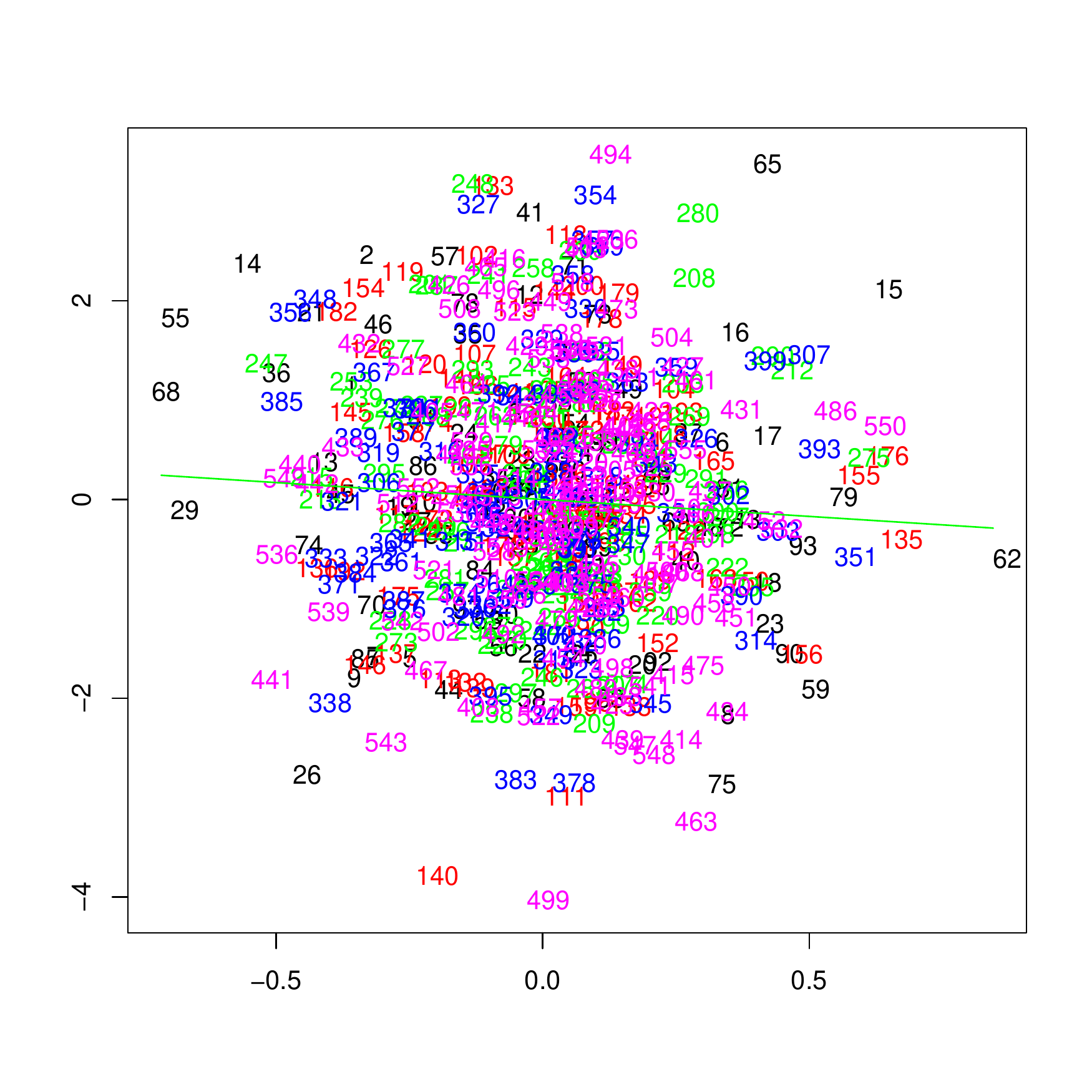}
                 \caption{FallTC2.}
                 \label{fig:RM_add_var3}
\end{subfigure}
 \caption{\textbf{Step 4} After the fit with covariates Elevation, Slope, SummerTC1, SpringTC2 and north-south quadratic trend; (b) to (d) are spectral-domain added variable plots.}\label{fig:step4_addvar}
         \end{figure}

                  \begin{table}[H]
\centering
  \begin{tabular}{ |c || c |c|c|}
    \hline
    Candidate covariates &  slope & p-value&pts with top 5 Cook's dist\\ \hline
    Elevation&  -- &--&--\\ \hline
    Slope & -- &--&--\\ \hline
    SpringTC2& --&--&--\\ \hline
    SpringTC3 & 0.87&0.02&2,15,280,506,7\\ \hline
    SummerTC1 &--&--&--\\ \hline
    \textbf{SummerTC3} & 0.99&0.002&2,499,518,489,12\\ \hline
    FallTC2 &-0.34&0.15&65,15,26,14,55\\ \hline  \hline
    \end{tabular}
  \caption{ \textbf{Step 4} Slopes of spectral-domain added variable plots after the fit with covariates Slope, Elevation, SpringTC2, SummerTC1 and north-south quadratic trend.  The covariate chosen to be added at the next step is in bold. The symbol ``--" denotes covariates already in the model.}\label{table:table_step4}
\end{table}

\begin{figure}[H]
         \centering
         \begin{subfigure}[b]{0.22\textwidth}
 \includegraphics[width=\textwidth]{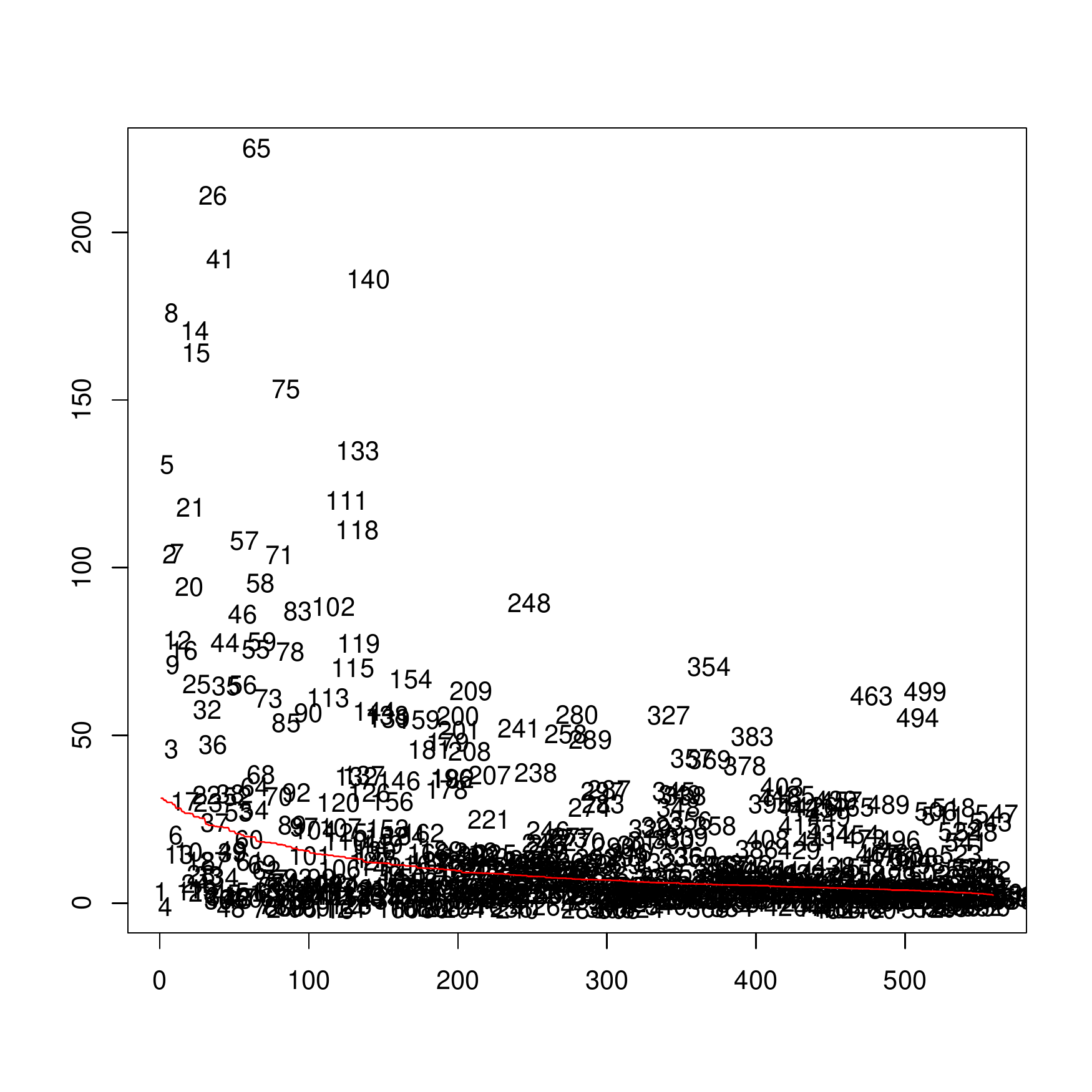}
                 \caption{$v_j^2$ vs $j$.}
                                 \label{fig:step5_vjsq}
\end{subfigure}
\quad
\begin{subfigure}[b]{0.22\textwidth}
                                                  \includegraphics[width=\textwidth]{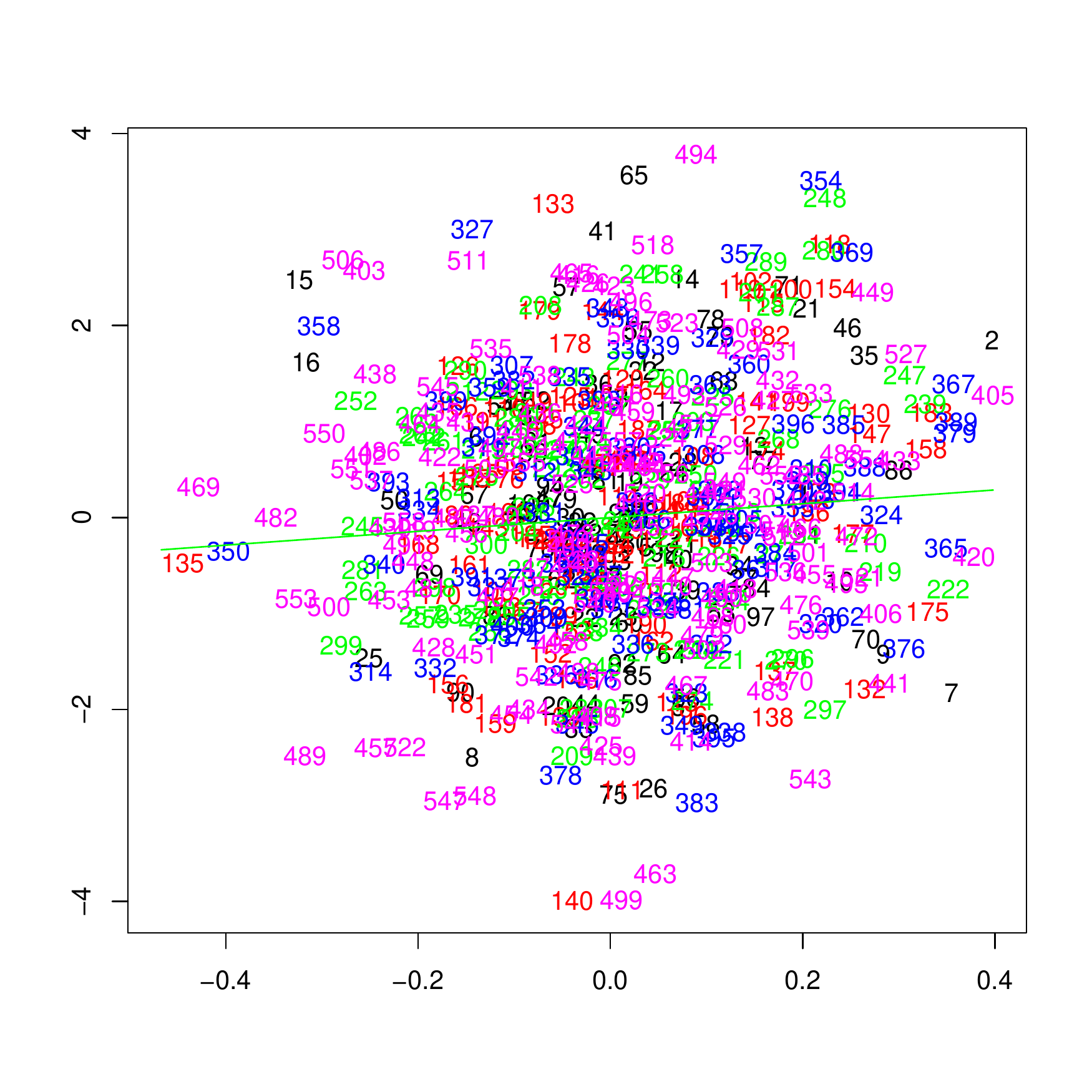}
                 \caption{SpringTC3.}
                 \label{fig:RM_add_var4}
\end{subfigure}
          \quad
\begin{subfigure}[b]{0.22\textwidth}
                                                  \includegraphics[width=\textwidth]{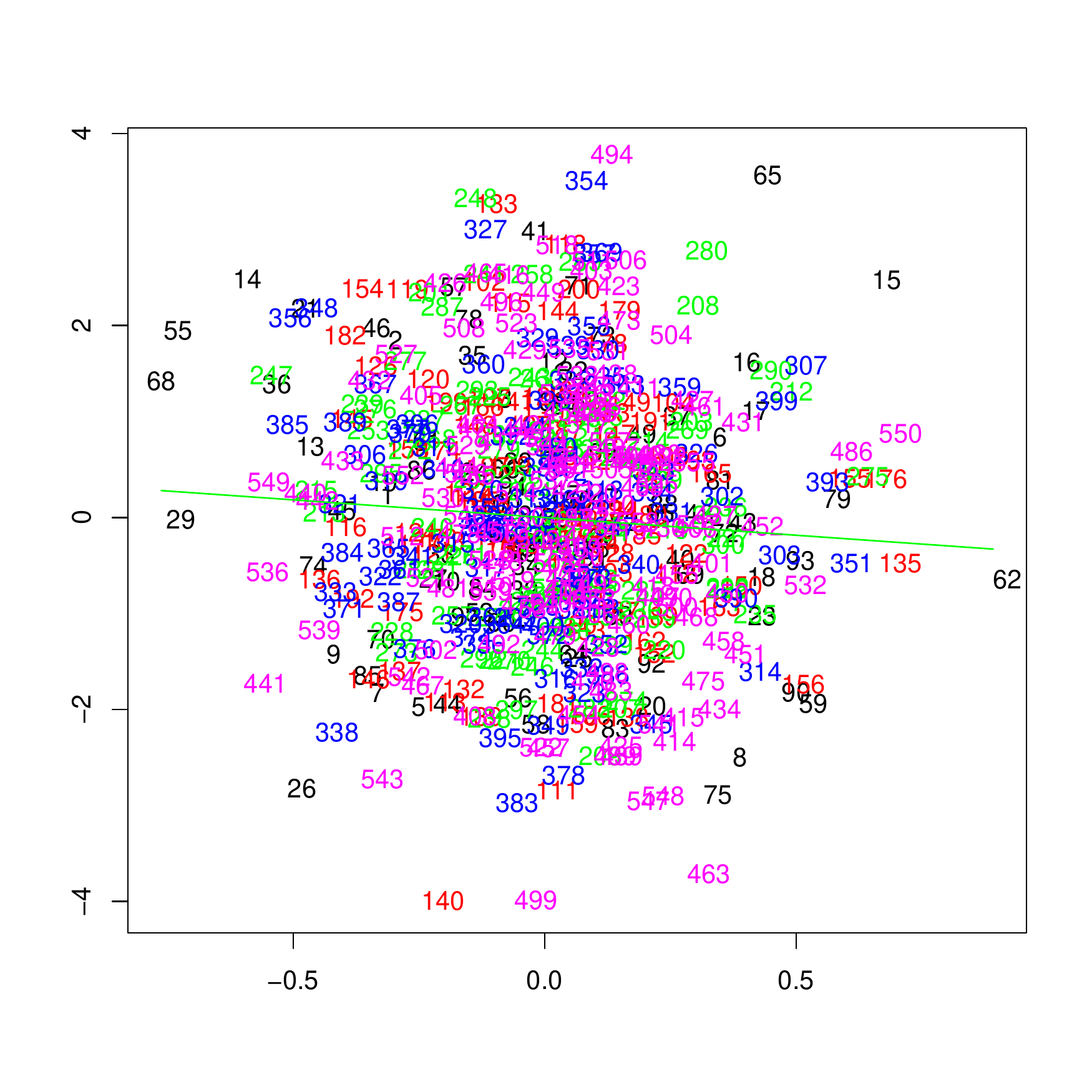}
                 \caption{FallTC2.}
                 \label{fig:RM_add_var3}
\end{subfigure}
 \caption{\textbf{Step 5} After the fit with covariates Elevation, Slope, SummerTC1, SpringTC2, north-south quadratic trend, and SummerTC3;  (b) to (c) are spectral-domain added variable plots. }\label{fig:step5_addvar}
         \end{figure}

                  \begin{table}[H]
\centering
  \begin{tabular}{ |c || c |c|c|}
    \hline
    Candidate covariates &  slope & p-value&pts with top 5 Cook's dist\\ \hline
    Elevation&  -- &--&--\\ \hline
    Slope & -- &--&--\\ \hline
    SpringTC2& --&--&--\\ \hline
    SpringTC3 & 0.72&0.06&15,506,403,489,7\\ \hline
    SummerTC1 &--&--&--\\ \hline
    SummerTC3 & --&--&--\\ \hline
    FallTC2 &-0.37&0.11&15,65,26,14,55\\ \hline  \hline
    \end{tabular}
  \caption{\textbf{Step 5}  Slopes of spectral-domain added variable plots after the fit with covariates Slope, Elevation, SpringTC2, SummerTC1, north-south quadratic trend, and SummerTC3. The symbol ``--" denotes covariates already in the model, in addition a north-south quadratic trend is also in the model.}\label{table:table_step5}
\end{table}

\section{Web Appendix I. \hspace*{3mm} Observation-domain added variable plots}\label{sec:real_data_yaddvar}
\begin{figure}[H]
         \centering \begin{subfigure}[b]{0.22\textwidth}
 \includegraphics[width=\textwidth]{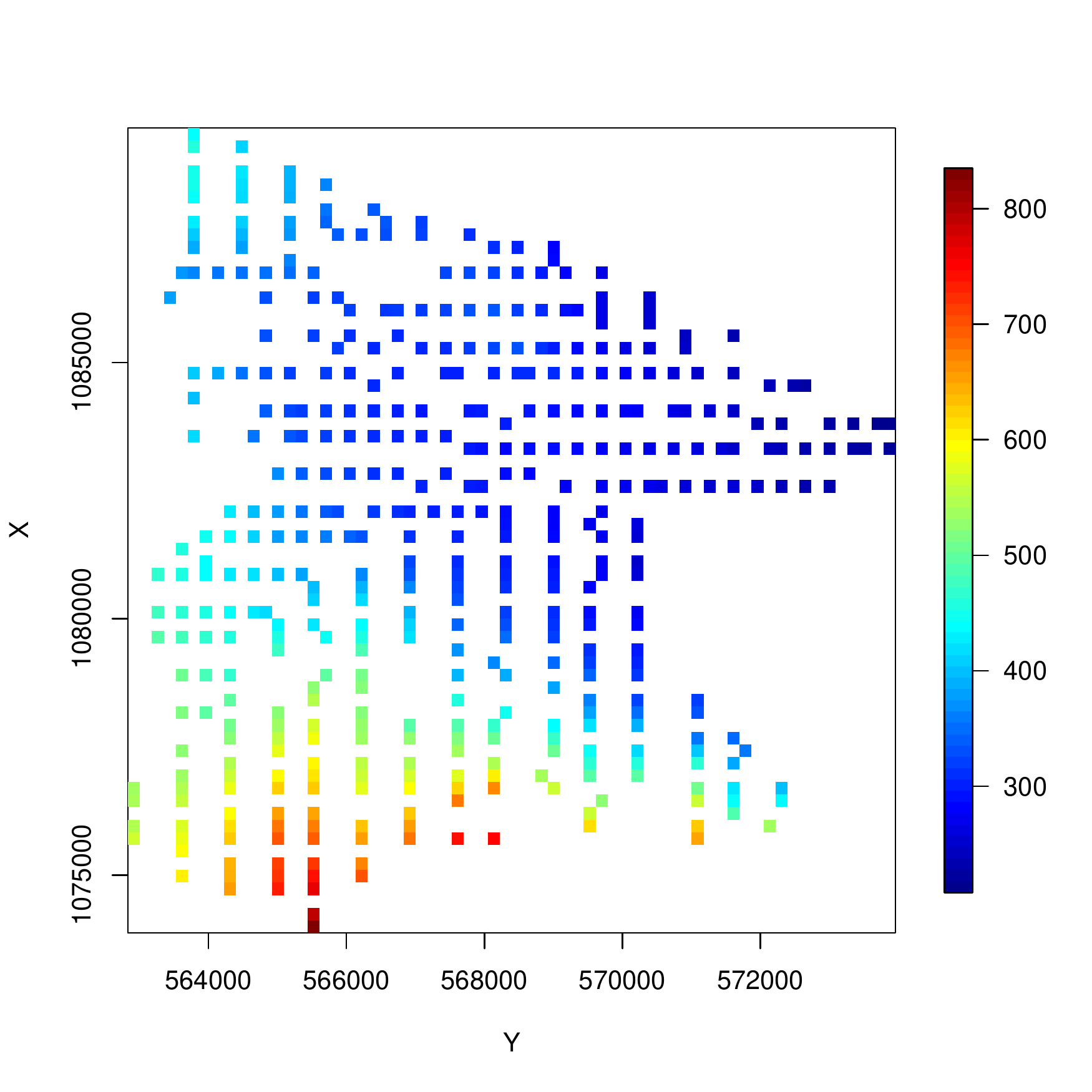}
                 \caption{Elevation.}
                                  \label{fig:RM_pred1}
\end{subfigure}
         \quad
\begin{subfigure}[b]{0.22\textwidth}
                                                  \includegraphics[width=\textwidth]{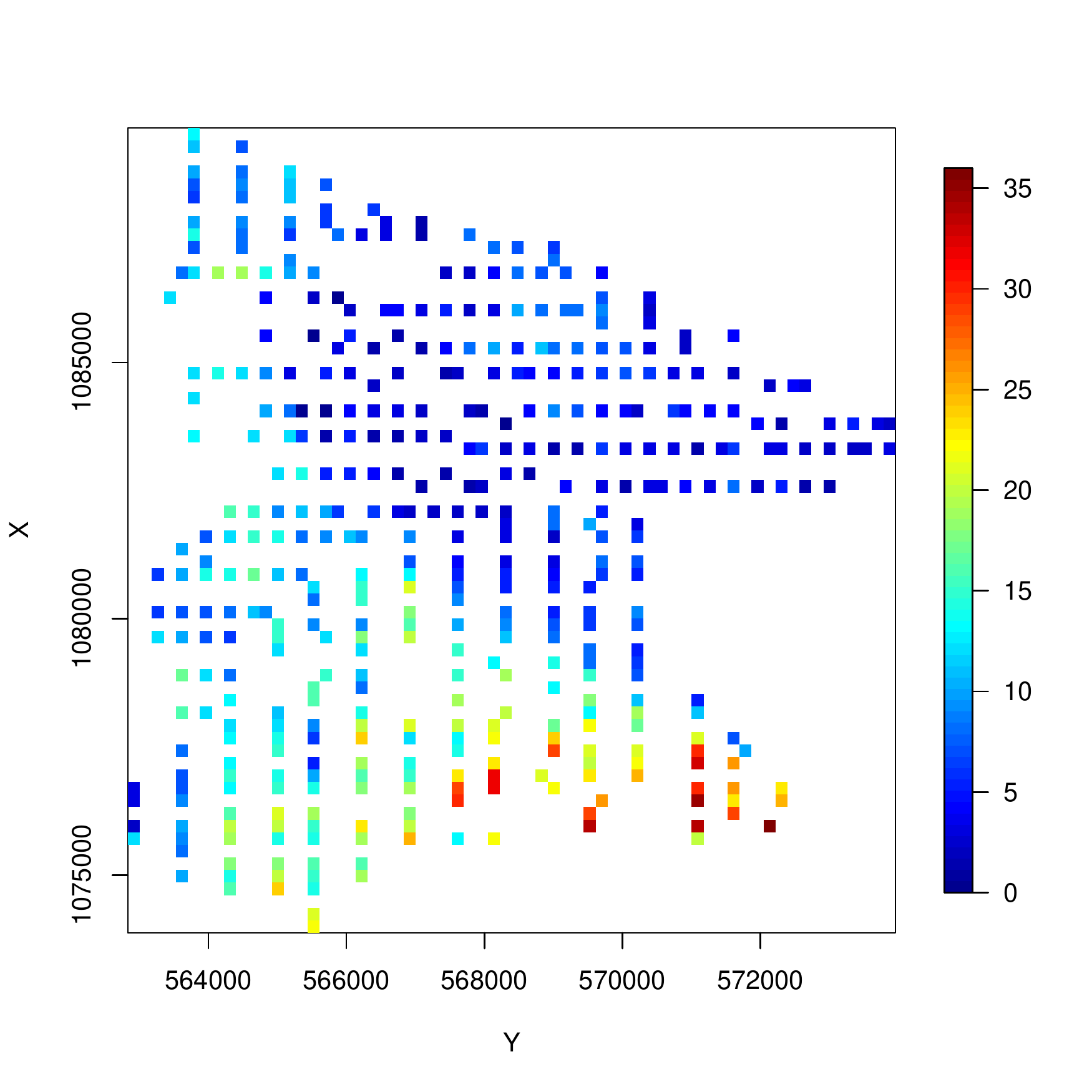}
                 \caption{Slope. }
                 \label{fig:RM_pred2}
\end{subfigure}
         \quad
\begin{subfigure}[b]{0.22\textwidth}
                                                  \includegraphics[width=\textwidth]{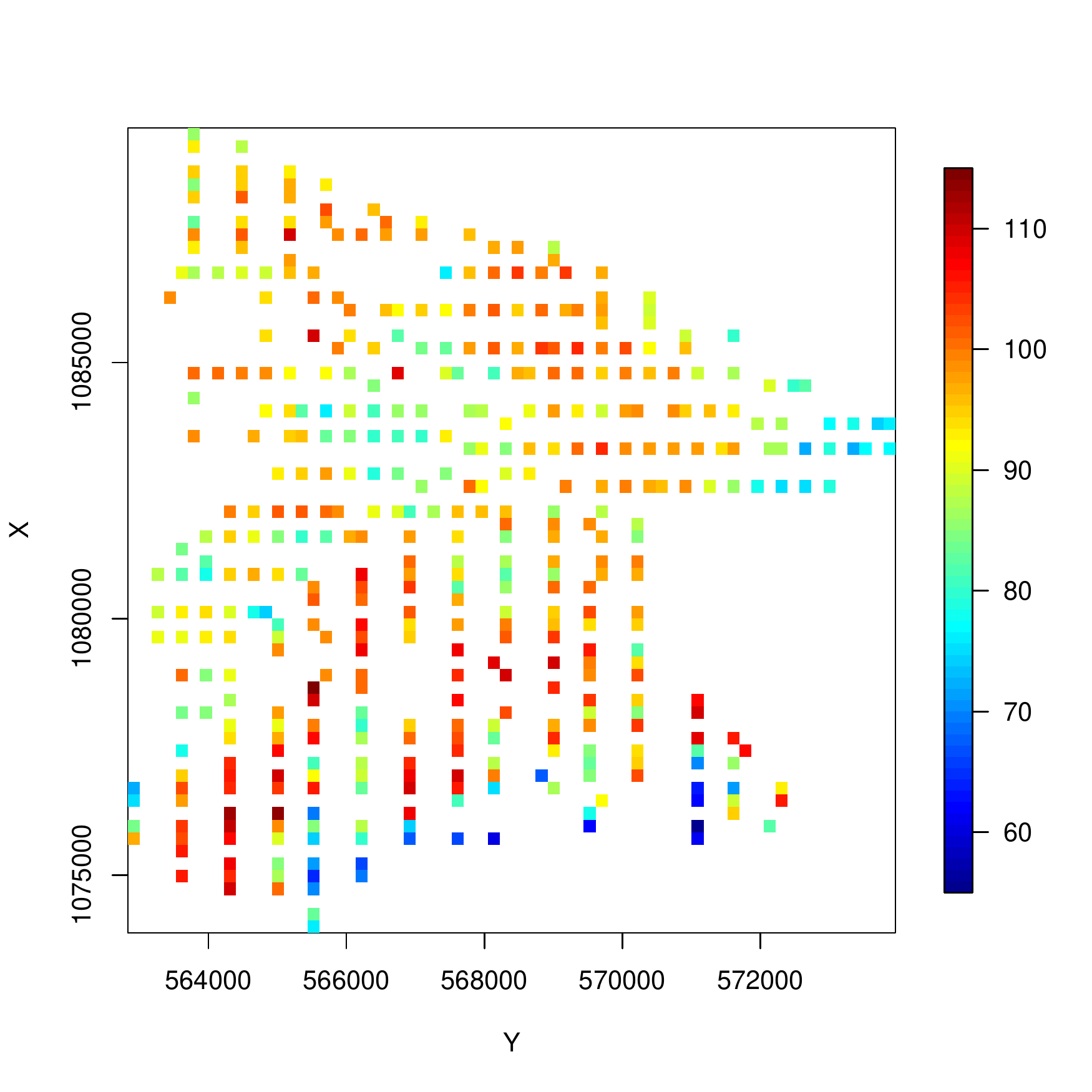}
                 \caption{SummerTC1.}
                 \label{fig:RM_pred3}
\end{subfigure}
\quad
\begin{subfigure}[b]{0.22\textwidth}
                                                  \includegraphics[width=\textwidth]{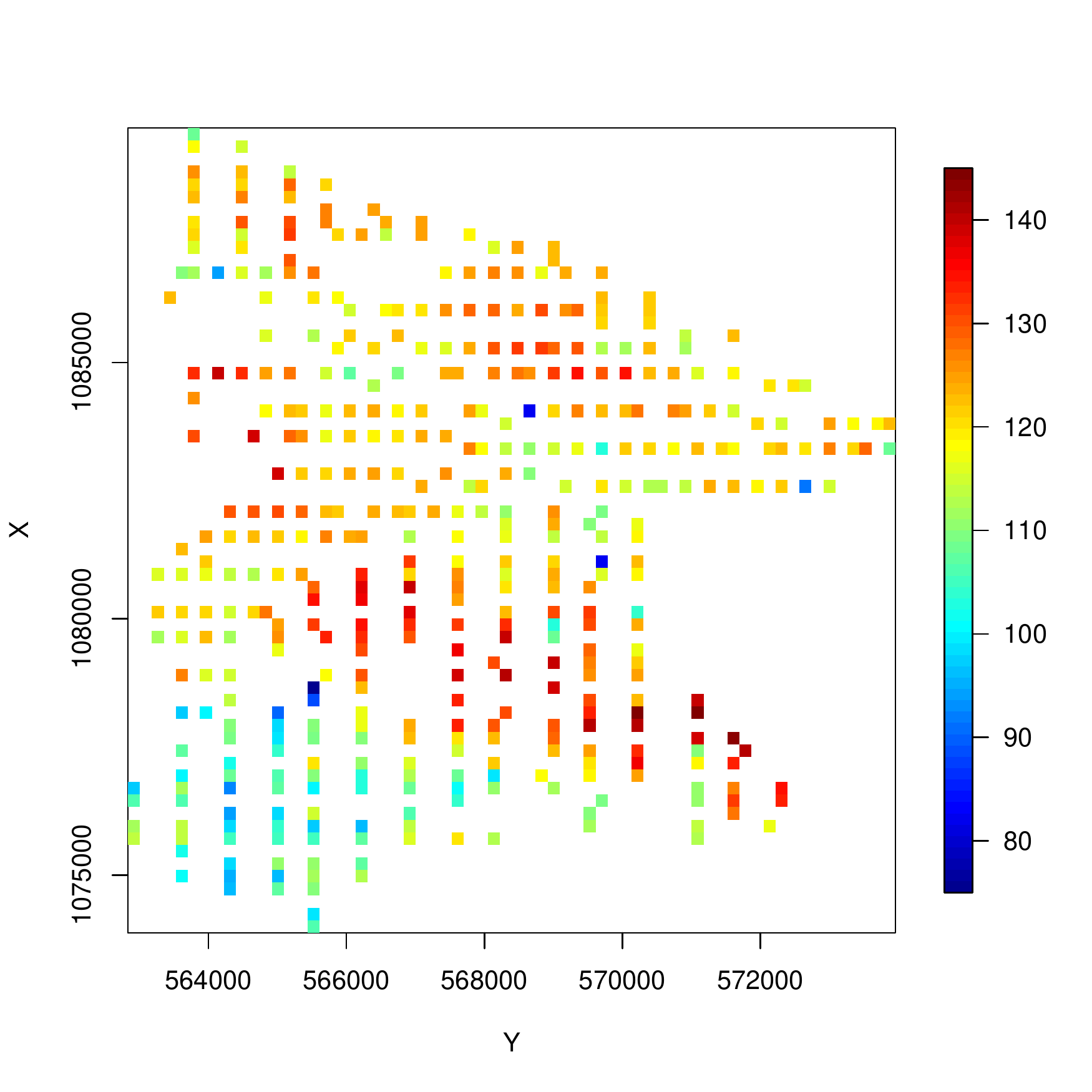}
                 \caption{ SpringTC2. }
                 \label{fig:RM_pred4}
\end{subfigure}
 \caption{ Plots of four of the covariates.}\label{fig:add_var}
         \end{figure}

\begin{figure}[H]
         \centering \begin{subfigure}[b]{0.22\textwidth}
 \includegraphics[width=\textwidth]{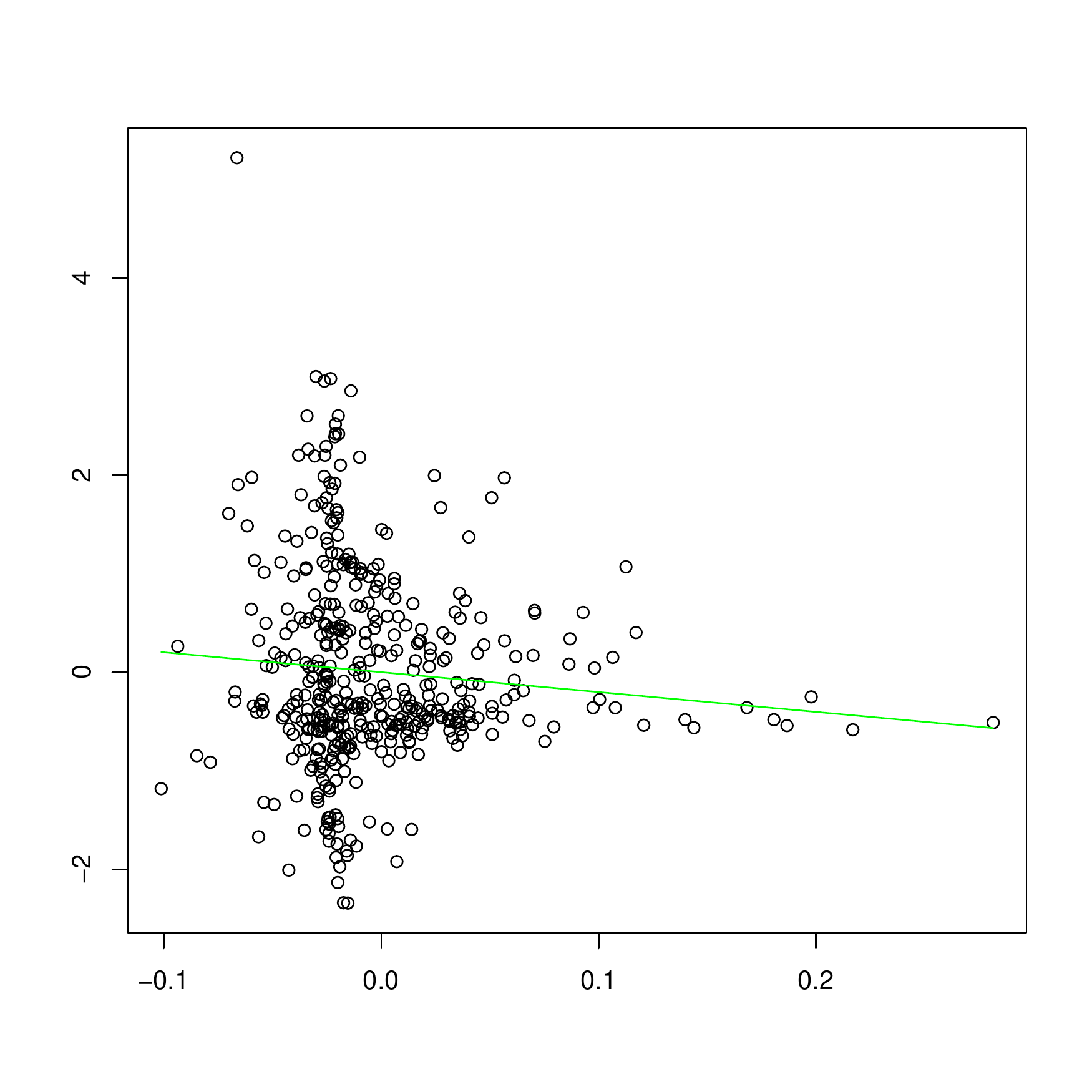}
                 \caption{Elevation.}
                                  \label{fig:RM_add_var1}
\end{subfigure}
         \quad
\begin{subfigure}[b]{0.22\textwidth}
\includegraphics[width=\textwidth]{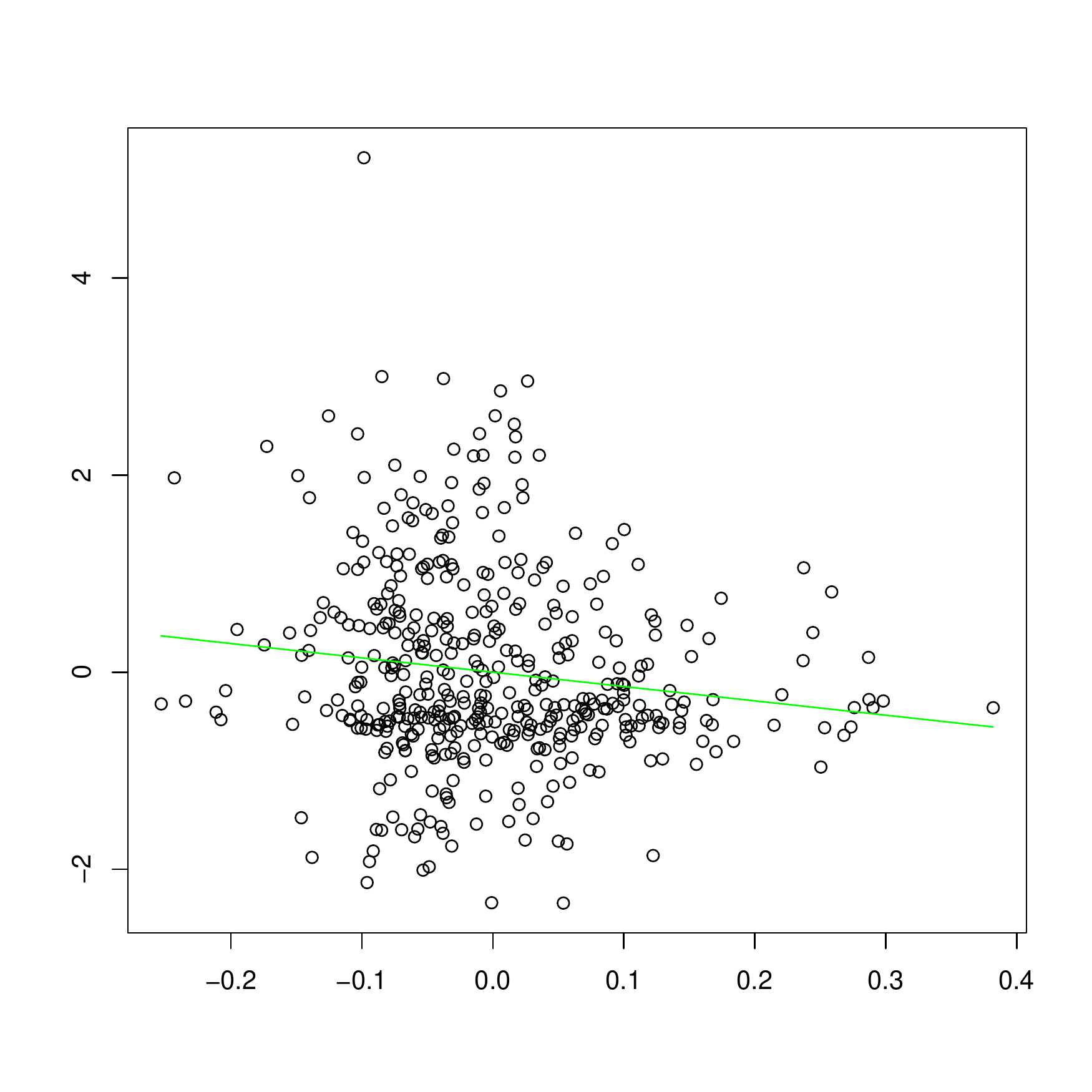}
                 \caption{Slope. }
                 \label{fig:RM_add_var2}
\end{subfigure}
         \quad
\begin{subfigure}[b]{0.22\textwidth}
 \includegraphics[width=\textwidth]{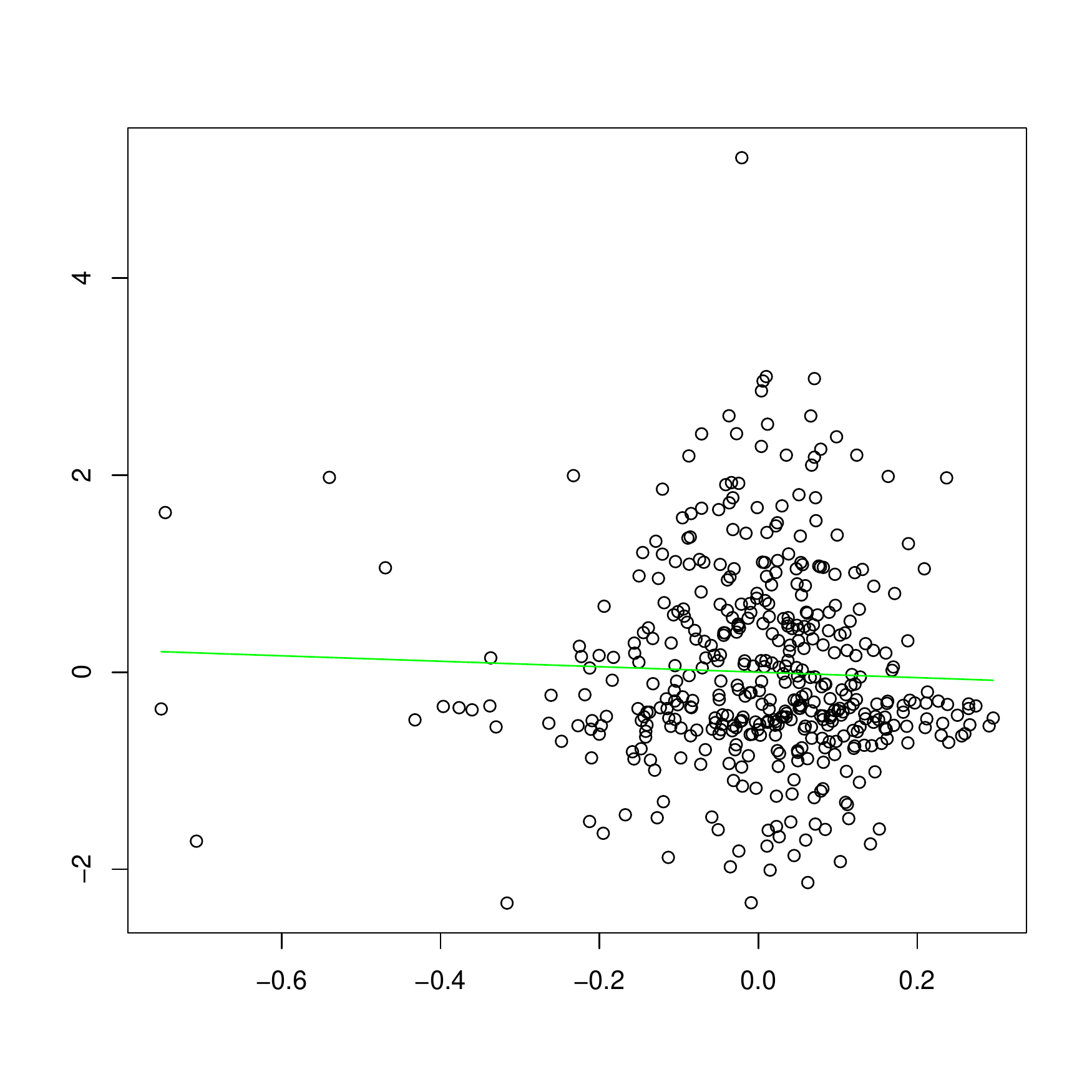}
                 \caption{SpringTC2.}
                 \label{fig:RM_add_var3}
\end{subfigure}
\quad
\begin{subfigure}[b]{0.22\textwidth}
\includegraphics[width=\textwidth]{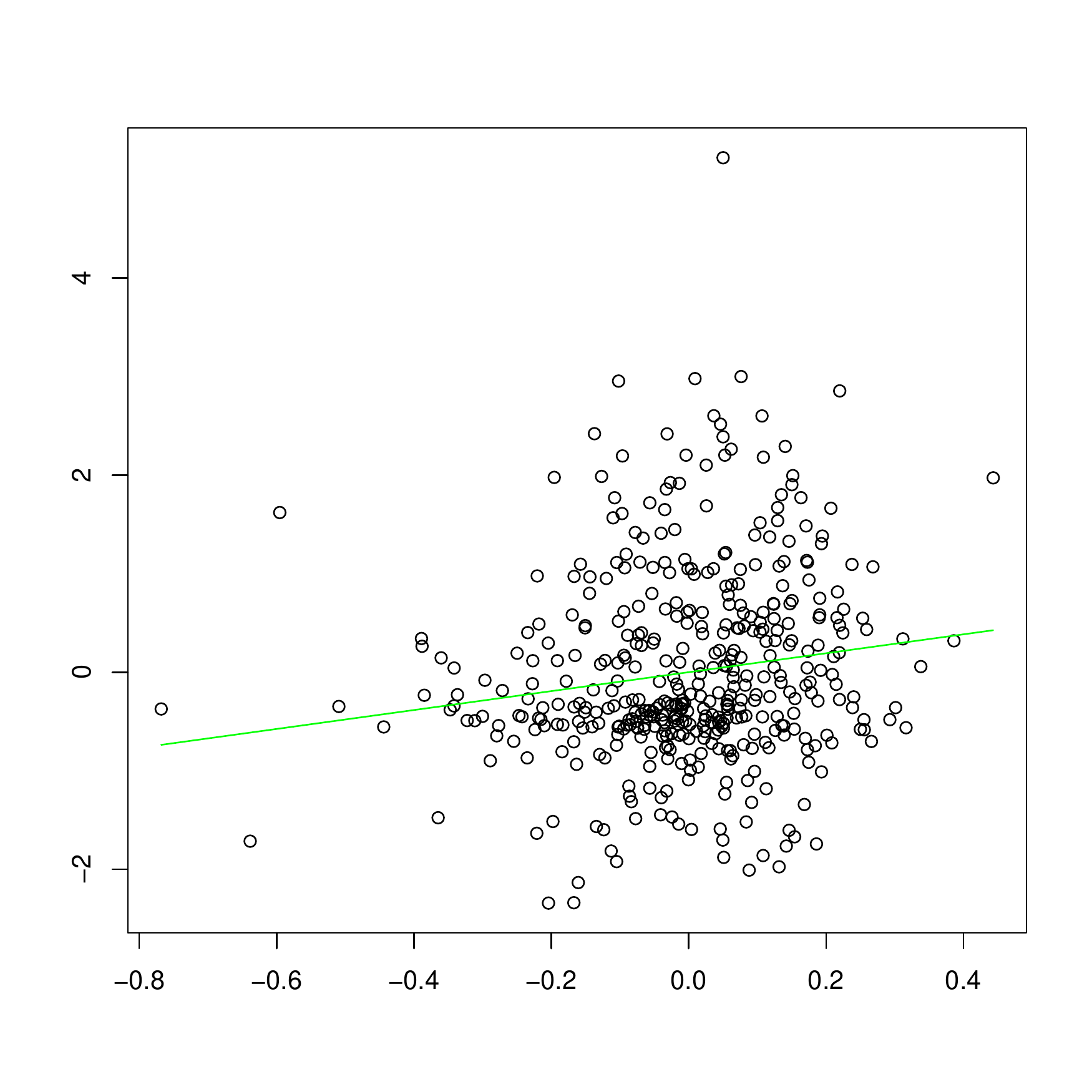}
                 \caption{SpringTC3.}
                 \label{fig:RM_add_var4}
\end{subfigure}
         \quad
         \centering \begin{subfigure}[b]{0.22\textwidth}
 \includegraphics[width=\textwidth]{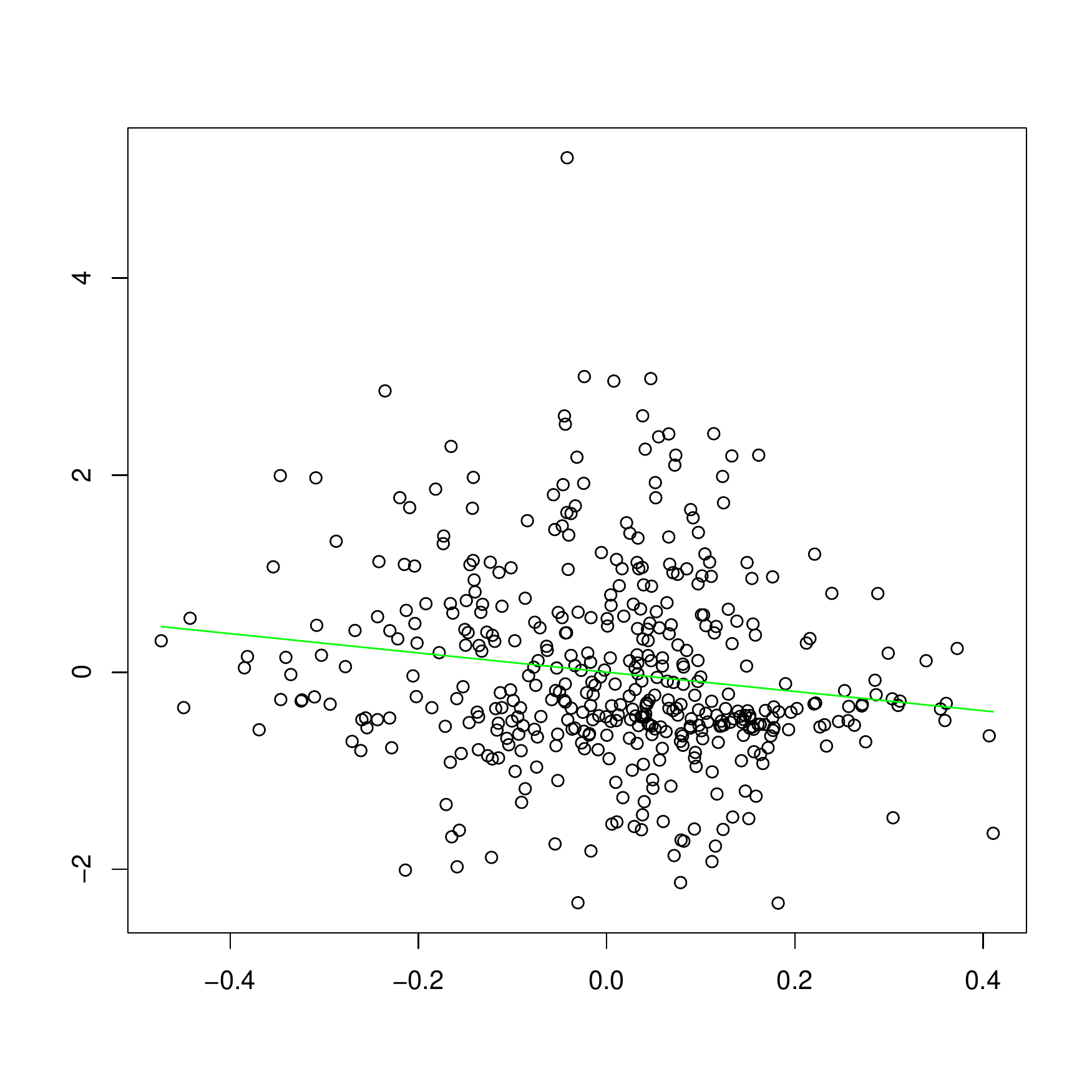}
                 \caption{SummerTC1.}
                                  \label{fig:RM_add_var1}
\end{subfigure}
         \quad
\begin{subfigure}[b]{0.22\textwidth}
 \includegraphics[width=\textwidth]{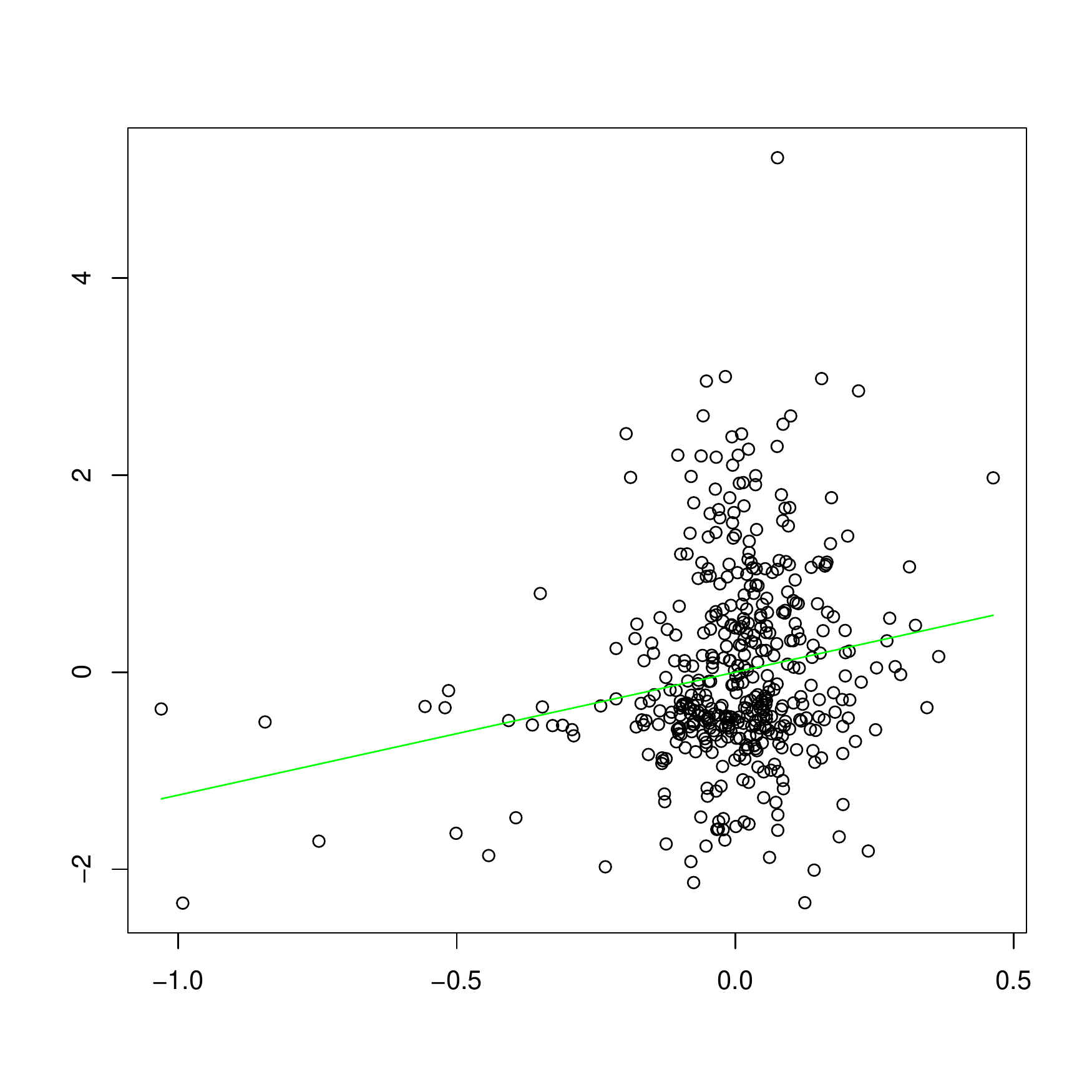}
                 \caption{SummerTC3. }
                 \label{fig:RM_add_var2}
\end{subfigure}
         \quad
\begin{subfigure}[b]{0.22\textwidth}
\includegraphics[width=\textwidth]{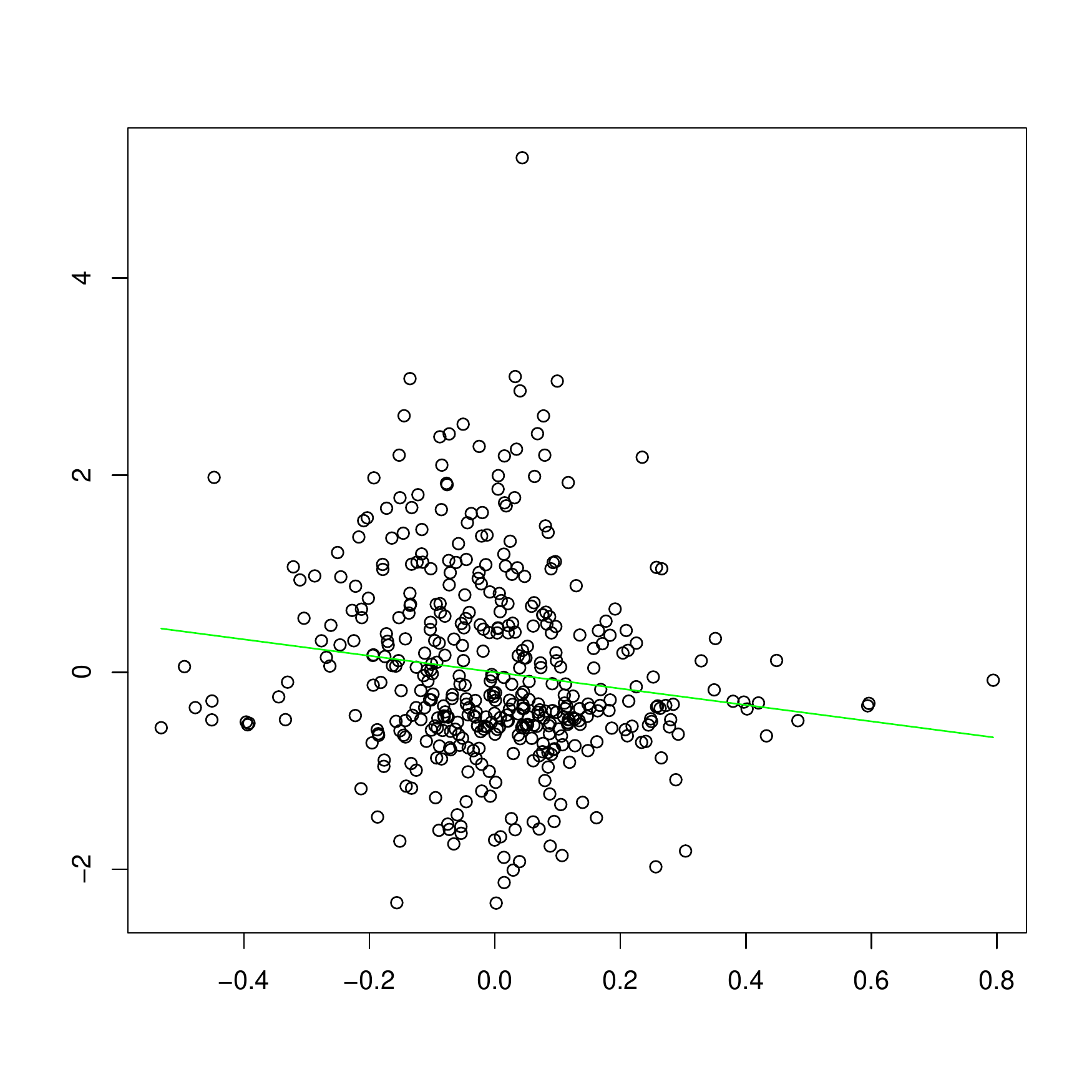}
                 \caption{FallTC2.}
                 \label{fig:RM_add_var3}
\end{subfigure}
 \caption{ \textbf{Step 1} Observation-domain added variable plots after the intercept-only fit.}\label{fig:add_var_obs1}
         \end{figure}

This Appendix shows the observation-domain added variable plots, with five steps adding covariates in the same order as they were added when considering the spectral domain in Appendix H.  For each step, a figure and table below summarize the step's observation-domain added variable plots;  following are comments comparing these added variable plots to those in the spectral domain.  Elevation and Slope have strong low-frequency components, as shown in plots of these covariates in Figures \ref{fig:RM_pred1} and \ref{fig:RM_pred2}).  Thus, their signal is weaker in the observation-domain added variable plots for the intercept-only model (Figure~\ref{fig:add_var_obs1}) than in their earlier spectral-domain added variable plots (Figure~\ref{fig:step1_addvar2});  compare slopes of spectral-domain added variable plots in Step 1, Table \ref{table:table_step1} with the observation-domain analogs, Table \ref{table:table_step1_obs}.  Compare also the spectral-domain added variable plots for Step 2, Table \ref{table:table_step2} with the observation-domain analogs, Table \ref{table:table_step2_obs}. The p-values for Elevation and Slope are much smaller in the spectral-domain added variable plots than in the observation-domain plots;  because these two covariates have relatively large power at low frequencies, they are more easily detectable in the spectral domain. SpringTC2 also has relatively large power in low frequencies so it behaves similarly:  the spectral-domain added variable plot has greater power to detect its slope (compare the spectral-domain added variable plots in Step 3, Table \ref{table:table_step3} with the observation-domain analogs, Table \ref{table:table_step3_obs}). SummerTC1 has both low- and high-frequency features so it gives a rather strong signal in the added variable plots in both domains:  compare the spectral-domain added variable plots in Step 2, Table \ref{table:table_step2} with the observation domain analogs, Table \ref{table:table_step2_obs}.

         \begin{table}[H]
\centering
  \begin{tabular}{ |c || c |c|}
    \hline
    Candidate covariates & slope&p-value \\ \hline
     \textbf{Elevation} &-2.02 &0.07\\ \hline
    Slope &-1.45 & 0.004\\ \hline
    SpringTC2&-0.28 &0.42\\ \hline
    SpringTC3 &0.96 &0.002\\ \hline
    SummerTC1  &-0.98&0.002\\ \hline
    SummerTC3  & 1.25&$10^{-5}$\\ \hline
    FallTC2  &-0.83&0.004\\ \hline  \hline
    \end{tabular}
  \caption{\textbf{Step 1} Slopes of observation-domain added variable plots after the intercept-only fit. Covariate selected to be added on the basis of these plots is in bold.  }\label{table:table_step1_obs}
\end{table}

         \begin{figure}[H]
         \centering \begin{subfigure}[b]{0.22\textwidth}
 \includegraphics[width=\textwidth]{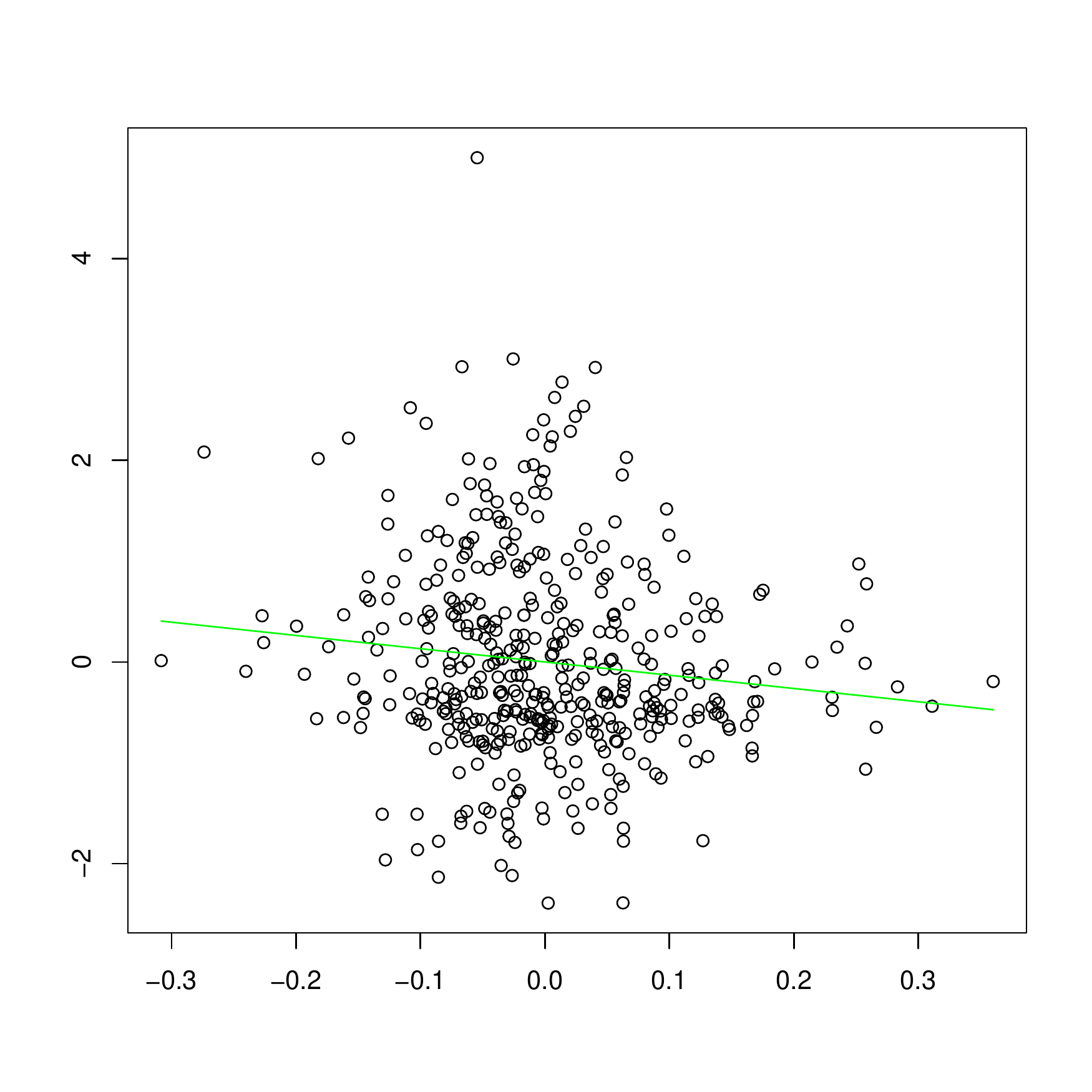}
                 \caption{Slope.}
                                  \label{fig:RM_add_var1}
\end{subfigure}
                  \quad
\begin{subfigure}[b]{0.22\textwidth}
\includegraphics[width=\textwidth]{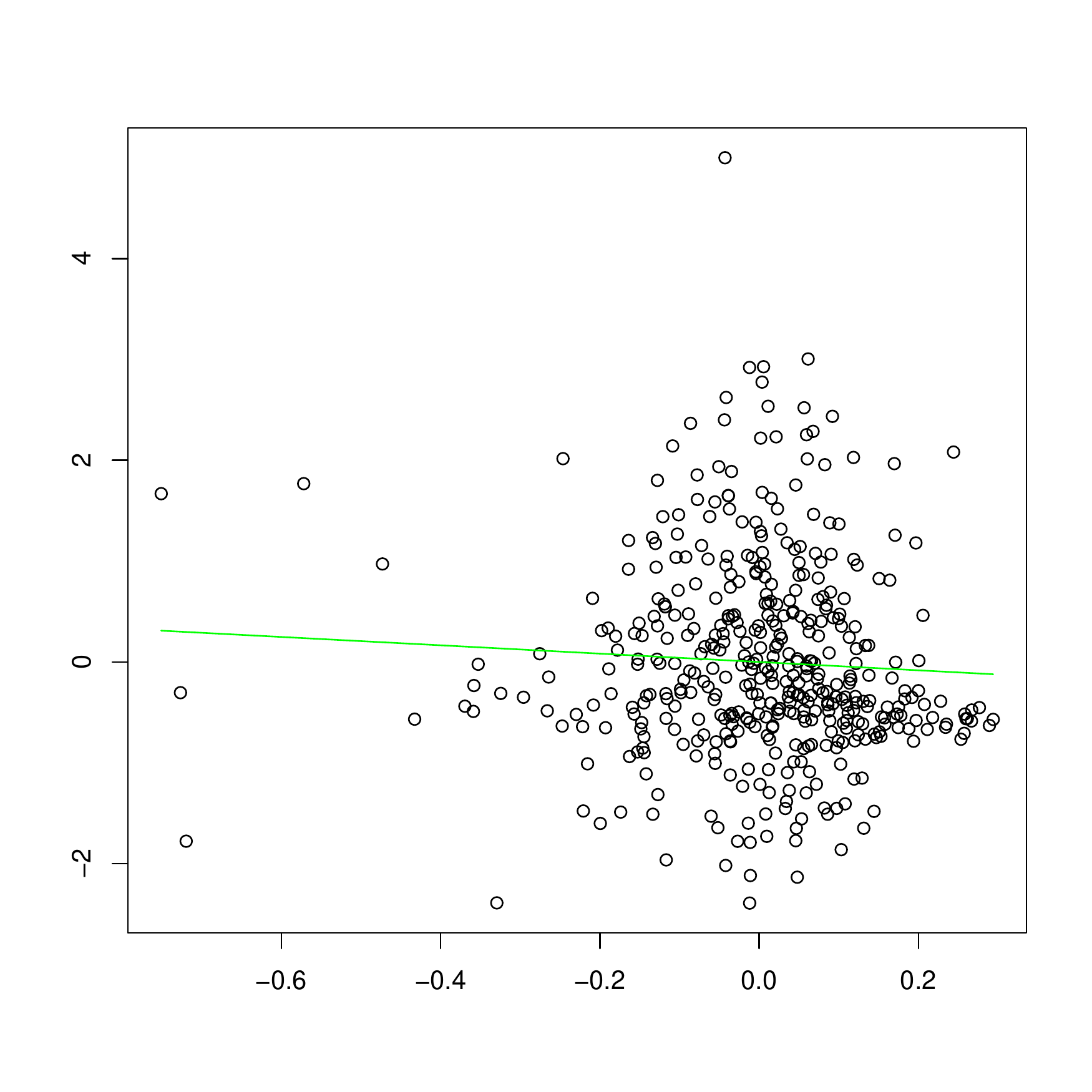}
                 \caption{SpringTC2.}
                 \label{fig:RM_add_var3}
\end{subfigure}
\quad
\begin{subfigure}[b]{0.22\textwidth}
 \includegraphics[width=\textwidth]{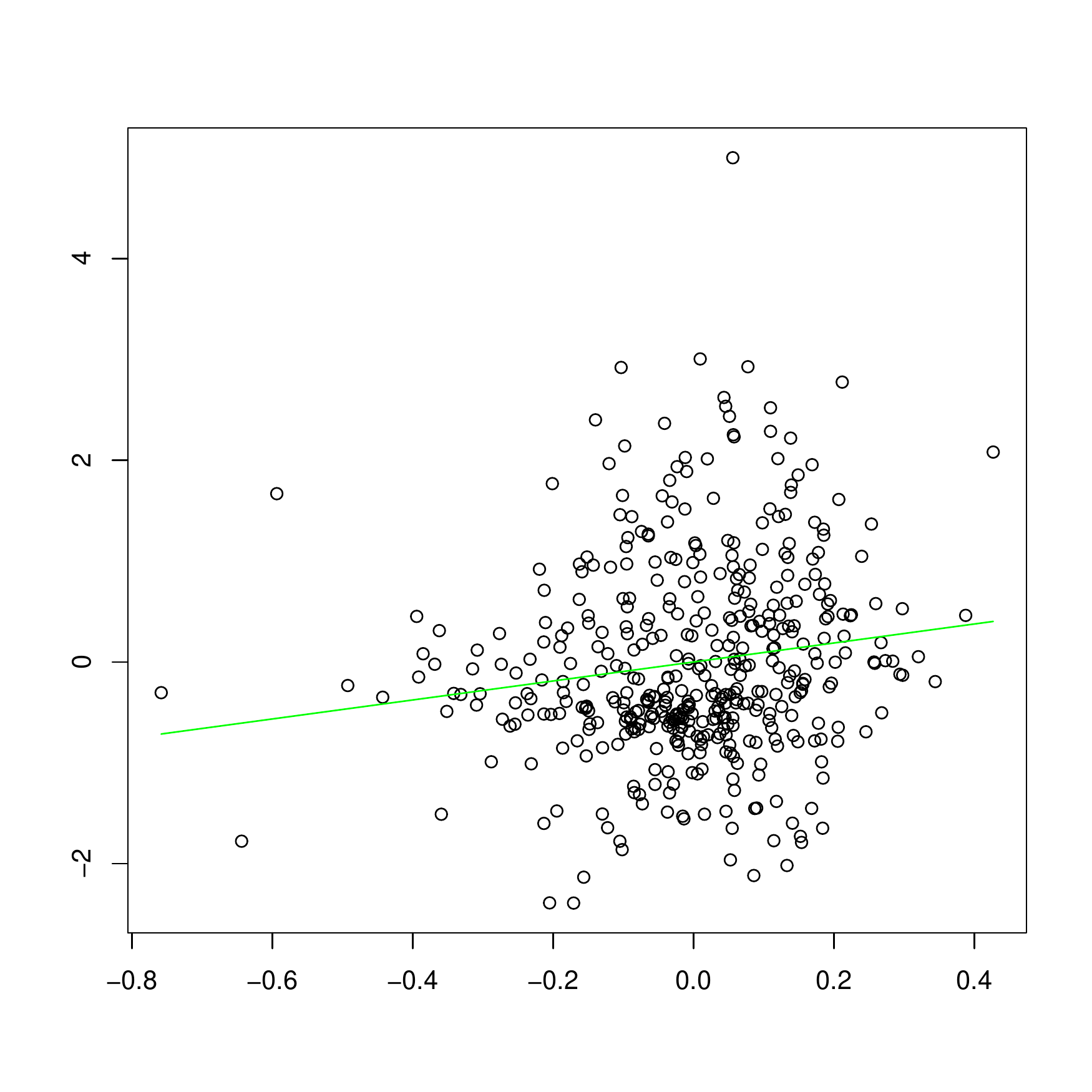}
                 \caption{SpringTC3.}
                 \label{fig:RM_add_var4}
\end{subfigure}
\quad
         \begin{subfigure}[b]{0.22\textwidth}
 \includegraphics[width=\textwidth]{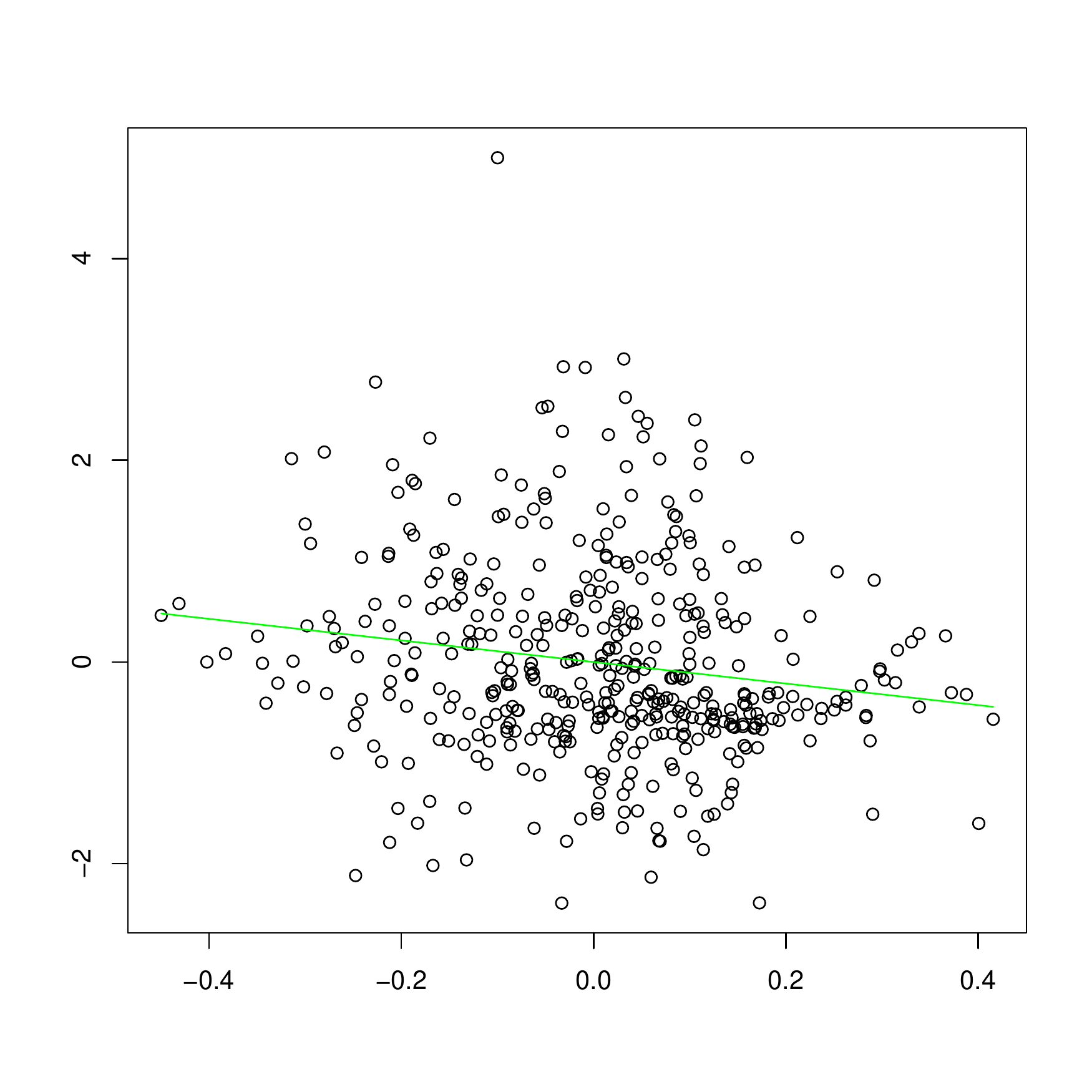}
                 \caption{SummerTC1.}
                                  \label{fig:RM_add_var1}
\end{subfigure}
         \quad
\begin{subfigure}[b]{0.22\textwidth}
\includegraphics[width=\textwidth]{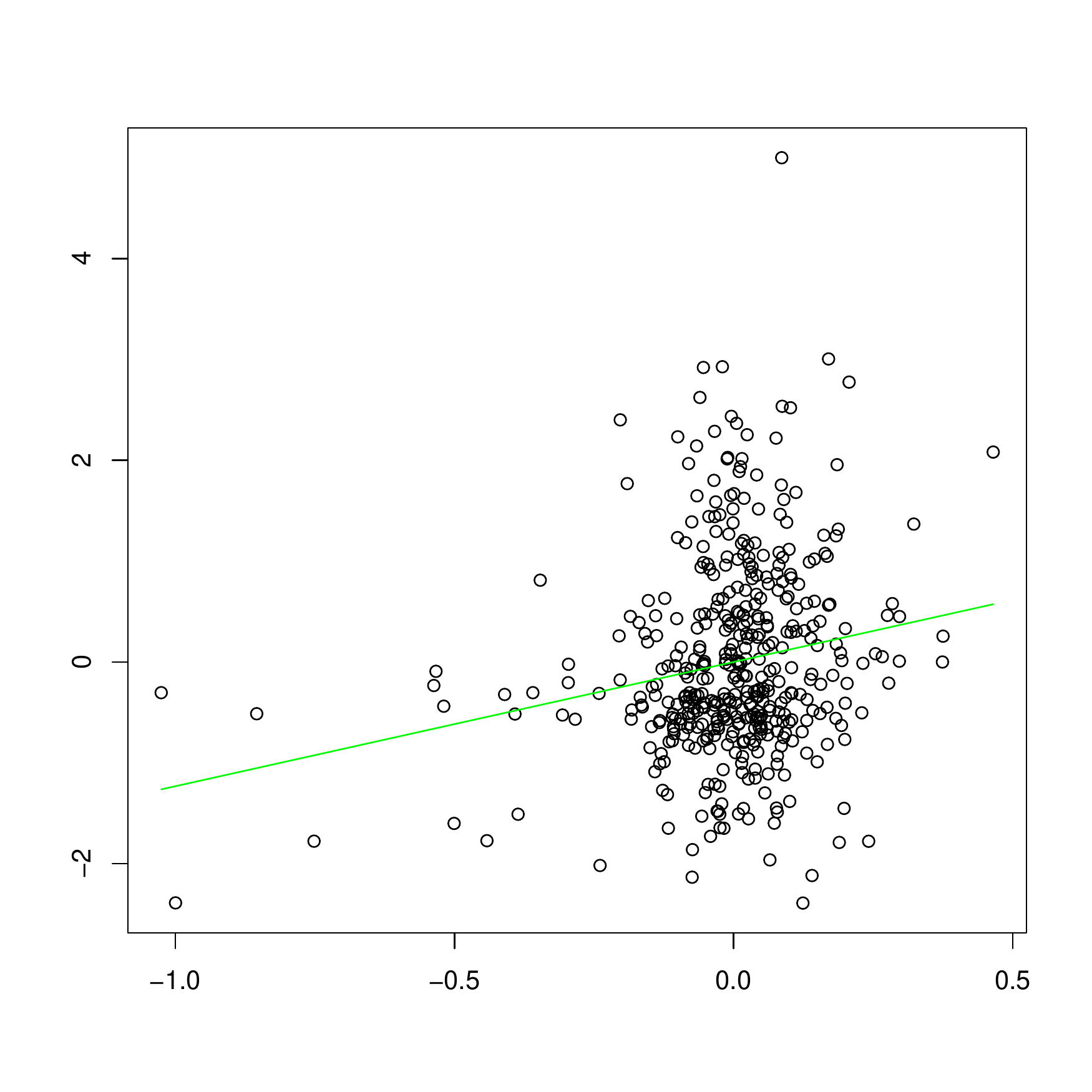}
                 \caption{SummerTC3. }
                 \label{fig:RM_add_var2}
\end{subfigure}
         \quad
\begin{subfigure}[b]{0.22\textwidth}
\includegraphics[width=\textwidth]{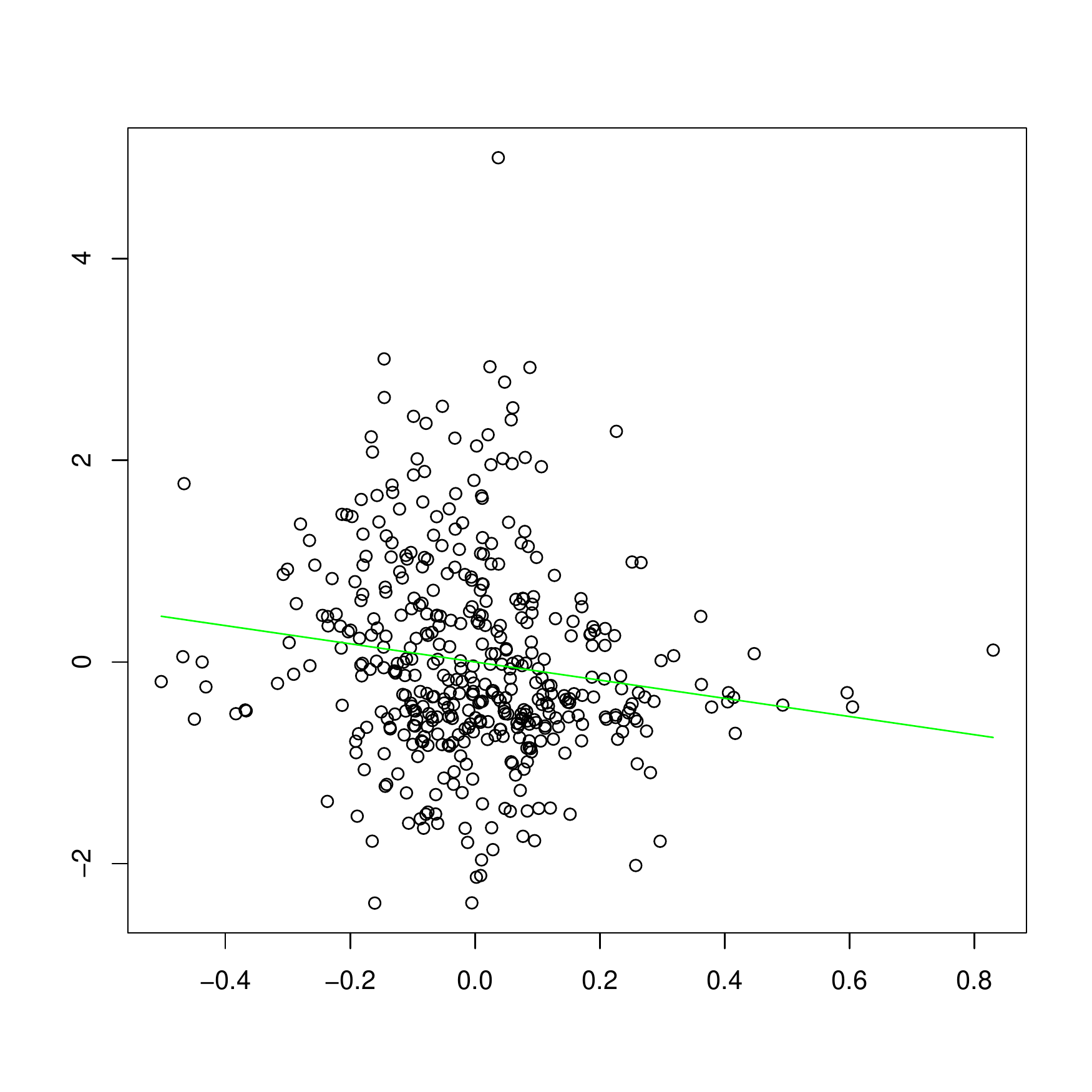}
                 \caption{FallTC2.}
                 \label{fig:RM_add_var3}
\end{subfigure}
 \caption{ \textbf{Step 2} Observation-domain added variable plots after the fit with covariate Elevation.}\label{fig:add_var}
         \end{figure}

         \begin{table}[H]
\centering
  \begin{tabular}{ |c || c |c|}
    \hline
    Candidate covariates &slope&p-value \\ \hline
      $\downarrow$ & & \\ \hline \hline
    Elevation& --&-- \\ \hline
    \textbf{Slope}  & -1.31&0.01 \\ \hline
    SpringTC2 &-0.41 &0.23\\ \hline
    SpringTC3  & 0.94&0.002\\ \hline
    \textbf{SummerTC1} &-1.07 &0.0008\\ \hline
    SummerTC3  &1.23&$10^{-5}$\\ \hline
    FallTC2 &-0.90&0.002\\ \hline  \hline
    \end{tabular}
  \caption{ \textbf{Step 2} Slopes of observation-domain added variable plots after the fit with Elevation. covariates selected to be added on the basis of these plots are in bold. The symbol ``--" denotes covariates already in the model.}\label{table:table_step2_obs}
\end{table}

         \begin{figure}[H]
         \centering
\begin{subfigure}[b]{0.22\textwidth}
\includegraphics[width=\textwidth]{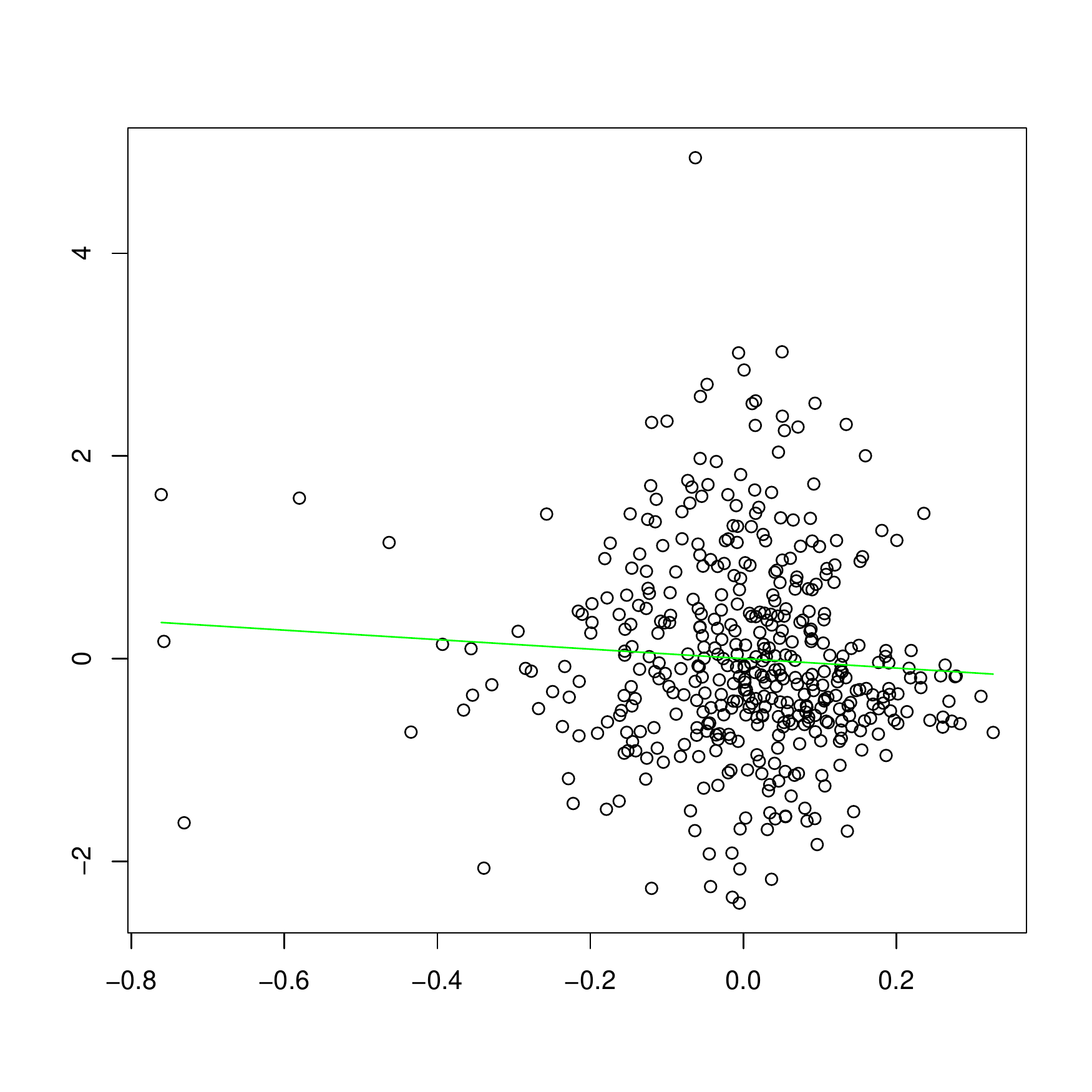}
                 \caption{SpringTC2.}
                 \label{fig:RM_add_var3}
\end{subfigure}
\quad
\begin{subfigure}[b]{0.22\textwidth}
\includegraphics[width=\textwidth]{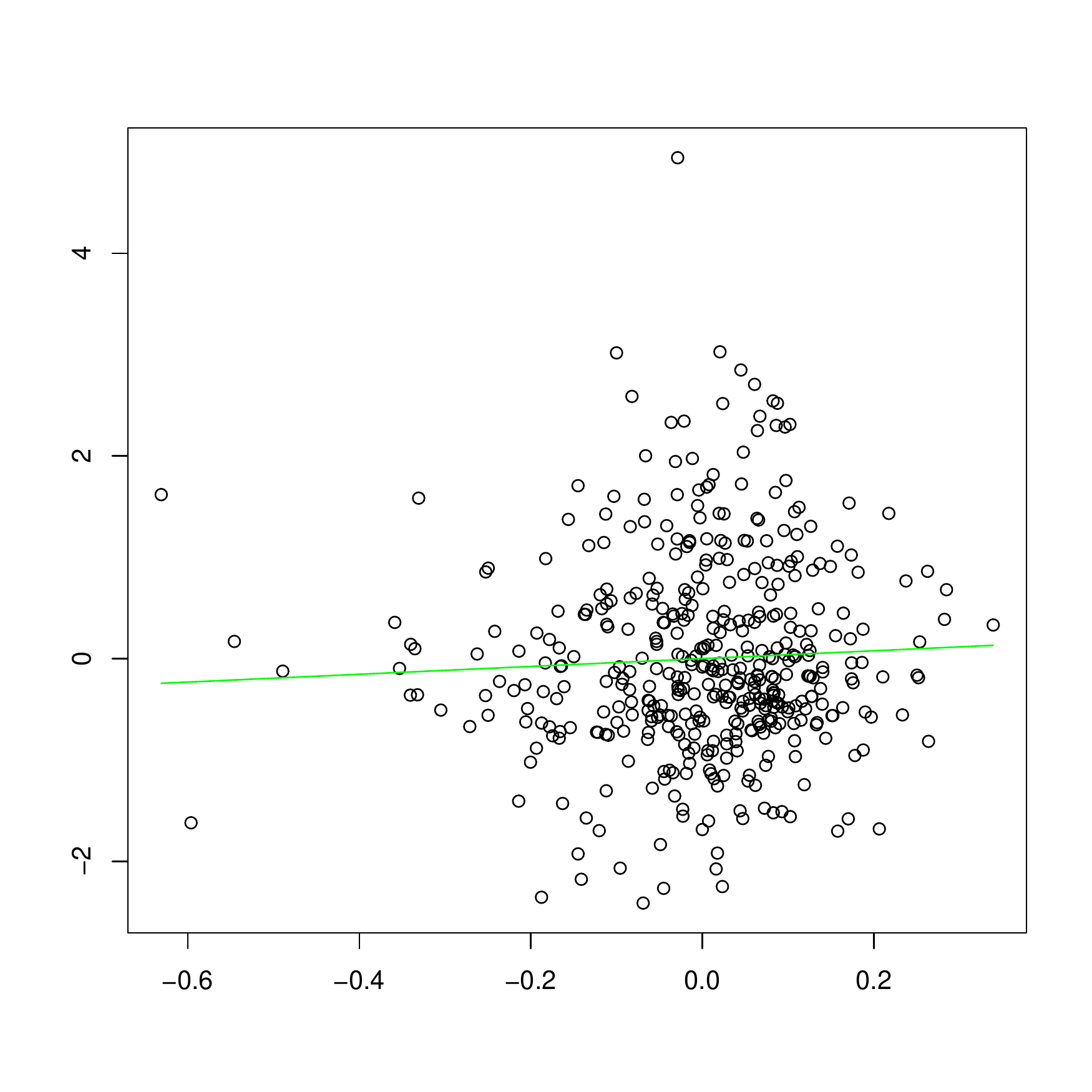}
                 \caption{SpringTC3.}
                 \label{fig:RM_add_var4}
\end{subfigure}
 \quad
 \begin{subfigure}[b]{0.22\textwidth}
 \includegraphics[width=\textwidth]{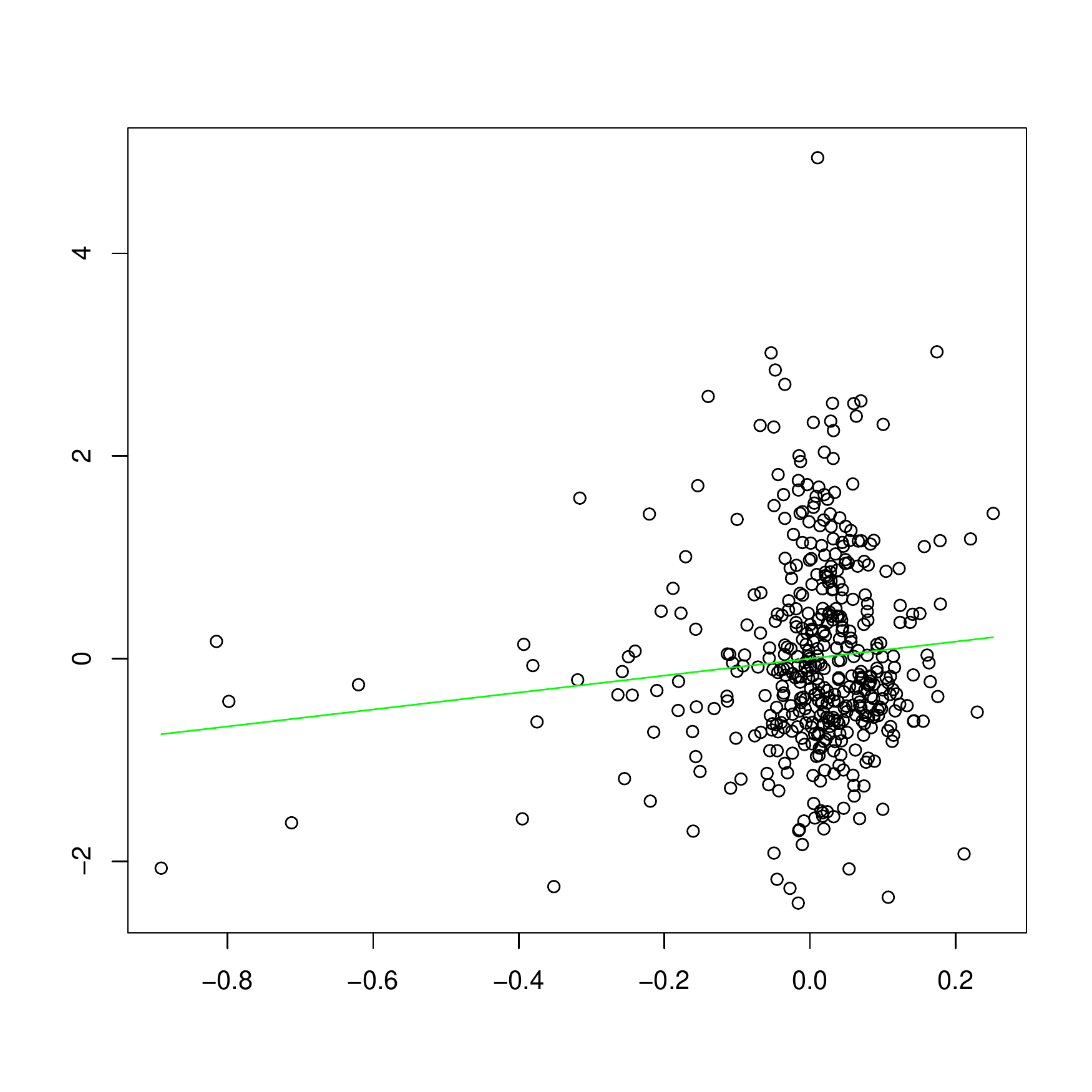}
                 \caption{SummerTC3.}
                                  \label{fig:RM_add_var1}
\end{subfigure}
         \quad
\begin{subfigure}[b]{0.22\textwidth}
\includegraphics[width=\textwidth]{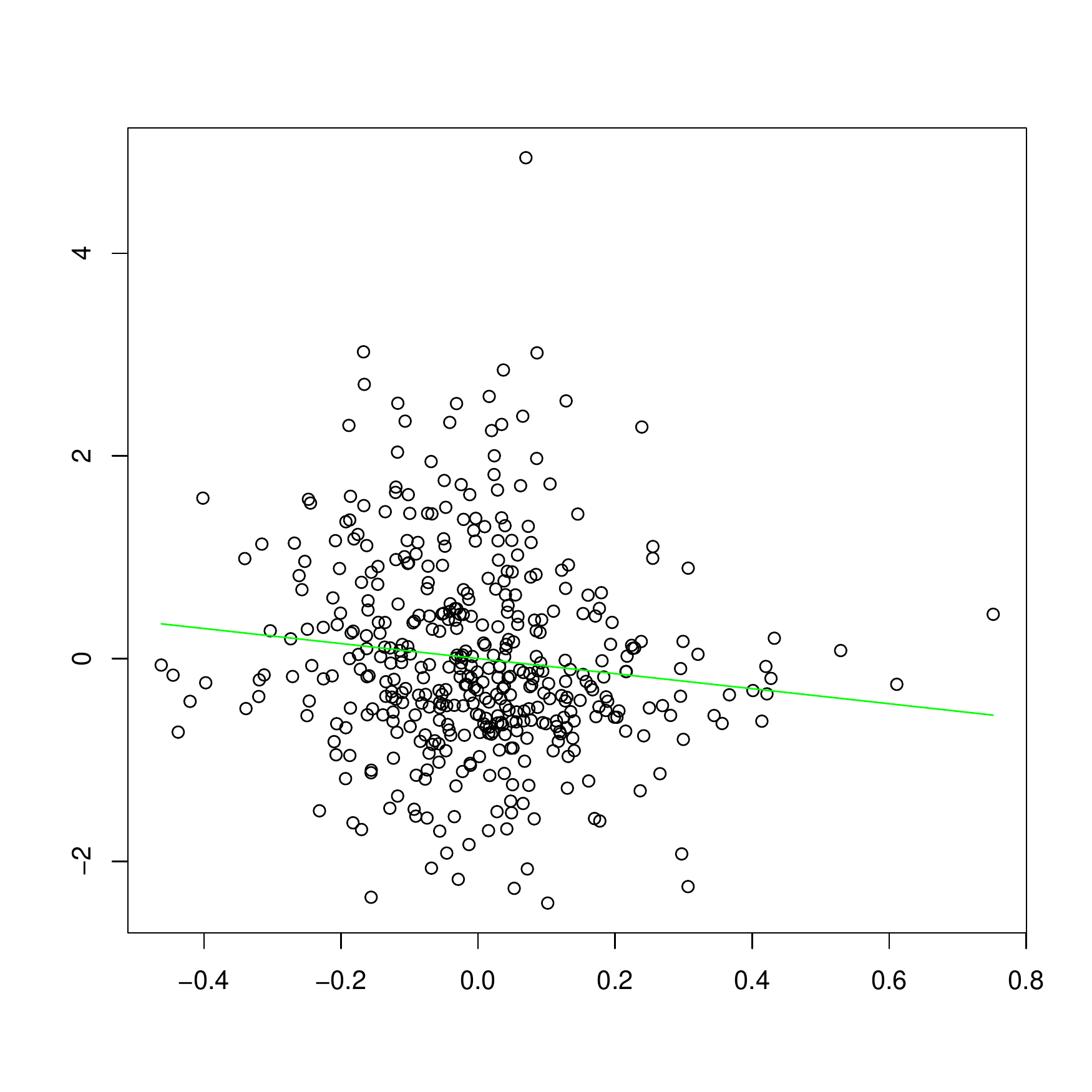}
                 \caption{FallTC2. }
                 \label{fig:RM_add_var2}
\end{subfigure}
         \caption{ \textbf{Step 3} Observation-domain added variable plots after the fit with covariates Elevation, Slope, and SummerTC1.}\label{fig:add_var}
         \end{figure}

         \begin{table}[H]
\centering
  \begin{tabular}{ |c || c |c|}
    \hline
    Candidate covariates &  slope&p-value \\ \hline
      $\downarrow$ & & \\ \hline \hline
    Elevation& --&-- \\ \hline
    Slope &-- &--\\ \hline
    \textbf{SpringTC2}&-0.47 &0.16\\ \hline
    SpringTC3  & 0.39&0.30\\ \hline
  SummerTC1  &-- &--\\ \hline
    SummerTC3 &0.84&0.03\\ \hline
    FallTC2  &-0.74&0.01\\ \hline  \hline
    \end{tabular}
  \caption{ \textbf{Step 3} Slopes of observation-domain added variable plots after the fit with Elevation, Slope, and SummerTC1. covariate selected to be added on the basis of these plots is in bold. The symbol ``--" denotes covariates already in the model.  }\label{table:table_step3_obs}
\end{table}

          \begin{figure}[H]
         \centering

\begin{subfigure}[b]{0.22\textwidth}
 \includegraphics[width=\textwidth]{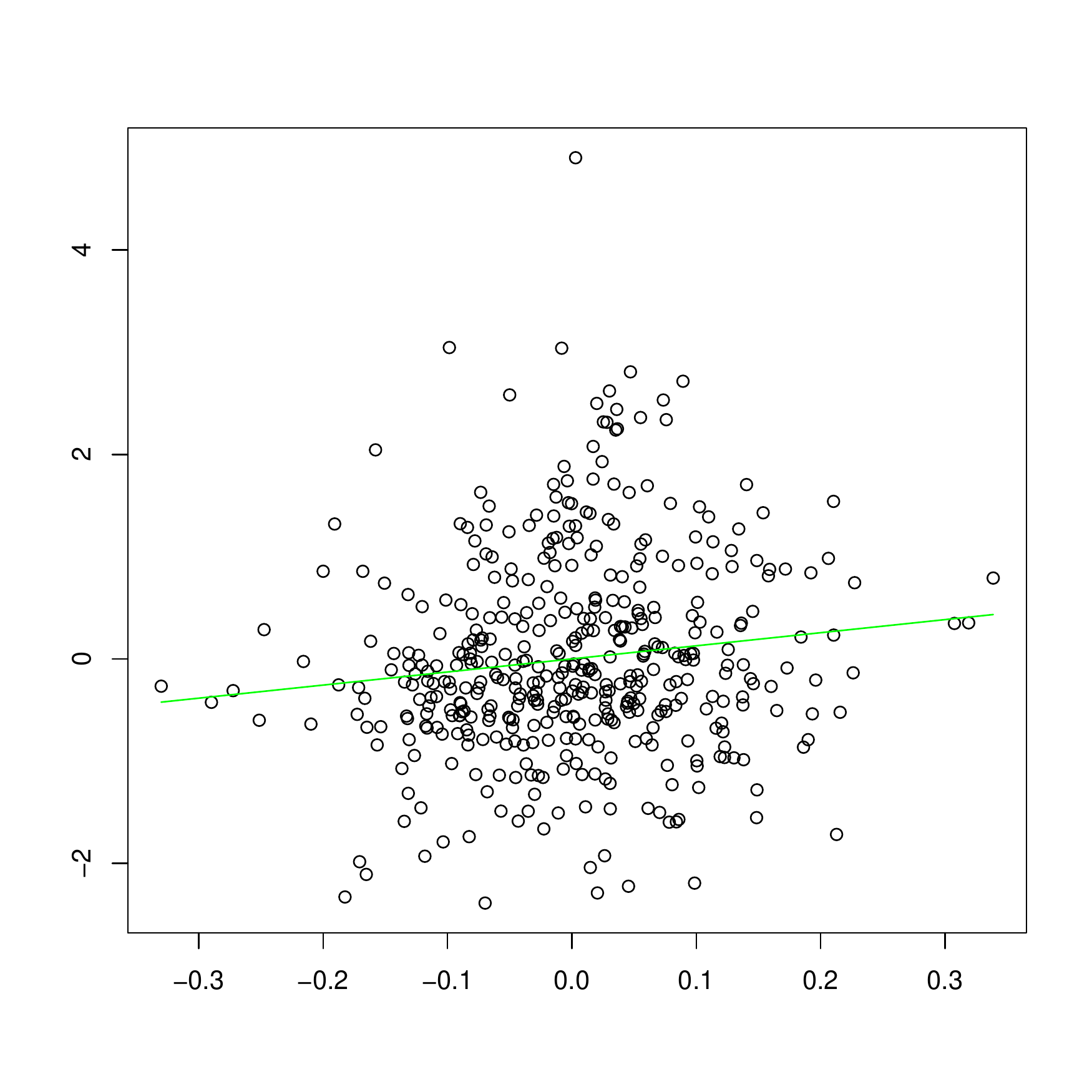}
                 \caption{SpringTC3.}
                 \label{fig:RM_add_var4}
\end{subfigure}
\quad
 \begin{subfigure}[b]{0.23\textwidth}
 \includegraphics[width=\textwidth]{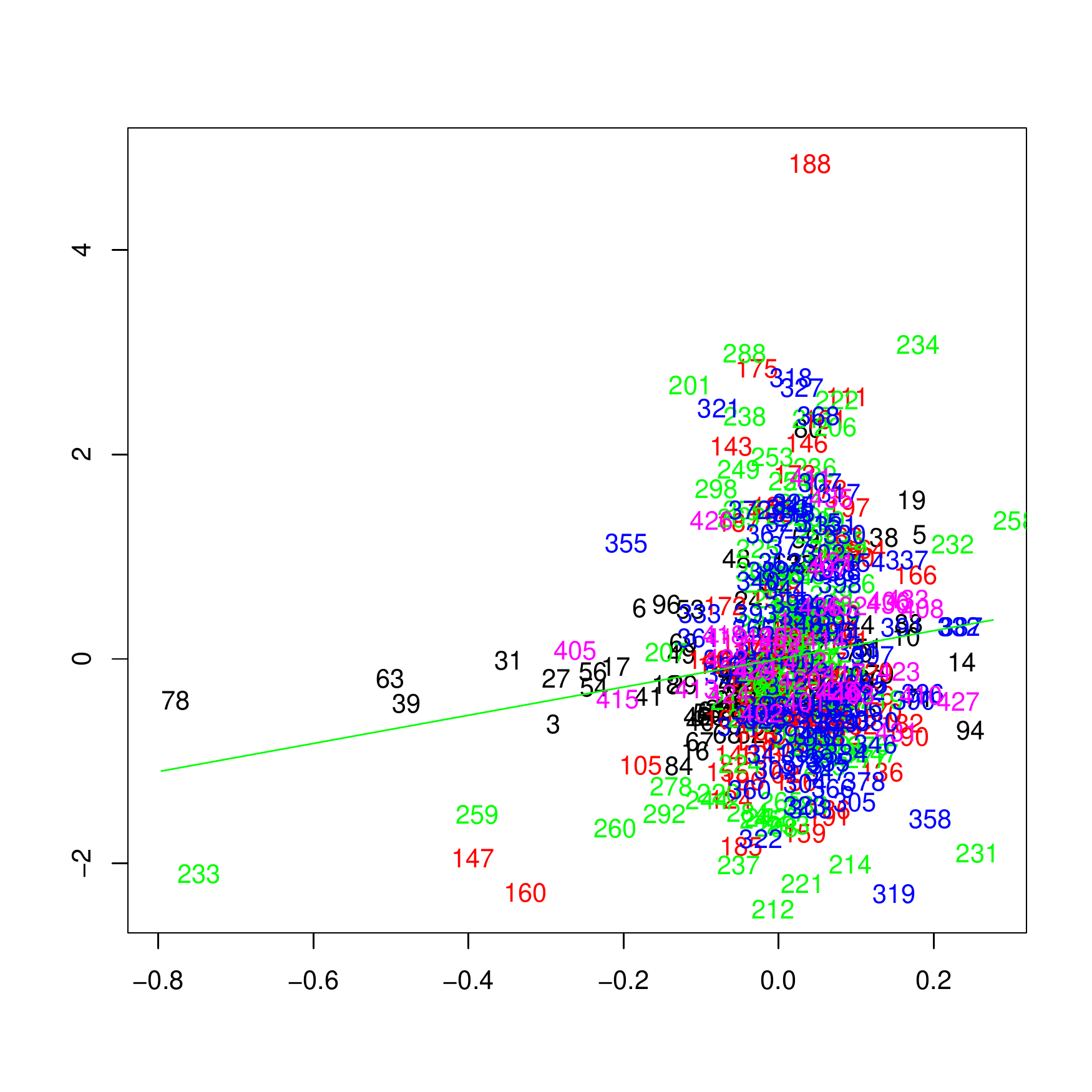}
                 \caption{SummerTC3.}
                                  \label{fig:RM_add_var1}
\end{subfigure}
         \quad
\begin{subfigure}[b]{0.22\textwidth}
\includegraphics[width=\textwidth]{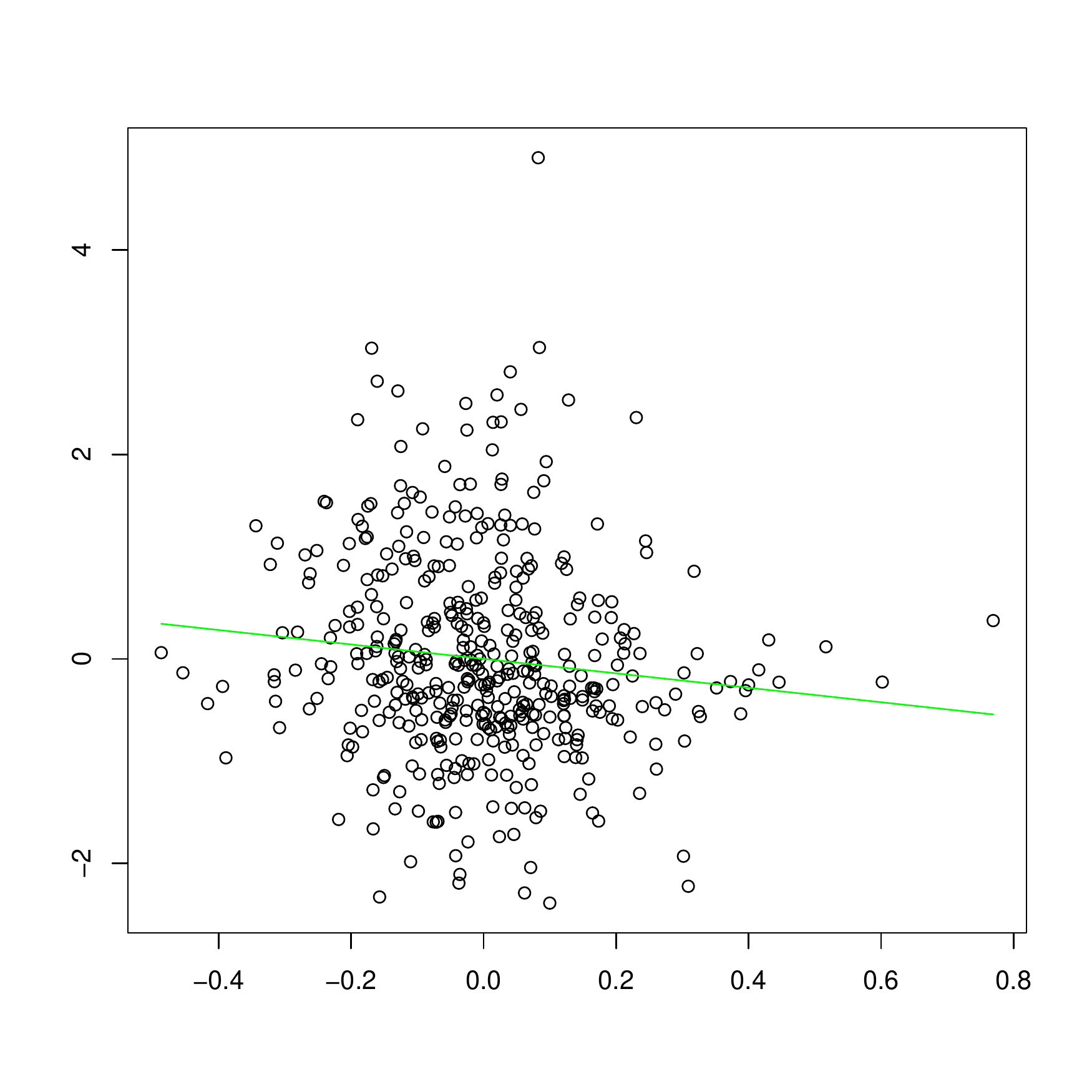}
                 \caption{FallTC2. }
                 \label{fig:RM_add_var2}
\end{subfigure}

 \caption{
         \textbf{Step 4} Observation-domain added variable plots after the fit with covariates Elevation, Slope, SummerTC1, SpringTC2, and north-south quadratic trend.}\label{fig:add_var}
         \end{figure}

\begin{table}[H]
\centering
  \begin{tabular}{ |c || c |c|}
    \hline
    Candidate covariates &  slope & p-value \\ \hline
      $\downarrow$ &&\\ \hline \hline
    Elevation& -- &--\\ \hline
    Slope & --&--\\ \hline
    SpringTC2&-- &--\\ \hline
    SpringTC3  &1.28  &0.009\\ \hline
  SummerTC1  &--&--\\ \hline
    \textbf{SummerTC3} &1.38 &0.001\\ \hline
    FallTC2   &-0.71  &0.02\\ \hline  \hline
    \end{tabular}
  \caption{
         \textbf{Step 4} Slopes of observation-domain added variable plots after the fit with covariates Elevation, Slope, SummerTC1, SpringTC2, and north-south quadratic trend. covariate selected to be added on the basis of these plots is in bold. The symbol ``--" denotes covariates already in the model. }\label{table:table_step4_obs}
\end{table}

          \begin{figure}[H]
         \centering
\begin{subfigure}[b]{0.22\textwidth}
 \includegraphics[width=\textwidth]{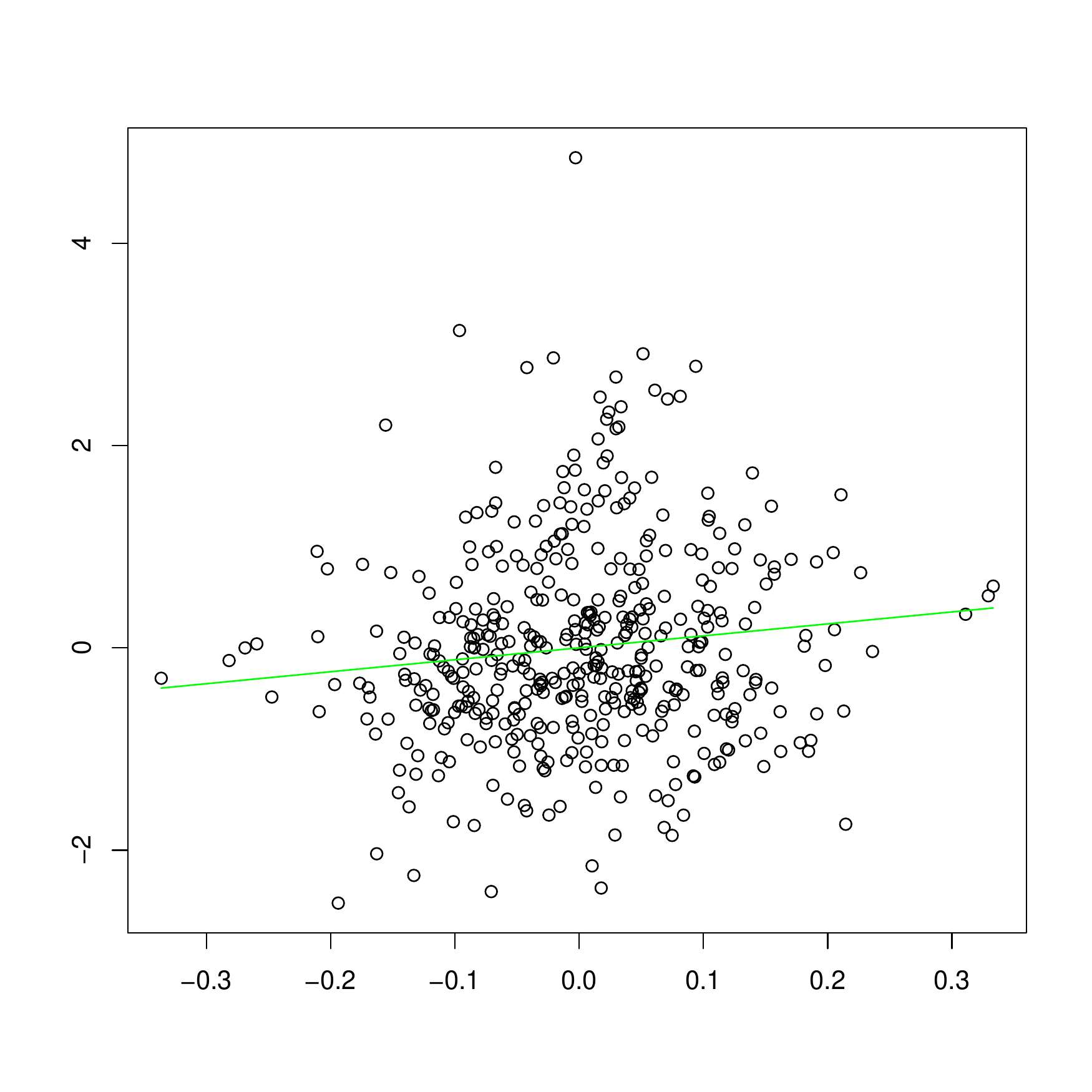}
                 \caption{SpringTC3.}
                 \label{fig:RM_add_var4}
\end{subfigure}
         \quad
\begin{subfigure}[b]{0.22\textwidth}
\includegraphics[width=\textwidth]{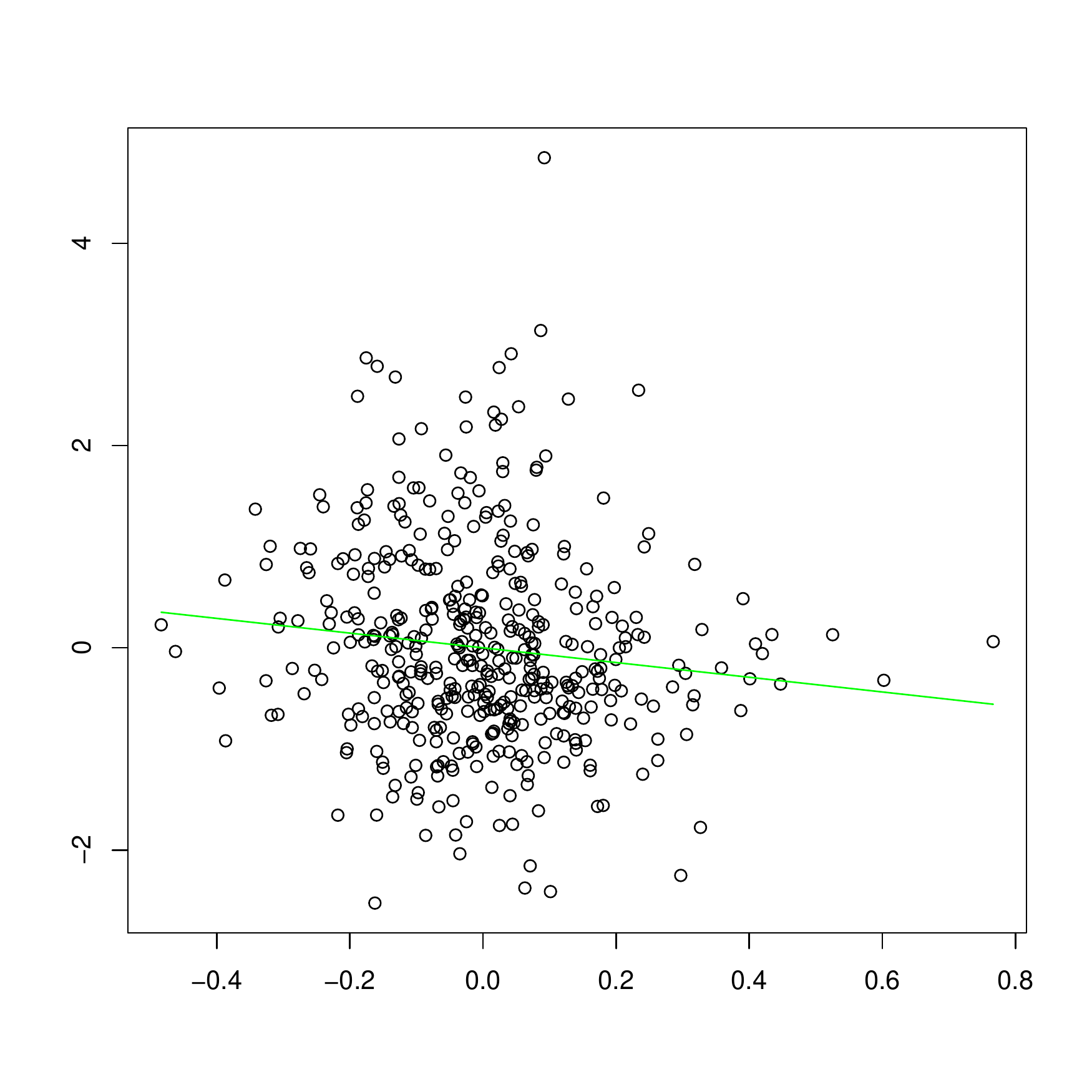}
                 \caption{FallTC2. }
                 \label{fig:RM_add_var2}
\end{subfigure}

 \caption{\textbf{Step 5} Observation-domain added variable plots after the fit with covariates Elevation, Slope, SummerTC1, SpringTC2, SummerTC3, and north-south quadratic trend.}\label{fig:add_var}
         \end{figure}

\begin{table}[H]
\centering
  \begin{tabular}{ |c || c |c|}
    \hline
    Candidate covariates &  slope & p-value \\ \hline
      $\downarrow$ &&\\ \hline \hline
    Elevation& -- &--\\ \hline
    Slope & --&--\\ \hline
    SpringTC2&-- &--\\ \hline
    SpringTC3  &1.16  &0.02\\ \hline
  SummerTC1  &--&--\\ \hline
    SummerTC3 &-- &--\\ \hline
    FallTC2   &-0.73  &0.01\\ \hline  \hline
    \end{tabular}
  \caption{\textbf{Step 5}  Slopes of observation-domain added variable plots after the fit with covariates Elevation, Slope, SummerTC1, SpringTC2, SummerTC3, and north-south quadratic trend. The symbol ``--" denotes covariates already in the model. }\label{table:table_step5_obs}
\end{table}

\section{Web Appendix J. \hspace*{3mm} Fits of the raw data using the exact restricted likelihood}\label{sec:real_data_exact}
\begin{figure}[H]
         \centering
 \includegraphics[width=0.22\textwidth]{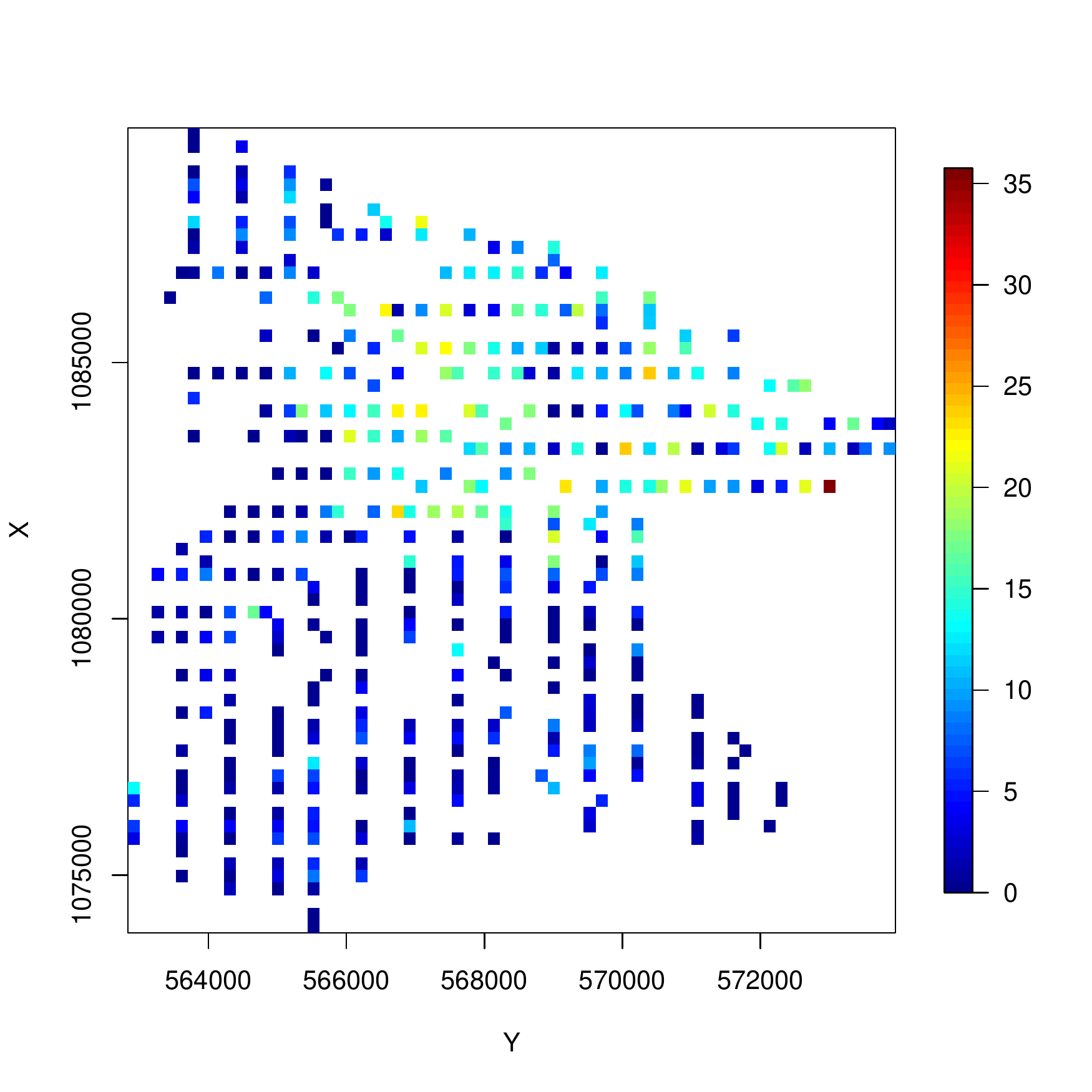}
                 \caption{ Observed data on the observation domain.}
                                  \label{fig:RM_obsdata}
\end{figure}

\begin{figure}[H]
         \centering \begin{subfigure}[b]{0.17\textwidth}
 \includegraphics[width=\textwidth]{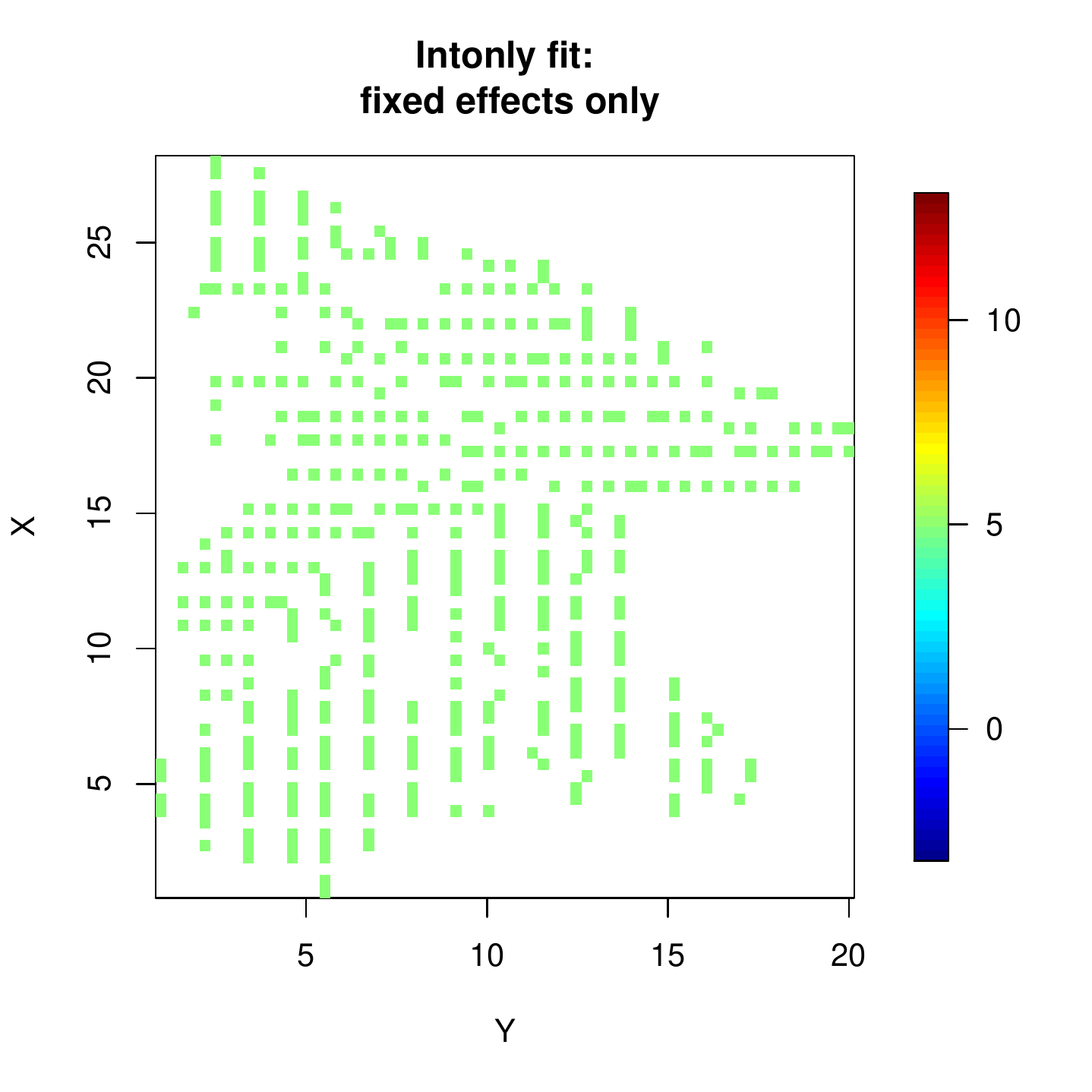}
                 \caption{\\Intercept-only \\fit.\\}
                                  \label{fig:RM_add_var1}
\end{subfigure}
         \quad
\begin{subfigure}[b]{0.17\textwidth}
 \includegraphics[width=\textwidth]{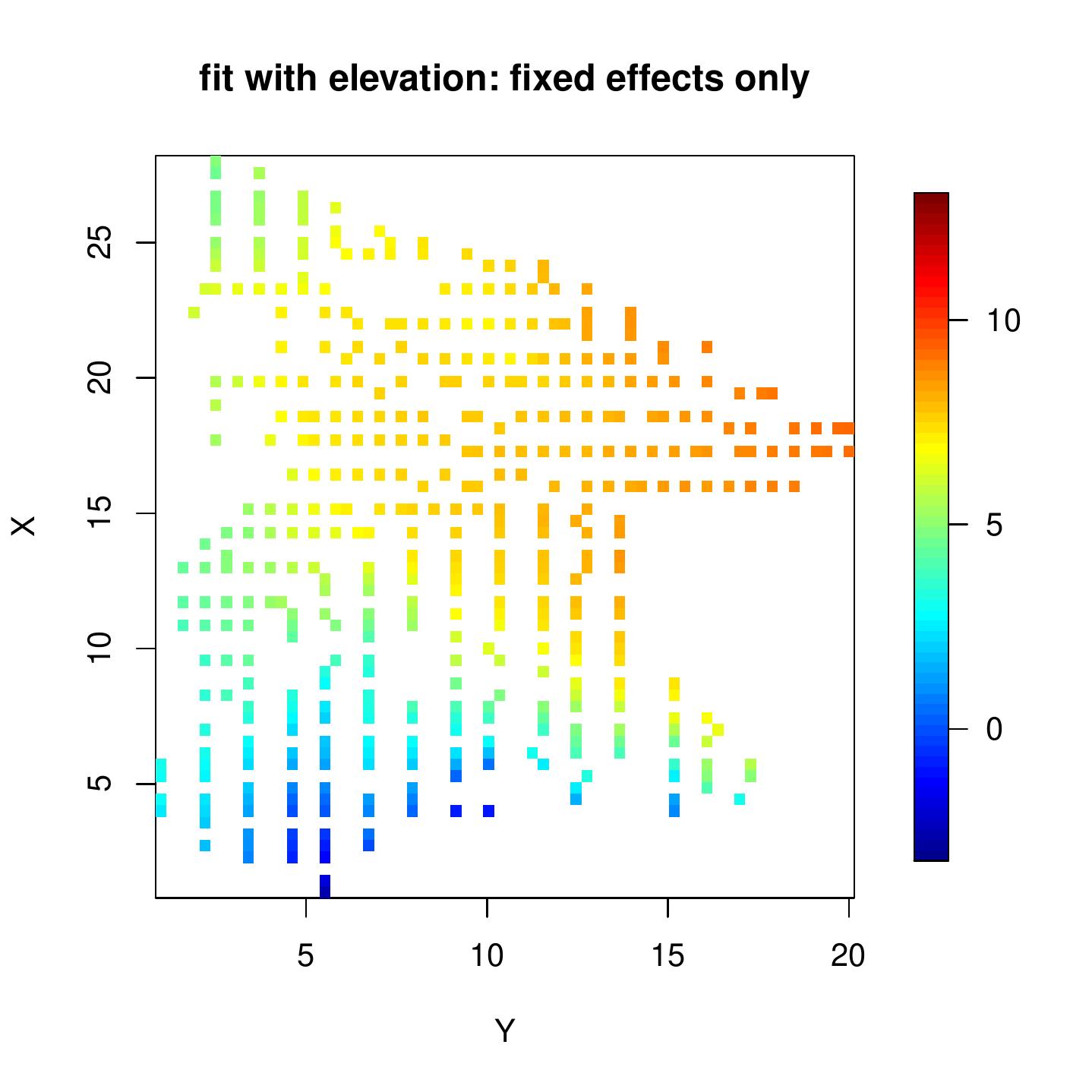}
                 \caption{\\Fit \\with Elevation.\\ }
                 \label{fig:RM_add_var2}
\end{subfigure}
         \quad
\begin{subfigure}[b]{0.17\textwidth}
\includegraphics[width=\textwidth]{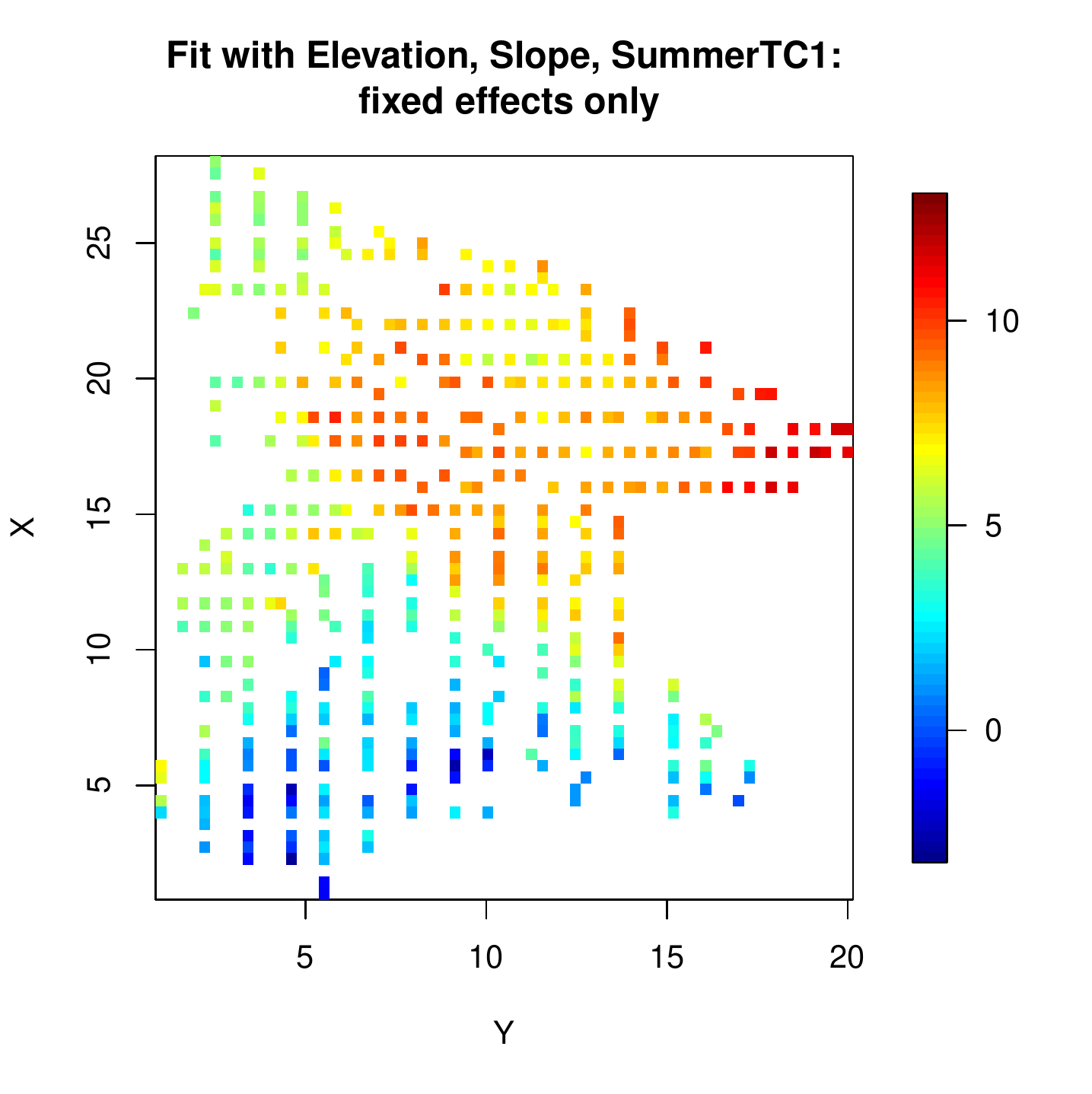}
                 \caption{\\Fit with Elevation, Slope, SummerTC1.}
                 \label{fig:RM_add_var3}
\end{subfigure}
\quad
\begin{subfigure}[b]{0.17\textwidth}
\includegraphics[width=\textwidth]{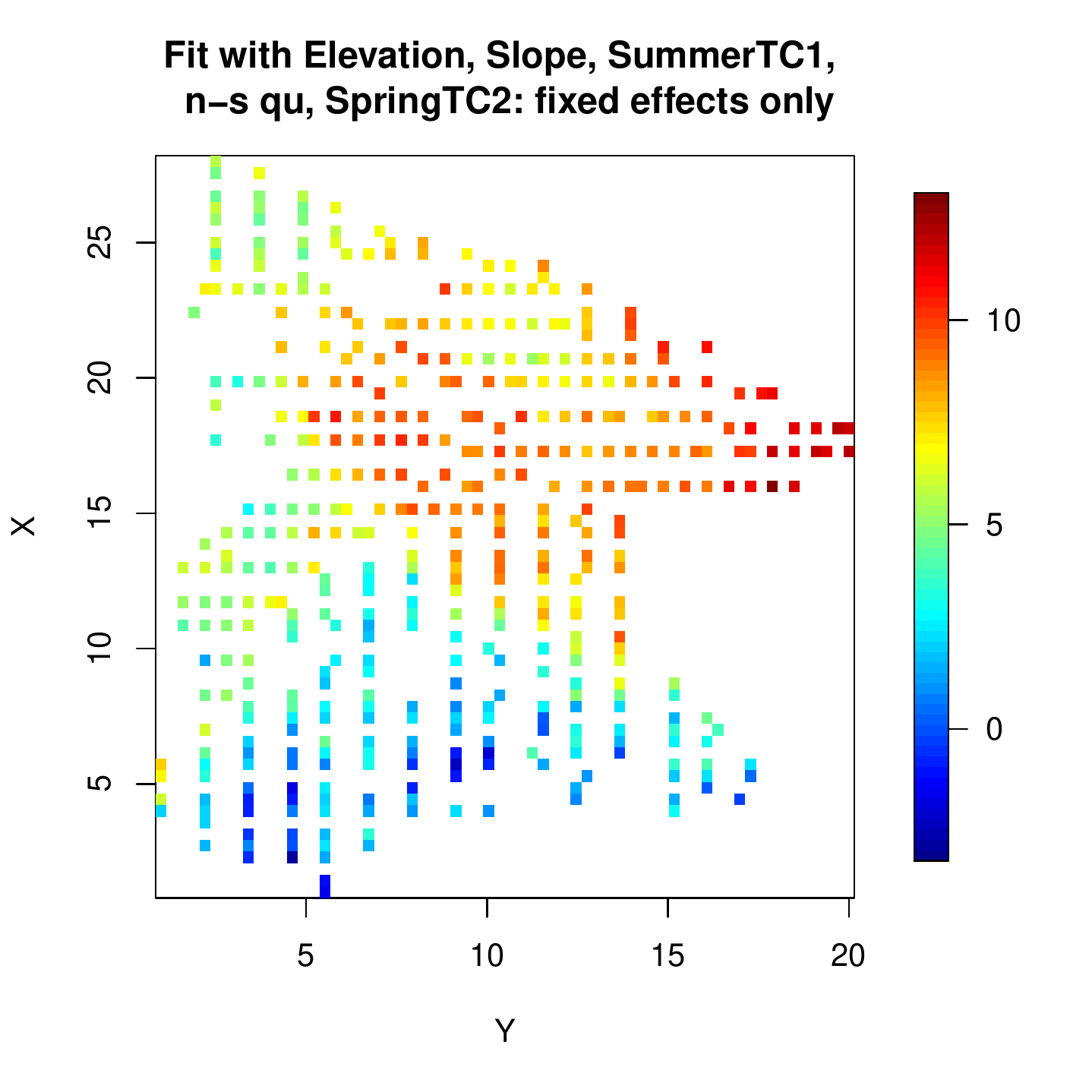}
                 \caption{Fit with Elev, Slope, SumTC1, SprTC2, north-south quad trend. }
                 \label{fig:RM_add_var4}
\end{subfigure}
\quad
\begin{subfigure}[b]{0.17\textwidth}
\includegraphics[width=\textwidth]{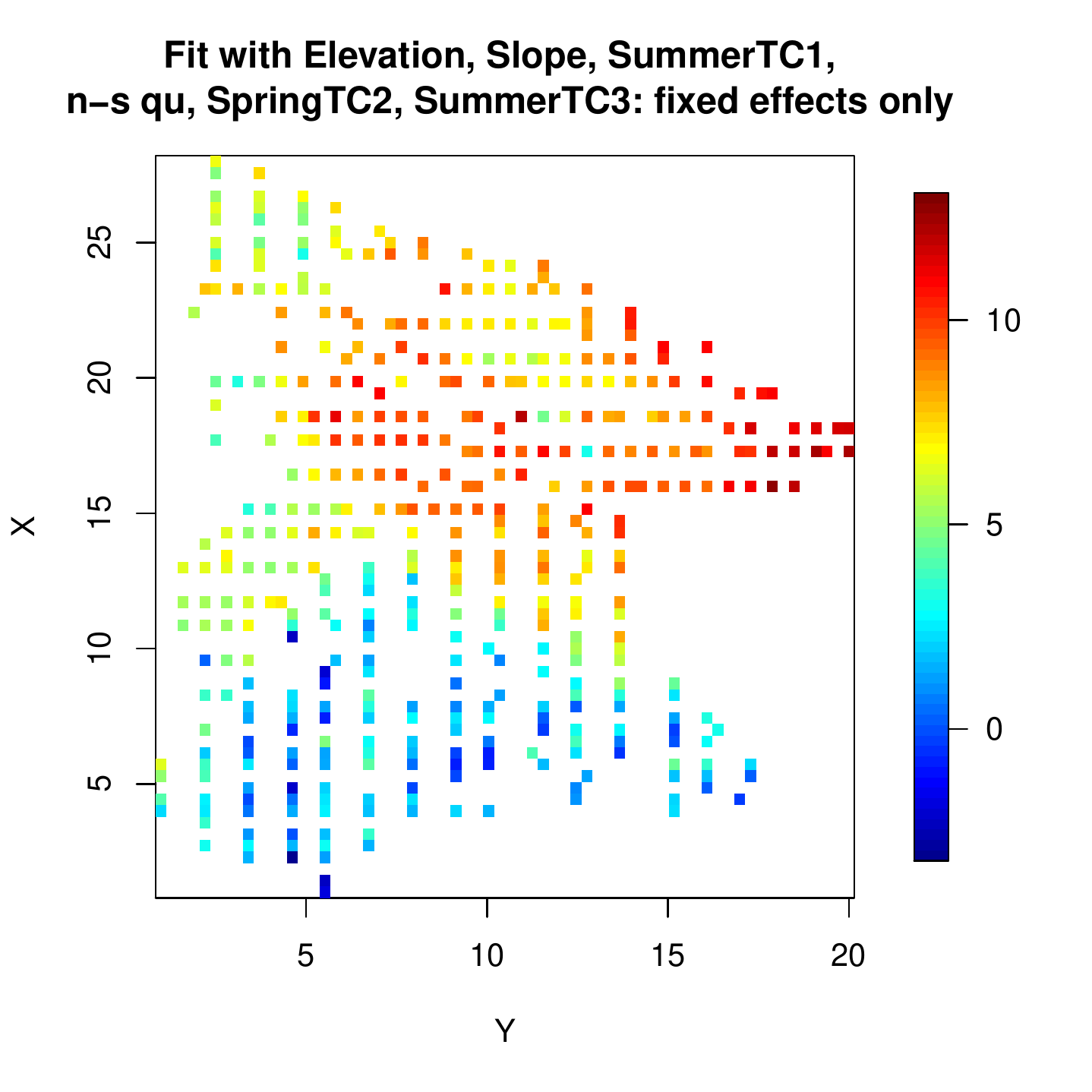}
                 \caption{Fit with Elev, Slope, SumTC1, SprTC2, n-s quad trend, SumTC3. }
                 \label{fig:RM_add_var4}
\end{subfigure}
 \caption{ Estimated fixed effects part of the fit for each of the five steps. }\label{fig:exact_fits}
         \end{figure}

\begin{figure}[H]
         \centering \begin{subfigure}[b]{0.17\textwidth}
 \includegraphics[width=\textwidth]{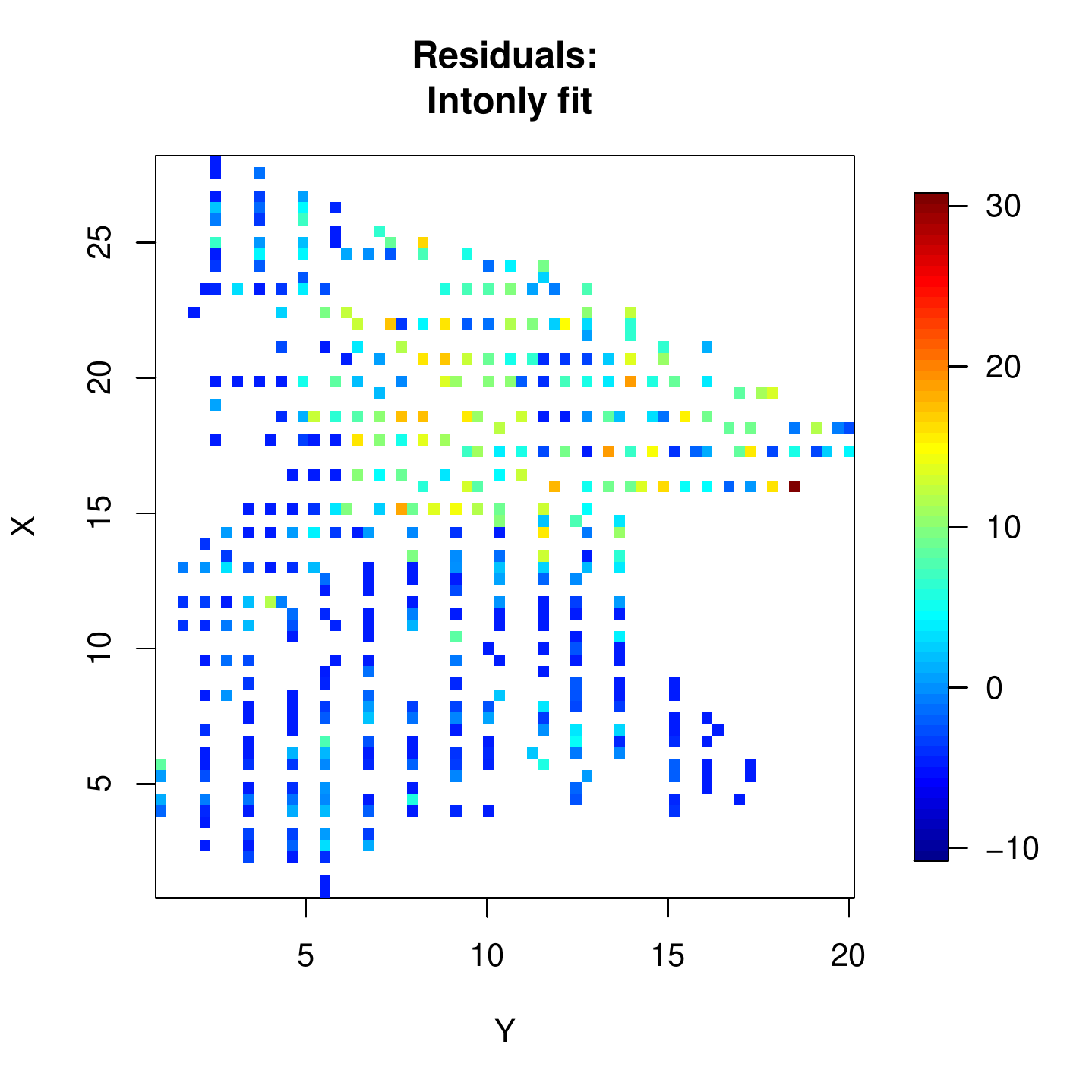}
                 \caption{\\ \\ Residuals from intercept-only fit.\\}
                                  \label{fig:RM_add_var1}
\end{subfigure}
         \quad
\begin{subfigure}[b]{0.17\textwidth}
\includegraphics[width=\textwidth]{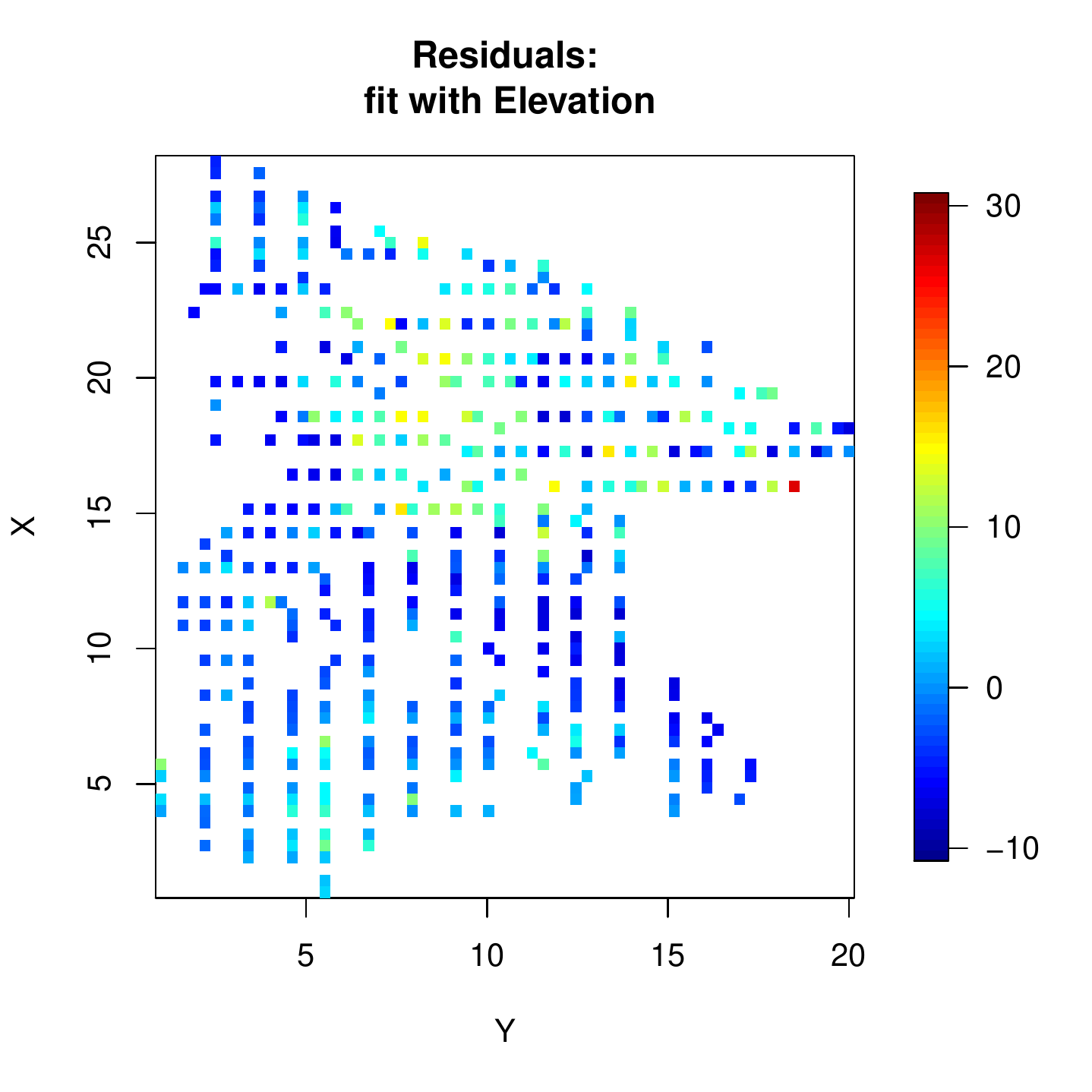}
                 \caption{\\ \\ Residuals from \\fit with Elevation.\\}
                 \label{fig:RM_add_var2}
\end{subfigure}
         \quad
\begin{subfigure}[b]{0.17\textwidth}
\includegraphics[width=\textwidth]{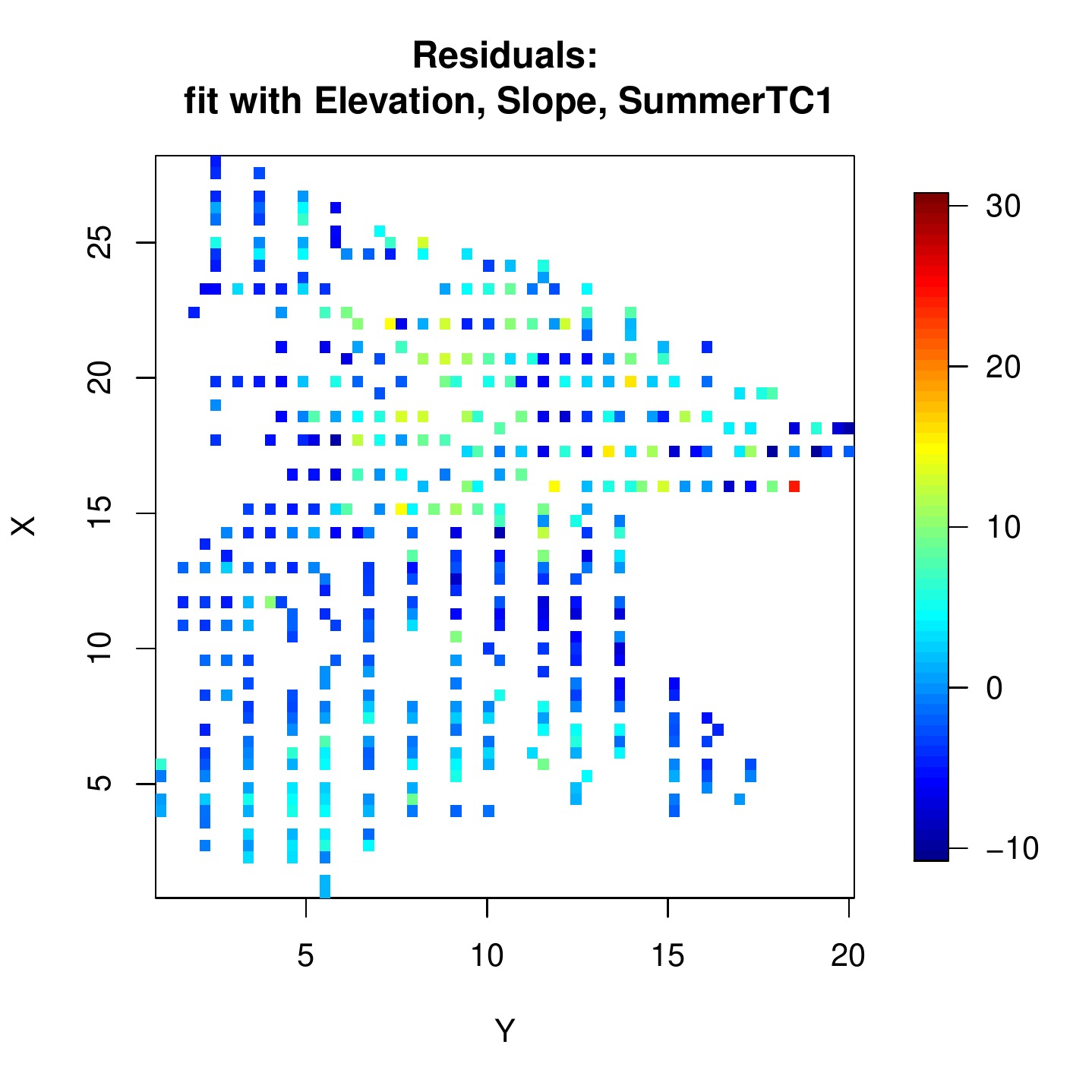}
                 \caption{Residuals from fit with\\ Elevation, Slope, SummerTC1.\\}
                 \label{fig:RM_add_var3}
\end{subfigure}
\quad
\begin{subfigure}[b]{0.17\textwidth}
\includegraphics[width=\textwidth]{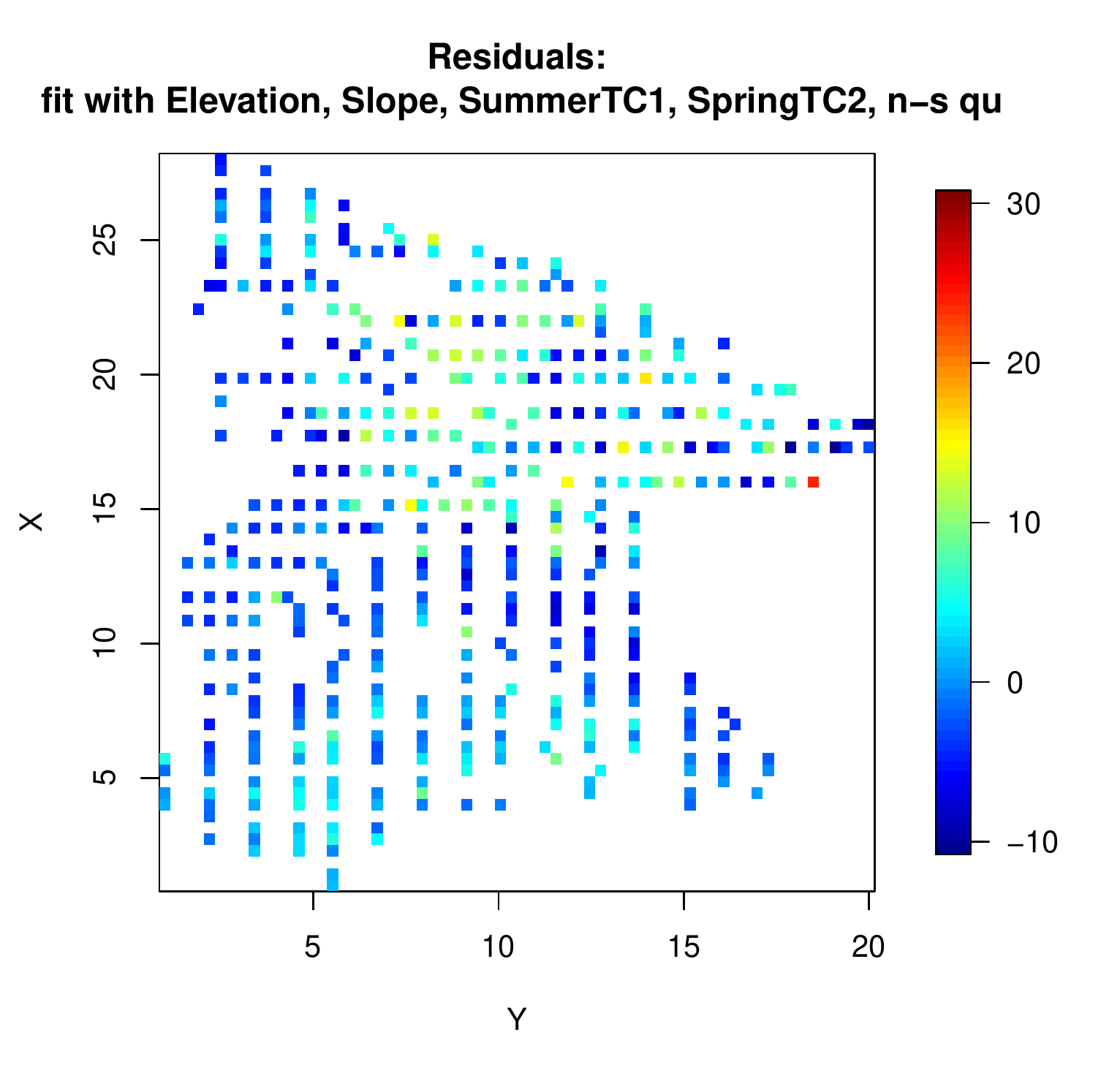}
                 \caption{Residuals from fit with Elev, Slope, SumTC1, SprTC2, north-south quad trend.}
                 \label{fig:RM_add_var4}
\end{subfigure}
\quad
\begin{subfigure}[b]{0.17\textwidth}
\includegraphics[width=\textwidth]{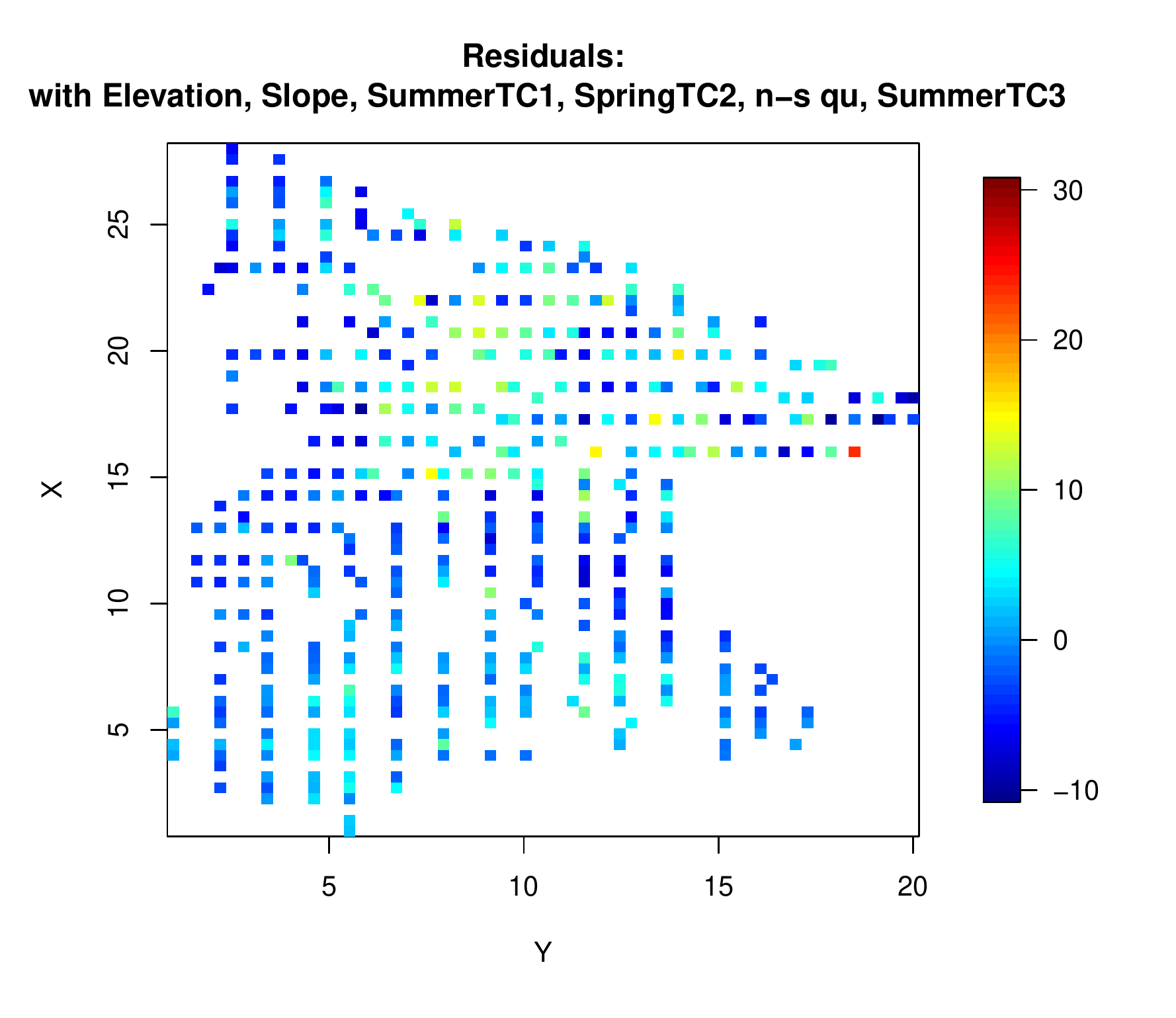}
                 \caption{Residuals from fit with Elev, Slope, SumTC1, SprTC2, n-s quad trend, SumTC3.}
                 \label{fig:RM_add_var4}
\end{subfigure}
 \caption{ Residuals, i.e., data minus estimated fixed effects, for each of the five steps. }\label{fig:exact_resids}
         \end{figure}

The spectral-domain added variable plots used the data smoothed on a grid to select covariates. This appendix shows the improvement of the fit to the actual (not smoothed) data, when fits are made using the exact restricted likelihood.  We repeat:  this Appendix involves no approximations of any kind.  Figure \ref{fig:RM_obsdata} shows the actual data, while Figure \ref{fig:exact_fits} shows the estimated fixed-effect fits at each of the five steps discussed above, on the same color scale as the observed data.  Figure \ref{fig:exact_resids} shows the residuals obtained by subtracting the estimated fixed effects from the data at each step. Table \ref{table:quantiles} shows quantiles of absolute values of the residuals with respect to the fixed effects, which become somewhat smaller as we add covariates, with most of the change arising from adding Elevation, Slope, and SummerTC1. Changes in the residuals are small and thus not very visible in the residual plots (Figure \ref{fig:exact_resids}).  However, it is clear from Table \ref{table:quantiles} that the fit to the data has indeed improved.

Tables \ref{table:dumbfit_first} and \ref{table:dumbfit_second} contain estimated coefficients of the covariates when they are added to the model and the model is re-fit at each step. The estimated coefficients are almost exactly equal to the slopes of the corresponding \textit{observation}-domain added variable plots in Appendix I. However the p-values from ordinary Wald tests of these coefficients (taking as known the estimates of $\sigma_\mathfrak{s}^2$, $\sigma_e^2$, and $\rho$, as is typical in non-Bayesian software) are not similar to the p-values for significance of the added variable plot slopes in either the observation or spectral domain. Table~\ref{table:exact_ests} shows the estimates of $\sigma_\mathfrak{s}^2$, $\sigma_e^2$, and $\rho$ at each step.  The covariates Elevation, Slope, and Summer TC1 have the largest effect on $\hat{\sigma}_\mathfrak{s}^2$, reducing it by almost half, as we might expect given the relatively strong low-frequency components in these covariates.  Adding Elevation to the model had the largest effect on $\hat{\rho}$, with modest reductions from adding further covariates, and adding Elevation produced most of the reduction in $\hat{\sigma}_e^2$.

         \begin{table}[H]
\centering
\hspace*{-2cm}
  \begin{tabular}{ |l || c |c|c|c|c|}
    \hline
    covariates in model&  minimum & 25\%&50\%&75\%&maximum \\ \hline
    \textbf{Step 1}: Intercept-only&  0.01 &2.69&4.95&5.40&30.81\\ \hline
    \textbf{Step 2}: Elevation & 0.00 &2.10&4.20&6.70&26.79\\ \hline
    \textbf{Step 3}: Elev, Slope, SumTC1& 0.00&1.76&3.93&5.96&24.48 \\ \hline
    \textbf{Step 4}: Elev, Slope, SumTC1,& & & & & \\
    SprTC2, n-s quad trend& 0.03&1.71&3.77&5.91&24.09\\ \hline
    \textbf{Step 5}: Elev, Slope, SumTC1, & & & & &\\SprTC2, n-s quad, SumTC3& 0.00&1.68&3.63&5.84&23.73\\ \hline
       \end{tabular}
  \caption{ Quantiles of absolute residuals, i.e., data minus estimated fixed effects. }\label{table:quantiles}
\end{table}

\begin{table}[H]
\centering
  \begin{tabular}{ |c ||c|c||c|c|}
    \hline
     & \textbf{Step 1}  &\textbf{Step 1}&  \textbf{Step 2} &\textbf{Step 2}\\ \hline
    Candidate covariates &  slope &p-value&  slope &p-value\\ \hline
    Elevation&-2.52&$<10^{-12}$ &-- &-- \\ \hline
    Slope  &-1.63&$10^{-6}$  &-1.37&$10^{-5}$ \\ \hline
    SpringTC2 &-0.28 &0.16 &-0.44 &0.04 \\ \hline
    SpringTC3 &0.99 &$10^{-5}$ &0.97&$10^{-5}$\\ \hline
    SummerTC1  &-1.03&$10^{-5}$& -1.09&$10^{-6}$ \\ \hline
    SummerTC3  &1.25&$10^{-7}$ &1.23&$10^{-7}$ \\ \hline
    FallTC2  &-0.87&0.0001  & -0.94&$10^{-5}$\\ \hline  \hline
    \end{tabular}
  \caption{ Coefficients for the covariates obtained by putting them into the model, and p-values from Wald tests. The symbol ``--" denotes covariate already in the model.}\label{table:dumbfit_first}
\end{table}

\begin{table}[H]
\centering
  \begin{tabular}{ |c || c |c|| c |c|| c |c|}
    \hline
    &  \textbf{Step 3}  &\textbf{Step 3}&  \textbf{Step 4} &\textbf{Step 4}&  \textbf{Step 5} &\textbf{Step 5}\\ \hline
    Candidate covariates &  slope&p-value&  slope&p-value&  slope&p-value \\ \hline
    Elevation& --&-- & --&--  & --&--\\ \hline
    Slope & --&-- & --&--& --&--\\ \hline
    SpringTC2 &-0.49 &0.03&-- &--&-- &--\\ \hline
    SpringTC3  &0.39 &0.10&1.30 &0.0005&1.19 &0.0008 \\ \hline
   SummerTC1 &-- &--&-- &-- &-- &--\\ \hline
    SummerTC3  &0.84 &0.003&1.39 &$10^{-6}$ &--&--\\ \hline
    FallTC2 &-0.76 &0.0007 &-0.72 &0.001&-0.74 &0.0005 \\ \hline  \hline
    \end{tabular}
  \caption{Coefficients for the covariates obtained by putting them into the model, and p-values from Wald tests. The symbol ``--" denotes covariates already in the model.}\label{table:dumbfit_second}
\end{table}

\begin{table}[H]
\small
\centering
  \begin{tabular}{ |l ||c|c|c|c|}
    \hline
    covariates in the model& \textbf{$\sigma_\mathfrak{s}^2$}& \textbf{$\sigma_e^2$}& \textbf{$\rho$}\\ \hline
    Step 1: Intercept-only &29.62&16.20&5.96\\ \hline
    Step 2: Elev &21.96&13.82&2.85  \\ \hline
    Step 3: Elev, Slope, SumTC1&15.98&15.13&2.65 \\ \hline
    Step 4: Elev, Slope, SumTC1, & & &\\SprTC2, north-south quad trend &16.04&14.37&2.28\\ \hline
    Step 5: Elev, Slope, SumTC1, SprTC2, & & &\\SumTC1, n-s quad, SumTC3  &16.35&12.89&1.92\\ \hline \hline
    \end{tabular}
  \caption{Estimates of $\sigma_\mathfrak{s}^2$, $\sigma_e^2$, $\rho$ using the exact restricted likelihood at each step.}\label{table:exact_ests}

\end{table}

\section{Web Appendix K. \hspace*{3mm} Approximation Effects and Accuracy}\label{sec:approx}

This appendix focuses on the effects of smoothing data observed at irregular locations to a regular grid using IDW, and approximating the restricted likelihood (RL) using the spectral approximation to the GP-distributed random effect.  The appendix mainly summarizes a simulation experiment examining the effect of smoothing data to the grid and the combined effect of smoothing to a grid and the spectral approximation.  Subsection~\ref{intuition} discusses the intuition about IDW that motivated the simulation experiment.  The spectral approximation has a large literature, which we do not attempt to summarize.  Subsection~\ref{simResults} gives and interprets the simulation experiment results.

\subsection{Intuition about smoothing to the grid using IDW}\label{intuition}

In the pseudo-data for a given grid point, IDW gives each actual observation a weight that is inversely proportional to the distance from the grid point to that observation's location, raised to the power $\lambda$.  A small $\lambda$ gives weight to many observations and thus produces smooth pseudo-data;  as $\lambda$ grows, weight is concentrated on fewer observations and with very large $\lambda$, most grid points are effectively assigned the nearest observation.

The clearest intuition is about smoothing's effect on the estimated error variance, $\hat{\sigma}^2_e$:  in general, $\hat{\sigma}^2_e$ will be biased downward because the weighted averaging at grid points dampens local or high-frequency variation.  The power $\lambda$ affects this most directly:  small $\lambda$ suppresses local or high-frequency variation more than large $\lambda$, so small $\lambda$ should produce greater bias than large $\lambda$.  As for grid density ($M_1$ and $M_2$), the downward bias of $\hat{\sigma}^2_e$ should be greatest at the extremes of very coarse and very dense grids even with large $\lambda$, though for different reasons at the two extremes.  For coarse grids, with considerably fewer grid points than observations, each grid point's weighted average will be affected only by the observations closest to it;  as the grid becomes coarser, the most discrepant observations are more likely to influence no grid points, so the downward bias should worsen as the grid coarsens.  For dense grids, with the number of grid points approaching or greater than the number of observations, more and more individual observations will be used for more than one grid point;  this repeated use will make the grid-smoothed data less variable than the actual data, so again the downward bias should worsen as the grid becomes increasingly dense.  This suggests a moderate grid density is best for bias in $\hat{\sigma}^2_e$.  Finally, many real datasets have measurements on non-rectangular regions, so the rectangular grid will have grid points in regions of the map that are empty of observations.  Pseudo-data for such grid points are necessarily averages of observations at or very near the edge of the empty regions so they will tend to be less variable than the actual data because of repeated use of the same observations.  Thus, larger empty regions should tend to have more downward bias in $\hat{\sigma}^2_e$.

Intuition about the effects of grid smoothing on the estimate of the GP's range, $\hat{\rho}$, largely follows from intuition about $\hat{\sigma}^2_e$.  Smoothing to a grid should have little effect on low-frequency features of the data but will suppress power at high frequencies, so that power will tend to decline as frequency increases {\it faster} in the grid smoothed data than in the actual data.  Using intuition about estimates based on the spectral approximation, because $\hat{\rho}$ is determined by this decline, $\hat{\rho}$ will tend to be biased high in grid-smoothed data.  As for the GP variance $\sigma_\mathfrak{s}^2$, in the same scheme of intuition, its estimate adjusts $\sigma_\mathfrak{s}^2 a_j(\rho) + \sigma^2_e$ to ``go through the middle of" the $v^2_j$ for small $j$, and it is not clear that the aforementioned effects will induce any consistent effect on $\hat{\sigma}_\mathfrak{s}^2$.

\subsection{Simulation experiment}\label{simResults}

In most simulation experiments done by statisticians, the purpose is to precisely estimate operating characteristics (e.g., type I error), so the experimental designs consider few conditions (simulation scenarios) and simulate many artificial datasets for each.  Our purpose is different --- to examine trends in estimates over many IDW settings --- so our experimental design is like one we would recommend to our collaborators, with relatively few simulated datasets.  (Our collaborators rarely need designs with 1000 replications per design cell.)  Specifically, we simulated 20 datasets with observations at locations in the unit square;  from each simulated dataset, we created 3 analysis datasets differing in the size of the empty region on the spatial map;  for each analysis dataset, we considered 5 grid sizes;  and for each grid size, we considered three values of the IDW smoothing parameter $\lambda$.  The experimental design is thus entirely within-subject, where a ``subject" is a simulated dataset (20 levels), and the factors are amount of blank space (3 levels), and estimator (16 levels, the estimates from maximizing the exact RL on the actual data plus 15 estimates for data smoothed to a grid, where 15 = 5 grid sizes $\times$ 3 $\lambda$ values).  Analysis of variance gave p $< 0.05$ for many effects, i.e., adequate power to detect trends of the sort described in the preceding section;  below, we focus on broad trends and do not report significance tests.

\underline{\bf Experimental settings.}  Twenty datasets were simulated, each with observations at 400 locations in the unit square, iid draws from a uniform distribution.  The same locations were used for all 20 datasets.  Simulated observations were drawn from an intercept-only linear mixed model with true $\beta$ = 0 and true $\sigma_\mathfrak{s}^2$ = 12, $\rho$ = 0.1 (i.e., 1/10 of each dimension of the unit square), and $\sigma_e^2$ = 5.  \underline{Amount of blank space.}  Blank space was created by omitting observations at locations within specified regions.  The levels of this factor were:  no blank space;  1/8 of the unit square blank;  and 1/4 of the unit square blank.  The omitted regions were wedges opening upward with apex at the point (0.5, 0.5), specifically:  to omit 1/8 of the unit square, omit points having $s_2 > -2s_1 + 1.5$ and $s_2 > 2s_1 - 0.5$;  to omit 1/4 of the unit square, omit points having $s_2 > -s_1 + 1$ and $s_2 > s_1$.  \underline{Grid size.}  All grids were $M \times M$, where $M$ was 12, 14, 16, 18, or 20.  With no omitted data, a simulated dataset had 400 observations, so these grids had, respectively, 36\%, 49\%, 64\%, 81\%, and 100\% as many points.  \underline{Smoothing parameter $\lambda$.}  We considered values of 5, 10, and 100.  Preliminary experiments considered $\lambda$ = 1 with catastrophic results.  The analyses in the main paper's Section 7 used $\lambda$ = 7 for $y$ and $\lambda$ = 9 for candidate covariates by minimizing the sum over $\sigma_\mathfrak{s}^2$, $\rho$, and $\sigma_e^2$ of the squared relative difference between the estimate obtained by maximizing the exact RL for the real data and the data smoothed to the grid.

For each of the 60 analysis datasets (20 simulated datasets $\times$ 3 levels of blank space), we produced estimates of $(\sigma_\mathfrak{s}^2, \rho, \sigma_e^2)$ by (a) maximizing the exact RL using the ``real" data;  then (b) maximizing the exact RL using pseudo-data smoothed to the grid;  then (c) maximizing the approximate RL, based on the spectral approximation to the GP, using pseudo-data smoothed to the grid.  Comparing the estimates from (b) to the exact estimates (a) isolates the effect of smoothing the data to a grid;  comparing the estimates from (c) to the exact estimates (a) shows the combined effect of smoothing the data to a grid and using the spectral approximation.  (We do not recommend obtaining estimates by maximizing the approximate restricted likelihood;  this is simply a compact way to capture the effect of smoothing to the grid followed by the spectral approximation.)  The following sections present these two comparisons;  for both comparisons, preliminary analyses indicated that the logarithms of the estimates were more appropriate dependent variables for ANOVA than untransformed values.

\underline{\bf Effect of smoothing to a grid.}

\begin{figure}[H]
         \centering
 \includegraphics[clip=true, trim= 0 0 0 0, width=\textwidth]{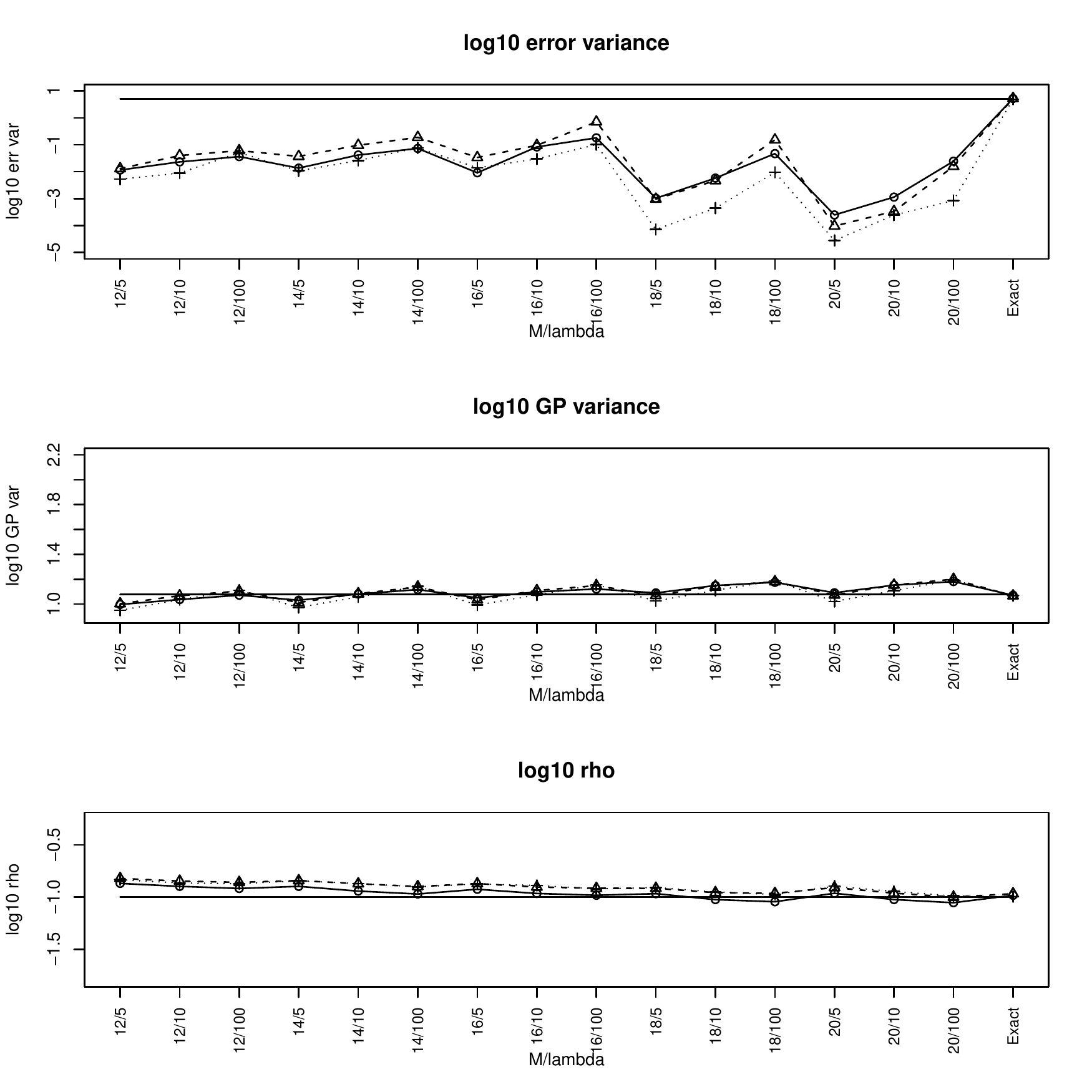}
                 \caption{Estimates from maximizing the exact RL;  average over 20 simulated datasets.  In each panel:  No empty space, solid line with circles;  1/8 empty space, dashed line with triangles;  1/4 empty space, dotted line with crosses;  the solid horizontal line is the true value used to simulate the data.  ``Exact" is the estimate from maximizing the exact RL for the actual data.  Vertical axes are the same as in Figure~\ref{fig:ApEsts}. Standard errors for individual estimates are 0.51 for log10 error variance, 0.023 for log10 GP variance, and 0.032 for log10 $\rho$.  Standard errors for comparing pairs of estimates are similar.  }
                                  \label{fig:ExEsts}
\end{figure}

\begin{figure}[H]
         \centering
 \includegraphics[clip=true, trim= 0 0 0 0, width=\textwidth]{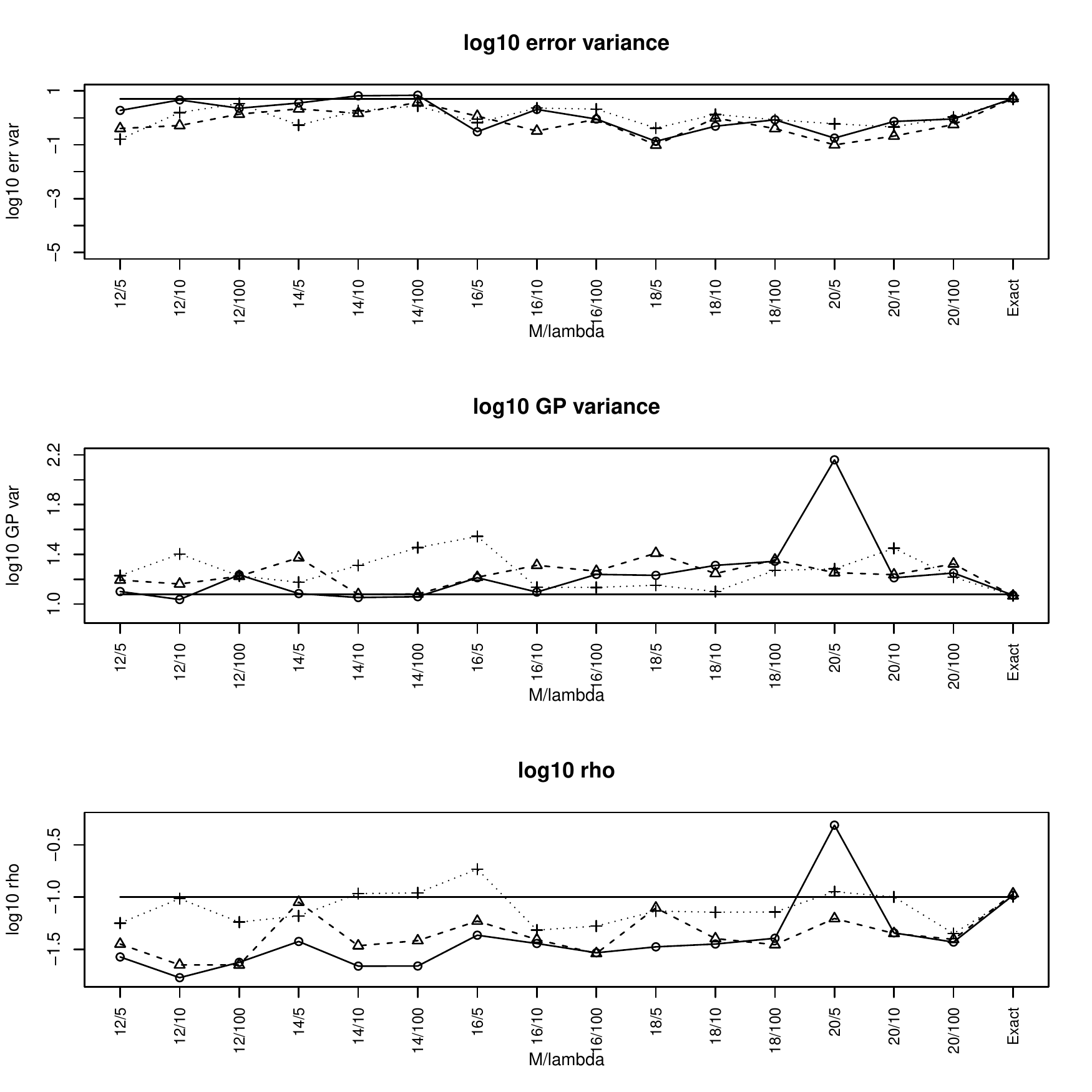}
                 \caption{Estimates from maximizing the approximate RL;  average over 20 simulated datasets.  In each panel:  No empty space, solid line with circles;  1/8 empty space, dashed line with triangles;  1/4 empty space, dotted line with crosses;  the solid horizontal line is the true value used to simulate the data.  ``Exact" is the estimate from maximizing the exact RL for the actual data.  Vertical axes are the same as in Figure~\ref{fig:ExEsts}.  Standard errors for individual estimates are 0.39 for log10 error variance, 0.19 for log10 GP variance, and 0.22 for log10 $\rho$.}
                                  \label{fig:ApEsts}
\end{figure}

Figure~\ref{fig:ExEsts} shows the estimates obtained by maximizing the exact likelihood for the actual data (``Exact" on the horizontal axis) or the pseudo-data smoothed to the grid (labelled as ``$M$/$\lambda$" on the horizontal axis).  Figures~\ref{fig:ExEsts} and~\ref{fig:ApEsts} have the same vertical axes to facilitate comparison.

Broadly, biases in the estimates can be summarized as follows.
\begin{itemize}
\item Error variance:  Estimates are biased downward, least so for moderate $M$ and large $\lambda$ though even for these settings the bias is substantial.  More empty space makes the bias worse for denser grids.
\item GP variance:  Estimates are affected much less than error variance.  The bias has no consistent direction of bias but is sensitive to both $M$ and $\lambda$.
\item GP range $\rho$:  Estimates are generally biased high, more so for coarser grids (small $M$) and for more omitted data.
\end{itemize}

Regarding choices for the IDW procedure:
\begin{itemize}
\item Grid size:  Moderate $M$ is the best compromise:  it is best for the error and GP variance, while large $M$ is best for $\rho$ though not by much.
\item Lambda:  100 is best for error variance;  for the GP variance, the best value depends on $M$, with moderate $\lambda$ best for moderate $M$;  and for $\rho$, $\lambda$ has little effect.
\end{itemize}

\underline{\bf Combined effect of smoothing to a grid and using the approximate RL.}

Figure~\ref{fig:ApEsts} shows estimates obtained by maximizing the exact RL for the actual data (``Exact" on the horizontal axis) and the approximate RL for pseudo-data smoothed to the grid (labelled ``$M$/$\lambda$" on the horizontal axis).  Figures~\ref{fig:ApEsts} and~\ref{fig:ExEsts} have the same vertical axes to facilitate comparison.

For data smoothed to the grid, the estimate of error variance is markedly less biased and less sensitive to the IDW settings when obtained by maximizing the approximate RL than when obtained by maximizing the exact RL.  Coarser grid sizes $M$ give modestly less bias compared to denser grids.  This may depend on the true parameter values used to simulate the data;  further experiments would be required to determine this.  Compared to estimates obtained using the exact RL, estimates of GP variance and $\rho$ are more biased on average using the approximate RL and also more variable, as reflected in the standard errors associated with each estimate, listed in the Figure captions.  (An outlier strongly affects the average estimates of $(\sigma_\mathfrak{s}^2$ and $\rho$ for $M$ = 20, $\lambda$ = 5).  Estimates of the GP variance are generally biased high but show no particular dependence on the IDW settings, while estimates of $\rho$ are generally biased downward but again show no particular dependence on the IDW settings.

\section{Web Appendix L. \hspace*{3mm} Spatial Confounding}\label{sec:confounding}

This brief discussion of spatial confounding, as it relates to the spectral approximation used here, is based on Chapter 15 of Hodges (2014), especially Section 15.2.3.

A conventional non-Bayesian fit of a linear mixed model has two steps:  (1) Maximize the restricted likelihood (RL) to estimate unknown parameters in the random-effect and error covariance matrices $G$ and $R$, and (2) insert those estimates into $G$ and $R$ as if they were known to be true, and estimate fixed effect coefficients $\beta$ and random effects $u$.  (A Bayesian analysis implicitly replaces Step 1 by computing the marginal posterior of the unknowns in $G$ and $R$, and replaces Step 2's plug-in estimates with the integral of the conditional posterior of $(\beta,u)$ against the marginal posterior of the unknowns in $G$ and $R$, which gives the marginal posterior of $(\beta,u)$.  If the fixed effect vector $\beta$ has $\pi(\beta) \propto 1$, the following paragraph still holds;  if $\beta$ has a multivariate normal prior with covariance matrix $\Sigma$ having a finite determinant, the following paragraph no longer holds.)

In Step 1, the RL attributes to the fixed effects {\it all} variation in $y$ that lies in the column space of the fixed effect design matrix $X$, so that estimates of the unknowns in $G$ and $R$ are determined entirely by variation in $y$ lying in the orthogonal complement to the column space of $X$.  Thus, spatial confounding has no effect whatsoever on RL-maximizing estimates of unknowns in $G$ and $R$;  our approximate analysis preserves this property.  In Step 2, the unknowns in $G$ and $R$ are set to their estimated values and the fixed effects $X\beta$ and random effects $Zu$ then compete to explain variation in $y$.  They compete to an extent determined by Step 1's estimates:  spatial confounding can occur in Step 2 {\it only} if shrinkage or smoothing is not marked;  this occurs if error variation ($R$) is not large relative to random-effect variation ($G$).  Our approximate analysis preserves this property as well.

The foregoing, applied to the spectral approximation, may help explain spatial confounding for linear mixed models including GP-distributed random effects, the best published treatment of which, as far as we know, is Hanks et al (2015).  Broadly, fixed effects with strong low-frequency components can be spatially confounded {\it only} if Step 1 produces an estimate of error variation that is not large relative to random-effect variation.  If this condition holds, then such fixed effects are spatially confounded by low-frequency components of the GP-distributed random effect to a degree determined by the size of their projection on the columns of $Z$ capturing those low frequency components, and by the extent to which the latter's coefficients in $u$ are in fact shrunk.  Fixed effects with {\it weak} low-frequency components are generally not subject to spatial confounding:  high-frequency components of the random-effect fit are shrunk a great deal and thus cannot confound fixed effects.  The exception to the latter generalization occurs when error variation is estimated to be very small relative to random-effect variation, in which case little shrinkage occurs and fitting fixed effects of any kind is hazardous because the model is barely identified.

\end{document}